\documentclass[epjc3,twocolumn,a4paper,fleqn]{svjour3} 
\usepackage[fleqn]{amsmath}
\usepackage{graphicx} 
\usepackage{epstopdf}
\usepackage{fix-cm}
\usepackage{txfonts}
\usepackage{listings}
\usepackage{lineno}
\usepackage{mdwlist}
\usepackage{pennames}
\usepackage[british]{babel}
\usepackage{color}
\usepackage{subfigure}
\usepackage[binary-units]{siunitx}
\DeclareSIUnit \clight	{\textit{c}}

\expandafter\def\csname ver@subfig.sty\endcsname{}

\usepackage{import}
\usepackage{amssymb}
\usepackage{epsfig}
\usepackage{cite}
\usepackage{hyperref}
\hypersetup{
colorlinks=true,
linkcolor=blue,
citecolor=blue,
urlcolor=blue,
}

\usepackage{pdflscape}

\DeclareGraphicsExtensions{.eps,.pdf,.jpeg,.png,.gif}
\newcommand{\meg}{\ifmmode\mathrm{\mu \to e \gamma}\else$\mathrm{\mu \to e \gamma}$\fi}
\newcommand{\megc}{\ifmmode\mathrm{\mu^+ \to e^+ \gamma}\else$\mathrm{\mu^+ \to e^+ \gamma}$\fi}
\newcommand{\michel}{\ifmmode\mathrm{\mu^+ \to e^+ \nu\bar{\nu}}\else$\mathrm{\mu^+ \to e^+ \nu\bar{\nu}}$\fi}
\newcommand{\radiative}{\ifmmode\mathrm{ \mu^+ \to e^+\nu\bar{\nu}\gamma} \else$\mathrm{\mu^+ \to e^+ \nu\bar{\nu}\gamma}$\fi}
\newcommand{\conv}{\ifmmode\mathrm{\mu^-N \to e^-N}\else$\mathrm{\mu^-N \to e^-N}$\fi}
\newcommand{\convme}{\ifmmode\mathrm{\mu^- \to e^-}\else$\mathrm{\mu^- \to e^-}$\fi}
\newcommand{\mute}{\ifmmode\mathrm{\mu \to 3e}\else $\mathrm{\mu \to 3e}$\fi}
\newcommand{\mutec}{\ifmmode\mathrm{\mu^+ \to e^+e^+e^-}\else $\mathrm{\mu^+ \to e^+e^+e^-}$\fi}
\newcommand{\aif}{\ifmmode\mathrm{e^+ e^- \to \gamma\gamma} \else$\mathrm{e^+ e^- \to \gamma \gamma}$\fi}

\newcommand*{\tmueg}{\mathrm\tau \to \ell \gamma}
\newcommand*{\tautl}{\mathrm\tau \to 3\ell}

\newcommand*{\muonp}          {\ifmmode\mathrm{\mu^+}\else$\mathrm{\mu^+}$\fi}
\newcommand*{\muon}          {\ifmmode\mathrm{\mu}\else$\mathrm{\mu}$\fi}
\newcommand*{\tauon}          {\ifmmode\mathrm{\tau}\else$\mathrm{\tau}$\fi}

\newcommand*{\egammapair}     {\mathrm{e^+\gamma}}
\newcommand*{\egamma}         {E_{\mathrm{\gamma}}}
\newcommand*{\positron}       {\ifmmode{\mathrm{e^+}}\else${\mathrm{e^+}}$\fi}
\newcommand*{\epositron}      {E_\mathrm{e^+}}
\newcommand*{\ppositron}      {p_\mathrm{e^+}}
\newcommand*{\tpositron}      {t_\mathrm{e^+}}
\newcommand*{\tgamma}         {t_{\mathrm{\gamma}}}
\newcommand*{\tegamma}        {t_{\mathrm{e^+ \gamma}}}
\newcommand*{\Thetaegamma}    {\Theta_{\mathrm{e^+ \gamma}}}

\newcommand*{\thetaegamma}    {\theta_{\mathrm{e^+ \gamma}}}
\newcommand*{\phiegamma}      {\phi_{\mathrm{e^+ \gamma}}}
\newcommand*{\thetae}         {\theta_\mathrm{e^+}}
\newcommand*{\phie}           {\phi_\mathrm{e^+}}

\newcommand*{\BR}     { {\cal B} }

\newcommand*{\ypos}          {y_\mathrm{e^+}}
\newcommand*{\zpos}          {z_\mathrm{e^+}}

\newcommand*{\ugamma}         {u_{\gamma}}
\newcommand*{\vgamma}         {v_{\gamma}}
\newcommand*{\wgamma}         {w_{\gamma}}

\newcommand*{\mathtentative}{}
\def\mathtentative#1#{\mathcoloraux{#1}}
\newcommand*{\mathcoloraux}[3]{%
  \protect\leavevmode
  \begingroup
    \color#1{#2}#3%
  \endgroup
}

\journalname{Eur. Phys. J. C} 

\begin{document}


\title{The design of the MEG~II experiment}

\date{Received: date / Accepted: date}

\newcommand*{\INFNPi}{INFN Sezione di Pisa$^{a}$; Dipartimento di Fisica$^{b}$ dell'Universit\`a, Largo B.~Pontecorvo~3, 56127 Pisa; Scuola Normale Superiore$^{c}$, Piazza dei Cavalieri, 56127 Pisa, Italy}
\newcommand*{\INFNGe}{INFN Sezione di Genova$^{a}$; Dipartimento di Fisica$^{b}$ dell'Universit\`a, Via Dodecaneso 33, 16146 Genoa, Italy}
\newcommand*{\INFNPv}{INFN Sezione di Pavia$^{a}$; Dipartimento di Fisica$^{b}$ dell'Universit\`a, Via Bassi 6, 27100 Pavia, Italy}
\newcommand*{\INFNRm}{INFN Sezione di Roma$^{a}$; Dipartimento di Fisica$^{b}$ dell'Universit\`a ``Sapienza'', Piazzale A.~Moro, 00185 Roma, Italy}
\newcommand*{\INFNLe}{INFN Sezione di Lecce$^{a}$; Dipartimento di Matematica e Fisica$^{b}$ dell'Universit\`a del Salento, Via per Arnesano, 73100 Lecce, Italy}
\newcommand*{\ICEPP} {ICEPP, The University of Tokyo, 7-3-1 Hongo, Bunkyo-ku, Tokyo 113-0033, Japan }
\newcommand*{\UCI}   {University of California, Irvine, CA 92697, USA}
\newcommand*{\KEK}   {KEK, High Energy Accelerator Research Organization 1-1 Oho, Tsukuba, Ibaraki 305-0801, Japan}
\newcommand*{\PSI}   {Paul Scherrer Institut PSI, 5232 Villigen, Switzerland}
\newcommand*{\BINP}  {Budker Institute of Nuclear Physics of Siberian Branch of Russian Academy of Sciences, 630090 Novosibirsk, Russia}
\newcommand*{\JINR}  {Joint Institute for Nuclear Research, 141980 Dubna, Russia}
\newcommand*{\ETHZ}  {Swiss Federal Institute of Technology ETH, 8093 Zurich, Switzerland}
\newcommand*{\NOVS}  {Novosibirsk State University, 630090 Novosibirsk, Russia}
\newcommand*{\NOVST} {Novosibirsk State Technical University, 630092 Novosibirsk, Russia}
\newcommand*{\Siena}{Dipartimento di Scienze Fisiche, della Terra e dell'Ambiente dell'Universit\`a, Via Roma 56, 53100, Siena, Italy}

\author{
The MEG II collaboration\\\\
       A.~M.~Baldini\thanksref{addr4}$^a$ \and
       E.~Baracchini\thanksref{addr3} \and
       C.~Bemporad\thanksref{addr4}$^{ab}$ \and
       F.~Berg\thanksref{addr1,addr2} \and
       M.~Biasotti\thanksref{addr5}$^{ab}$ \and
       G.~Boca\thanksref{addr7}$^{ab}$ \and
       P.~W.~Cattaneo\thanksref{addr7}$^{a,}$\thanksref{e1}  \and
       G.~Cavoto\thanksref{addr8}$^{ab}$ \and
       F.~Cei\thanksref{addr4}$^{ab}$ \and
       M.~Chiappini\thanksref{addr16,addr4}$^{a}$ \and
       G.~Chiarello\thanksref{addr6}$^{ab}$ \and
       C.~Chiri\thanksref{addr6}$^{a}$ \and
       G.~Cocciolo\thanksref{addr6}$^{ab}$ \and
       A.~Corvaglia\thanksref{addr6}$^{a}$ \and
       A.~de~Bari\thanksref{addr7}$^{ab}$ \and
       M.~De~Gerone\thanksref{addr5}$^{a}$ \and
       A.~D'Onofrio\thanksref{addr4}$^{ab}$ \and
       M.~Francesconi\thanksref{addr4}$^{ab}$ \and
       Y.~Fujii\thanksref{addr3}  \and
       L.~Galli\thanksref{addr4}$^{a}$ \and
       F.~Gatti\thanksref{addr5}$^{ab}$ \and
       F.~Grancagnolo\thanksref{addr6}$^{a}$ \and
       M.~Grassi\thanksref{addr4}$^{a}$ \and
       D.~N.~Grigoriev\thanksref{addr12,addr14,addr15} \and
       M.~Hildebrandt\thanksref{addr1} \and
       Z.~Hodge\thanksref{addr1,addr2} \and
       K.~Ieki\thanksref{addr3}  \and
       F.~Ignatov\thanksref{addr12,addr15} \and
       R.~Iwai\thanksref{addr3}  \and
       T.~Iwamoto\thanksref{addr3}  \and
       D.~Kaneko\thanksref{addr3}  \and
       K.~Kasami\thanksref{addr9}  \and
       P.-R.~Kettle\thanksref{addr1} \and
       B.~I.~Khazin\thanksref{addr12,addr15,e2} \and
       N.~Khomutov\thanksref{addr13} \and
       A.~Korenchenko\thanksref{addr13,e2}  \and
       N.~Kravchuk\thanksref{addr13}  \and
       T.~Libeiro\thanksref{addr11} \and
       M.~Maki\thanksref{addr9}  \and
       N.~Matsuzawa\thanksref{addr3} \and
       S.~Mihara\thanksref{addr9}  \and
       M.~Milgie\thanksref{addr11} \and
       W.~Molzon\thanksref{addr11} \and
       Toshinori~Mori\thanksref{addr3}  \and
       F.~Morsani\thanksref{addr4}$^{a}$ \and
       A.~Mtchedilishvili\thanksref{addr1}  \and
       M.~Nakao\thanksref{addr3}  \and
       S.~Nakaura\thanksref{addr3}  \and 
       D.~Nicol\`o\thanksref{addr4}$^{ab}$ \and
       H.~Nishiguchi\thanksref{addr9}  \and
       M.~Nishimura\thanksref{addr3}  \and 
       S.~Ogawa\thanksref{addr3}  \and
       W.~Ootani\thanksref{addr3}  \and
       M.~Panareo\thanksref{addr6}$^{ab}$ \and
       A.~Papa\thanksref{addr1} \and
       A.~Pepino\thanksref{addr6}$^{ab}$ \and
       G.~Piredda\thanksref{addr8}$^{a,}$\thanksref{e2} \and
       A.~Popov\thanksref{addr12,addr15} \and
       F.~Raffaelli\thanksref{addr4}$^{a}$ \and
       F.~Renga\thanksref{addr8}$^{a}$ \and
       E.~Ripiccini\thanksref{addr8}$^{ab}$ \and
       S.~Ritt\thanksref{addr1} \and
       M.~Rossella\thanksref{addr7}$^{a}$ \and
       G.~Rutar\thanksref{addr1,addr2} \and
       R.~Sawada\thanksref{addr3}  \and
       G.~Signorelli\thanksref{addr4}$^{a}$ \and
       M.~Simonetta\thanksref{addr7}$^{ab}$  \and
       G.~F.~Tassielli\thanksref{addr6}$^{ab}$ \and
       Y.~Uchiyama\thanksref{addr3} \and
       M.~Usami\thanksref{addr3} \and
       M.~Venturini\thanksref{addr4}$^{ac}$ \and
       C.~Voena\thanksref{addr8}$^{a}$ \and
       K.~Yoshida\thanksref{addr3} \and
       Yu.~V.~Yudin\thanksref{addr12,addr15}  \and
       Y.~Zhang\thanksref{addr11} 
}

\institute{\INFNPi \label{addr4}
           \and
              \ICEPP \label{addr3}
           \and
             \PSI \label{addr1} 
           \and
              \ETHZ \label{addr2}
           \and
             \INFNGe \label{addr5}
           \and
             \INFNPv \label{addr7}
           \and
             \INFNRm \label{addr8}
           \and
             \Siena   \label{addr16}
           \and
             \INFNLe \label{addr6} 
           \and
             \BINP   \label{addr12}
           \and
             \NOVST  \label{addr14}
           \and
             \NOVS   \label{addr15}
           \and
             \KEK    \label{addr9}
           \and
             \JINR   \label{addr13}
           \and
             \UCI    \label{addr11}
}


\thankstext[*]{e1}{Corresponding author: paolo.cattaneo@pv.infn.it} 
\thankstext[$\dagger$]{e2}{Deceased } 
\maketitle

\begin{abstract}
The MEG experiment, designed to search for the \megc\ decay at a \num{e-13} sensitivity level, completed data-taking in 2013.
In order to increase the sensitivity reach of the experiment by an order of magnitude to the level of \num{6e-14} 
for the branching ratio, a total upgrade, involving substantial changes to the experiment, has been undertaken, known as MEG~II. 
We present both the motivation for the upgrade and a detailed 
overview of the design of the experiment and of the expected detector performance.
\end{abstract}

\tableofcontents

\newpage
\section{Introduction}
\label{sec:Introduction}

\subsection{Status of the MEG experiment in the framework of charged Lepton Flavour Violation (cLFV) searches}
\label{sec:Status_of_MEG}

\begin{figure}
\centering
\includegraphics[width=\columnwidth]{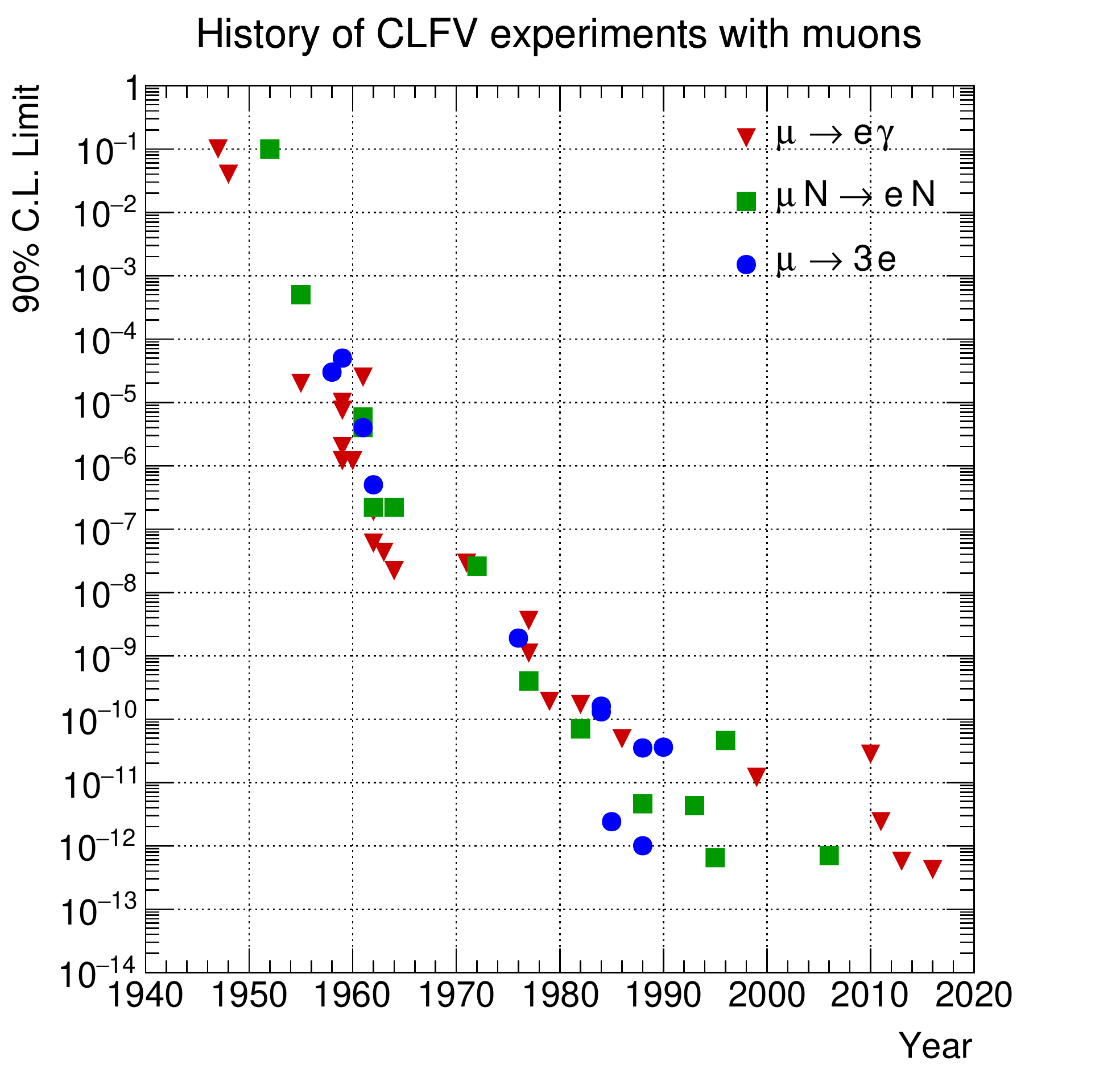}
\caption{Chronology of upper limits on cLFV processes. \label{fig:LFVlimits}}
\end{figure}

The experimental upper limits established in searching for cLFV processes with muons, including the \megc\ decay, 
are shown in Fig.~\ref{fig:LFVlimits} versus the year of the result publication. 
Historically, the negative results of these experiments led to 
the formulation of the Standard Model (SM) of elementary particles interactions, 
in which lepton flavour conservation was empirically included. 
During the past 35 years the experimental sensitivity to the \megc\ decay has improved by almost three orders of magnitude, mainly due
to improvements in detector and beam technologies. 
In particular,
\lq{}surface\rq{} muon beams (i.e. beams of muons originating from stopped $\pi^+$s decay in the surface layers of the pion production target)
with virtually
monochromatic momenta of \SI[per-mode=symbol]{\sim29}{\MeV\per\clight}, 
offer the highest muon stop densities obtainable at present 
in low-mass targets,
allowing ultimate resolution in positron momentum and emission angle 
and suppressing the photon background production.

The signal of the two-body \megc\ decay at rest 
can be distinguished from the background by measuring the photon energy $\egamma$, 
the positron momentum $\ppositron$, their relative angle $\Thetaegamma$  
and timing $\tegamma$ with the best possible resolutions.

The background comes either from radiative muon decays (RMD) $\radiative$ 
in which the neutrinos carry away a small amount of energy or from an accidental coincidence of an energetic positron
from Michel decay $\michel$ with a photon coming from RMD, bremsstrahlung or positron annihilation-in-flight (AIF) $\aif$. 
In experiments using high intensity beams, such as MEG, this latter background is dominant.

The keys for \megc\ search experiments achieving high sensitivities can be summarised as 
\begin{enumerate}
\item A high intensity continuous surface muon beam to gain the data statistics with minimising the accidental background rate (cf. Eq.~(\ref{eq:bacc}) below).
\item A low-mass positron detector with high rate capability to deal with the abundant positrons from muon decays.
\item A high-resolution photon detector, especially in the energy measurement, to suppress the high-energy random photon background.
\end{enumerate}

\begin{figure*}[t]
\centering
\includegraphics[width=\textwidth, clip, trim=0 0 0 2pc]{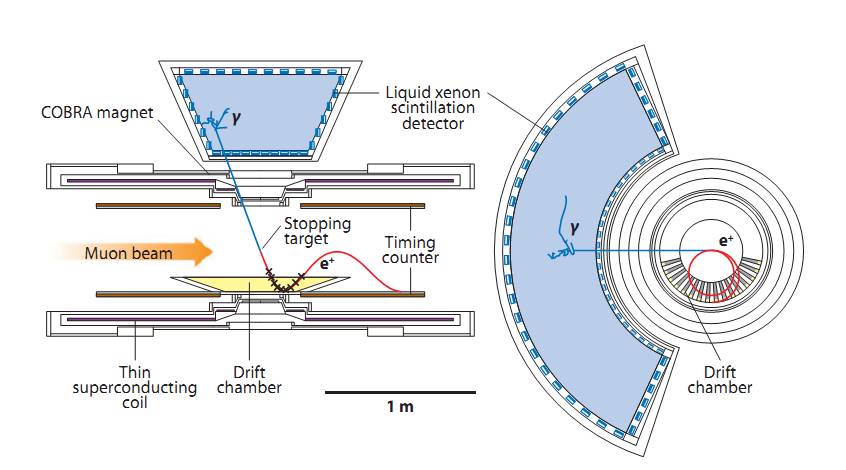}
\caption{Schematic of the MEG experiment.
\label{fig:MEG}
}
\end{figure*}

The MEG experiment \cite{megdet} at the Paul Scherrer Institute (PSI, Switzerland) uses one of the world's most intense 
(maximum rate higher than \SI{e8}{\muonp\per\second} continuous surface muon beams, 
but, for reasons explained in the following, the stopping intensity is limited to
\SI{3e7}{\muonp\per\second} . 
The muons are stopped in a thin (\SI{205}{\um}) polyethylene target, 
placed at the centre of the experimental set-up which includes a positron spectrometer and a photon detector, 
as shown schematically in Fig.~\ref{fig:MEG}. 

The positron spectrometer consists of a set of drift chambers 
and scintillating timing counters located inside a superconducting solenoid 
COBRA (COnstant Bending RAdius)
with a gradient magnetic field along the beam axis, 
ranging from \SI{1.27}{\tesla} at the centre to \SI{0.49}{\tesla}  at either end, that guarantees
a bending radius of positrons weakly dependent on the polar angle. 
The gradient field is also designed to remove quickly spiralling positrons sweeping
them outside the spectrometer to reduce the track density inside the tracking volume.

The photon detector, located outside of the solenoid, is a homogeneous volume (\SI{900}{\litre}) 
of liquid xenon (LXe) viewed by 846 UV-sensitive photomultiplier tubes (PMTs) submerged 
in the liquid, that read the scintillating light from the LXe.
The spectrometer measures the positron momentum vector and timing, 
while the LXe photon detector measures the photon energy 
as well as the position and time of its interaction in LXe. 
The photon direction is measured connecting the interaction vertex in the LXe photon detector 
with the positron vertex in the target obtained by extrapolating the positron track.
All the signals are individually digitised by in-house designed waveform digitisers (DRS4) 
\cite{Ritt2010486}.

The number of expected signal events for a given branching ratio $\BR$ 
is related to the rate of stopping muons $R_\muonp$, the measurement time $T$,
the solid angle $\Omega$ subtended by the photon and positron detectors, 
the efficiencies of these detectors ($\epsilon_\mathrm{\gamma}, \epsilon_\positron$) 
and the efficiency of the selection criteria $\epsilon_\mathrm{s}$:\footnote{An usual selection criterion is to choose 90\% efficient cuts on each of the variables 
($\egamma$, $\ppositron$, $\Thetaegamma$, $\tegamma$) 
around the values expected for the signal: 
this criterion defines the selection efficiency to be $\epsilon_\mathrm{s} = (0.9)^4$. 
This kind of analysis in which one counts the number of events within some selection cuts 
and compares the number found with predictions for the background is named \lq\lq box analysis\rq\rq. 
MEG/MEG~II adopt more refined analyses 
which take into account the different distributions of 
($\egamma$,  $\ppositron$, $\Thetaegamma$,  $\tegamma$) for background and signal type events by using maximum likelihood methods. }
\begin{eqnarray}
\label{eq:signal}
N_\mathrm{sig} = R_\muonp \times T \times \Omega \times \BR \times \epsilon_\gamma \times \epsilon_\positron \times \epsilon_\mathrm{s}.
\end{eqnarray}
The single event sensitivity (SES) is defined as the $\BR$ for which the experiment would see one event. 
In principle the lowest SES, and therefore the largest possible $R_\muonp$, 
is desirable in order to be sensitive to the lowest possible $\BR$. 
%
The number of accidental coincidences $N_\mathrm{acc}$, for given selection criteria, 
depends on the experimental resolutions (indicated as $\Delta$ in Eq.~\ref{eq:bacc}) 
with which the four relevant quantities 
($\egamma$, $\ppositron$, $\Thetaegamma$, $\tegamma$) are measured. 
By integrating the RMD photon and Michel positron spectra over respectively the photon energy
and positron momentum resolution intervals, it can be shown that:
\begin{eqnarray}
\label{eq:bacc}
N_\mathrm{acc} \propto R_{\mu^+}^2 \times {\Delta\egamma}^2 \times {\Delta\ppositron} \times  \Delta\Thetaegamma^2 \times \Delta\tegamma \times T.
\end{eqnarray}
Due to the quadratic dependence on $R_\muonp$, 
the accidental coincidences largely dominate over the background coming from RMD 
(which is linearly dependent on $R_\muonp$). 
It is clear from Eqs.~(\ref{eq:signal}) and (\ref{eq:bacc}) that, 
for fixed experimental resolutions, the muon stopping rate cannot be increased arbitrarily 
but must be chosen in order to keep a reasonable signal to background ratio.


The current published MEG limits of $\BR(\megc) < \num{4.2e-13}$ at 90\%\ confidence level (CL), based on the full data-set \cite{meg2016} is currently the most stringent limit on this decay.
Given that the background (accidental) extends into the signal region, only a limited gain in sensitivity could be achieved with further statistics, hence data-taking ceased in 2013, allowing the upgrade program to proceed with full impetus.

Other cLFV channels, complementary to \megc\ and being actively pursued are: 
\conv, \mute, $\tmueg$ and $\tautl$ ($\ell = \mathrm{e}$ or $\mathrm{\mu}$).  
In the \conv\ conversion experiments, negative muons are stopped in a thin target and form muonic atoms. 
The conversion of the muon into an electron in the field of the nucleus results 
in the emission of a monochromatic electron of momentum \SI{\sim100}{\MeV/\clight}, depending on the target nucleus used.
Here the backgrounds to be rejected are totally different from the \megc\ case. 
The dominant examples are muon decay-in-orbit and those correlated with the presence of beam
impurities, such as pions. In order to reduce these backgrounds the experiments planned at Fermilab (Mu2e) \cite{Car2008,mu2e2014}
and J-PARC (COMET \cite{Bry2007,cometTDR} and DeeMe \cite{deemme2010}) will use pulsed proton beams to produce their muons.

Since muonic atoms have lifetimes ranging from hundreds of nanoseconds up
to the free muon lifetime at low Z,
the conversion electrons are therefore searched for in the intrabunch intervals. 

The COMET collaboration plans to start the first phase of the experiment in 
2018 with a sensitivity reach better than \num{e-14} followed by the second 
phase aiming for a goal sensitivity of \num{7e-17}, while
the Mu2e experiment is foreseen to start in 2021 with a first phase 
sensitivity goal of \num{7e-17}. 
These experiments can in principle
reach sensitivities below \num{e-17}~\cite{PRISM2006,Knoepfel:2013}. 

The $\mute$ decay search is being pursued in a new experiment, 
proposed at PSI: Mu3e \cite{Blondel:2013ia}.
This plans a staged approach to reach its target a sensitivity of \num{e-16}. 
The initial stage involves sharing part of the MEG beam line and seeks a
three orders-of-magnitude increase in sensitivity over the current limit,
its goal being \num{e-15}.
The final stage foresees muon stopping rates of the order of \SI{e9}{\muonp\per\second}.

$\tmueg$ and $\tautl$ will be explored by the Belle~II experiment at SuperKEKB \cite{BelleII_TDR,PhysicsAtSuperKEKB} and a proposed experiment at the super Charm-Tau factory \cite{Bondar2013,Hao2016}
where sensitivities of the order of \num{e-9} to the branching ratios for these channels are expected.

A comparison between the sensitivity planned for MEG~II and that envisaged for the other above mentioned cLFV processes 
will be discussed in the next section after a very short introduction to cLFV predictions in theories beyond the SM.


\subsection{Scientific merits of the MEG~II experiment}
Although the SM has proved to be extremely successful in explaining a wide variety of phenomena in the energy scale from sub-\si{\electronvolt} to $O(\SI{1}{\TeV})$, it is widely considered a low 
energy approximation of a more general theory. 
One of the attractive candidates for such theory is the grand-unified theory (GUT) 
\cite{Georgi:1974} 
which unifies all the SM gauge groups into a single group as well as quarks and
 leptons into common multiplets of the group.
In particular, the supersymmetric version (SUSY-GUT) has received a great amount of
 attention after the LEP experiments showed that a proper unification of the forces 
 can be achieved at around a scale $M_\mathrm{GUT}\SI{\sim e16}{\GeV}$ 
if SUSY particles exist at a scale ${\cal O}(\SI{1}{\tera\electronvolt})$ \cite{Amaldi:1991}.  
The search for TeV-scale SUSY particles has been one of the goals
of the LHC program. Results so far have been negative for masses up to \SIrange[range-units = single,range-phrase = --]{1}{2}{\tera\electronvolt}
\cite{atlassusy2017,cmssusy2017}.

The experimentally measured phenomenon of neutrino oscillations~\cite{SK:1998,SNO:2002,PDBook_2016}
 requires an extension of the SM. 
It demonstrates that lepton flavour is violated, and neutrinos have masses 
but they are orders of magnitude smaller than those of quarks and charged leptons. 	
%
An appealing extension of the SM consists in introducing Majorana masses for neutrinos 
to naturally account for the tiny neutrino masses via the seesaw mechanism 
\cite{Minkowski:1977,Yanagida:1979,Gell-Mann:1979,Mohapatra:1980}.
This approach predicts the existence of heavy right-handed Majorana neutrinos\footnote{%
	This is called Type-I seesaw. Other types of the seesaw mechanism have also been invented;
	 see \cite{Mohapatra:2007} for a review.} 
 in the range of \SIrange[range-units = single,range-phrase = --]{e9}{e15}{\GeV}.
This ultra-high mass scale may be indicative of their connection to SUSY-GUT 
(e.g. all the SM fermions plus the right-handed neutrino in a generation can fit into a single multiplet in SO(10) GUT).
The Majorana neutrinos violate the lepton number, and may account for 
the matter--antimatter asymmetry in the Universe \cite{Fukugita:1986}.

It is generally difficult to detect, even indirectly, the effects of 
such ultra-high energy scale physics. 
However, the situation changes with SUSY, 
and cLFV signals provide a general test of SUSY-GUT and SUSY-seesaw as discussed below.

It is well known that cLFV is sensitive to SUSY~\cite{Ellis:1982,Campbell:1983,Lee:1984}; 
in fact the parameter space for the minimal SUSY extension of the SM (MSSM) 
has largely been constrained by flavour- and CP-violation processes involving 
charged leptons and quarks~\cite{Gabbiani:1996,Altmannshofer:2010,Arana-Catania:2013,Altmannshofer:2013}.
These experimental observations lead to considering special mechanisms 
of SUSY breaking, 
requiring e.g. the universal condition of SUSY particles\rq{} masses at some high scale.
It was however shown that mixing in sleptons emerges unavoidably at low energy
 in SUSY-GUT~\cite{Hall:1986} and SUSY-seesaw~\cite{Borzumati:1986} models
 even if the lepton flavour is conserved at high scale.
This is  because flavour-violation sources, i.e. at least the quark and/or neutrino Yukawa interactions, 
do exist in the theory and radiatively contribute to the mass-squared matrices of 
sleptons during the evolution of the renormalisation-group equation.\footnote{%
	This effect is enhanced by large Yukawa couplings. The large top Yukawa coupling 
	does it in SUSY-GUT models. 
	The neutrino Yukawa couplings can be the same order as those for quarks and 
	charged leptons in the seesaw mechanism. 
	In particular, in SO(10) GUT, neutrino Yukawa couplings are related to up-type ones
	and at least one of them should be as large as the top one~\cite{Masiero:2003}.}
As a result, $\BR(\meg)$ is predicted at an observable level  \numrange[range-phrase = --]{e-11}{e-14}
\cite{barbieri1994,barbieri,Ciafaloni:1996,Hisano:1997,
Hisano:1995,hisano_1996_prd,Hisano:1999,Casas:2001}.
This theoretical framework motivated the MEG and MEG~II experiment.

In order to appreciate this, we recall that the SM, even introducing massive neutrinos, 
practically forbids any observable rate of cLFV ($\BR(\meg) < \num{e-50}$) \cite{Petcov:1977,Cheng:1980}. 
Processes with cLFV are therefore clean channels 
to look for possible new physics beyond the SM, for which
a positive signal would be unambiguous evidence.

Over the last five years, two epoch-making developments took place in particle physics: 
the discovery of Higgs boson \cite{ATLAS:2012,CMS:2012} and the measurement of the last 
unknown neutrino mixing angle $\theta_{13}$ \cite{DayaBay,Reno,Abe2011,Abe2012}.
The mass of Higgs boson at \SI{125}{\GeV}~\cite{ATLAS-CMS:2015}, rather light, 
on one hand 
supports the SUSY-GUT scenario since it is actually in the predicted region~\cite{Akula:2011}.
On the other hand, it is relatively heavy in MSSM and suggests, together with the null results
in the direct searches at LHC, that the SUSY particles would be heavier than expected.
This implies that a smaller $\BR(\meg)$ is expected because of the approximate dependence
 $\propto 1/M_\mathrm{SUSY}^4$.
This might explain why MEG was not able to detect the signal 
as well as why other flavour observables, particularly 
$\mathrm{b} \to \mathrm{s}\gamma$~\cite{HFAG:2016}
and $B_\mathrm{s} \to \mu^+ \mu^-$~\cite{CMS-LHCb:2015}, 
have been measured to be consistent with the SM so far.
In contrast, the observed large mixing angle 
$\theta_{13}\sim \ang{8.5}$~\cite{PDBook_2016} suggests 
higher $\BR(\meg)$ in many physics scenarios such as SUSY-seesaw.

Updated studies of SUSY-GUT/seesaw models taking those recent experimental results 
into account show that $\BR(\meg)\sim \num{e-13}$--$\num{e-14}$ is possible
up to SUSY particles\rq{} masses around \SIrange[range-units = single,range-phrase = --]{5}{10}{\tera\electronvolt}
\cite{Calibbi:2012gr,Hirsh:2012,Cannoni:2013,Dutta:2013,Chowdhury:2013,Moroi:2014,Goto:2015,Bora:2015,Fukuyama:2016}, 
well above the region where LHC (including HL-LHC) direct searches can reach.
In addition, cLFV searches are sensitive to components which do not strongly interact 
(e.g. sleptons and electroweakinos in MSSM) and thus are not much constrained by the LHC results.
Considering these situations, further exploration of the range $\BR(\meg)\sim O(\num{e-14})$
in coincidence with the 14-\si{\tera\electronvolt} LHC run provides a unique and powerful probe, 
complementary and synergistic to LHC,  to explore new physics.

So far, we discussed SUSY scenarios, the main motivation of MEG~II, 
but many other scenarios, such as models with 
extra-dimensions~\cite{Agashe:2006,Iyer:2012,Beneke:2016}, 
left-right symmetry~\cite{Cirigliano:2004,Akeroyd:2007,Lee:2013,Perez:2017},
 leptoquarks~\cite{Benbrik:2008,Arnold:2013,Gripaios:2015,Varzielas:2015}, 
and little Higgs~\cite{Choudhury:2007,Blanke:2010,delAguila:2011,Wang:2012}, also predict observable rates of \meg\ within the reach of MEG~II.

Comparison between different $\mu \to \mathrm{e}$ transition processes can be done 
model independently by an effective-field-theory approach.
Considering new physics, cLFV processes are generated by higher-dimensional operators; 
the lowest one that directly contributes to \meg\ is the following 
dimension-six (dipole-type) operator,
\begin{align}
{\cal O}^D_{L(R)} =  \langle H \rangle \left( \bar{e}_{R(L)} \sigma_{\alpha \beta} \mu_{L(R)}\right)   F^{\alpha \beta},
\label{eq:dipole}
\end{align}
where $\langle H \rangle$ is the vacuum expectation value of the Higgs field and $F^{\mu \nu}$ is the field-strength tensor of photon.
This operator also induces \mute\ and \conv\ via the propagation of a virtual photon.
There are several other dimension-six operators which cause the $\mu \to \mathrm{e}$ transitions, 
and their amplitudes to each of the three processes are model-dependent.\footnote{%
	Recent  effective-field-theory analyses have shown that those operators valid at some 
	high scale mix at the low energy scale where the experiments take place via the evolution 
	of renormalisation-group equation \cite{Crivellin:2017}.
	Due to this mixing effect as well as higher order contributions, the limit on $\BR(\megc)$ 
	provides severe constraints also on operators other than (\ref{eq:dipole}).}

In many models, especially most of SUSY models including the above mentioned 
SUSY-GUT/seesaw models, the operator (\ref{eq:dipole}) dominates the 
$\mu \to \mathrm{e}$ transitions.
In such a case, the following relations hold independently of the parameters in the models
\cite{kuno_2001,Czarnecki:1998}:
\begin{align}
&\frac{\BR(\mute)}{\BR(\megc)}\approx \num{6e-3},  \label{eq:ratio1}\\
&\frac{\BR(\conv)}{\BR(\megc)}\approx \num{2.6e-3} \quad \mathrm{(for~N=Al)}. \label{eq:ratio2}
\end{align}
Therefore, a search for \megc\ with a sensitivity of \num{\sim 6e-14}, 
which is the target of MEG~II, with a much shorter timescale and a far lower budget than other 
future projects, is competitive not only with the second phase of the Mu3e experiment \cite{Blondel:2013ia}
but also with the COMET \cite{Bry2007} and Mu2e \cite{mu2e2014} experiments. 
On the other hand, in case of discovery, we can benefit from a synergistic effect by the results from these experiments, 
providing a strong model-discriminant power;
any observations of discrepancy from the relations (\ref{eq:ratio1}) (\ref{eq:ratio2}) 
would suggest the existence of the contributions from operators other than  (\ref{eq:dipole}).

The comparison between $\mu$ and $\tau$ processes is more model dependent.
In the SUSY-seesaw models with and without GUT relations, the ratio 
$\BR(\tau \to \mu\gamma)/\BR(\meg)$ roughly ranges from \numrange{1}{e4}.\footnote{%
	$\BR(\tau \to \mathrm{e}\gamma) $ and $\BR(\tautl)$ are typically orders of 
	magnitude smaller than  $\BR(\tau \to \mu\gamma)$ in these models.}
Therefore, the present MEG bound on \megc\ already sets strong constraints on $\tmueg$ 
to be measured in the coming experiments \cite{PhysicsAtSuperKEKB}.
If $\tmueg$ will be detected in these experiments without a discovery of \megc\ in MEG~II, 
such models will be strongly disfavoured.

 We finally note that MEG~II will represent the best effort to address the search of the $\megc$ rare decay with the
 available detector technology coupled with the most intense continuous muon beam in the world. 
 Experience shows that to achieve any significant improvement in this field several years are required (more than one decade was necessary to pass from MEGA to MEG) 
 and therefore we feel committed to push the sensitivity of the search to the ultimate limits.


\subsection{Overview of the MEG~II experiment}

The MEG~II experiment plans to continue the search for the $\megc$ decay, 
aiming for a sensitivity enhancement of one order of magnitude compared to the final MEG result, i.e. down to \num{6e-14} for $\BR(\megc)$.
Our proposal for upgrading MEG \cite{Baldini:2013ke} was approved by the PSI research committee in 2013
and then, the details of the technical design has been fixed after intensive R\&D and is reported in this paper.

\begin{figure*}[tb]
\centering
\includegraphics[width=0.8\textwidth, clip, trim=0 1pc 0 2pc]{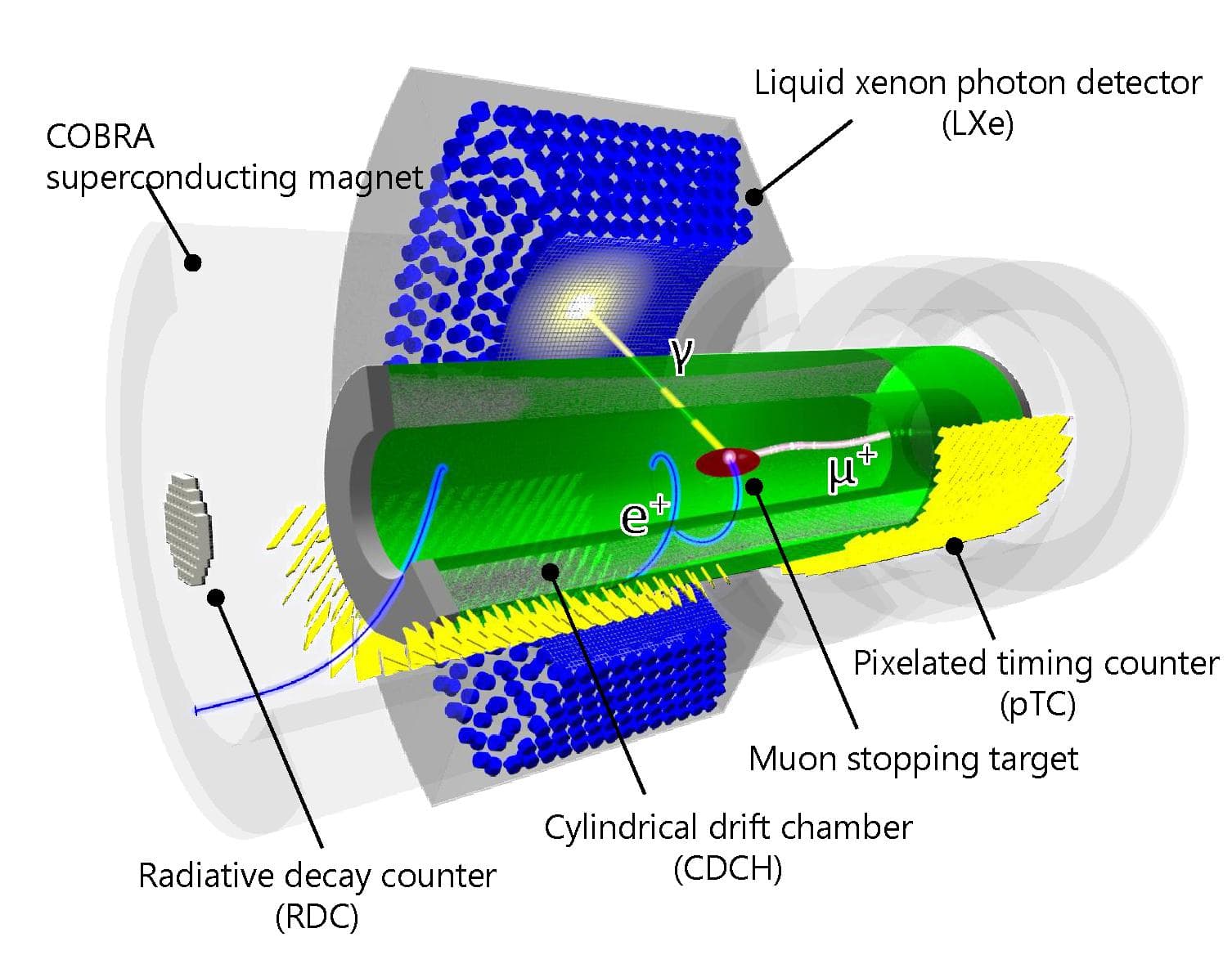}
\caption{\label{fig:MEGII}A schematic of the MEG~II experiment
}
\end{figure*}

The basic idea of the MEG~II experiment is to achieve the highest possible sensitivity by making maximum use of
the available muon intensity at PSI with the basic principle of the MEG experiment but with improved detectors.
A schematic view of  MEG~II is shown in Fig.~\ref{fig:MEGII}.
\label{sec:OverView}

A beam of surface \muonp\ is extracted from the $\pi$E5 channel of the PSI 
high-intensity proton accelerator complex, as in MEG, but the intensity is increased to the maximum. 
After the MEG beam transport system, the muons are stopped in a target, 
which is thinner than the MEG one to reduce both multiple Coulomb scattering 
of the emitted positrons and photon background generated by them.
The stopping rate becomes $R_\muonp= \SI{7e7}{\per\second}$, more than twice that of MEG (see Sect.~\ref{sec:Beam_Line}).

The positron spectrometer uses the gradient magnetic field to sweep away the low-momentum 
\positron. The COBRA magnet is retained from MEG, while the positron detectors inside are replaced by new ones.
Positron tracks are measured by a newly designed single-volume cylindrical drift chamber (CDCH) 
 able to sustain the required high rate. 
The resolution for the \positron\ momentum vector is improved with more hits per track by the high density of drift cells (see Sect.~\ref{sec:Cylindrical_Chamber}).
The positron time is measured with improved accuracy by a new pixelated timing counter (pTC) based on scintillator tiles read out by SiPMs (see Sect.~\ref{sec:Pixelated_Timing_Counter}).
The new design of the spectrometer increases the signal acceptance by more than a factor 2 due to the reduction of inactive materials between CDCH and pTC.

The photon energy, interaction point position and time are measured by an upgraded LXe photon detector. The energy and position resolutions are improved with a more uniform collection of scintillation light achieved by replacing the PMTs on the photon entrance face with new vacuum-ultraviolet (VUV) sensitive \SI[product-units=single]{12 x 12}{\mm\squared} SiPMs 
(see Sect.~\ref{sec:LXe_Calorimeter}).

A novel device for an active background suppression is newly introduced:
the Radiative Decay Counter (RDC) which employs plastic scintillators for timing and scintillating crystals for energy measurement in order to identify low-momentum \positron\ associated to high-energy RMD photons (see Sect.~\ref{sec:RDC}). 

The trigger and data-acquisition system (TDAQ) is also upgraded to meet the stringent requirements of an increased number of read-out channels and to cope with the required bandwidth by integrating the various functions of analogue signal processing, biasing for SiPMs, high-speed waveform digitisation, and trigger capability into one condensed unit (see Sect.~\ref{sec:tdaq}).

In rare decay searches the capability of improving the experimental sensitivity depends on the use of intense beams and high performance detectors, accurately calibrated and monitored. This is the only way to ensure that the beam characteristics and the detector performances are reached and maintained over the experiment lifetime. To that purpose several complementary approaches have been developed with some of the methods requiring dedicated beams and/or auxiliary detectors. 
Many of them have been introduced and commissioned in MEG and will be inherited by MEG~II with some modifications to match the upgrade. In addition new methods are introduced to meet the increased complexity of the new experiment. 

Finally, the sensitivity of MEG~II with a running time of three years is estimated in Sect.~\ref{sec:sensitivity}.


\newpage
\section{Beam line }
\label{sec:Beam_Line}

\subsection{MEG beam line layout}

The main beam requirements for a high rate, high sensitivity,
ultra-rare decay coincidence experiment such as MEG are:
\begin{itemize}
\item high stopping intensity ($R_\muonp = \SI{7e7}{\per\second}$) 
on target with high transmission optics,
\item small beam spot to minimise the stopping target size,
\item large momentum-byte $\Delta p_\muonp/p_\muonp \num{\sim7}\%$ (FWHM) with an
achromatic final focus, yielding an almost monochromatic
beam with a high stop density for a thin target,
\item minimal and well separated beam-correlated
backgrounds such as positrons from Michel decay
or $\pi^0$-decay in the production target or decay
particles from along the beam line and
\item minimisation of material budget along the beam
line to suppress multiple scattering and photon
production, use of vacuum or helium environments
as far as possible.
\end{itemize}

\begin{figure*}
\centering
\includegraphics[width=1.0\textwidth]{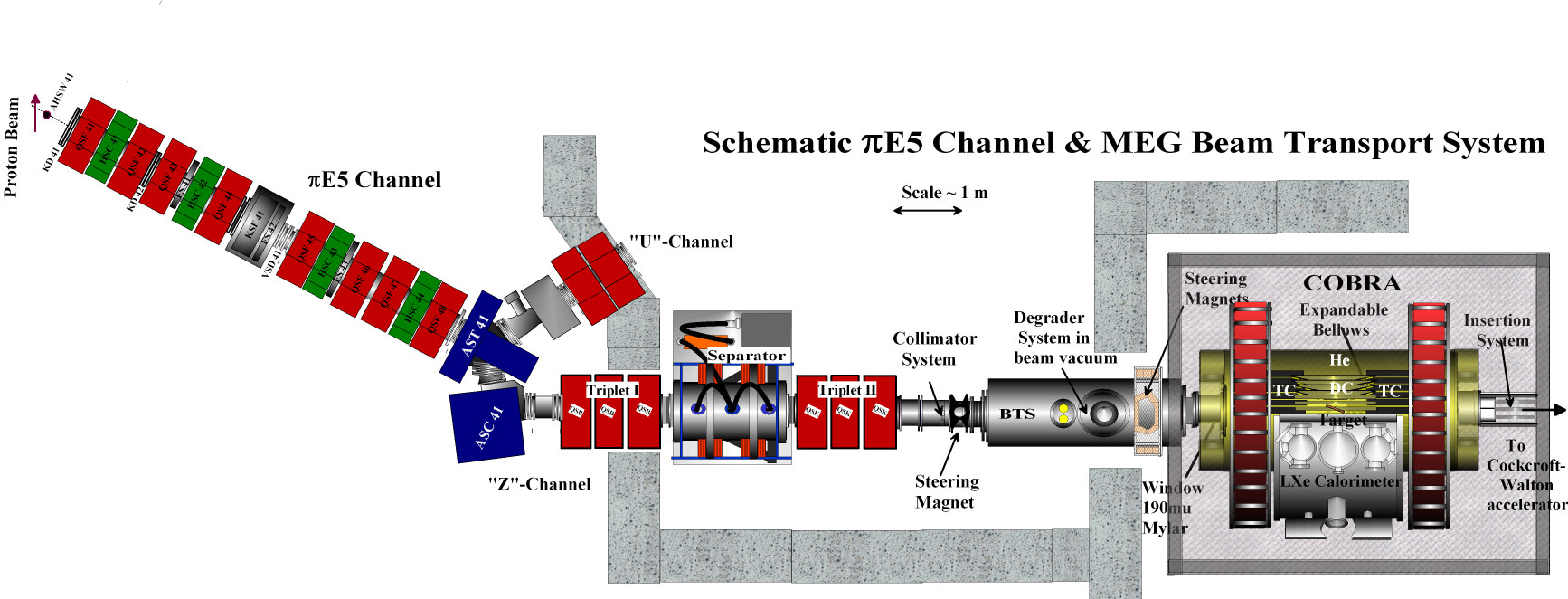}
\caption{MEG Beam line with the $\pi$E5 channel and MEG detector system incorporated in and around the COBRA magnet. \label{fig:beamline}}
\end{figure*}

\begin{figure}
\centering
\includegraphics[width=0.49\textwidth]{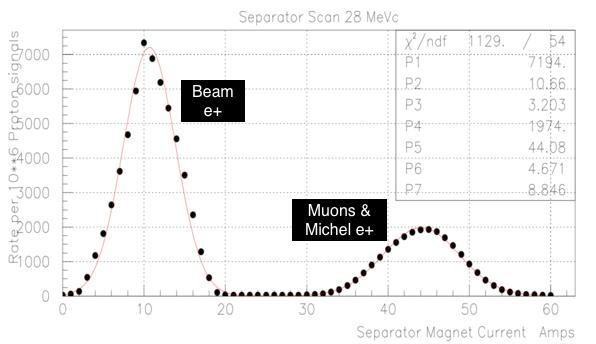}
\caption{Measurement of the separation quality with the
Wien-filter during the 2015 Pre-Engineering Run. \label{fig:separation}}
\end{figure}

Coupling the MEG COBRA spectrometer and LXe photon
detector to the $\pi$E5 channel, which ends with the last dipole
magnet ASC41 in the shielding wall, is achieved
with a Wien-filter (cross-field separator) and two sets of
quadrupole triplet magnets, as shown in Fig.~\ref{fig:beamline}. These
front-elements of the MEG beam line allow a maximal
transmission optics through the separator, followed by
an achromatic focus at the intermediate collimator system.
Here an optimal separation quality between surface
muons and the eight-fold higher beam positron
contamination from Michel positrons or positrons
derived from $\pi^0$-decay in the target and having the correct
momentum, can be achieved (see Fig.~\ref{fig:separation}) \cite{megdet}. 
The muon range-momentum adjustment is made at the centre
of the superconducting beam transport solenoid BTS
where a Mylar\textsuperscript{\textregistered} degrader system is placed at the central
focus to minimise multiple scattering. The degrader
thickness of \SI{300}{\um} takes into account the remaining
material budget of the vacuum window at the entrance
to the COBRA magnet and the helium atmosphere inside,
so adjusting the residual range of the muons to
stop at the centre of a \SI{205}{\um} thick polyethylene target
placed at \ang{20.5} to the axis.

The residual polarisation of the initially 100\%
polarised muons at production has been estimated by
considering depolarising effect at production, during
propagation and due to moderation in the stopping target.
The net polarisation is seen in the asymmetry of the
angular distribution of decay Michel positrons from the
target. The estimate is consistent with measurements
made using Michel positrons at the centre of the COBRA
spectrometer \cite{Baldini:2015lwl}, where the energy-dependent angular
distributions were analysed. A high residual polarisation
of $P_{\mu^+} = -0.86\pm 0.02~\mathrm{(stat.)} + 0.06  - 0.05~\mathrm{(syst.)}$ was
found, with the single largest depolarising contribution
coming from the cloud muon content of the beam.
These are muons derived from pion decay-in-flight
in and around the target and inherently have a low polarisation
due to the widely differing acceptance kinematics.
The cloud muon content in the \SI{28}{\MeV/\clight} surface
muon beam was derived from measurements where the
muon momentum spectrum was fitted with a constant
cloud muon content over the limited region of the kinematic
edge of the spectrum at \SI{29.79}{\MeV/\clight}. This was
cross-checked against measurements at \SI{28}{\MeV/\clight} using
a negative muon beam. In this case, there are no
such surface muons (due to the formation of pionic atoms
on stopping) and hence a clear cloud muon signal
can be measured. When comparing the cross-sections
and the kinematics of pions of both charge signs consistency
is found, with a ratio of $\sim1.2$\% of negative
cloud muons to surface muons at \SI{28}{\MeV/\clight}. This situation
is not expected to change significantly for MEG~II,
apart from the slightly higher divergences expected due
to the increased $\Delta p_\muonp/p_\muonp$ and a possible difference in the
polarisation quenching properties of the target material
in a magnetic field \cite{Buhler1}, which is still under investigation.

\subsection{Upgrade concept}

The increased sensitivity sought in MEG~II will partially
be realised by the full exploitation of the available
beam intensity and partially by the increased detector
performances, allowing the most significant contribution
to the background from overlapping accidental
events, to be managed, at the level of an order of magnitude
higher sensitivity for the experiment. As outlined
in Sect.~\ref{sec:Status_of_MEG} the accidental background has a quadratic
dependence on the muon beam stopping rate,
whereas the signal is directly proportional to the stopping
rate. This puts stringent limits on the material
budget and the suppression of beam-correlated backgrounds
in the beam line, while having to allow for the
flexibility and versatility of different beam modes required
for calibration purposes. The three main modes required are:
\begin{itemize}
\item stopped surface muon beam for normal data-taking
at \SI{28}{\MeV/\clight}, 
\item stopped negative pion beam of \SI{70.5}{\MeV/\clight} for
charge-exchange $\pi^-\mathrm{p}\to \pi^0\mathrm{n}$ (CEX) and radiative
capture $\pi^-\mathrm{p}\to \gamma \mathrm{n}$ (RC) photons (see Sect.~\ref{sec:LXe_Calorimeter_Calibration}) and
\item a monochromatic positron beam of \SI{53}{\MeV/\clight} 
for Mott scattering calibrations (see Sect.~\ref{sec:Mott}).
\end{itemize}

For MEG~II, the beam line components and optics
will stay the same, apart from the introduction of extra
beam monitoring tools (cf. Sect.~\ref{sec:CsI_luminophore}). However, the
increased muon rate for MEG~II, while maintaining the
high transmission optics, can only be achieved by an
increase in the momentum-byte $\Delta p_\muonp/p_\muonp$ i.e. by means of
opening the $\pi$E5 channel momentum slits to their full
extent. An increased $\Delta p_\muonp$ however, implies an increased
range straggling of the beam. A study undertaken for
the MEG~II upgrade proposal \cite{Baldini:2013ke} looked at various
beam/target scenarios comparing the use of a surface
muon beam of \SI{28}{\MeV/\clight} (mean range 
\SI{\sim 125}{\mg\per\cm\squared}) to
that of a sub-surface beam of \SI{25}{\MeV/\clight} (mean range 
\SI{\sim 85}{\mg\per\cm\squared}). As the name implies, these are muons with
a unique momentum of \SI{29.79}{\MeV/\clight} from stopped pion
decay, which are selected from deeper within the target
and lose some of their energy on exiting.

The potential advantage of such a sub-surface
beam is then the reduced range straggling which is
comprised of two components (cf. Eq.~(\ref{eq:range})). The first factor
from energy-loss straggling of the intervening material,
which at these momenta amounts to about 9\% (FWHM) of the
range \cite{Pifer} and the second from the momentum-byte
$\Delta p_\muonp/p_\muonp$. However, the
range and the straggling vary most strongly with momentum,
being proportional to $a\times p^{3.5}$, where \lq{}$a$\rq{} is a material
constant,
\begin{align}
\Delta R_\mathrm{TOT} = a\sqrt{(0.09)^2 + (3.5\Delta p_\muonp/p_\muonp)^2}\times p_\muonp^{3.5}.
\label{eq:range}
\end{align}Therefore, the most efficient way to reduce
the range straggling is by reducing the momentum
rather than the $\Delta p_\muonp/p_\muonp$.

\begin{figure}
\centering
\includegraphics[width=0.49\textwidth]{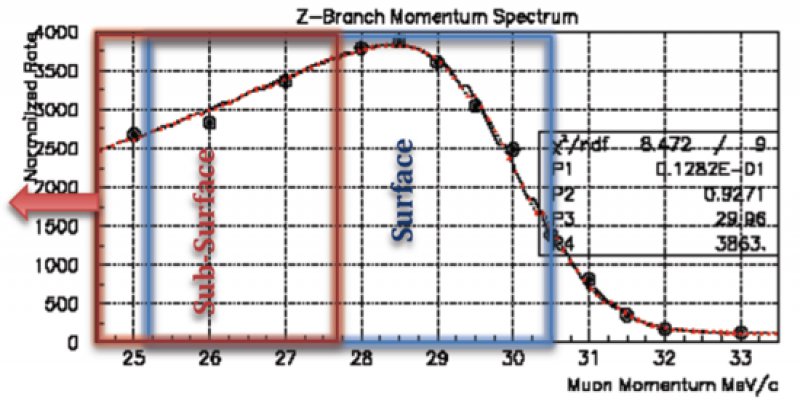}
\caption{Shows the $\pi$E5 measured momentum spectrum with full momentum-byte. 
The red curve is a fit to the data with a $p^{3.5}$ power law, folded with a Gaussian
momentum resolution corresponding to the momentum byte
as well as a constant cloud muon contribution. \label{fig:Lpspect}}
\end{figure}

A momentum change has a direct impact on the
target thickness, which is a balance between maximising
the stop density and minimising the multiple scattering
of the out-going Michel positrons and the photon
background produced in the target. Furthermore, the
surface muon rate also decreases with $p^{3.5}$ and therefore
ultimately limits how low one can go down in momentum.
This behaviour is shown in Fig.~\ref{fig:Lpspect}, where the
measured muon momentum spectrum is fitted with a
$p^{3.5}$ power-law, folded with a Gaussian momentum resolution
equivalent to the momentum-byte, plus a constant cloud muon content. 
The blue and the red (truncated) 
boxes show the $\pm 3\sigma_{p_\muonp}$ momentum acceptance
for the surface/sub-surface beams, corresponding
respectively to (\num{\pm 2.7}/\num{\pm 2.5})~\si{\MeV/\clight}. The optimal
momentum yielding the highest intensity within the
full momentum-byte is centred around \SI{28.5}{\MeV/\clight}. For
each data-point the whole beam line must be optimised.
The upgrade study \cite{Baldini:2013ke} investigated various
combinations of beam momentum and target parameters
such as thickness which varied between \SIrange[range-units=single,range-phrase=--]{100}{250}{\um}
and orientation angle varying between \SIrange[range-phrase=--]{15.0}{20.5}{\degree}.
This resulted in only one really viable solution that could yield
the required muon stopping intensity of \SI{7e7}{\muonp\per\second}
suitable for achieving the goal sensitivity within
a measuring period of \num{\sim 3} years:
a surface muon beam of \SI{28}{\MeV/\clight} with a polyethylene
target of \SI{140}{\um} thickness, placed at an angle of 
\ang{15.0} to the axis.

A sub-surface beam solution was only able to meet the criteria
by scaling-up the target thickness to \SI{160}{\um}, which negated 
the principle.
Hence the baseline solution chosen for MEG~II was the 
surface muon beam solution due to the thinner target and higher 
achievable rate as well as its beneficial impact on the resolutions and
background.

\subsection{Beam monitoring }

Two new detectors have been developed to measure the beam profile and rate: 
the sampling scintillating fibre beam monitoring (sampling SciFi) mounted at 
the entrance to the spectrometer and the luminophore foil detector (CsI on a 
Mylar support) coupled with a CCD camera installed at the intermediate focus 
collimator system. 

\subsubsection{The sampling SciFi beam monitoring detector}

This detector is a quasi non-invasive, high rate sustainable beam monitoring tool, 
able to provide beam rate, profile measurements and particle identification in real time.
It is based on scintillating fibres (SciFi) coupled to SiPMs;
the usage of SiPMs allows for a detector able to work in high magnetic fields.

It consists of a grid of two orthogonal fibre layers: one with the fibres running along the $x$-axis
and the other with the fibres along the $y$-axis.
The detector is expected to be located at the end of the vacuum beam line, just in front of the spectrometer. A movable configuration allows the remote removal/insertion of the detector into the beam.

\begin{figure}
\centering
\includegraphics[width=0.99\linewidth, clip, trim=2pt 8pc 2pt 7pc]{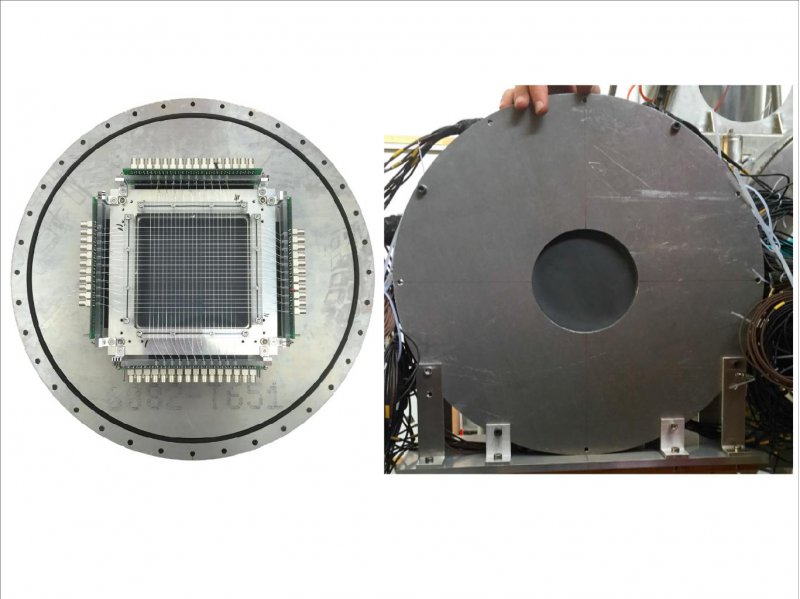}
\caption{The orthogonal double layer scintillating fibre prototype (left) and the 
front view of the detector assembly (right). A \SI{25}{\um} thick Tedlar foil is used as a detector entrance window.}
\label{fig:SamplingSciFiPrototype}
\end{figure}

Figure{~\ref{fig:SamplingSciFiPrototype}} shows the built and tested full scale prototype. We used Saint-Gobain BCF-12, \SI[product-units=single]{250x250}{\um\squared} double-cladding fibres~\cite{SaintGobainFibres}, each one independently coupled at both ends to Hamamatsu S13360-1350CS SiPMs (with an active area of \SI[product-units=single]{1.3 x 1.3}{\mm\squared} and a pixel size of \SI[product-units=single]{50 x 50}{\um\squared})~\cite{hamamatsu-MPPC-SiFi}. The relative distance between adjacent fibres mounted in the same layer is equal to \SI{4.75}{\mm}, a pitch which satisfies the requirements for a precise measurement of the beam profile and rate. Furthermore a large detector transparency $T > 92\%$ (where $1-T =$ particles hitting the fibres / total incident particles) is achieved with a relatively small number of channels ($\approx 100$). In fact for this prototype we mounted 21 fibres per layer giving a total number of 84 channels. The signals are sent to the TDAQ prototype (see Sect.~\ref{sec:tdaq}) that includes also the preamplifiers (with adjustable gain up to 100, which is what we used here) and the power supplies for the SiPMs (operated at \SI{\approx 55.6}{\volt}). The trigger used for the beam profile and rate measurements is the ``OR" of all the ``AND"s of the SiPMs coupled to the same fibre, with a common threshold for all channels \num{\geq 0.5} photoelectrons.

\begin{figure}
\centering
\includegraphics[width=0.99\linewidth, clip, trim=0 2pc 0 4pc]{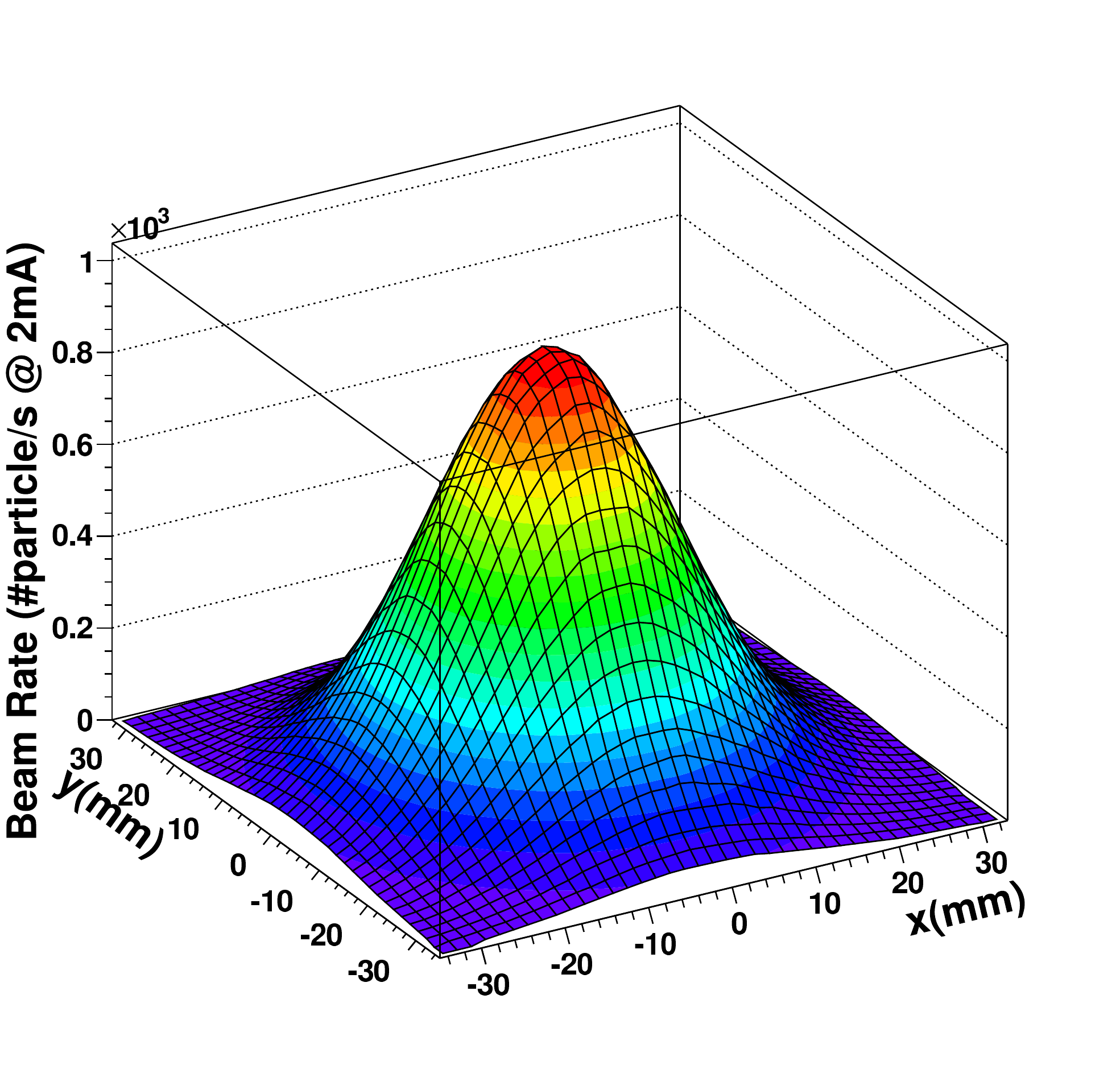}
\caption{Positive muon beam profile and rate as measured along the $\pi$E5 beam line.}
\label{fig:SciFiBeamProfiles}
\end{figure}

Figure~\ref{fig:SciFiBeamProfiles} shows the beam profile as measured with the detector 
mounted along the $\pi$E5 beam line. The incident particles are positive muons with an 
initial momentum of \SI{28}{\MeV/\clight}, after having left the \SI{190}{\um} Mylar 
window at the end of the vacuum beam line and travelling some \SI{15}{\cm} in air before 
traversing the \SI{25}{\um} of Tedlar\textsuperscript{\textregistered} used as a light tight shield. 
The corresponding total rate and beam profiles were 
$R_\muonp (\mathrm{at}~I_{\mathrm{p}}= \SI{2.2}{\milli\ampere}) = 
\SI[separate-uncertainty]{1.11(1)e8}{\muonp\per\second}$ and 
$(\sigma_x,\sigma_y)= (\num[separate-uncertainty]{18.1(1)}, \num[separate-uncertainty]{17.8(1)})~\si{\mm}$, 
respectively. These measured numbers are consistent to within 5\% or better 
with those provided by our ``standard" beam monitoring tools (methods based on a 2D x-y 
scanner using a large depletion layer APD or a pill scintillator coupled to a miniature PMT). 
One of the most attractive features of this detector is its capability of providing the 
full beam characterisation in just tens of seconds with all the associated benefits 
(faster beam tuning, real time feedback about a malfunctioning of the beam/apparatus, 
reduced systematic uncertainties etc.).

\begin{figure}
\centering
\includegraphics[width=0.99\linewidth, clip, trim=0 0 0 4pc]{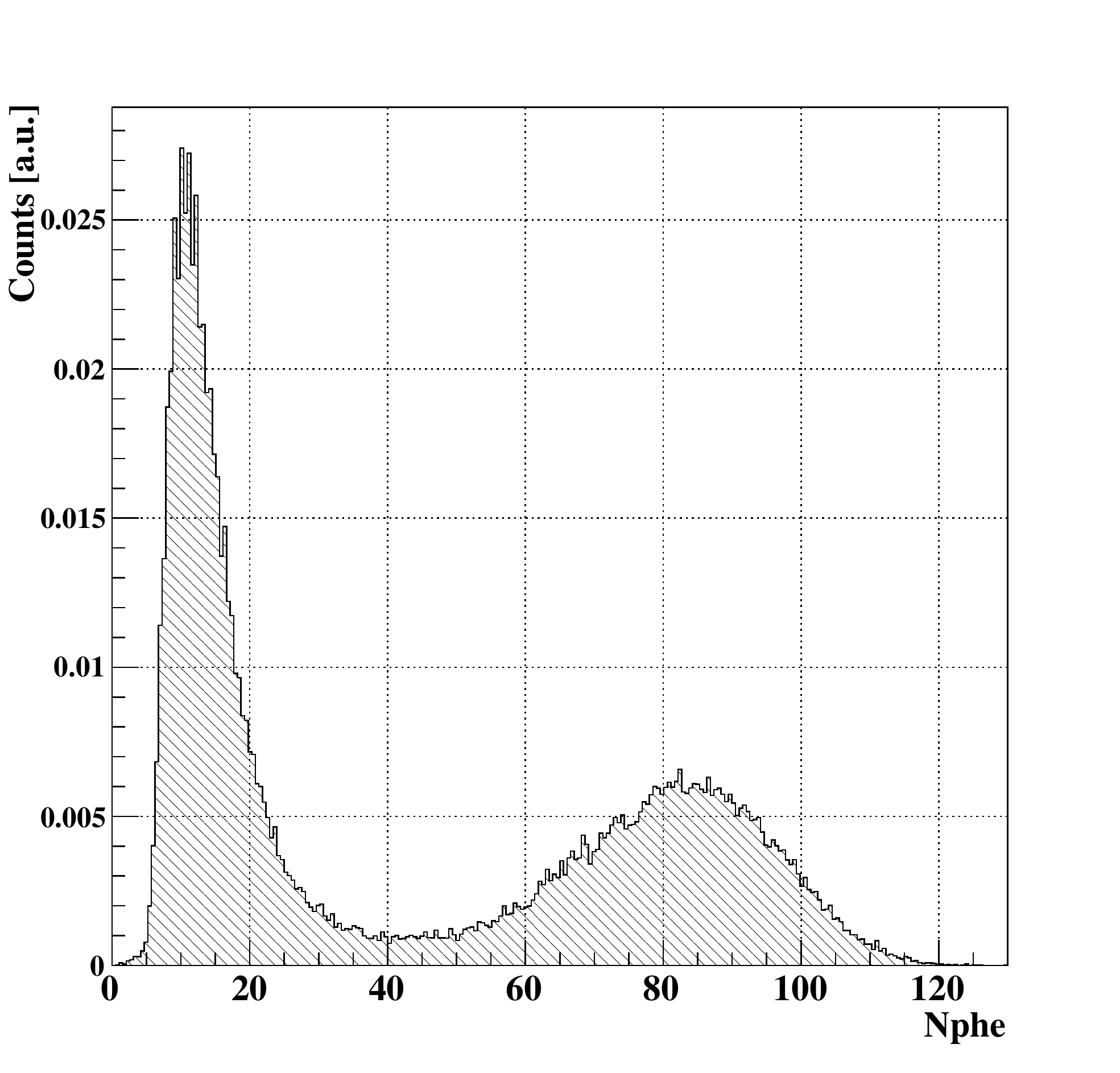}
\caption{Particle identification using the different energy deposited by positrons (m.i.p) (peak on the left) and muons (peak on the right) with an original momentum of $p=\SI{28}{\MeV/\clight}$.}
\label{fig:SciFiParticleID}
\end{figure}

Figure~\ref{fig:SciFiParticleID} shows the detected charge associated with positrons 
of \SI{28}{\MeV/\clight} and stopping muons in the fibres. A clear separation between 
the positrons (which are minimum ionising particles m.i.p.) and the low energy muons can be seen. 

\begin{figure}
\centering
\includegraphics[width=0.99\linewidth, clip, trim=0 0 0 3pc]{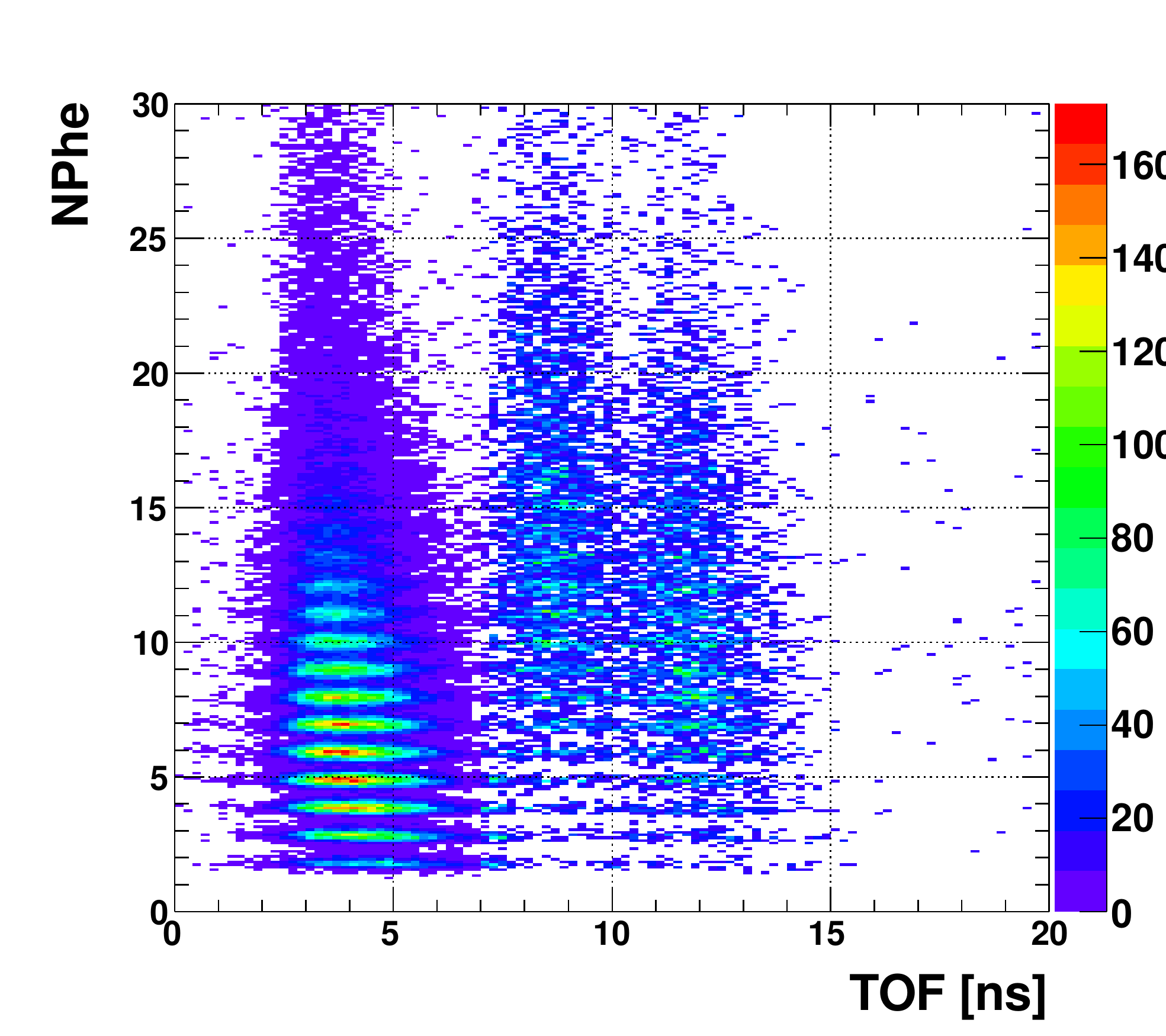}
\caption{Scatter plot of the measured charge versus the time difference between the arrival time of the particles (with momentum $p=\SI{115}{\MeV/\clight}$) and the radio frequency of the main accelerator. From left to right we have positrons, pions and muons.}
\label{fig:SciFi_TOF}
\end{figure}

Figure~\ref{fig:SciFi_TOF}, finally, shows the capability of the detector to distinguish 
between high momentum particles ($p=\SI{115}{\MeV/\clight}$) by plotting the measured 
charge associated to them versus their time-of-flight (the radio frequency of the main accelerator is 
used as a time reference). From left to right we have positrons, pions and muons. 

\subsubsection{An ultra-thin CsI(Tl) luminophore foil beam monitor}
\label{sec:CsI_luminophore}
A new in-situ, high rate and non-destructive beam monitoring system based on a thin CsI(Tl) scintillation foil (luminophore) and a CCD camera system has been developed for MEG~II. Initial tests as an external device able to measure both the beam intensity as well as giving a quantitative measure of the beam spot size have led to a permanent installation incorporated into the beam line vacuum at the MEG intermediate focus collimator system.

The advantages of such a system over the standard MEG pill-scintillator 2D x-y scanner system are four-fold:
in-situ, non-destructive measurement of the beam characteristics, no dismantling of beam line components necessary, as in the case of the pill-scintillator scanner system;
in vacuum measurement, no corrections needed for multiple scattering in the vacuum window or air;
comparatively fast measurement, multiple exposures each of \SIrange[range-phrase=--,range-units=single]{10}{100}{\second} compared with a pill-scintillator 2D ``cross-scan" of \SI{10}{\minute} or a 2D~``raster-scan" of \SI{90}{\minute};
continuous monitoring possible allowing online centring in the event of beam steering due to changes of the proton beam position on the muon production target E.

\paragraph{{Cs}{I}({Tl}) foils and CCD camera system}
CsI(Tl) is a well known and common inorganic scintillator with a relatively high light yield at 
more than \SI{5e4}{ph\per\MeV} of deposited energy. The peak emission of CsI(Tl) 
is approximately \SI{560}{\nm} and well suited for use in visible light imaging systems such as a CCD. 
The scintillation light decay constants (\SI{\sim1}{\us}) are rather long compared to fast organic 
scintillators though not problematic for this application due to the much longer exposure times.

Four foils were constructed using a Lavsan (Mylar\textsuperscript{\textregistered} equivalent) 
base structure, where a thin layer of CsI(Tl) was applied using chemical vapour deposition. 
The precise CsI(Tl) layer thickness was varied between \SI{3.0}{\um} and \SI{5.2}{\um}, allowing 
for the comparison and possible optimisation of layer thickness.

The imaging system used was a Hamamatsu ORCA FLASH4.0 camera providing 4.19 megapixels along with \SI{16}{bit} pixel depth. An internal Peltier cooling device as well as an external water cooling system allow the sensor temperature to be reduced to \SI{-30}{\degreeCelsius} and hence significantly reducing the thermal noise. The sensor's peak quantum efficiency matches well to the CsI(Tl) peak emission near \SI{560}{\nm}.

\paragraph{Beam image analysis}
Beam profile imaging consists of multi-frame (typically 10) exposures each of 
\SI{10}{\second} length together with an equivalent set of background exposures 
taken with the beam-blocker closed, enabling stray ambient light and the inherent 
thermal noise of the sensor to be eliminated on subtraction. 

\begin{figure}[tb]
\centering
   \includegraphics[width=\columnwidth]{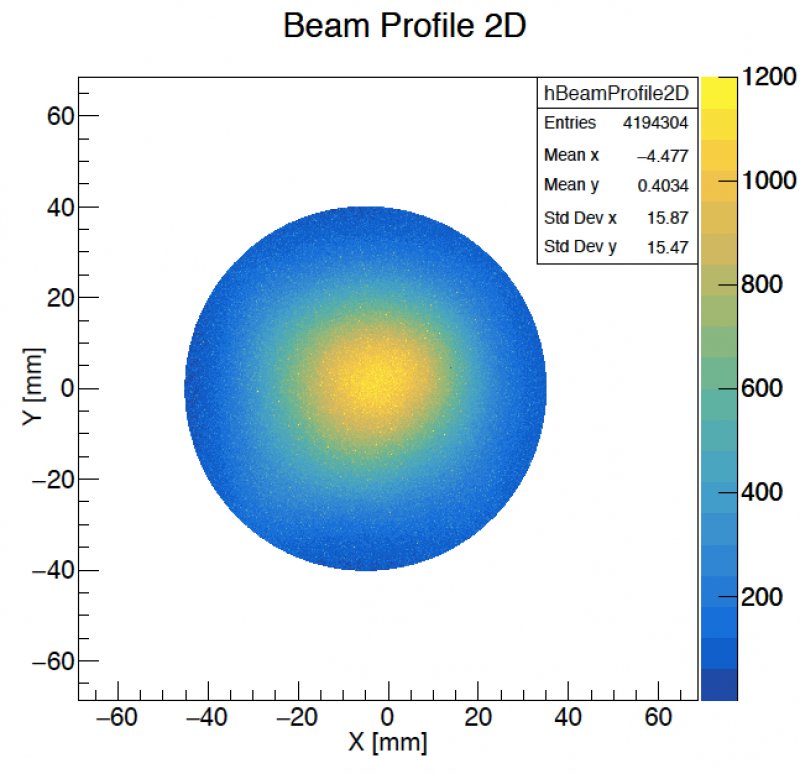} 
   \caption{Beam profile signal image after background subtraction, cut to a region of interest, and normalised to the proton current.}
   \label{fig:beamprofile-2D}
\end{figure}

All signal and background images are first summed and averaged and then subtracted 
to generate a calibrated signal image, from which a central region of interest is selected. 
This image is then fitted using a 2D correlated Gaussian function to obtain the beam 
position and widths in $x$ and $y$ as well as their correlations.
The summed image intensity is normalised by the total proton current during the exposure period. 
The current measurement is initiated by a simultaneous external trigger of the proton signal 
scalar and the camera shutter. A typical image after processing is shown in Fig.~\ref{fig:beamprofile-2D}.

\begin{figure}[tbp]
\centering

\subfigure[]{
   \includegraphics[width=0.47\linewidth]{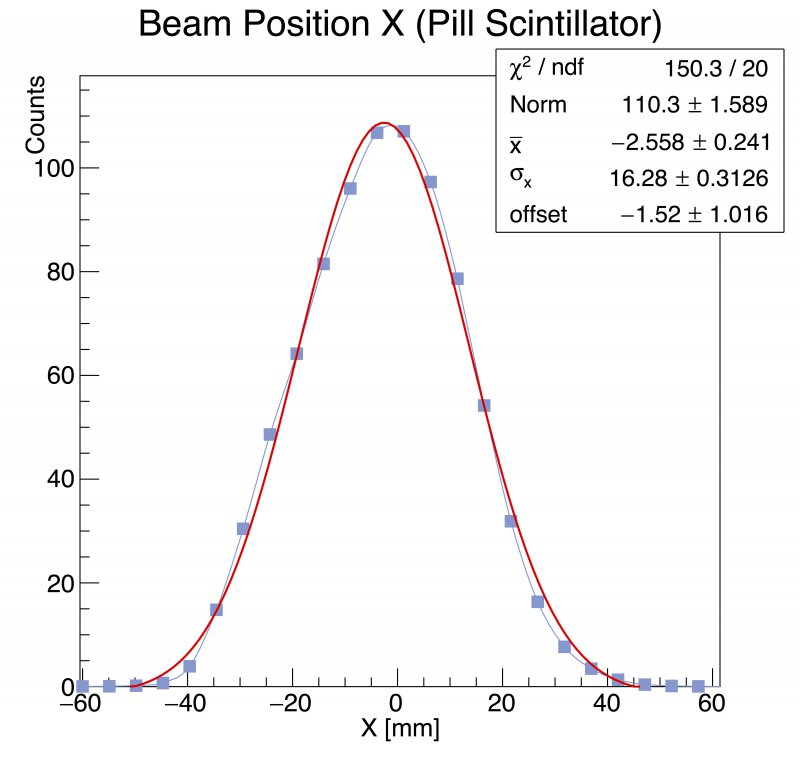}
    }
\subfigure[]{
   \includegraphics[width=0.47\linewidth]{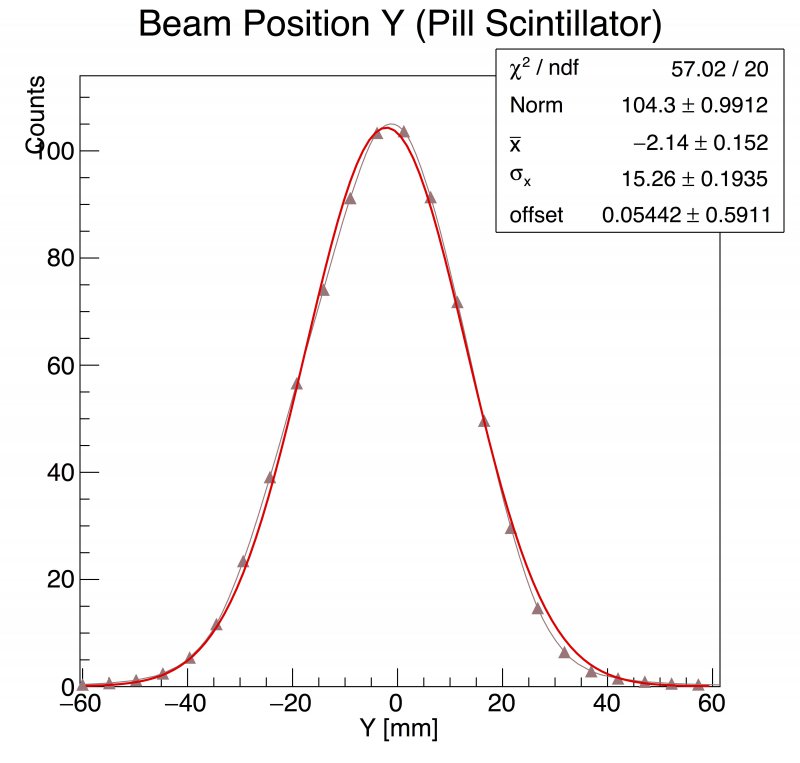}
    }
\subfigure[]{
   \includegraphics[width=0.47\linewidth]{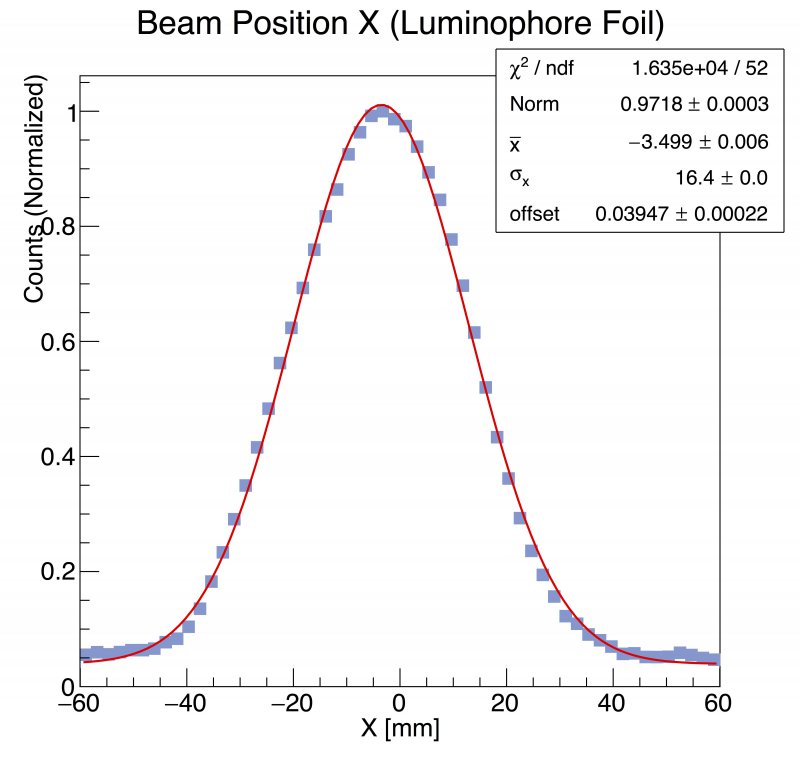}
    }
\subfigure[]{
   \includegraphics[width=0.47\linewidth]{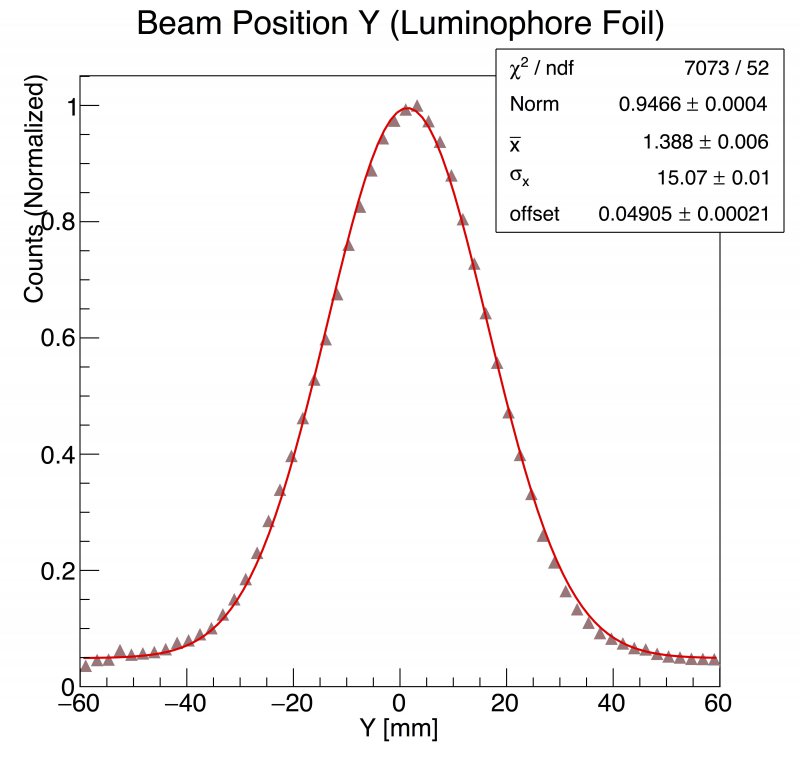}
    }

   \caption{The beam profiles in $x$ and $y$ measured with the pill-scintillator in (a) and (b) and projections from the luminophore foil in (c) and (d) fitted with Gaussian functions. Emphasis is on the beam widths, as differences in mean positions are attributed to alignment differences in the two setups.}
   \label{fig:beamprofiles}
\end{figure}

\paragraph{Beam width}
A comparison of the beam spots as measured by the pill-scintillator to those obtained 
from $x-y$ projections of the luminophore foil image are shown in Fig.~\ref{fig:beamprofiles} 
with good agreement within the fit widths.
The difference in centroids is due to the difference in alignment between the two setups. 

\begin{figure}[tbp]
   \centering
   \includegraphics[width=\columnwidth]{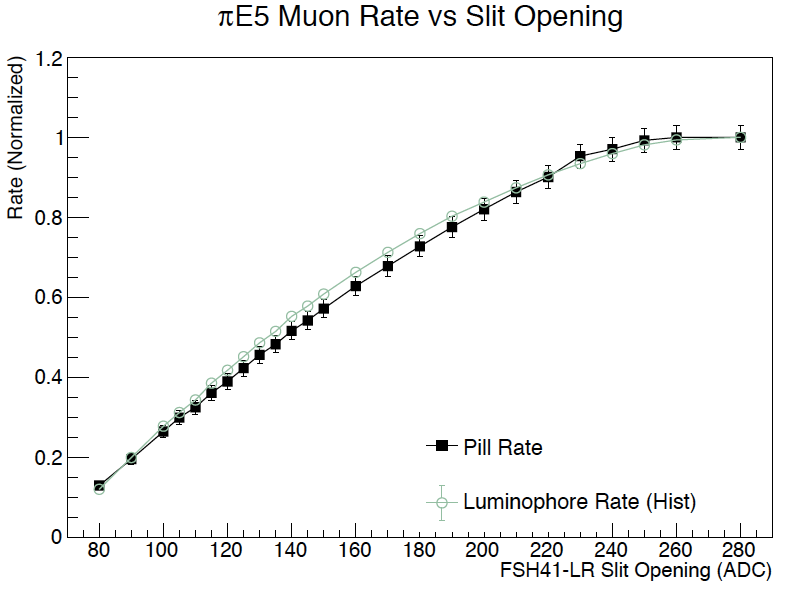} 
   \caption{The muon rate as a function of the beam line slit opening, measured using the pill-scintillator and luminophore foil.}
   \label{fig:beamrate_slitcurve}
\end{figure}

The spatial resolution of the luminophore foil system was determined by placing an Al grid just upstream of the foil, while irradiating with the muon beam. The grid edges of the resultant picture image, when fitted with a step-function convoluted with a Gaussian resolution function, yield an upper limit on the combined foil, camera and beam resolution of \SI{650}{\um} which includes beam divergence and range straggling effects, so that the intrinsic spatial resolution of the foil is much smaller. 
\paragraph{Beam intensity}
A beam intensity comparison between the luminophore system and the pill-scintillator system was made by symmetrically opening the $\pi$E5 FS41L/R slit system in small steps, so scanning the full beam intensity over an order of magnitude. The comparative plot of relative intensity normalised to the proton beam intensity is shown in Fig.~\ref{fig:beamrate_slitcurve}. Good agreement can be seen at the 5$\%$ level which can be understood as being due to the difference in technique. The pill-scintillator measurement samples only a \SI{2}{\mm} diameter portion of the beam on the beam-axis, whereas the luminophore samples the entire beam spot which changes in size with slit opening, at the 10$\%$ level over the entire range.

\paragraph{Beam line setup}

\begin{figure}[tbp]
\centering
  \includegraphics[width=0.99\columnwidth]{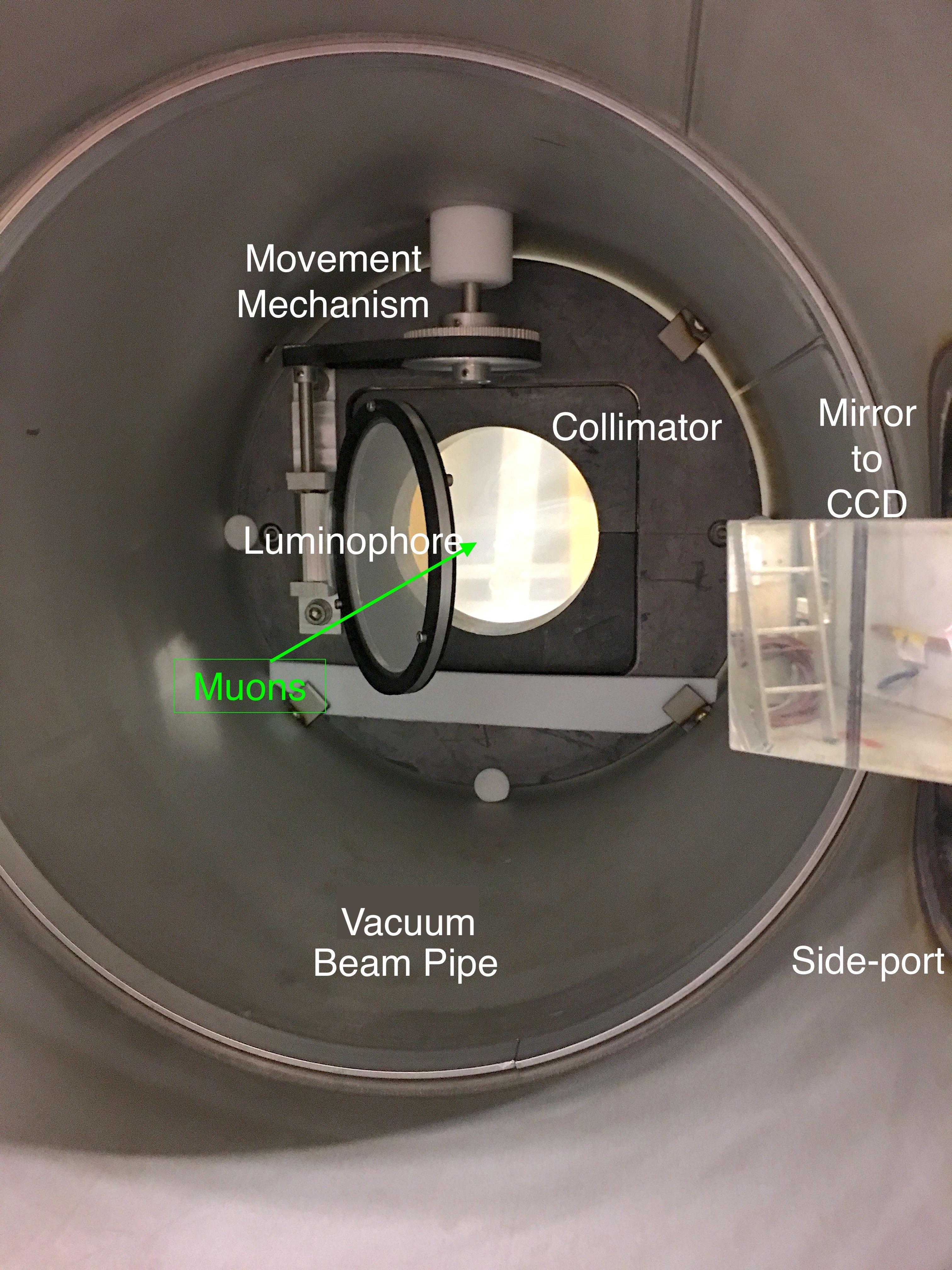}
   \label{fig:foilframe}
   \caption{The luminophore foil set-up at the collimator. The imaging is done via a mirror system through a side-port to a CCD camera outside the vacuum pipe.}
\end{figure}

The initially developed external system has since been incorporated into the beam line vacuum as part of the intermediate focus collimator system shown in Fig.~\ref{fig:foilframe}. The foil frame is attached to a drive shaft and pulley system that allows the foil to be rotated in and out of the beam while under vacuum. A calibration grid is attached to the surface of the frame to allow for a pixel-to-millimetre conversion. The foil and frame are viewed inside the beam pipe, under vacuum and imaged with the CCD camera via a mirror system and glass window on a side port. The interior of the vacuum pipe can be illuminated with a UV LED to conduct calibration measurements of the foil and CCD system within the light-tight region. 

\begin{figure}[tbp]
   \centering
   \includegraphics[width=\columnwidth]{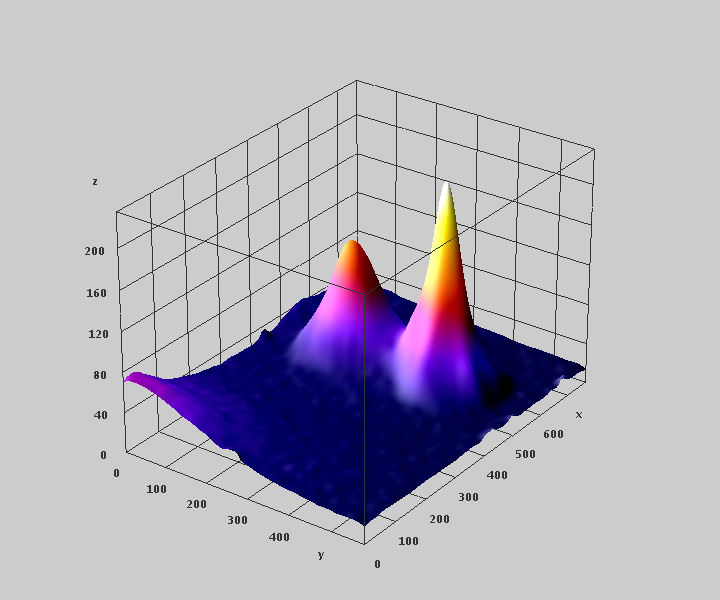} 
   \caption{A pseudo 3D light intensity plot showing the muon beam (small peak) and positron beam (large peak) spots together in one image. This is achieved by reducing the Wien filter (SEP41) separation power through reduced E and B fields.}
   \label{fig:BeamMuAndE_pseudo}
\end{figure}

An example of the usefulness of such a system can be seen in Fig.~\ref{fig:BeamMuAndE_pseudo}, which shows the separation quality between muon and positron beam spots imaged at the collimator system with the luminophore foil. The separation quality has purposely been reduced by adjusting the parameters of the Wien filter in order that both spots can be seen simultaneously on the picture. The use of the luminophore allows a calibration of the spatial separation to be made effectively online.

\paragraph{Conclusions}
Thin CsI(Tl) luminophore foils offer fast, in-situ beam monitoring possibilities, with negligible impact on beam rate and emittance. The foils combined with a cooled camera system with sufficient resolution reproduces beam profile and rate measurements conducted with the scanning pill-scintillator. Full 2D beam measurement can be made approximately ten times faster while providing long-term non-destructive beam information. Furthermore, it allows a direct measure of beam parameters without the need for multiple scattering corrections due to air or vacuum windows and allows direct feedback on external influences on the beam position or intensity.

\newpage
\section{Target }
\label{sec:Target}

The basic requirements for a MEG stopping target are six-fold:
\begin{itemize}
\item a high muon stopping density over a limited
axial region centred on the COBRA fiducial volume,
\item minimisation of multiple scattering for the outgoing positrons,
\item minimisation of photon conversions from RMD in the target,
\item minimisation of positron AIF or bremsstrahlung with photons 
entering the detector acceptance,
\item allow reconstruction of the positron decay vertex
and initial direction at the vertex, onto the target plane and
\item mechanically stable with good planarity and
remotely movable for compatibility with calibrations
requiring other targets.
\end{itemize}

\begin{figure}
\centering
\includegraphics[width=0.49\textwidth]{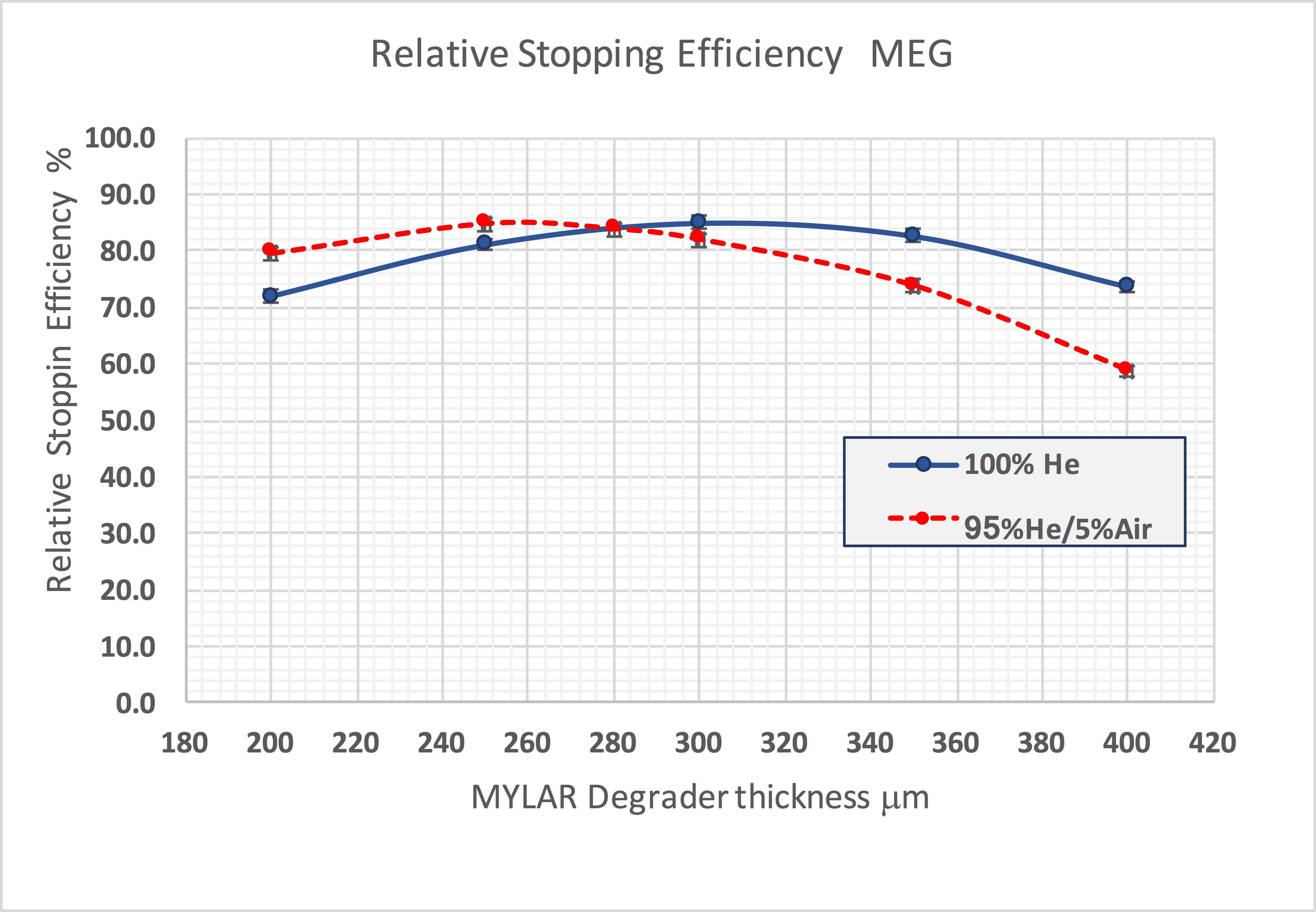}
\caption{Muon stopping efficiency versus degrader thickness for 
the MEG \SI{205}{\um} polyethylene (CH$_2$) target, for two different 
He-concentrations inside COBRA. \label{fig:stopeffI}}
\end{figure}

Owing to the thinner target, smaller angle for MEG~II and
the increased $\Delta p_\muonp/p_\muonp$, the remaining variable material
budget consisting of degrader and COBRA helium environment,
must then be matched to give an optimal residual range at
the target. Figure~\ref{fig:stopeffI} shows the simulation results for the
optimal stopping efficiency versus degrader thickness
for the previous MEG \SI{205}{\um} thick polyethylene target.
Two different He-concentrations are shown, from which can be seen 
that 1\%\ of air is equivalent to \SI{\sim 10}{\um} of Mylar.

For MEG~II a separate target study was also undertaken
to examine the material possibilities for a target
equivalent to the baseline \SI{140}{\um} polyethylene
(CH$_2$) target, placed at \ang{15.0} to the axis. The resulting set
of candidate targets are listed in Table 1 below. Since
the material thickness for each target is equivalent in
terms of the surface density \si{\gram\per\cm\squared}, the residual range
and hence the degrader thickness is therefore also the same.

\begin{table*}[tb]
  \caption{The candidate target parameters for an equivalent thickness 
  to the baseline solution of \SI{140}{\um} polyethylene (CH$_2$) target, 
  placed at \ang{15.0} to the axis.
  }
  \centering
  \label{beam:material}
   \begin{tabular}{cccccccccc}
    \hline
    Material   & Degrader   & Thickness  & Thickness & Inclination & Density                    & Stop Efficiency & \multicolumn{2}{c}{ Multiple Scattering} \\
               & (\si{\um}) & (\si{\um}) & (X$_0$)   &   (deg)     & (\si{\gram\per\cm\cubed})  & (\%)            & \multicolumn{2}{c}{ (\si{\milli\radian})} \\
               &            &            &           &             &                            &                 & \muonp[\SI{18}{\MeV}] &  \positron[\SI{52}{\MeV}]\\
    \hline
    CH$_2$     & 350 & 140 & \num{2.8e-4} & 15.0 & 0.893 & 83 & 52.0 & 3.0 \\
    \hline
    Be         & 350 &  90 & \num{2.6e-4} & 15.0 & 1.848 & 83 & 49.3 & 2.9 \\
    \hline
    Mylar      & 350 & 100 & \num{3.5e-4} & 15.0 & 1.390 & 84 & 58.5 & 3.4 \\
    \hline
    Scint. PVT & 350 & 130 & \num{3.1e-4} & 15.0 & 1.032 & 84 & 54.5 & 3.2 \\
    \hline
    Diamond    & 350 &  40 & \num{3.3e-4} & 15.0 & 3.515 & 81 & 56.8 & 3.3 \\
    \hline
   \end{tabular}
\end{table*}

\begin{figure}
\centering
\includegraphics[width=0.49\textwidth]{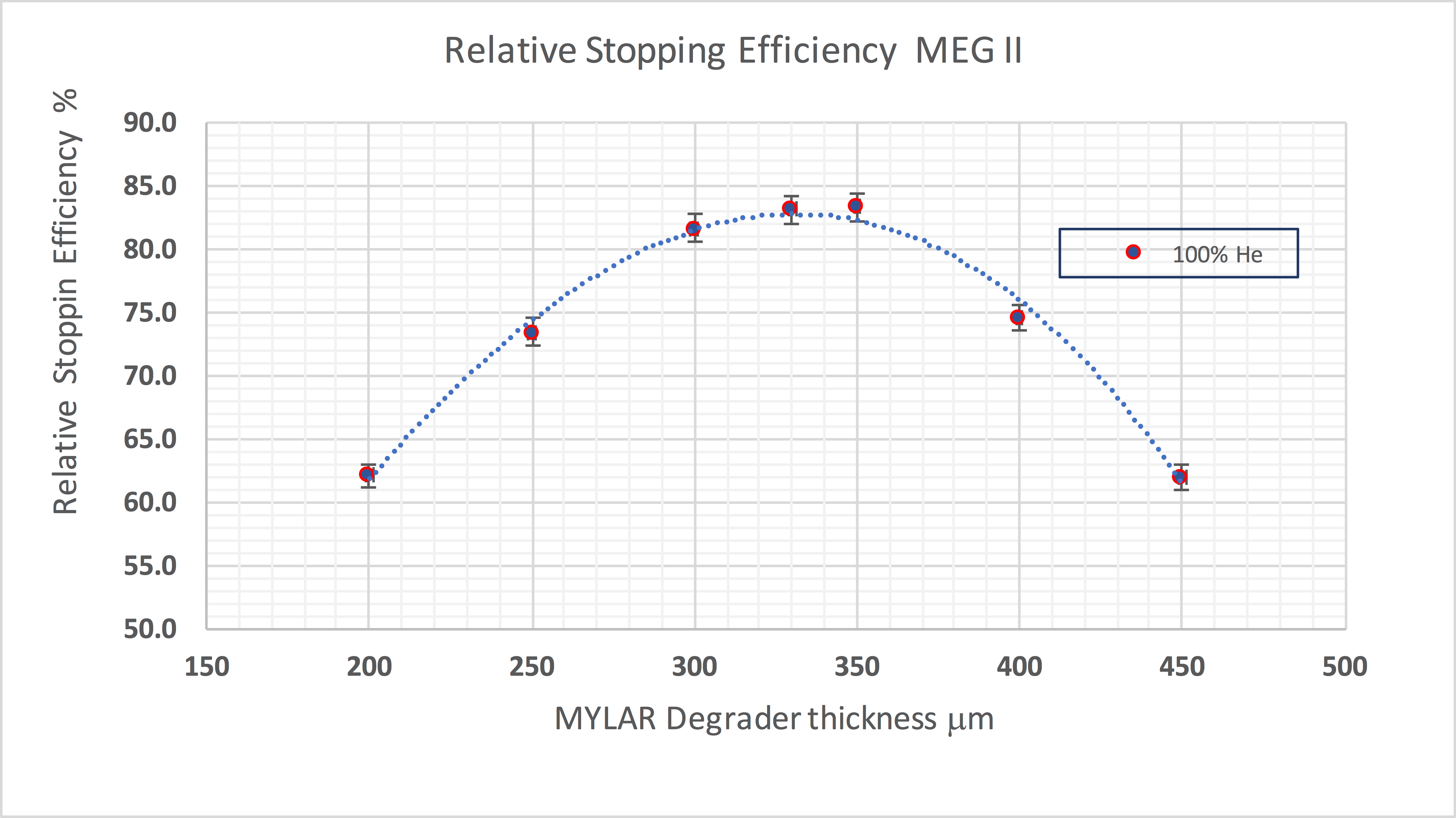}
\caption{Equivalent MEG~II case of muon stopping efficiency versus 
degrader thickness for a \SI{140}{\um} polyethylene (CH$_2$) target. \label{fig:stopeffII}}
\end{figure}

The main properties affecting tracking and
background production, as well as the target stopping
efficiency show that there are no dramatic differences
between the candidates, with multiple scattering estimates
varying less than 10\%\ from the average, while the
equivalent thickness in radiation lengths 
varies by about 15\%\ from the average. A separate
background study to estimate the number of background
photons with energy $\egamma > \SI{48}{\MeV}$ produced in
the fiducial volume of COBRA per incident muon and
entering the LXe photon detector gave values between
\SI[separate-uncertainty]{1.14(5)e-6}{\gamma\per\muonp}
for the scintillation target and
\SI[separate-uncertainty]{1.22(5)e-6}{\gamma\per\muonp}
for the Mylar target. The equivalent
simulated optimised stopping efficiency in the case
of the MEG~II polyethylene target is shown in Fig.~\ref{fig:stopeffII}.

Table~\ref{beam:material} shows that different materials outperform
each other in different categories. In general, the
beryllium target shows an overall good performance,
though from the thickness and size required, as well as
from the safety aspects it is not favoured. Diamond,
which is mechanically stable and known to be more radiation
tolerant has the smallest radiation length, as well
as having scintillation properties. However, it is currently
not commercially available in the size required for a
MEG~II target. The scintillation target (BC400B) from
Saint-Gobain lies in the mid-range of the performance
span, though with the lowest number of accepted background
photons per muon of all targets. A very important
and added advantage over the other non-scintillating
targets is, the possibility of non-destructive
beam intensity and profile measurements, using a CCD
camera and optical system. This would allow corrections,
caused by proton beam shifts on the main pion
production target, to be made to the beam centring on
the MEG muon target during data-taking. Two prototype
targets have so far been implemented for the Pre-Engineering
Runs 2015/16, a polyethylene (PE) and a polyvinyltoluene
(PVT) one. The prototype scintillation target
(PVT) is seen in Fig.~\ref{fig:PVTarget}.

\subsection{Scintillation target prototype}

\begin{figure}
\centering
\includegraphics[width=0.49\textwidth]{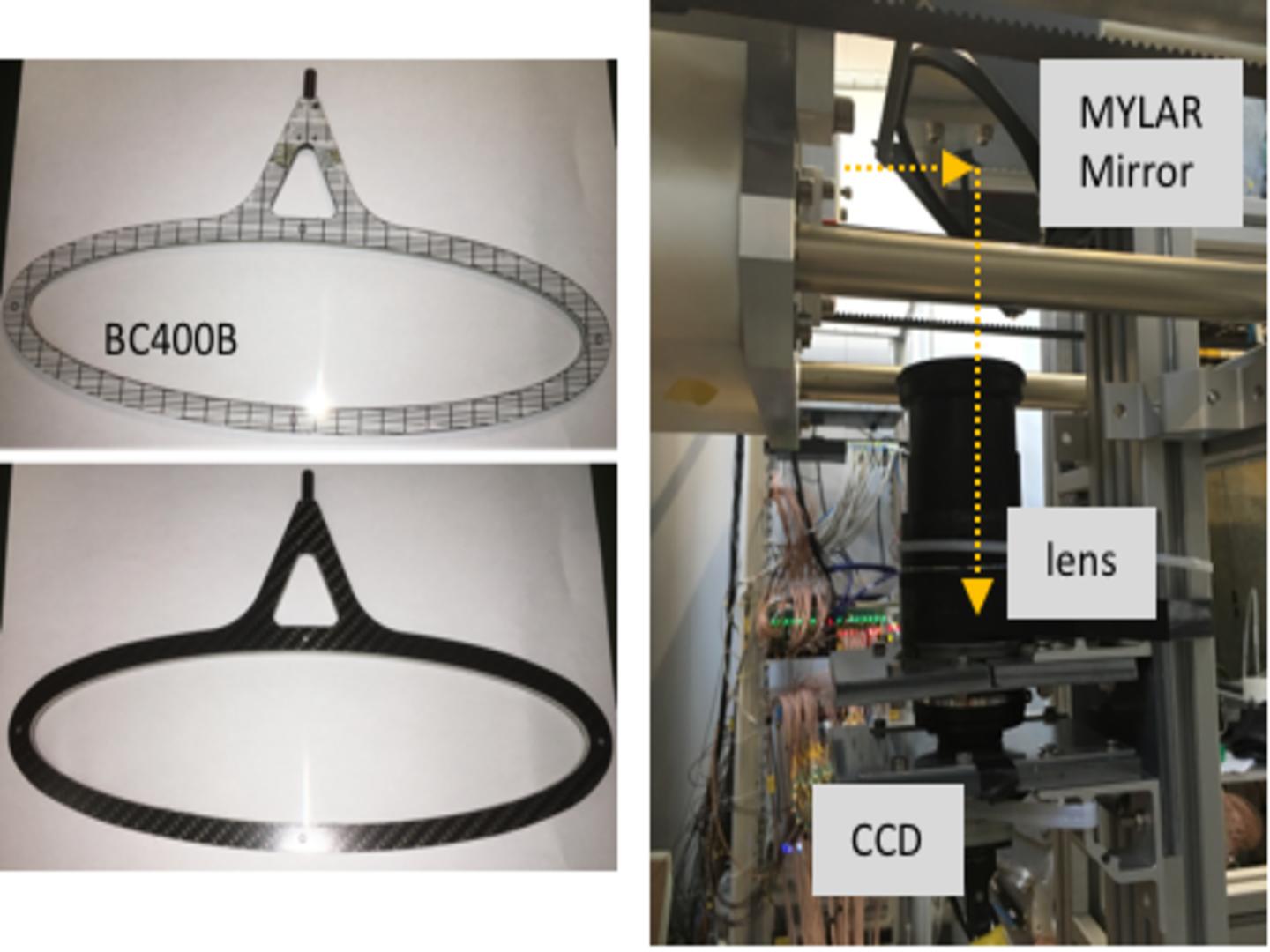}
\caption{(Left) shows two sides of the prototype PVT
target used during the 2016 Pre-Engineering Run. The
calibration grid is used for the perspective transformation.
The carbon-fibre/Rohacell\textsuperscript{\textregistered} foam frame can
be seen from the other side. (Right) shows the CCD
setup and Mylar mirror at the downstream side (DS)
of the COBRA magnet \SI{\sim 2.1}{\metre} DS of the target. \label{fig:PVTarget}}
\end{figure}

Figure~\ref{fig:PVTarget} shows the two sides of the prototype
target used in the 2016 Pre-Engineering Run, the downstream
CCD viewing side has a calibration grid as part
of the frame to ensure a correct perspective transformation
of the beam image. The frame is a sandwich of
carbon-fibre and Rohacell foam ensuring a lightweight
construction and strength, as can be seen from
the lower image in Fig.~\ref{fig:PVTarget} (left). The fiducial size of
the scintillator, excluding the frame is \SI[product-units=single]{260x70}{\mm\squared}.

\begin{figure}
\centering
\includegraphics[width=0.49\textwidth]{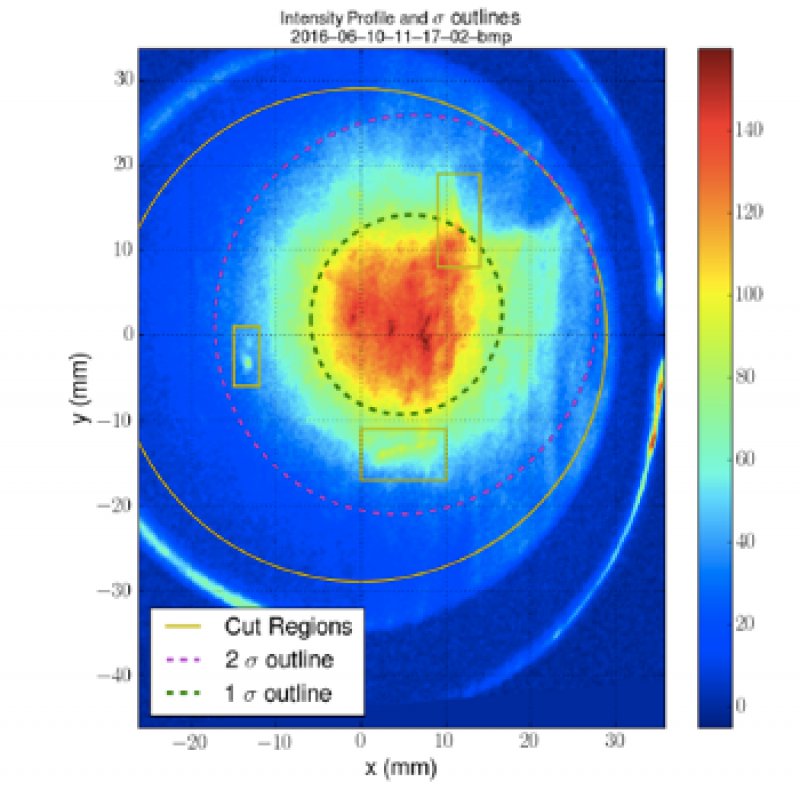}
\caption{Example of a perspective corrected target beam image viewed 
originally under \ang{15.0} to the target plane. The $1\sigma$- and 
$2\sigma$-contours from the 2D Gaussian fit 
are also shown. \label{fig:imegae}}
\end{figure}

\begin{figure}
\centering
\includegraphics[width=0.49\textwidth]{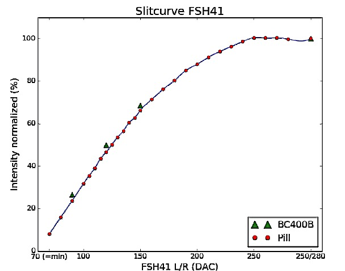}
\caption{Slit curve comparison measured with the scintillation
target (triangles) and 2D pill scintillator scanner
system (circles), showing an intensity variation of
the muon beam of a factor of $\sim 20$
. \label{fig:slitcurve}}
\end{figure}

The bare setup including CCD camera, lens and thin
Mylar mirror system placed \SI{\sim 2.1}{\metre} away from the
target, on the downstream-side (DS) of the COBRA
magnet is shown in Fig.~\ref{fig:PVTarget} (right).
Analysed background subtracted, perspective
corrected and 2D Gaussian fitted beam images (see Fig.~\ref{fig:imegae}) 
show that even with a non-ideal CCD camera (no cooling),
and exposures of \SI{100}{\second} in a strong gradient
magnetic field of several $\sim \si{\tesla}$, comparable results, at the
sub-millimetre level, to the usual 2D APD \lq\lq{}raster scans\rq\rq{}
performed at the centre of COBRA, can be obtained, in
a fraction of the time. Furthermore, it was demonstrated
that the beam intensity could be measured over the
range of a factor of 50 and reproduce results measured
independently with the \lq\lq{}pill scintillator\rq\rq{} scanner system
as shown in Fig.~\ref{fig:slitcurve}. The measurements were made
by adjusting the opening of the FS41L/R momentum
slits of the channel, so changing the intensity. Good
agreement is seen.

\begin{figure}
\centering
\includegraphics[width=0.49\textwidth]{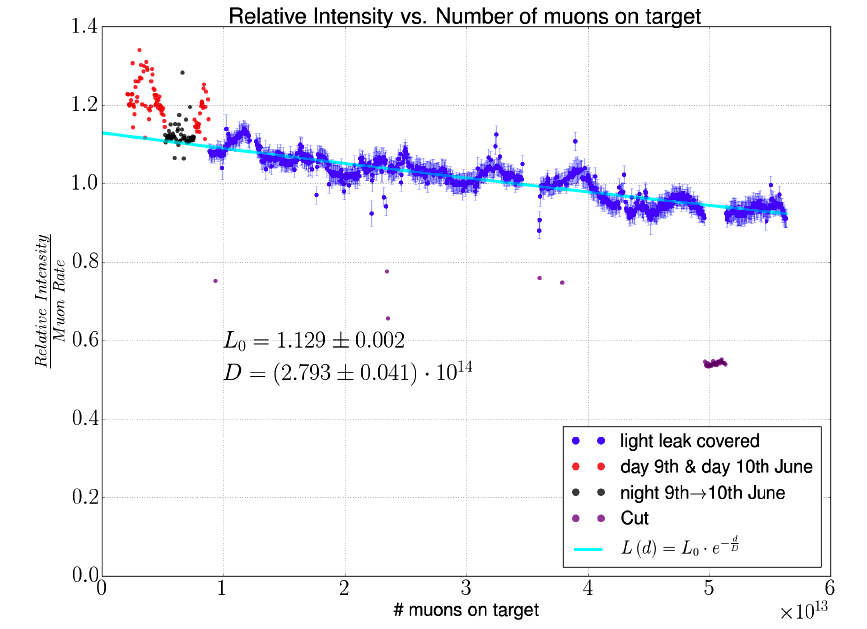}
\caption{Light yield curve for PVT exposed to \SI{\sim 30}{\kilo\gray} (\SI{3}{\mega\radian}) 
of integrated dose from the muon beam. An exponential fit to the data 
is shown with the resulting decay constant $D$. \label{fig:RadiationDamag}}
\end{figure}

Finally, a first radiation damage study was also
undertaken during the 2016 run with about \SI{5.5e13}{\muonp}
integrated, corresponding to an integrated dose of \SI{\sim 30}{\kilo\gray} (\SI{3}{\mega\radian}). 
A loss in light yield was seen, though less
than expected \cite{Protopopov1995}, which may be understood by the way
in which the scintillation light is collected namely,
through the very thin scintillator thickness thereby being
less sensitive to attenuation. A fit to the data with an exponential
decay law gives a decay constant of $D=\SI[separate-uncertainty]{2.793(41)e14}{\muonp}$
as shown in Fig.~\ref{fig:RadiationDamag}. Extrapolating
this to the longest MEG beam run of 2012 at the
MEG~II beam intensity as measured above, 
would lead to a light yield of \num{\sim 14}\%\ at the end of a 1-year period however,
still yielding measurable profiles and intensities as
demonstrated above. Normalising UV-LED measurements
would be however required for a corrected intensity measurement.
Furthermore, this would necessitate a new target for
each year. Further radiation tests are envisaged to
study the effect on the mechanical properties such as planarity, 
before a final decision on the target material is taken. A new CCD camera
system for imaging the beam on target has now been procured, including 
cooling and a mechanical shutter which should significantly improve the image 
quality and the analysis procedure.

\newpage

\subsubsection{Target alignment}

An important consideration for the target implementation is the accurate 
knowledge of the target position, in particular the knowledge of the target planarity 
and its perpendicular distance from its nominal position. 
Errors in this coordinate introduce a systematic error in the positron direction at the 
target due to the error in the path length of the curved positron trajectory projected on to 
the target plane. An offset of \SI{1}{\mm} in the target plane introduces a systematic error 
in the positron $\phi$-angle of \SIrange{7}{12}{\milli\radian}, comparable to the $\phi$ angular resolution 
achieved by MEG \cite{meg2016}. In MEG, this position was monitored by imaging 
small holes in the target foil. 
This monitoring was statistics limited in its ability to monitor deformation of the 
target foil during the run; lack of precise target position and shape information 
introduced a significant contribution to the systematic uncertainties in the positron angle 
measurement. With the anticipated improved angular resolution in MEG~II, improved 
monitoring of the target position and shape is required, with a goal of monitoring the target 
planarity and transverse position to a precision \SI{< 50}{\um} and the axial position 
to precision \SI{< 100}{\um}. 

\begin{figure}
\centering
\includegraphics[width=0.49\textwidth]{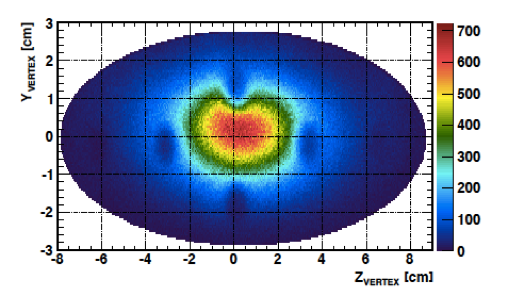}
\caption{Example reconstructed vertex positions on
the target plane for the MEG 2011 data. \label{fig:2011vertex}}
\end{figure}

It is envisaged, as in MEG, to implement both an optical survey for the determination 
of the target position, orientation, and shape 
and the software alignment method introduced above. 
The perpendicular distance of the target plane from the origin is 
determined by imaging the $y$-positions of a number of holes; there is a deficit of 
trajectories originating from the position of the holes. Any error in the perpendicular 
distance of the target from its nominal position results in the hole images varying 
in a systematic way depending on the value of $\phie$ 
(see \cite{megdet} for a full description of this technique). 
An example of a reconstructed vertex plot of the target is shown in Fig.~\ref{fig:2011vertex}
corresponding to the 2011 run data. As in MEG, this technique will be statistics limited and not allow continuous monitoring of the target position and planarity.

\begin{figure*}
\centering
\includegraphics[width=1.0\textwidth]{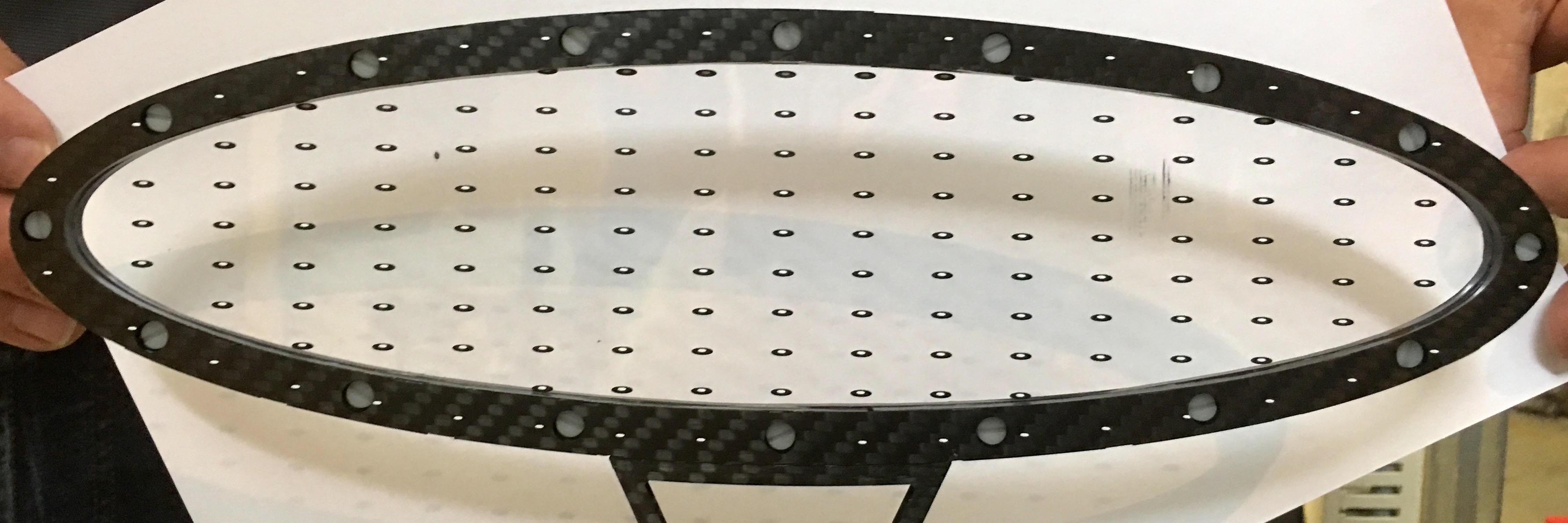}
\caption{The optical markings on the scintillator target used to test the
photogrammetric monitoring principle.
\label{fig:ScintTarget}}
\end{figure*}

A number of further improvements to the target and its optical imagery are planned and under study:
\begin{itemize}
\item a distortion-free/distortion minimising target suspension system allowing minimal 
impact of the target frame on the target foil;
\item further investigations to understand the origin of the previous MEG target distortion
(e.g. radiation damage, brittleness due to dry He-environment);
\item measurement of the target planarity both before and after exposure using a coordinate 
measuring machine with a precision better than \SI{50}{\um};
\item determination of the target frame position in the experiment to a precision of 
\SI{\sim 15}{\um} using a laser survey technique with low-mass corner-cube reflectors mounted 
on the target frame;
\item photogrammetric monitoring of target position, orientation and shape. A series of
printed patterns (dots) are optically monitored by CCD cameras viewing the target close
to axially. Preliminary studies show a precision of \SI{\sim 10}{\um} in the transverse
coordinate ($x$-$y$) and \SI{\sim 100}{\um} in the axial coordinate can be achieved. 
The current scintillator target with its printed pattern is shown in Fig.~\ref{fig:ScintTarget}.
\end{itemize}

\clearpage

\clearpage
\newpage
\section{Cylindrical drift chamber}
\label{sec:Cylindrical_Chamber}

\subsection{Cylindrical drift chamber overview}

The MEG~II Cylindrical Drift Chamber (CDCH) is a single volume detector, whose 
design was optimized to satisfy the fundamental requirements of high
transparency and low multiple scattering contribution for \SI{50}{\MeV} positrons, 
sustainable occupancy (at \SI{\sim 7e7}{\muonp\per\second} stopped on target) and fast 
electronics for cluster timing capabilities~\cite{Cascella:2014kca}. 
Despite the fact that in MEG~II the acceptance of the apparatus is dictated by 
the C-shaped LXe photon detector (see Sect.~\ref{sec:LXe_Calorimeter}), 
CDCH has full coverage ($2\pi$ in $\phi$),
to avoid non-homogeneous and asymmetric electric fields.

\begin{figure*}
	\centering
		\includegraphics[width=0.8\linewidth]{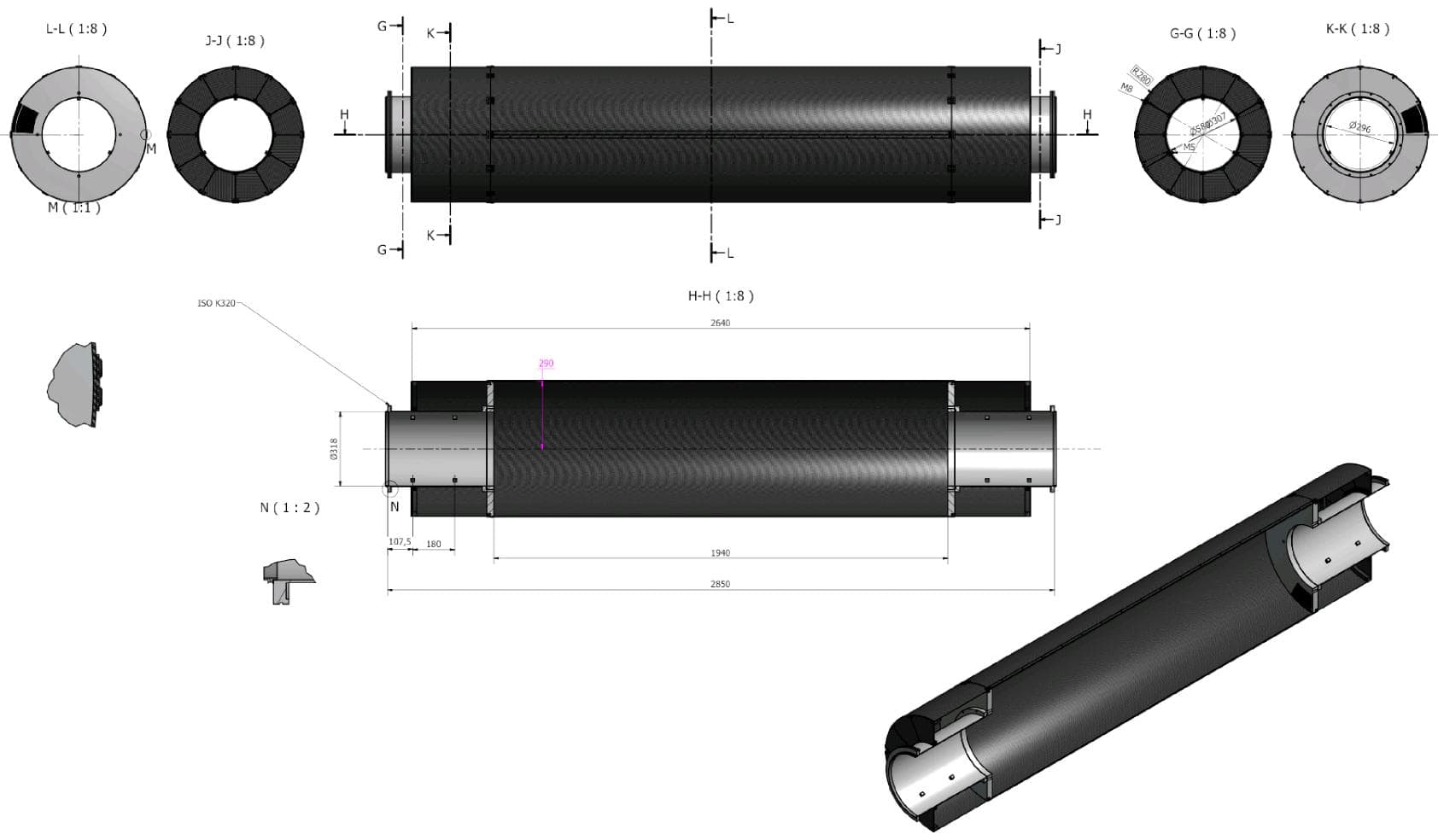}
		\caption{Cylindrical drift chamber structure.} \label{DC_mechanical_structure}
\end{figure*}

\begin{figure}[bt]
	\centering
	\includegraphics[width=0.50\textwidth]{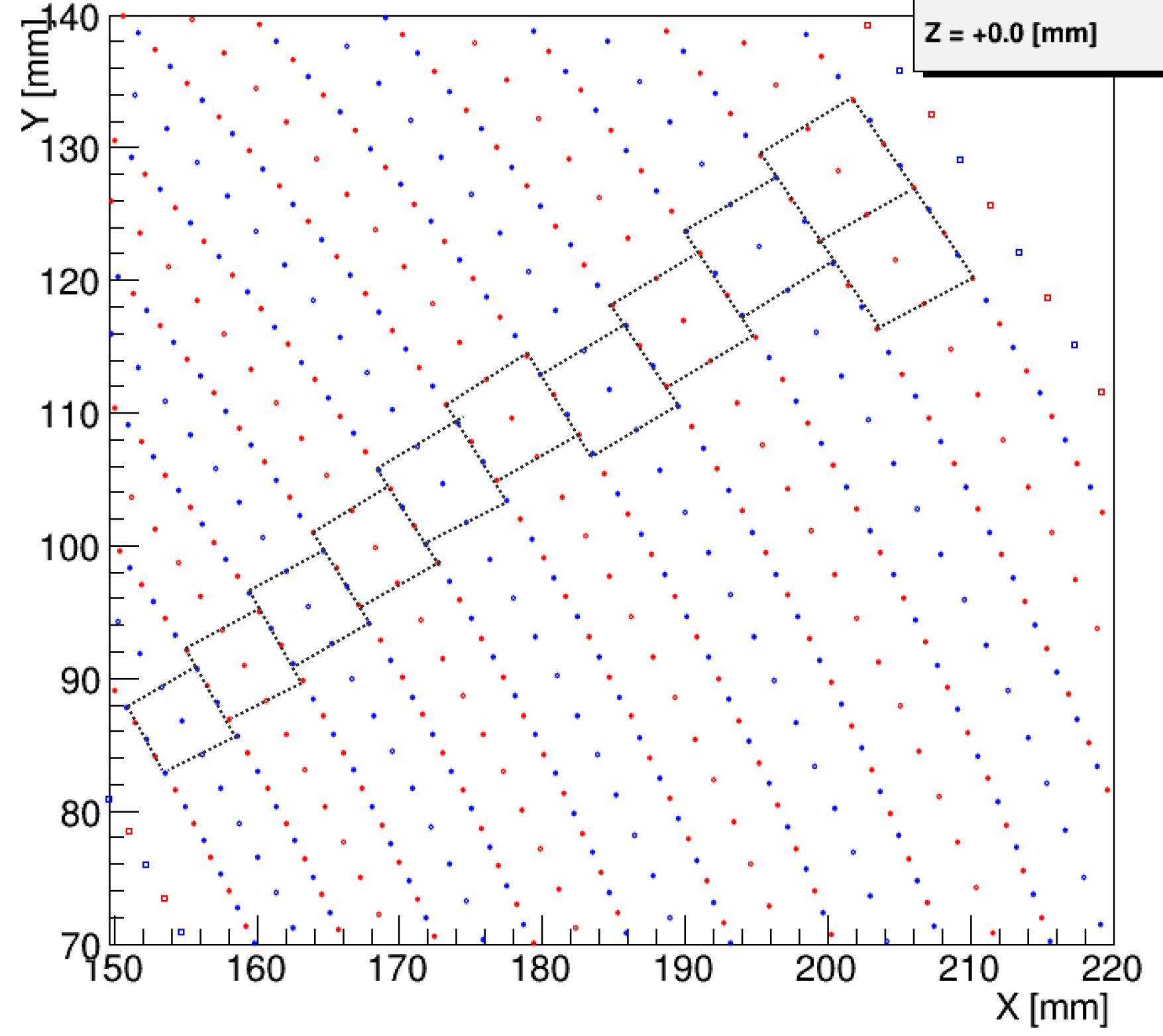}
	\caption{Drift cells configuration at the centre of CDCH.} \label{fig:cellsize}
\end{figure}

The mechanical structure, shown in Fig.~\ref{DC_mechanical_structure}, 
consists of a \SI{1.91}{\metre} long cylinder, inner radius of \SI{17}{\cm} 
and outer radius of \SI{29}{\cm}. It is composed of 10 concentric layers 
(see Fig.~\ref{fig:cellsize}), azimuthally divided in 12 identical \ang{30} 
sectors per layer, 16 drift cells wide. 
Each drift cell layer consists of two criss-crossing field wires planes enclosing a sense wires
plane at alternating signs stereo angles (approximately ranging from \ang{6.0} to \ang{8.5}
while radius increases) with respect to contiguous layers for a precise reconstruction 
of the $z$-longitudinal coordinate.

The double readout of the wires with the techniques of charge division 
and of time propagation difference, together with the ability to implement 
the cluster counting-timing technique \cite{Cascella:2014kca}, will further 
improve the longitudinal coordinate measurement. 

The stereo configuration of wires gives a hyperbolic profile to the active volume along the $z$-axis. 
The single drift cell (see Fig.~\ref{fig:cellsize}) is approximately square, 
\SI{6.6}{\mm} (in the innermost layer) to \SI{9.0}{\mm} (in the outermost one) wide, 
with a \SI{20}{\um} diameter gold plated W sense wire 
surrounded by \SI{40}{\um} diameter silver plated Al field wires in a ratio of 5:1. 
For equalising the gains of the innermost and outermost layers, two guard wires layers (\SI{50}{\um} 
silver-plated Al) have been added at proper radii and at appropriate  voltages. 
The total number of wires amounts to 13056 for an equivalent radiation length per track turn of about 
\num{1.58e-3}~X$_{0}$ when the chamber is filled with an ultra-low mass gas 
mixture of helium and isobutane (C$_4$H$_{10}$)
in the ratio 90:10 (compared with \num{2.0e-3}~X$_{0}$ in the MEG DCH \cite{megdet}). 
The drift chamber is built by overlapping along the radius, alternatively, 
PC Boards (PCB), to which the ends of the wires are soldered, and PEEK\textsuperscript{\textregistered} 
\footnote{PolyEther Ether Ketone, a colourless organic thermoplastic polymer.}
spacers, to set the proper cell width, in each of the twelve sectors, between the spokes of the helm shaped end-plate (see Fig.~\ref{fig:rocker}). A carbon fibre support structure guarantees the proper wire tension and encloses the gas volume. At the innermost radius, an Al Mylar foil separates the drift chamber gas volume from the helium filled target region.

Prototypes have been built~\cite{Baldini:2016rrk} to demonstrate that the design single 
hit resolution of the chamber ($\sigma_r\simeq$\SI{110}{\um}) can be reached 
and the detector can be operated in the high particle flux environment of
MEG-II without a significant ageing, as detailed in Sect.~\ref{sec:performance}.

\subsection{The choice of the filling gas}

CDCH uses a helium based gas mixture. The choice of helium is very advantageous, 
because of its large radiation length (X$_0 \SI{\sim 5300}{\metre}$ at STP), 
which ensures a small contribution in terms of multiple Coulomb scattering, 
a very important feature in low momentum measurements. 

A small amount ($10\%$) of isobutane is required as a quencher to avoid self-sustained discharge. 
Such a percentage is sufficient as it raises the number of primary ionisation pairs 
to \SI{\sim 13}{\per\cm}~\cite{kloe_155} though lowers the mixture radiation length to X$_0 \SI{\sim 1300}{\metre}$.
Unfortunately, the use of an organic quencher also results in additional 
problems after exposure to high radiation fluxes. The recombination of 
dissociated organic molecules results in the formation of solid or liquid 
polymers which accumulate on the anodes and cathodes, contributing to the ageing of the chamber.

The fairly constant drift velocity in helium based gas mixtures assures a linear time-distance 
relation, up to very close distance to the sense wire. On the other hand, the high helium 
ionisation potential of \SI{24.6}{\eV} is such that a crossing particle produces only
a small number of primary electron-ion pairs in helium based gas mixture.
In combination with the small size of the drift cells, it enhances the
contribution to the spatial resolutions coming from the statistical fluctuation of the 
primary ionisation along the track, if only the first arriving electrons are timed. 
An improvement can be obtained using the cluster timing technique, i.e. by 
timing all arriving ionisation clusters and so reconstructing their distribution along 
the ionisation track~\cite{Cascella:2014kca}.

\subsection{Electronics}
\label{sec:CDC_Electronics}
In order to permit the detection of single ionisation clusters, the electronic read-out interface has to process high speed
signals. For this purpose, a specific high performance {8-channels} front-end electronics (FE) has been designed with commercial 
devices such as fast operational amplifiers. This FE was designed for a gain which must produce a suitable read-out signal for further processing, 
low power consumption, a bandwidth adequate to the expected signal spectral density and a fast pulse rise time response, 
to exploit the cluster timing technique \cite{Chiarello:2016nbj,Chiarello:2015ypa}. 

\begin{figure}[!t]
	\centering{\includegraphics[width=1\linewidth]{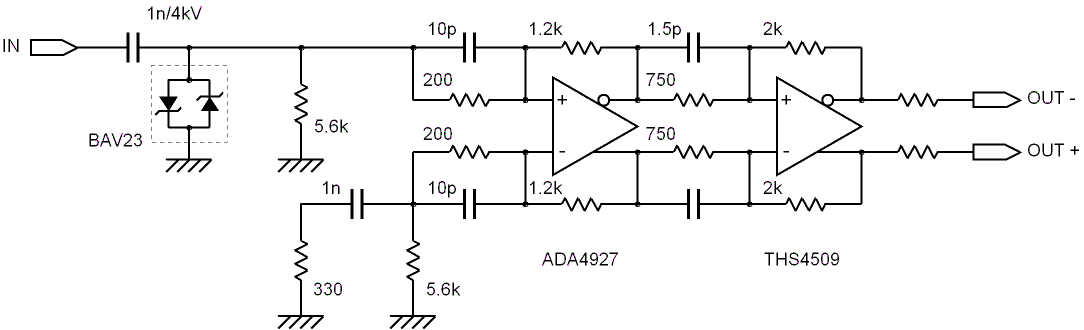}}
	\caption{Front-end single channel schematic.} \label{fe_schem}
\end{figure}
The FE single channel schematic is represented in Fig.~\ref{fe_schem}. The input network 
provides decoupling and protection, while signal amplification is realized with a double gain 
stage made from ADA4927 and THS4509. Analog Device's op-amp ADA4927 \cite{ADA4927} works as 
a first gain stage: it is a low noise, ultra-low distortion, high speed, current feedback 
differential amplifier. The current feedback architecture 
provides a loop gain that is nearly independent of the closed-loop gain, achieving wide bandwidth, 
low distortion, low noise (input voltage noise of only \SI{1.3}{\nano\volt}/$\sqrt{\si{\hertz}}$ 
at high gains) and lower power consumption than comparable voltage feedback amplifiers. 
The THS4509 \cite{THS4509} by Texas Instruments is used as a second gain stage and output driver. 
It is a wide-band, fully differential operational amplifier with a very low noise 
(\SI{1.9}{\nano\volt}/$\sqrt{\si{\hertz}}$), and extremely low harmonic distortion of \SI{-75}{\decibel}c $\mathrm{HD}_2$ 
and \SI{-80}{\decibel}c $\mathrm{HD}_3$ at \SI{100}{\MHz}. The slew-rate is \SI{6600}{\volt\per\micro\second} with a settling time of \SI{2}{\nano\second} to 1\% for a \SI{2}{\volt} step; it is ideal 
for pulsed applications. The output of the FE is differential, 
in order to improve the noise immunity and it is connected to the WaveDREAM Board \cite{Francesconi2014} through a custom cable \SI{5}{\metre} long, 
designed to have a stable, flat frequency response (Amphenol Spectra Strip SkewClear \cite{miniSaS}). This cable is made from shielded 
parallel pairs, each pair being individually shielded; an overall ground jacket is also present, giving a maximum attenuation of \SI{0.75}{\decibel\per\metre} at \SI{625}{\MHz}. 

In order to balance the attenuation of the output cable, a pre-emphasis on both gain stages has been implemented. The pre-emphasis introduces a high 
frequency peak that compensates the output cable losses resulting in a total bandwidth of nearly \SI{1}{\giga\hertz}.

\begin{figure}[!t]
	\centering
		\includegraphics[width=1\linewidth]{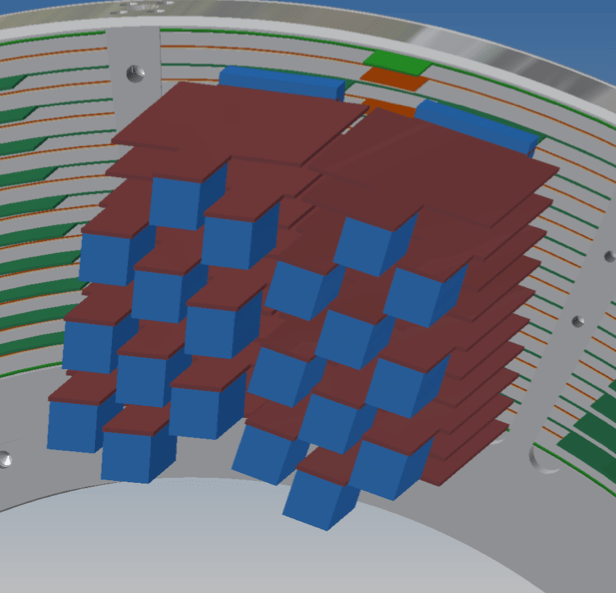}
		\caption{CDCH end-plate scheme.} \label{dc_endplate}
\end{figure}

The FE electronics boards are placed in each sector of CDCH; in Fig.~\ref{dc_endplate} 
the end-plate mechanical scheme, in which the boards will be inserted, is shown. 
Due to the area of the FE output connector socket and considering the available space between the layers, 
three different board versions have been designed, one with the output connector on the right, one in the centre and one on the left.

Pre-amplified differential signals are successively digitised by the WaveDREAM board at a (programmable) speed 
of 2 GSPS (Giga-samples per second) with an analogue bandwidth of \SI{1}{\giga\hertz} \cite{Francesconi2014}.

The current consumption for each channel is \SI{60}{\milli\ampere} at a voltage supply of \SI{\pm2.5}{\volt}; 
this correspond to a total power dissipation per end-plate of about \SI{300}{\watt}, therefore an appropriate 
cooling system relying both on recirculation of coolant fluid and on forced air is foreseen.   

\subsection{The wiring procedure}

\begin{figure*}[t]
	\centering
		\includegraphics[width=0.8\linewidth]{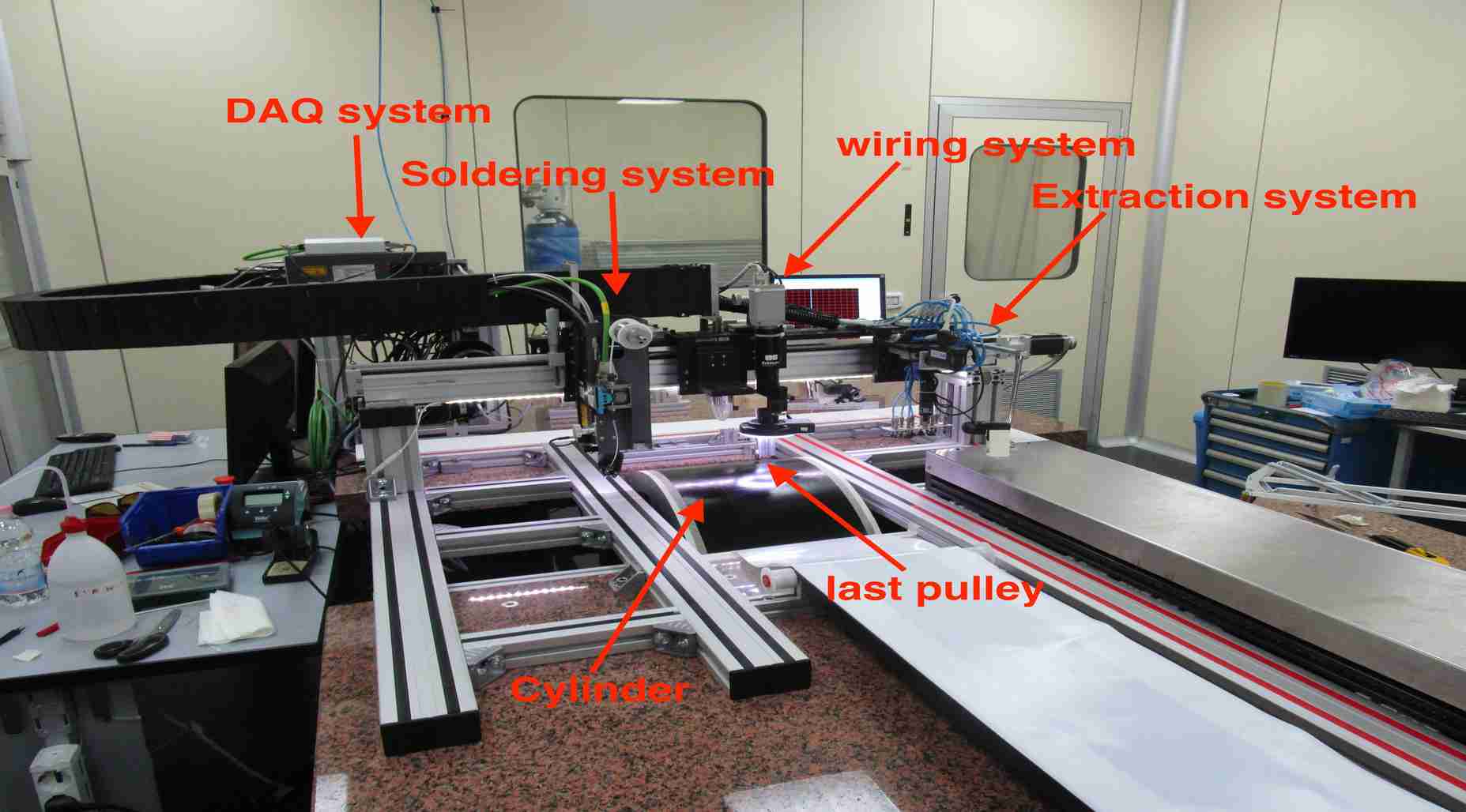}
		\caption{The wiring robot.} \label{wiring_robot}
\end{figure*}

A wiring system robot \cite{Chiarello:2016adb} has been designed and assembled 
in the clean room (see Fig.~\ref{wiring_robot}).
It allows to automatically stretch 
the wires on PCB frames, keeping under control the wire tension 
and pitch parameters; moreover the system fixes the wires on the PCB by a contact-less 
soldering. Since CDCH has a high wire density 
(\SI{12}{wires\per\cm\squared}), the classical feed-through technique, as a wire 
anchoring system, is hard to implement, therefore the development of a new 
wiring strategy was required.

The wiring robot has been designed with the following goals: 
\begin{itemize}
	\item[-] managing a very large number of densely spaced wires,
	\item[-] applying the wire mechanical tension and maintaining it constant and uniform throughout all the winding process,
	\item[-] monitoring the wire positions and their alignments within a few tens \si{\um},
	\item[-] fixing the wires on the PCB with a contact-less soldering system and
	\item[-] monitoring the solder quality of the wires to the supporting PCBs.
\end{itemize}
These requirements are satisfied by the following three systems:
\begin{enumerate}
	\item {\bf A wiring system} that uses a semi-automatic machine to 
	simultaneously stretch the multi-wire layer with a high degree of control on 
	the wire mechanical tension (better than \SI{0.2}{\g}) and on the wire position 
	(of the order of \SI{20}{\um}) . 
	\item {\bf A soldering system} composed of an infrared (IR) laser soldering 
	system and tin-feeder. 
	\item {\bf An automatic handling system} which extracts the multi-wire layers 
	from the wiring system and places them in a storage/transport frame.
\end{enumerate}
A dedicated LabView\textsuperscript{\textregistered} software \cite{Chiarello:2016adb}, based on a CompactRIO platform \cite{crio}, controls the three systems simultaneously, sequencing and synchronising all the different operations. 

\subsubsection{Wiring system}

\begin{figure}[bt]
	\centering
		\includegraphics[width=1\linewidth]{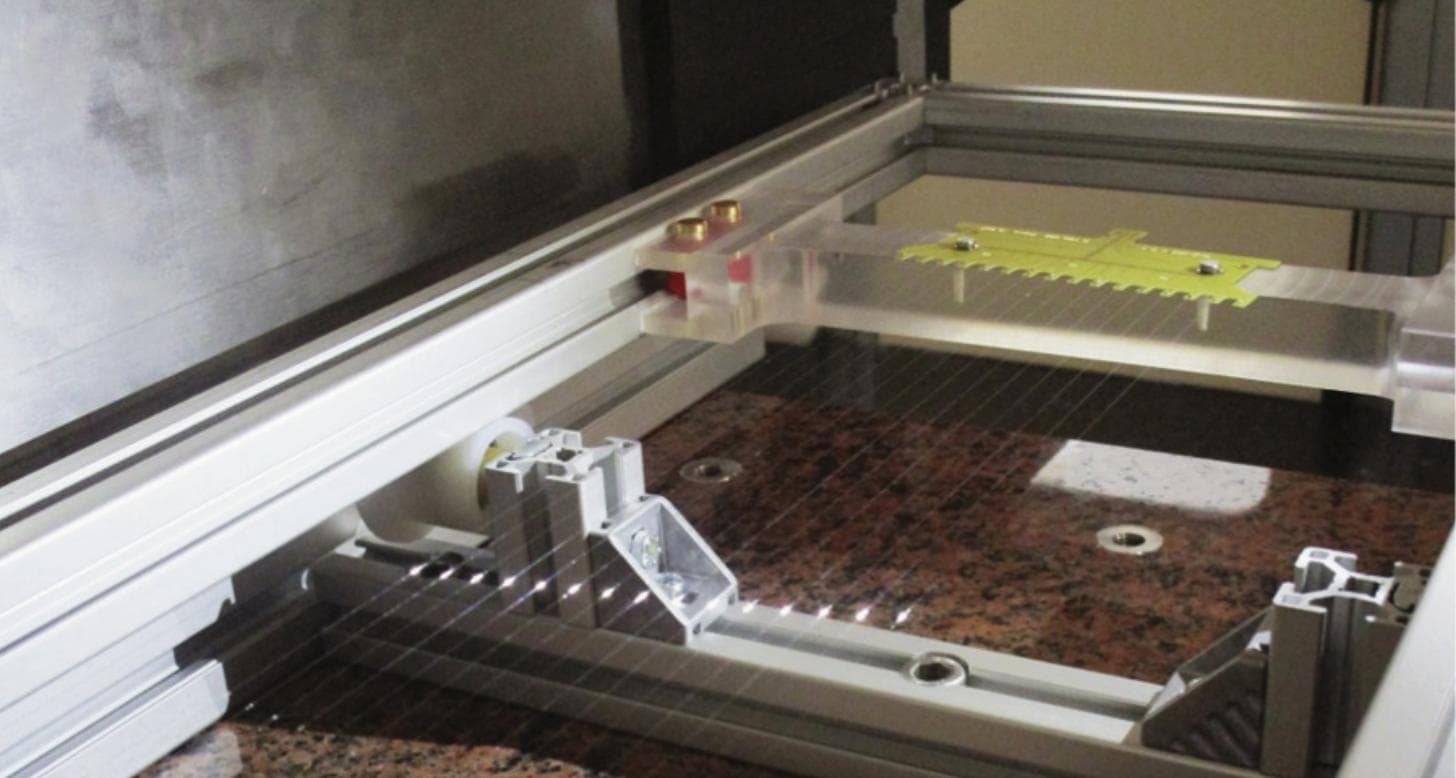}
		\caption{A multi-wire frame.} \label{frame}
\end{figure}

The purpose of the wiring system is the winding of a multi-wire layer consisting of 32 parallel 
wires at any stereo angle. In order to achieve a multi-wire layer (see Fig.~\ref{frame}), two PCBs, 
aligned and oriented at the proper stereo angle, are placed back-to-back on the winding cylinder. 
The multi-wire layer is obtained in a single operation by winding along a helical path the same 
wire 32 times around the cylinder with a pitch corresponding to the wire PCBs spacing. 
The correct pitch is achieved by a system of synchronised stepping motors, through the CompactRIO 
system and controlled by a digital camera with position accuracy of the order of \SI{20}{\um}. 
The wire mechanical tension is monitored by a high precision strain gauge and corrected with a 
real-time feedback system acting on the wire spool electromagnetic brake.

\begin{figure}[bt]
	\centering
		\includegraphics[width=1\linewidth]{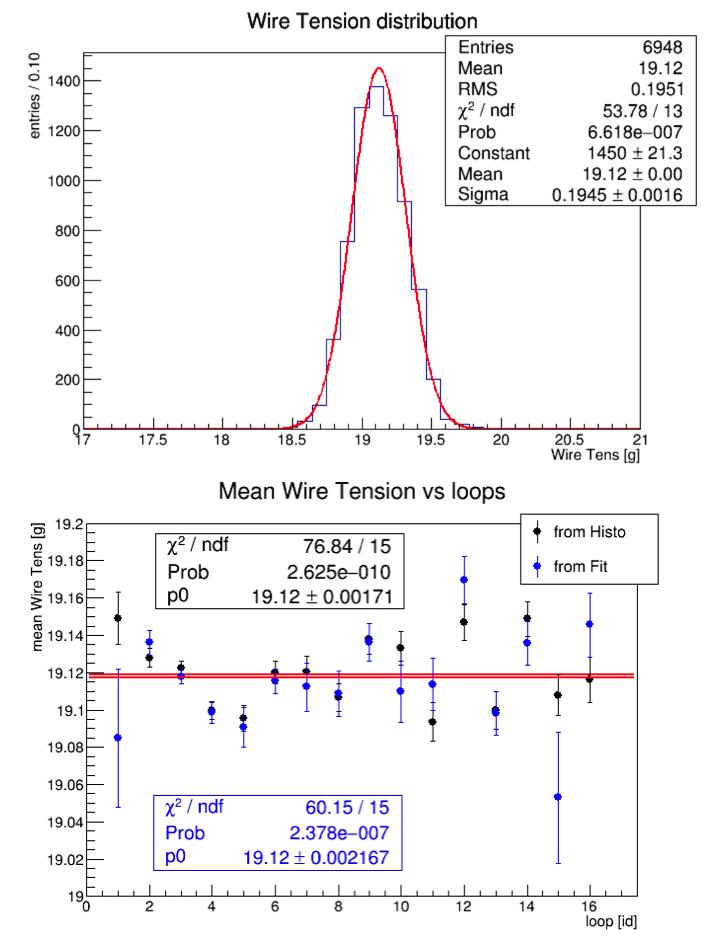}
		\caption{Top: the distribution of the wire tension during the winding. Bottom: average wire tension for each loop.} \label{wire_tension}
\end{figure}

The wire tension variations are of the order of \SI{\pm 1.5}{\gram}, without the feedback system, because of the mechanical tolerances. The feedback system reduces these variations to about \SI{\pm0.2}{\gram} (see Fig.~\ref{wire_tension}).

\subsubsection{Soldering system}

The soldering phase is accomplished by an IR laser soldering system (LASCON Hybrid with 
a solder wire feeder \cite{laser}). Each wire is fixed at both ends while still constrained 
around the winding cylinder under its own tension. The laser system is controlled by the 
CompactRIO and it is synchronised with the positioning system by using a pattern matching 
software to localise the soldering pad. All the soldering parameters (temperature, soldering time, 
solder wire length and feeding speed) are defined through a proper script. 

\subsubsection{Automatic handling system}

The wound layer of soldered wires around the cylinder is unrolled and detensioned for 
storage and transport. This is accomplished with an automatic device. The first wire PCB 
is lifted off from the cylinder surface with a linear actuator connected to a set of 
vacuum operated suction cups and placed on the storage and transport frame. 
The unrolling is accomplished by synchronising the cylinder rotation with the linear 
displacement of the frame. Once the layer of soldered wires is completely unrolled, 
the second wire PCB is lifted off from the cylinder, as the first one, and placed on the frame.
The frame hosts two supports made of polycarbonate, dedicated to holding the wire PCBs 
at the correct position by means of nylon screws.
One of the two supports can slide into the frame by adjusting the wire length, with a longitudinal threaded rod.
The wiring information relative to each frame is stored in a database. 
Then the wires on the frame are examined, stored and prepared for transportation 
to the CDCH assembly station.

\subsection{The assembling procedure}

The assembly of the drift chamber is as critical as the wiring phase and has to be performed under very carefully controlled 
conditions \cite{Venturini_thesis}. In fact, to reach the required accuracy on the drift chamber geometry and to avoid 
over-tensioning of wires, it is necessary to measure the position of the end-plates to better than
\SI{100}{\um}. For example, an error of \ang{1} on the twist angle can correspond to an extra elongation
of the wire of about \SI{1}{\mm}. It is therefore very important to have accurate position measurements 
over the chamber length of \SI{\sim 2}{\m}.		
For this reason, the assembly is performed by using a coordinate measuring machine; the machine, a DEA Ghibli \cite{DEA_Ghibli}, has 
a maximum machine travel distance of \SI{2500 x 1500 x 1000}{\mm} and a nominal accuracy of \SI{5}{\um}
with a contact measuring tool. The measurements of the positions of the PCBs are performed using 
an optical tool for the identification of the cross marks placed on the PCBs. 
The accuracy of the optical measurement is \SI{\sim 20}{\um}
in the horizontal plane and (making use of the focal distance of the optics) \SI{\sim 40}{\um} on the vertical axis.

The first test on the wire trays is a quick measurement of the elongation-tension curve in the 
proximity of the working point.
In this test the wire elongation is measured with the optical tool of the measuring 
machine and the wire tension is measured both by acoustic and electrical methods. 
In the acoustic method a periodic signal at a frequency close to the wire resonance is measured 
in the readout circuit by applying a HV difference between two adjacent wires and by using 
an acoustic source to excite the wires' oscillation. This system has the ability of measuring 
simultaneously up to 16 wires. In the electrical method the wire oscillation is forced by 
applying a HV signal at a known frequency. The mutual capacitance variation between two adjacent 
wires is then measured during a HV frequency scan on an external auto oscillating circuit connected to the wires.

\begin{figure}[!t]
	\centering
	\includegraphics[width=1\linewidth]{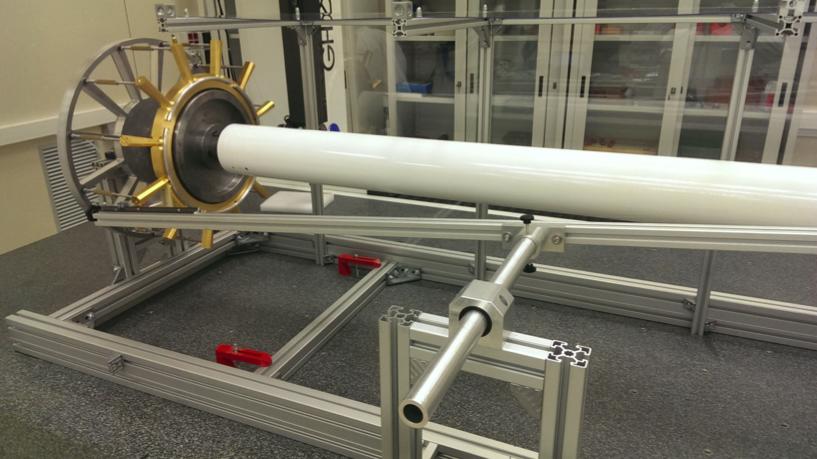}\quad\quad
	
	\includegraphics[width=1\linewidth]{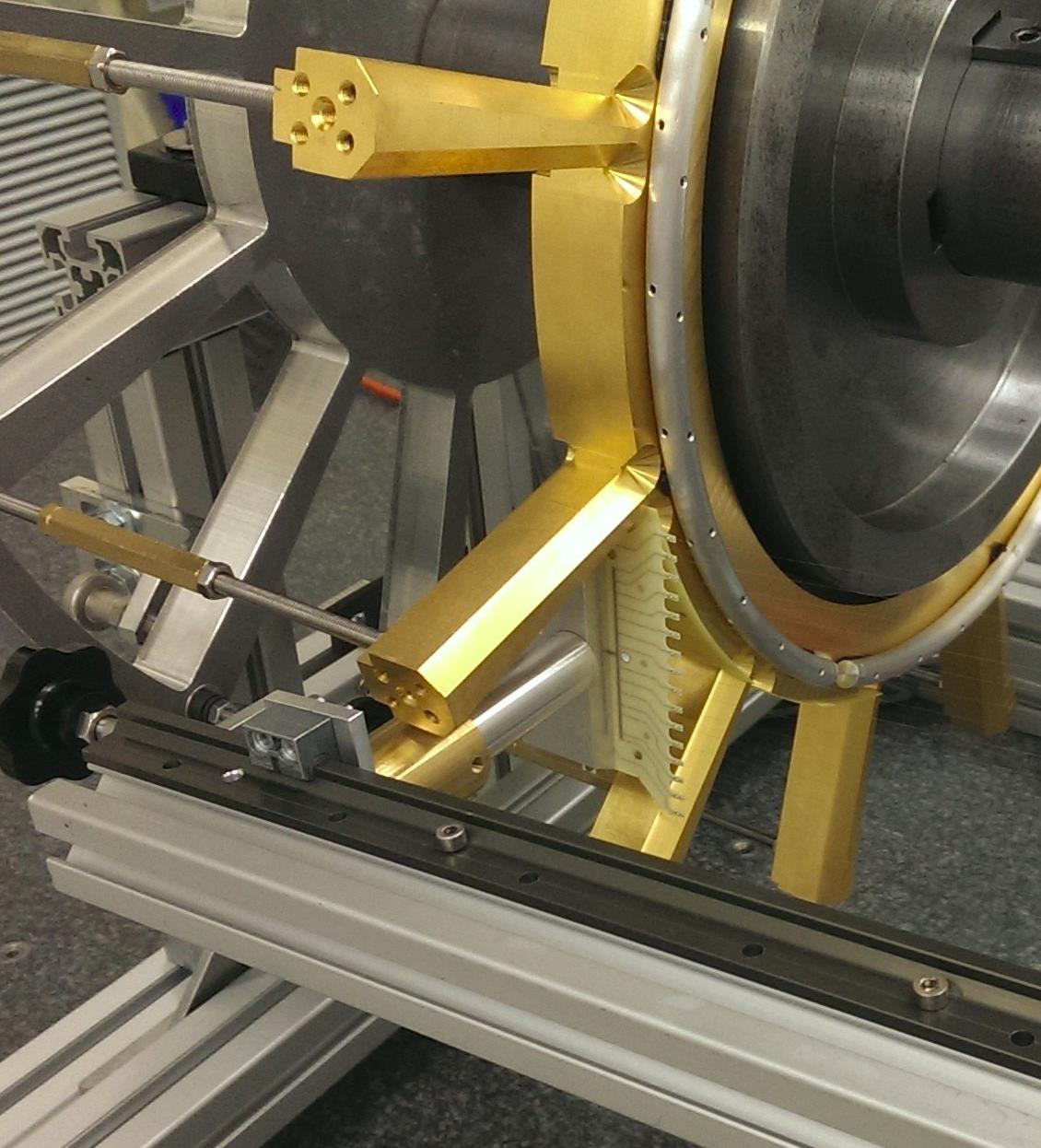}
	\caption{Pictures of the rocker arm during the tray-mounting procedure.}
	\label{fig:rocker}
\end{figure}
\begin{figure}[!t]
	\centering
	\includegraphics[width=1\linewidth]{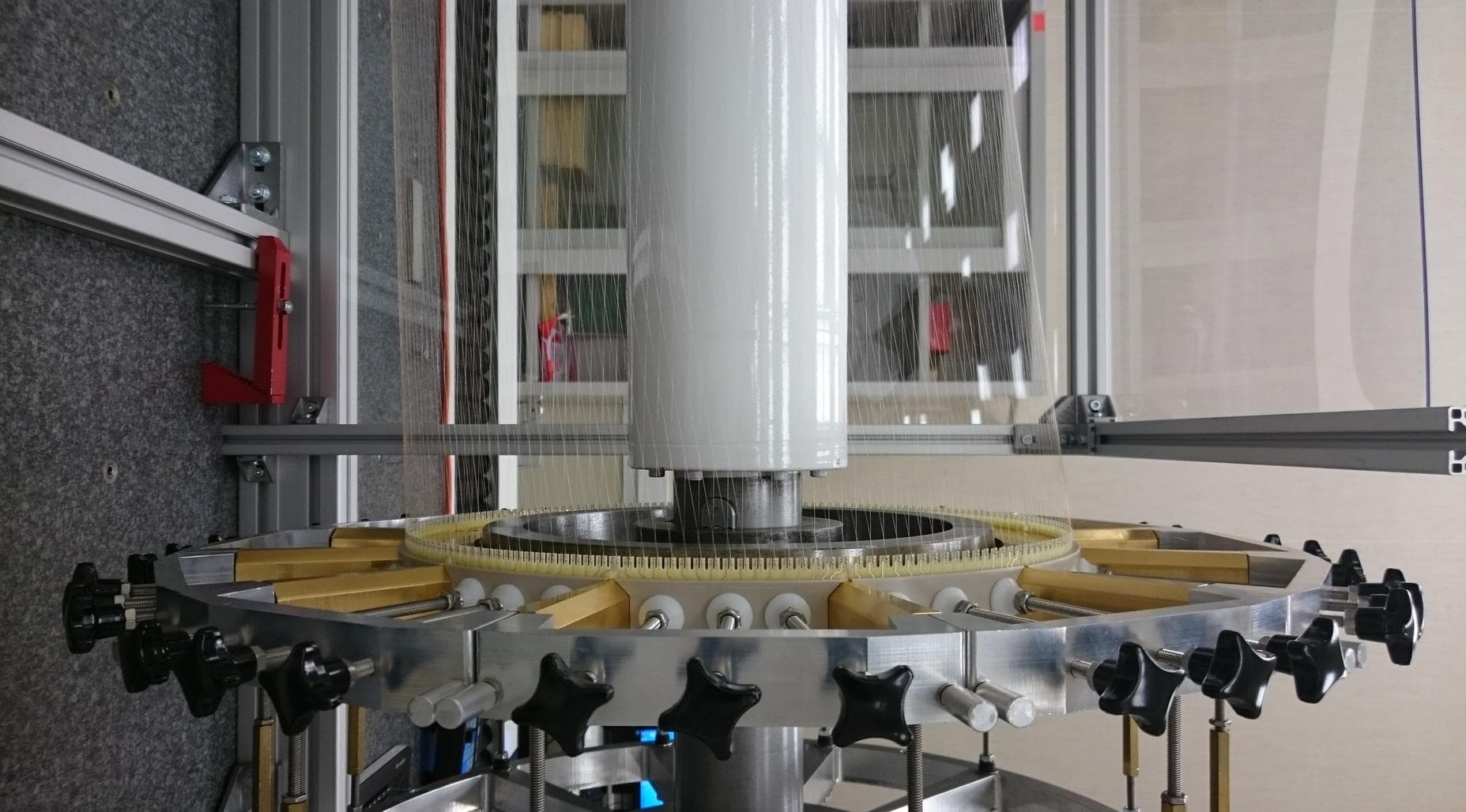}
	\caption{An end-plate after mounting about the 80\% of the wires.}
	\label{fig:AssembCdc}
\end{figure}
\begin{figure*}[t]
	\centering
	\includegraphics[width=0.8\linewidth]{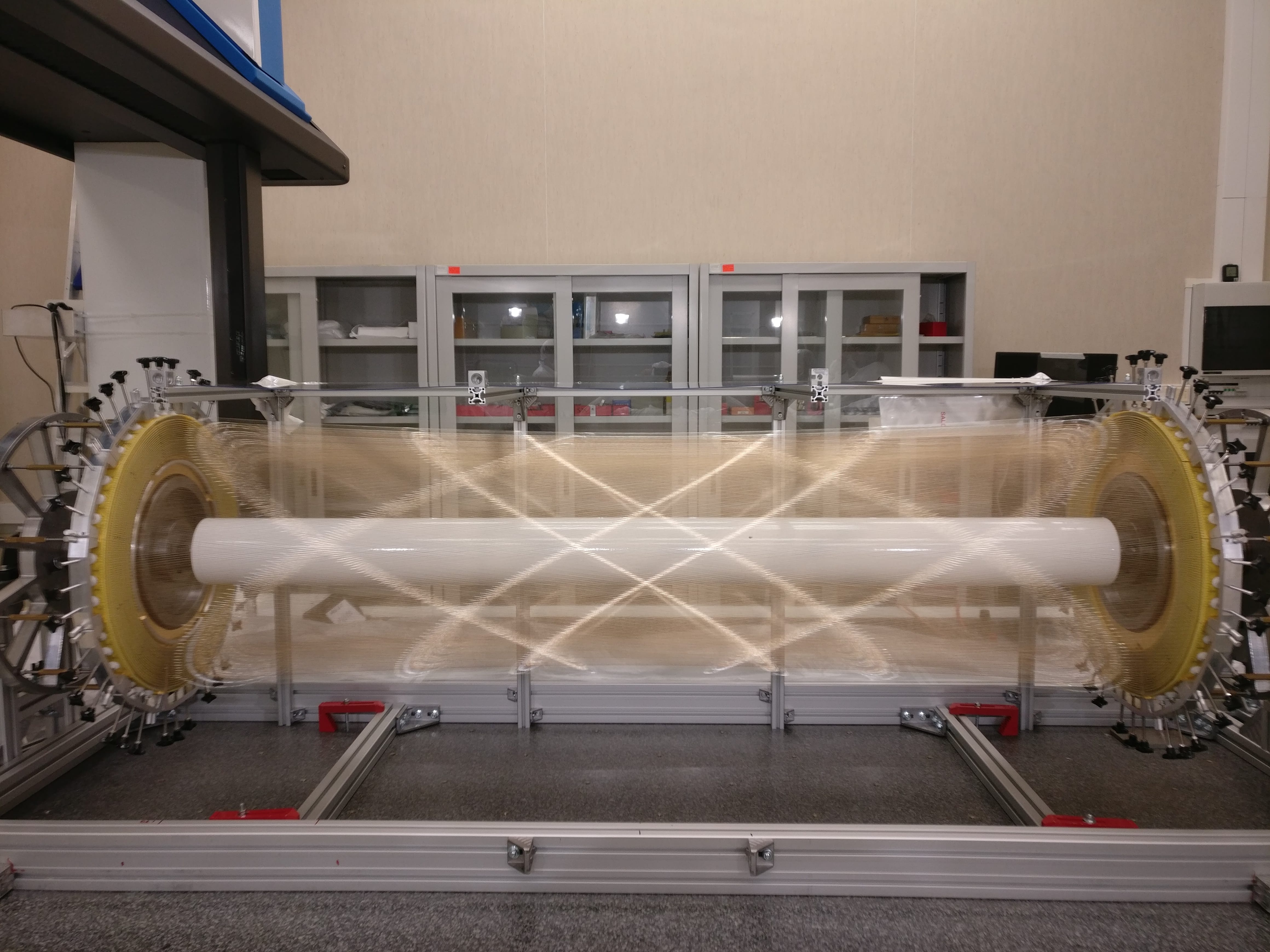}
	\caption{The entire drift chamber with all layers mounted. The hyperbolic profile of 
		the chamber is visible.}
	\label{fig:AssembCdcAll}
\end{figure*}

The drift chamber assembly is performed in safe conditions with unstretched wires: the 
distance between the end-plates is fixed at \SI{1906}{\mm}, \SI{6}{\mm} less than the nominal length (\SI{1912}{\mm}) and
\SI{2}{\mm} less than the untensioned wire length.
The positioning of the wire trays on the drift chamber is done in a well-constrained way using
a rocker arm, shown in Fig.~\ref{fig:rocker}. 

The wire tray is first engaged to the rocker arm by means of two precision pins fitting two PCB holes and a 
clip. The rocker arm is then engaged to a support that leaves it free to rotate and transfers the wire tray on the end-plates between two
spokes. The final positioning is driven by hand though dedicated nippers. The wire PCBs are glued on the PEEK spacers with double sided tape
previously applied on the inner layer. The PEEK spacers are needed to separate the layers at the right distance. Two pressing arches are used for ensuring a good adhesion of the tape.

In Fig.~\ref{fig:AssembCdc} and Fig.~\ref{fig:AssembCdcAll} we show two pictures of the drift chamber
after the completion of the two internal guard wire layers. In Fig.~\ref{fig:AssembCdc} the crossing 
of the layers in the two stereo views is visible, while Fig.~\ref{fig:AssembCdcAll} shows the hyperbolic 
profile of the drift chamber.

\subsection{Calibration and monitoring }

Michel events represent the natural way to continuously and fully characterise the spectrometer 
with dedicated pre-scaled triggers. The Michel positrons at the edge of the continuous energy spectrum 
are actually used to perform the alignment of the spectrometer, to define the energy scale of the 
detector and to extract all the positron kinematic variable resolutions (energy, time and angular variable resolutions).

\subsubsection{The Mott monochromatic positron beam }
\label{sec:Mott}

The continuous Michel positron spectrum makes the calibration difficult and subject to significant systematic errors, while delivering mono-energetic positrons would bring important advantages.

Positrons are an abundant component of the MEG/MEG~II beam (eight times 
more intense than the \muonp-surface component, but they are normally separated and rejected). 
Turning the muon beam into a positron beam line and tuning the positron 
momentum very close to the \megc\ signal energy ($\ppositron \SI{\sim 53}{\MeV/\clight}$), 
a quasi-monochromatic intense beam ($\sigma^\mathrm{beam}_{p_\mathrm{e^+}} 
\SI{\sim 250}{\keV/\clight}$, $I_\positron$ \SI{\sim e7}{\positron\per\second}) 
can be Mott scattered on the light nuclei present in the muon stopping target, 
providing a very useful \positron-line for a full understanding of the spectrometer 
from alignment to the positron kinematic variables' resolution.

The merits of the method, some of them unique, can be listed as 
\begin{itemize}
\item Spectrometer absolute energy scale determination.
\item Spectrometer alignment: the alignment is performed as an iterative procedure 
on the residuals of the expected and measured hits of the tracks. The alignment is 
executed with the detector under normal running conditions (i.e. with the magnetic 
field on) using curved tracks having monochromatic energy which simplify the procedure.
\item Spectrometer checks: the well known relative dependence of the Mott scattered 
positron-momentum on the angular variables $\phie$ and $\thetae$ makes possible
a detailed investigation of the spectrometer, any distortion would signal 
deviation from the expected detector behaviour.
\item Spectrometer acceptance: the well known Mott cross section permits the 
direct measurement of the spectrometer acceptance.
\item Independent check of the muon polarisation: the comparison of the Michel 
versus Mott $\thetae$-distribution, after taking into account the $\theta$ 
cross-section dependence of the Mott events, allows a cross-check of the muon 
polarisation at the Mott positron energy.
\item Positron momentum and angular resolutions: positron momentum and angular 
resolutions are extracted using double-turn track events. The double-turn 
track is divided in two independent tracks, the two tracks are propagated 
towards the target and the difference between the relevant observable 
(i.e. the $\ppositron$, $\phie$ or $\thetae$ variable) is computed.
\end{itemize}

As final remarks it should be noted that the high Mott positron rate enables for a fast calibration, 
the method does not require a dedicated target (i.e. the Mott target is the MEG~II muon stopping target) 
and does not need additional beam infrastructures.

\begin{figure}
\centering
\includegraphics[width=0.99\linewidth, clip, trim=0 0 0 3.5pc]{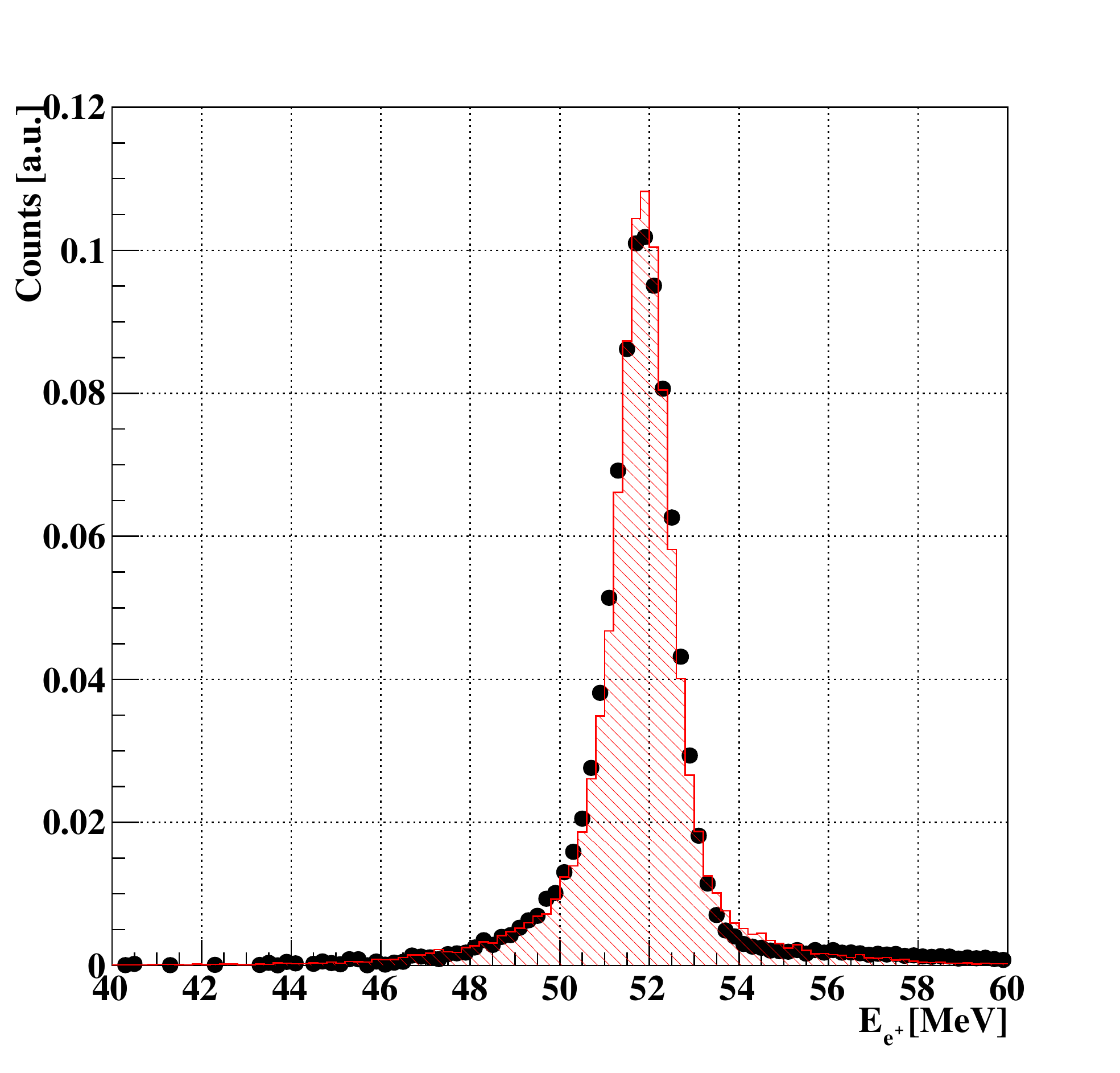}
\caption{The Mott scattered positron energy distribution in our spectrometer angular 
acceptance with a mean value at at $\hat{E}_{\positron} = \SI[separate-uncertainty]{51.840(3)}{\MeV}$. 
The comparison between data (black dot points) and MC simulation (red dashed area) is shown.}
\label{fig:Motteline_DatavsMC}
\end{figure}

The potential of this method has been proven using dedicated beam tests performed 
at the $\pi$E5 beam line (i.e. the MEG~II beam line) with the MEG spectrometer in 2012. 
Figure~\ref{fig:Motteline_DatavsMC} shows the good agreement between the Mott \positron-line 
(black dot points) and the Monte Carlo (MC) simulation prediction (red dashed area). 
The data are fitted with a double Gaussian function: one taking into account the core 
of the distribution and one the low energy tail. With the beam momentum slits virtually 
``fully closed" we get a line centred at $\hat{E}_\positron = \SI[separate-uncertainty]{51.840(3)}{\MeV}$ 
with a width $\sigma_{\epositron}^\mathrm{core} = \SI[separate-uncertainty]{412(10)}{\kilo\eV}$.

\begin{table*}
\caption{The Mott \positron energy spectrum: comparison between reprocessed data based on Michel vs. Mott alignment for 2013 data.}
\centering
\begin{tabular}{lccc}
\hline
Parameter & Michel Alignment & Mott Alignment& Difference (Mott--Michel)\\
\hline
Number of events & \num{80339} & \num{79059} & \num{-1.6}\% \\
$\hat{E}_\positron$ (\si{\MeV}) & $51.793 \pm 0.003$ & $51.762 \pm 0.003$ & \num{-0.031}\\
$\sigma_{\positron}^\mathrm{core}$(\si{\kilo\eV}) & $491\pm 2$ & $507\pm3$ & 16\\ 
\hline
\end{tabular}
\label{tab:MottvsMichelalignment}
\end{table*}%

The ability of performing the spectrometer alignment and obtaining consistent results 
can be seen in Table~\ref{tab:MottvsMichelalignment} which shows a reconstructed set 
of Mott data taken in 2013 based on the Michel alignment versus Mott alignment: 
both the mean energy and width are compared. The two data sets are in good agreement. 
The two different methods allow different systematic errors to be identified.

\begin{figure}
\centering
\includegraphics[width=0.99\linewidth, clip, trim=0 1pc 0 1pc]{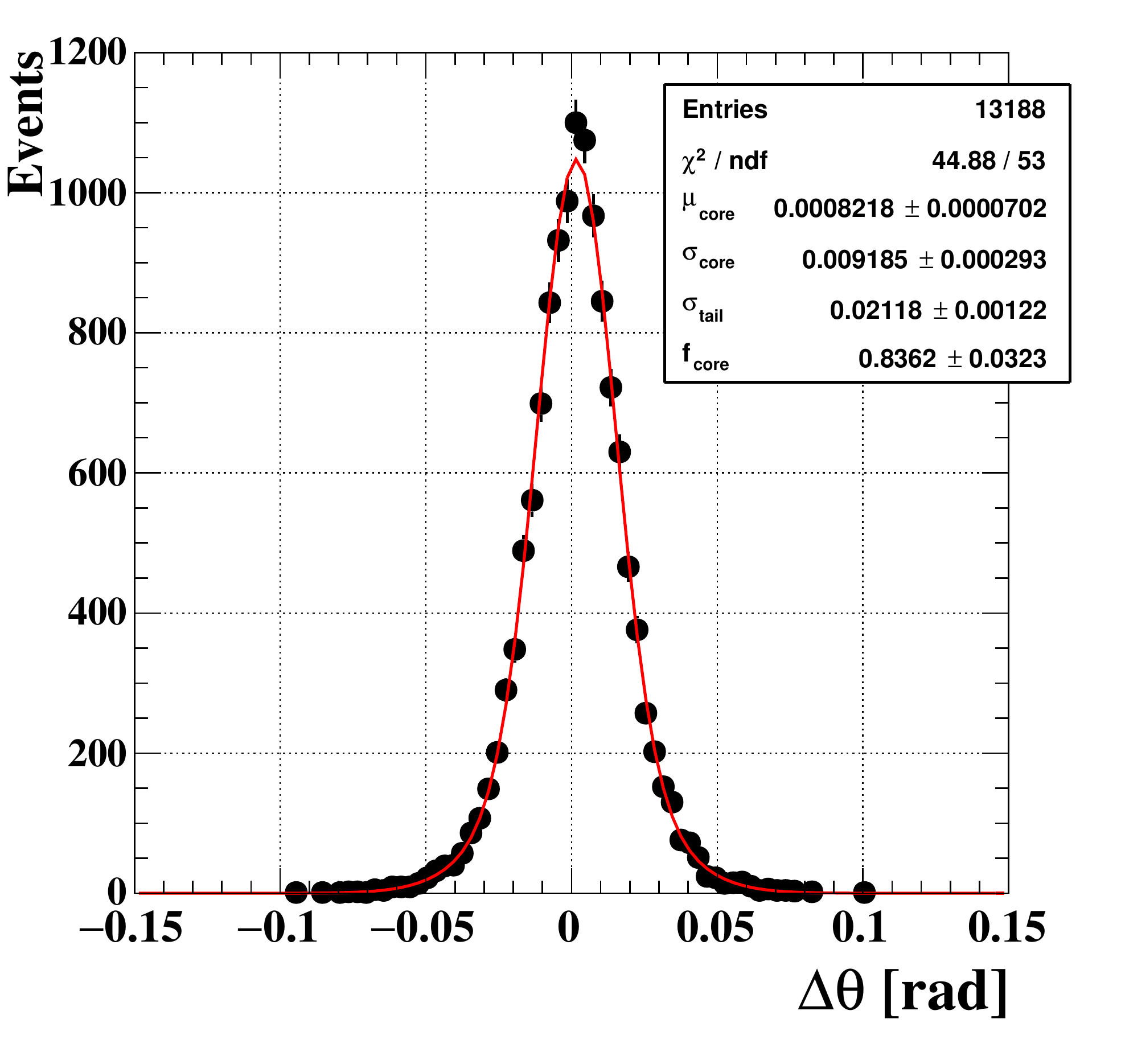}
\caption{The distribution of $\mathrm{\Delta} \theta = \theta_1 - \theta_2$ as obtained using 
the Mott data sample and the double-turn method, where $\theta_1$ and $\theta_2$ are 
the reconstructed $\theta$-angles associated with the first and second part a 
double-turn track, respectively. The distribution is fitted with a double Gaussian function.}
\label{fig:MottDoubleTurn_Theta}
\end{figure}

Similarly a comparison between the $\ppositron$ and angular variable resolutions 
extracted using the double-turn track method applied to the Mott sample and the Michel 
sample has also been performed. An example of the $\thetae$-angular distribution obtained 
using the Mott sample and applying the double-turn method is shown in Fig.~\ref{fig:MottDoubleTurn_Theta}. 
Actually the double turn resolutions on all positron variables measured with the Mott 
sample were found to be similar or even better (up to $20\%$) than that measured in the Michel data. 
The difference has been understood in terms of the different pile-up conditions in which 
the spectrometer works in the two cases. This is another example in which independent 
methods complement each other for a better understanding of the detector.

\begin{figure}
\centering
\includegraphics[width=0.99\linewidth, clip, trim=0 3pc 0 3pc]{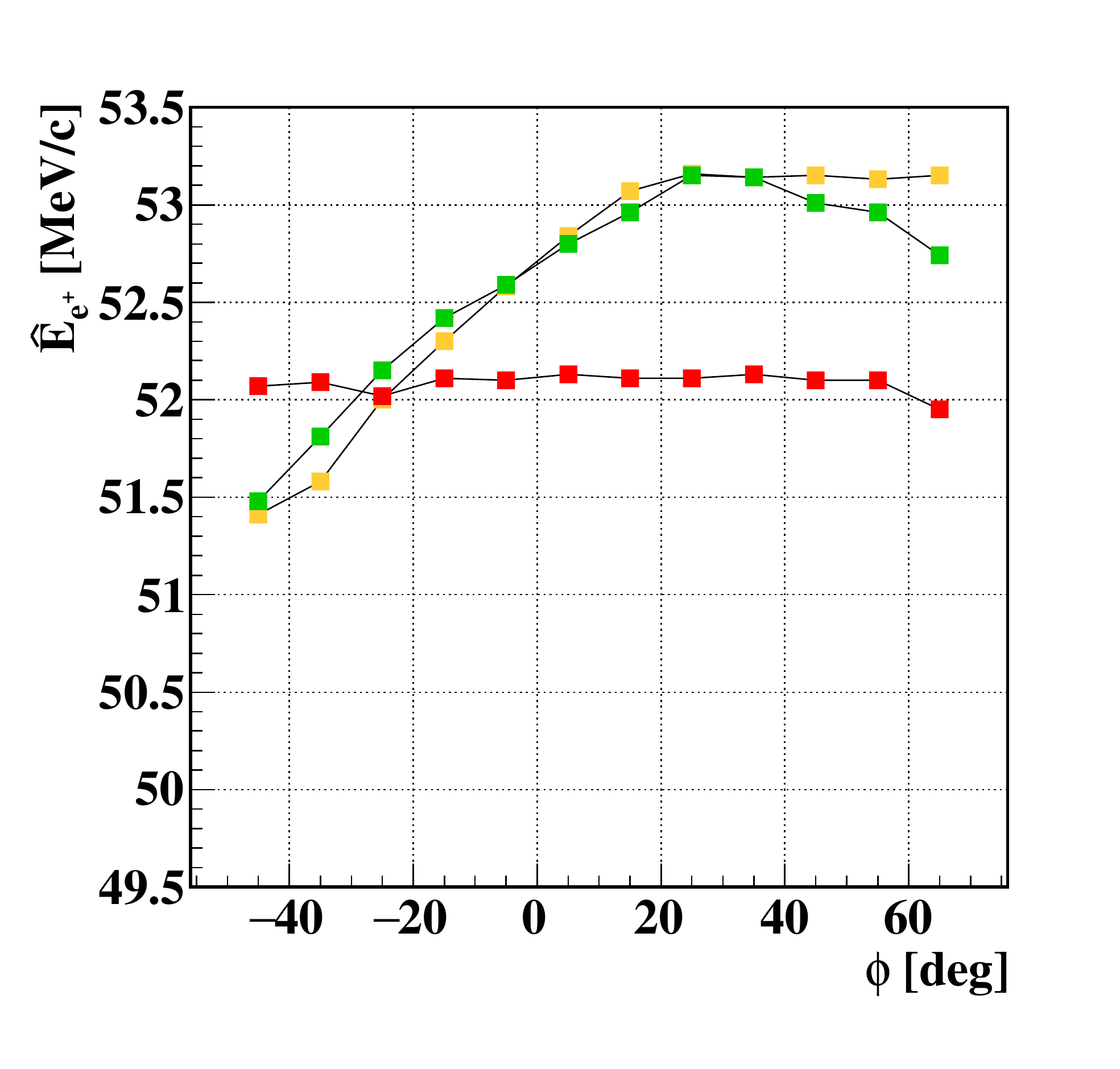}
\caption{Reconstructed Mott positron energy versus reconstructed $\phi$-angle. 
Under normal functioning conditions the trend of energy versus the $\phi$-angle is flat (red points). 
If some distortions are present, deviations are observed, as shown in the case of green and yellow points. 
See the text for more details. }
\label{fig:AlignmentCheck}
\end{figure}

Figure~\ref{fig:AlignmentCheck} shows how the method is very sensitive to misalignment. 
The red points show the expected 
dependence of the reconstructed $\epositron$ versus the reconstructed $\phie$; 
the green points show the same measurement in presence of an erroneous set of survey data
used as input to the alignment procedure; the plot highlights unambiguously the problem.
It is also possible to reproduce the plot in the simulation when using inconsistent alignment data
(see the yellow points).

These results validate the method as a standard calibration tool for MEG~II.

\subsection{Expected performances}
\label{sec:performance}

\begin{figure}
	\centering
	\includegraphics[width=1\linewidth]{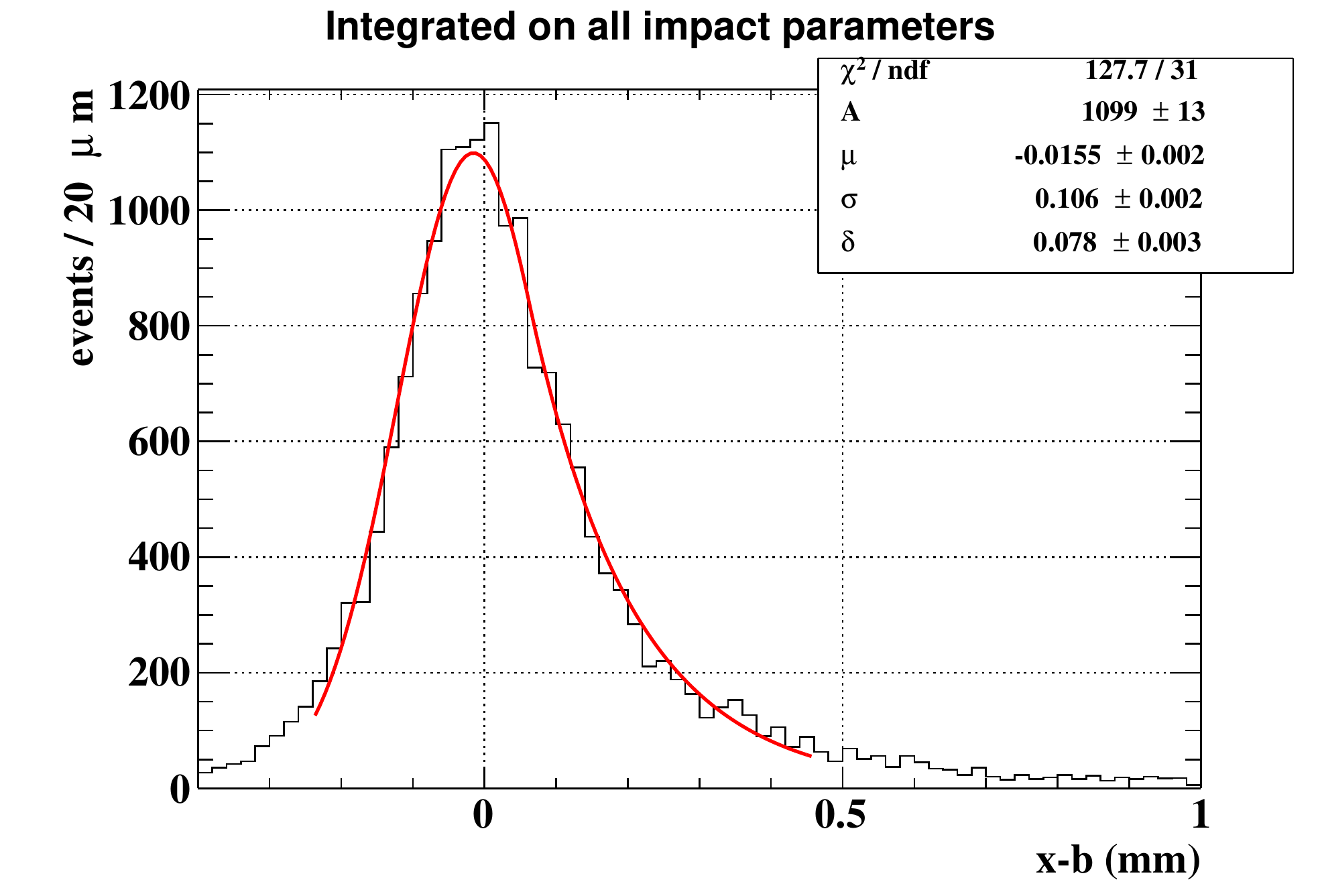}
	\caption{CDCH single hit resolution function, measured on a prototype in a cosmic ray facility, as the difference
      between the measured drift distance $x$ and the particle's impact parameter $b$. 
A fit is performed with a Gaussian core function of mean $\mu$ and width $\sigma$, analytically 
matched with an exponential tail starting at $\mu + \delta$ (see \cite{Baldini:2016rrk} for more details).}
	\label{fig:cdc_single_hit}
\end{figure}

As preliminary tests, the spatial resolution and the ageing properties of the chamber have been measured on prototypes.

For a precise measurement of the single-hit resolution, several drift chamber prototypes were tested in a 
cosmic ray facility set-up \cite{Galli:2015ufa,Baldini:2016rrk}, and an example result is 
shown in Fig.~\ref{fig:cdc_single_hit}.
Expected biases and resolution tails are observed, due to the poor ionisation statistics in 
the very light helium-based gas mixture. Despite the presence of these tails, the bulk of the resolution function has a 
Gaussian shape, with a width of $\sigma_r\simeq\SI{110}{\um}$, averaged over a large range 
of angles and impact parameters. Since the longitudinal coordinate of hits is determined 
by exploiting the stereo angle, the corresponding resolution is then expected to be 
$\sigma_z=\sigma_r/\sin\theta_\mathrm{s}\simeq\SI{1}{\mm}$. 
However in the final chamber further improvements are expected due to the new
front-end electronics with a \SI{1}{\giga\hertz} bandwidth allowing for
the exploitation of the cluster timing technique.

\begin{figure}[bt]
	\centering
	\includegraphics[width=0.50\textwidth]{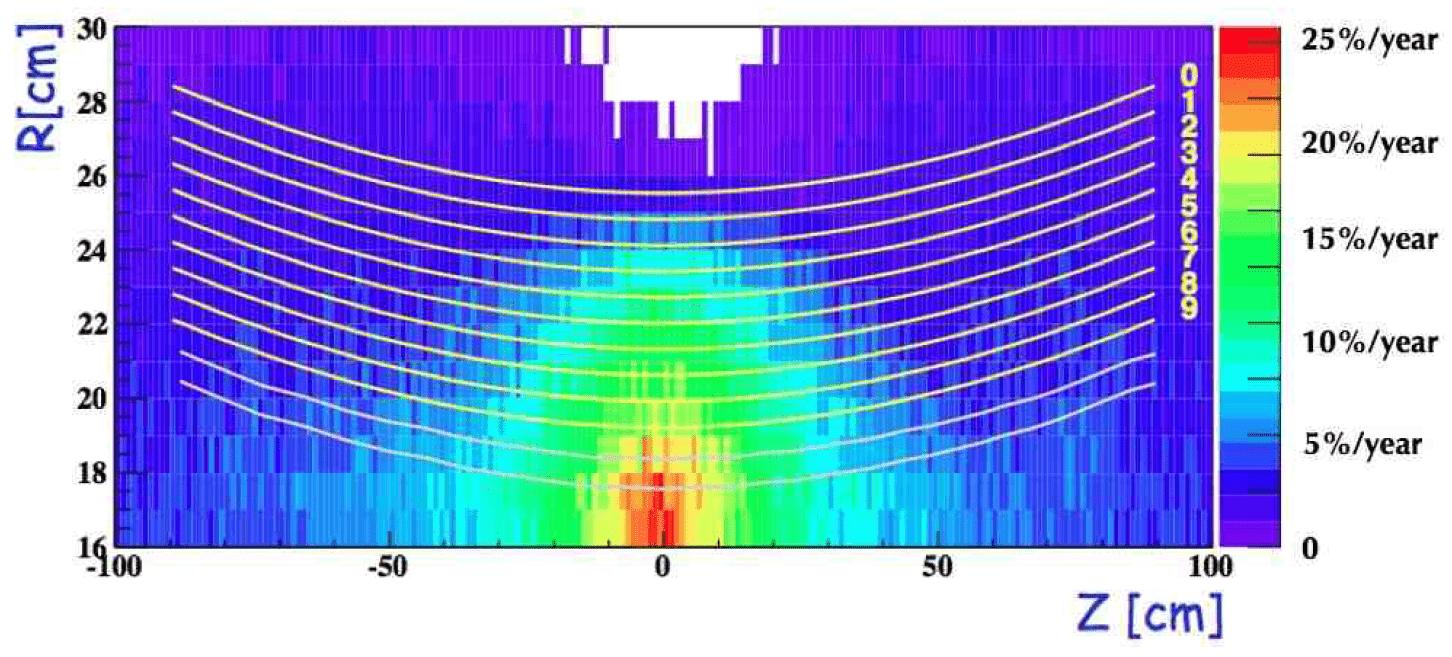}
	\caption{Gain drop in one year of DAQ time.} \label{fig:gaindrop}
\end{figure}

The operation and performance of the chamber will also be affected by the extremely high positron 
rate in CDCH (up to \SI{\sim 30}{\kilo\hertz\per\cm\squared}),
which will induce a huge amount of charge collected in the hottest portion of the innermost wire (\SI{\sim 0.5}{\coulomb\per\cm}). 
Since at such values of collected charge wire chambers can present inefficiencies and loss of gain, laboratory tests on prototypes 
in a dedicate irradiation facility set-up were performed \cite{Baldini:2016rrk}. Tests returned sustainable gain losses of less than
20\%\ per DAQ year (see Fig.~\ref{fig:gaindrop}) in the hottest few centimetres of the innermost wires, an effect which can be easily compensated for by increasing the 
voltage of the affected wires.

\begin{table}[bt]
	\caption{Expected MEG~II CDCH performances compared with MEG DCH (core resolutions).}
	\centering
	\label{tab:ex_DC_P}
	\begin{tabular}{ l c c } 
		\hline
		                                           &    MEG           &     MEG~II   \\ 
		\hline
		$\sigma^\mathrm{core}_{\ppositron}$ (\si{\kilo\eV})                 & 306              &   130  \\
		$\sigma^\mathrm{core}_{\thetae}$ (\si{\milli\radian})               & 9.4              &    5.3  \\
		$\sigma^\mathrm{core}_{\phie}$ (\si{\milli\radian})                 & 8.7              &    3.7   \\
		Tracking efficiency (\%)           & 65               &   78   \\
		CDCH-pTC matching efficiency (\%)   & 45               &   90   \\
		\hline
	\end{tabular}
\end{table}

The expected CDCH performance compared to the MEG DCH system is summarised in 
Table~\ref{tab:ex_DC_P}. The resolutions are obtained using the results of tests with prototypes 
as input for the simulation of the detector (under the assumption of Gaussian single hit resolutions), 
and cross-checked with a full simulation of the detector response (which also accounts for non-Gaussian tails). 
Non-Gaussian tails are observed in the resolution functions, as expected from Coulomb 
scattering at large angles and from energy-loss fluctuations. Core resolutions are shown in 
Table~\ref{tab:ex_DC_P}, but the full resolution functions have been used for the estimate 
of the MEG~II sensitivity. 

In the table we quote separately the efficiency for tracking a signal positron
and the probability that such a positron reaches the pTC in a place that can be geometrically matched to the reconstructed track 
(\emph{matching efficiency}). In MEG, the matching efficiency was limited by the positron scattering on service materials 
(electronics, cables, etc.) in the volume between the drift chambers and the timing counter. 
The new design significantly reduces this loss of efficiency, and the estimated transparency 
toward the pTC is doubled. The preliminary estimate of the tracking efficiency in 
MEG~II is expected to improve with further developments of the reconstruction algorithms.

\clearpage
\newpage
\section{Pixelated timing counter}
\label{sec:Pixelated_Timing_Counter}
Precise measurement of the time coincidence of $\egammapair$ pairs is 
one of the important features of the MEG~II experiment in order to suppress the predominant accidental background.
The positron time $\tpositron$ must be precisely measured by the pixelated timing counter (pTC), succeeding the MEG timing counter,
with a resolution $\sigma_{\tpositron}\sim\SI{30}{\ps}$ at a hit rate \SI{\sim5}{\MHz}.
In addition, it also generates trigger signals by providing prompt timing and direction information on the positron.

\subsection{Limitations of the MEG timing counter}
In the past decades, timing detectors based on scintillation counters with PMT read-out have been built and operated successfully. The best achievements with this technique gave time resolutions slightly better than \SI{50}{\ps} for a minimum ionising particle (e.g. \cite{sugitate_1986, shikaze_2000}). 
One of them was the MEG timing counter consisting of 30 scintillator bars (BC-404 with \SI[product-units=power]{80 x 4 x 4}{\cm} dimensions) each of which read out by fine mesh PMTs at both ends \cite{DeGerone:2011te}.
It showed a good intrinsic time resolution of \SI{40}{\ps} in beam tests,
but the operative time resolution on the experimental floor was measured 
to be $\sigma_{\tpositron}\SI{\sim70}{\ps}$. The main causes of the degradation were:
\begin{enumerate}
\item a large variation of the optical photon paths originating from the large size of the 
scintillator (long longitudinal propagation and incident-angle dependence due to its thickness),
\item a degradation of the PMT performance in the MEG magnetic field,
\item the error of timing alignment among the bars (time calibration) and
\item the electronic time jitter.
\end{enumerate}
The sum of all these contributions accounted for the above mentioned operating timing resolution. 

Furthermore, a positron crossing a bar sometimes impinged on the same bar 
again while moving along its approximately helical trajectory. Such double-hit events 
produced a tail component in the timing response function.

Finally, since the PMTs worked at the far edge of its performance versus single event 
rate (\SI{1}{\MHz} per PMT), the designed increase of the muon stopping would have required 
a segmentation of at least the same factor with respect to the present configuration 
in order to preserve the proper PMT working point. 

\subsection{Upgrade concept}
\label{sec:Upgrade_concept}
We plan to overcome such limitations by a detector based on a new concept: a highly segmented 
scintillation counter. 
In the new configuration, the 30 scintillator bars are replaced by 512 small scintillation tiles; 
we call the new detector pixelated timing counter (pTC).
There are several advantages in this design over the previous one:
\begin{enumerate}
\item The single counter can easily have a good time resolution due to the small dimensions.
\item The hit rate of each counter is under control to keep the pile-up probability 
as well as the \lq{}double-hit\rq{} probability negligibly low. 
\item Each particle\rq{}s time is measured with many counters to significantly improve the total time resolution.
\item A flexible detector layout is possible to maximise the detection efficiency and the hit multiplicity.
\end{enumerate} 
The third point is of particular importance:
by properly combining the times measured by $N_\mathrm{hit}$ counters, the total time resolution is expected to improve as
\begin{align}
	\sigma_{\tpositron}(N_\mathrm{hit}) = \frac{\sigma_{\tpositron}^\mathrm{single}}{\sqrt{N_\mathrm{hit}}}
	= \frac{
              \sigma_{\tpositron}^\mathrm{counter} \oplus \sigma_{\tpositron}^\mathrm{inter\mathchar`-counter} \oplus \sigma_{\tpositron}^\mathrm{elec} } {\sqrt{N_\mathrm{hit}}
        },
	  \label{eq:resolution}
\end{align}
where $\sigma_{\tpositron}^\mathrm{single}$ is the total time resolution of a 
single-counter measurement which includes the counter intrinsic resolution $\sigma_{\tpositron}^\mathrm{counter}$, 
the error in time alignment over the counters $\sigma_{\tpositron}^\mathrm{inter\mathchar`-counter}$ 
and the electronics jitter $\sigma_{\tpositron}^\mathrm{elec}$.
The contribution of the multiple Coulomb scattering, which does not scale linearly with $\sqrt{N_\mathrm{hit}}$, 
is negligible.
Therefore, the multi-hit-measurement approach overcomes most of the limitations mentioned above 
and is superior to pursuing the ultimate time resolution of a single device.
Note that to properly combine the hit times, the positron propagation times between the counters 
have to be well known; the trajectory extrapolated from CDCH is used as well
as refinement of it by the reconstructed counter hit positions. 

This pixelated design became possible by using a new type of solid state photo-sensor: the silicon photomultiplier (SiPM),
that is a valuable replacement of the conventional PMT because of its excellent properties as listed below:
\begin{itemize}
   \item compact size,
   \item sensitivity to single photons,
   \item high internal gain (\numrange[range-phrase = --]{e5}{e6}),
   \item high photon detection efficiency peaked at $\lambda \SI{\sim 450}{\nm}$,
   \item good time resolution (\SI{<100}{\ps} for a single photon),
   \item immunity to magnetic fields,
   \item low bias voltage (\SI{<100}{\volt}) and low power consumption,
   \item no avalanche fluctuation (excess noise factor \numrange[range-phrase = --]{1}{1.5}) and
   \item low cost.
\end{itemize}
The compactness and low cost of SiPMs allows a high segmentation and together with the high immunity to magnetic fields enables flexible design of the counter layout without deterioration of the performance in the COBRA field.
A high time resolution of a SiPM-based scintillation counter was demonstrated in \cite{StoykovNDIP11} prior to designing MEG~II. 
It should also be the best solution for the read-out of the pixel module in this detector.

\subsection{Design}
\begin{figure}[tb]
\centering
\includegraphics[width=1\linewidth]{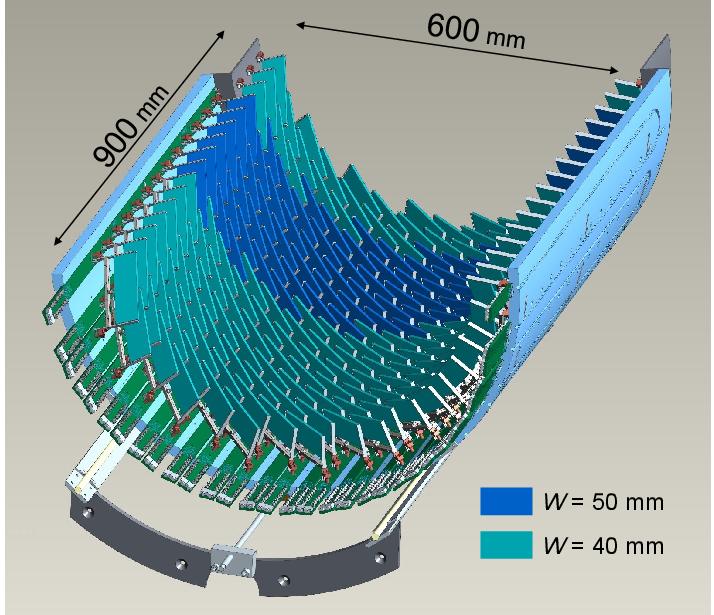}
\caption{Design of the downstream pTC super-module.}
\label{fig:TCdesign}
\end{figure}
The pTC consists of two semi-cylindrical super-modules like the previous ones, mirror symmetric to each other and placed upstream and downstream in the COBRA spectrometer.
Figure~\ref{fig:TCdesign} shows one of the super-modules composed of 256 counters fitted to the space between the CDCH and the COBRA magnet. The volume is separated from the CDCH, with the pTC modules placed in air.

Each counter is a small ultra-fast scintillator tile with SiPM read-out described in detail in Sect.~\ref{sec:pTC_Single_counter_design}.
Sixteen counters align in the longitudinal ($z$) direction at a \SI{5.5}{\cm} interval, and 16 lines are cylindrically arranged at a \ang{10.3} interval, alternately staggered by a half counter.
The counters are tilted at \ang{45} to be approximately perpendicular to the signal \positron\ trajectories.
The total longitudinal and $\phi$ coverages are $23.0<|z|<\SI{116.7}{\cm}$ and $\ang{-165.8}<\phi < \ang{+5.2}$, respectively, which fully cover the angular acceptance of the \positron\ from $\megc$ decays when the photon points to the LXe photon detector. 
This counter configuration was determined via a MC study to maximise the experimental sensitivity (given by the detection efficiency and the total time resolution) within the constraint of a limited number of electronics read-out channels (1024 channels in total).

\subsubsection {Counter module design}
\label{sec:pTC_Single_counter_design}
The single counter is composed of a scintillator tile and multiple SiPMs. The counter dimensions are defined by the length ($L$), width ($W$), and thickness ($T$) of the scintillator tile and described as $L\times W\times T$ below.
Multiple SiPMs are optically coupled to each $W\times T$ side of the scintillator. 
The signals from the SiPMs on each end are summed up and fed to one readout channel.
The \positron\ impact time at each counter is obtained by averaging the times measured at both ends.

We performed an extensive study to optimise the single counter design,
starting from a comparative study of scintillator material, SiPM models, number of SiPMs per counter, and connection scheme. 
Then, an optimisation of the scintillator geometry was performed to find the best compromise between the total resolution, detection efficiency and required number of channels.
The results are reported in detail in \cite{ootani-nima,metcjinst2014,Cattaneo:2014uya,Nishimura:2015qev} and summarised below.

\paragraph{Scintillator}
The choice of the scintillator material is crucial to optimise the time resolution.
The candidates selected from the viewpoint of light yield, rise- and decay-times, and emission spectrum are the ultra-fast plastic scintillators from Saint-Gobain listed in Table \ref{table:Saint-Gobain PS}. Note that the smaller counter dimensions allow the use of such very short rise time scintillators, which typically have short attenuation lengths.
\begin{table*}[!t]
\caption{Properties of ultra-fast plastic scintillators from Saint-Gobain. The properties of BC-404, which was used in the previous timing counter bar, is also shown for comparison.}
\label{table:Saint-Gobain PS}
\centering
  \begin{minipage}{0.65\linewidth}
   \renewcommand{\thefootnote}{\alph{footnote})}	
   \renewcommand{\thempfootnote}{\alph{mpfootnote})}	
\centering
\begin{tabular}{@{}lccccc}
\hline
\textbf{Properties}  & \textbf{BC-418} & \textbf{BC-420} & \textbf{BC-422} & \textbf{BC-422Q} & \textbf{BC-404} \\
\hline \hline
 Light Output\footnotemark[1] (\% Anthracene) & $67$ & $64$ & $55$ & $19$ & $68$\\
 Rise Time\footnotemark[1]\footnotemark[2] (\si{\ns}) & $0.5$ & $0.5$ & $0.35$ & $0.11$ & $0.7$\\
 Decay Time\footnotemark[1] (\si{\ns}) & $1.4$ & $1.5$ & $1.6$ & $0.7$ & $1.8$\\
 Peak Wavelength\footnotemark[1] (\si{\nm}) & $391$ & $391$ & $370$ & $370$ & $408$\\
 Attenuation Length\footnotemark[1] (\si{\cm}) & $100$ & $110$ & $8$ & $8$ & $140$\\
 Time Resolution\footnotemark[3] (\si{\ps}) & $48\pm2$ & $51\pm2$ & $43\pm2$ & $66\pm3$ & --\\
\hline
\end{tabular}
  \vspace{-0.2cm}
  \footnotetext[1]{From Saint-Gobain catalogue \cite{Saint-Gobain-BC4XX}.}
  \footnotetext[2]{Those values are dominated by the measurement setup. The intrinsic values are much faster. For example, a BC-422 rise time of \SI{<20}{\ps} was reported in \cite{BC422RiseTime}.}
  \footnotetext[3]{Measured value in \cite{Cattaneo:2014uya} with \SI[product-units=power]{60 x 30 x 5}{\mm} sized counter read-out with 3 HPK SiPMs (S10362-33-050C) at each end.}
\end{minipage}
\end{table*}
The time resolutions were measured for all the types of scintillator and different sizes.
BC-422 was found to always give the highest time resolution for each size (tested up to \SI[product-units=power]{120 x 40 x 5}{\mm}) and
therefore was chosen.

Different type of reflectors such as no reflector, Teflon\textsuperscript{\textregistered} tape, aluminised Mylar\textsuperscript{\textregistered} and enhanced specular reflector (ESR) from 3M were tested to improve the light collection and hence the time resolution.
The best time resolution was obtained with ESR film, while a small worsening was observed with Teflon tape (diffuse reflector) compared to no reflector \cite{ootani-nima}.

\paragraph{SiPM}
The photo-sensors must be sensitive to the scintillation light in the near-ultraviolet (NUV) range.
Recently, several manufacturers have developed such NUV-sensitive SiPMs based on \lq{}p-on-n\rq{} diode structures.
Therefore, we tested a number of such NUV-sensitive SiPMs available as of 2013 from AdvanSiD (ASD), Hamamatsu Photonics (HPK), KETEK, and SensL.

\begin{figure}[tb] 
\centering
\includegraphics[width=1\linewidth, clip, trim=0 0 0 2pc]{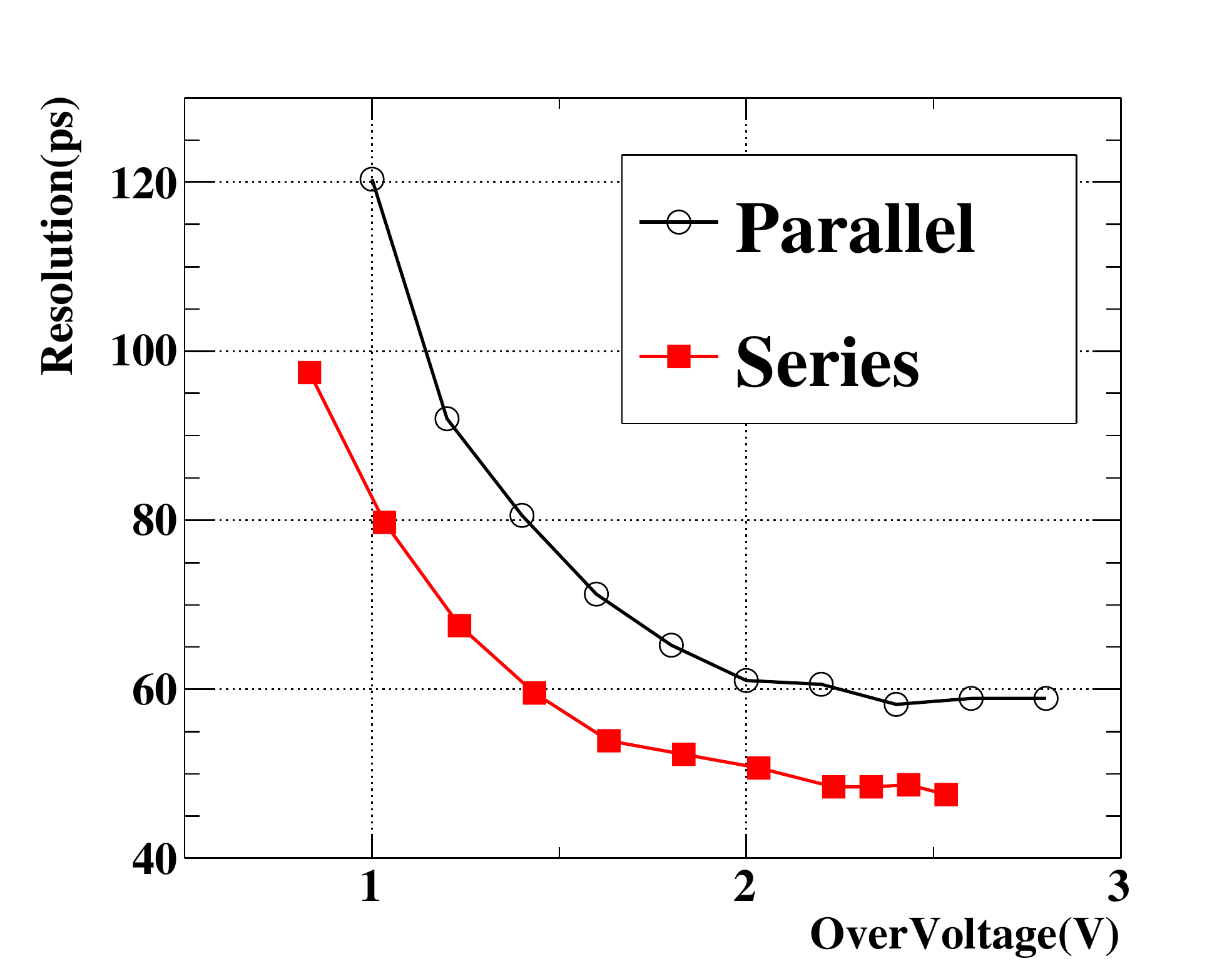}
\caption{Comparison of time resolutions between series and parallel connections measured with 
\SI[product-units=power]{60 x 30 x 5}{\mm} sized counter read-out with 3 HPK SiPMs (S10362-33-050C) at each end (from \cite{Nishimura:2015qev}).}
\label{fig:SeriesParallelResolution}
\end{figure}

Before the decision of SiPM models, we first examined the schemes of SiPM connection.
In order to compensate the small active area of SiPMs, multiple SiPMs are connected in parallel for read-out.
However, performance issues for the parallel connection are: increase in the signal rise time and width and increase in the parallel and series noise; both originate from the larger sensor capacitance and negatively affect the time resolution.
We have examined an alternative connection: series connection of multiple SiPMs ($N_\mathrm{SiPM}=3-6$).\footnote{Series connection of avalanche photodiodes was proposed and tested in \cite{apdSeries} and the first application to SiPMs is found in \cite{sipmSeries}.}
Figure~\ref{fig:SeriesParallelResolution} shows a comparison of time resolutions between series and parallel connections.
Series connection gives better time resolutions at all over-voltages.\footnote{The over-voltage is the excess bias voltage over the SiPM breakdown voltage. In the series connection case, it quotes over-voltage per SiPM.}
This is due to the narrower output pulse shape because of the reduced total sensor capacitance in the series circuit.
Although the total charge (gain) is reduced to $1/N_\mathrm{SiPM}$ of that of a single SiPM, the signal amplitude (pulse height) is kept comparable (compensated by the $N_\mathrm{SiPM}$ times faster decay time). 
Thus, we conclude that series connection is better for the pTC application. We simply connect SiPMs in series on a custom print circuit board (PCB) while we adopt a more complex way for the MPPCs used in LXe photon detector (see Sect.~\ref{sec:largeMPPC}).

\begin{figure}[tb] 
\centering
\includegraphics[width=1\linewidth]{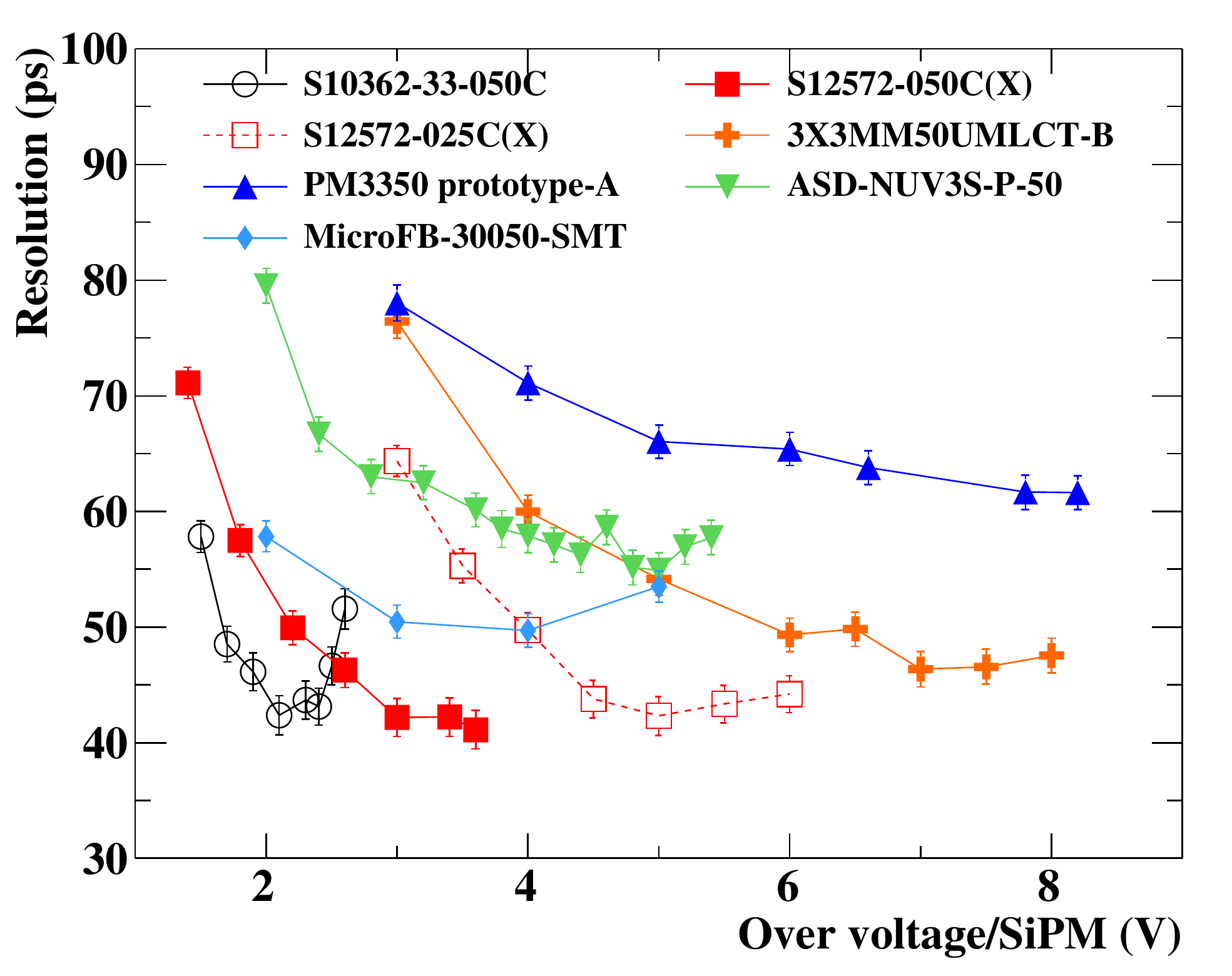}
\caption{Time resolutions measured with different types of SiPMs (3 SiPMs at each end) and \SI[product-units=power]{60 x 30 x 5}{\mm} scintillator (from \cite{Cattaneo:2014uya}).}
\label{fig:SiPM_Comparison}
\end{figure}

For each type of SiPM, we measured the device characteristics (such as dark count rate, cross-talk probability, PDE, and temperature dependence) and the time resolution when coupled to a scintillator.
The main results are shown in Fig.~\ref{fig:SiPM_Comparison}.
The best time resolution is obtained with SiPMs from HPK, which have the highest PDE.
This result indicates that the time resolution of our counter is predominantly limited by the photon statistics and increasing the number of detected photons is the most important and straightforward way of improving the time resolution.
Using higher PDE SiPMs is one way. 

\begin{figure}[tb] 
\centering
\includegraphics[width=1\linewidth]{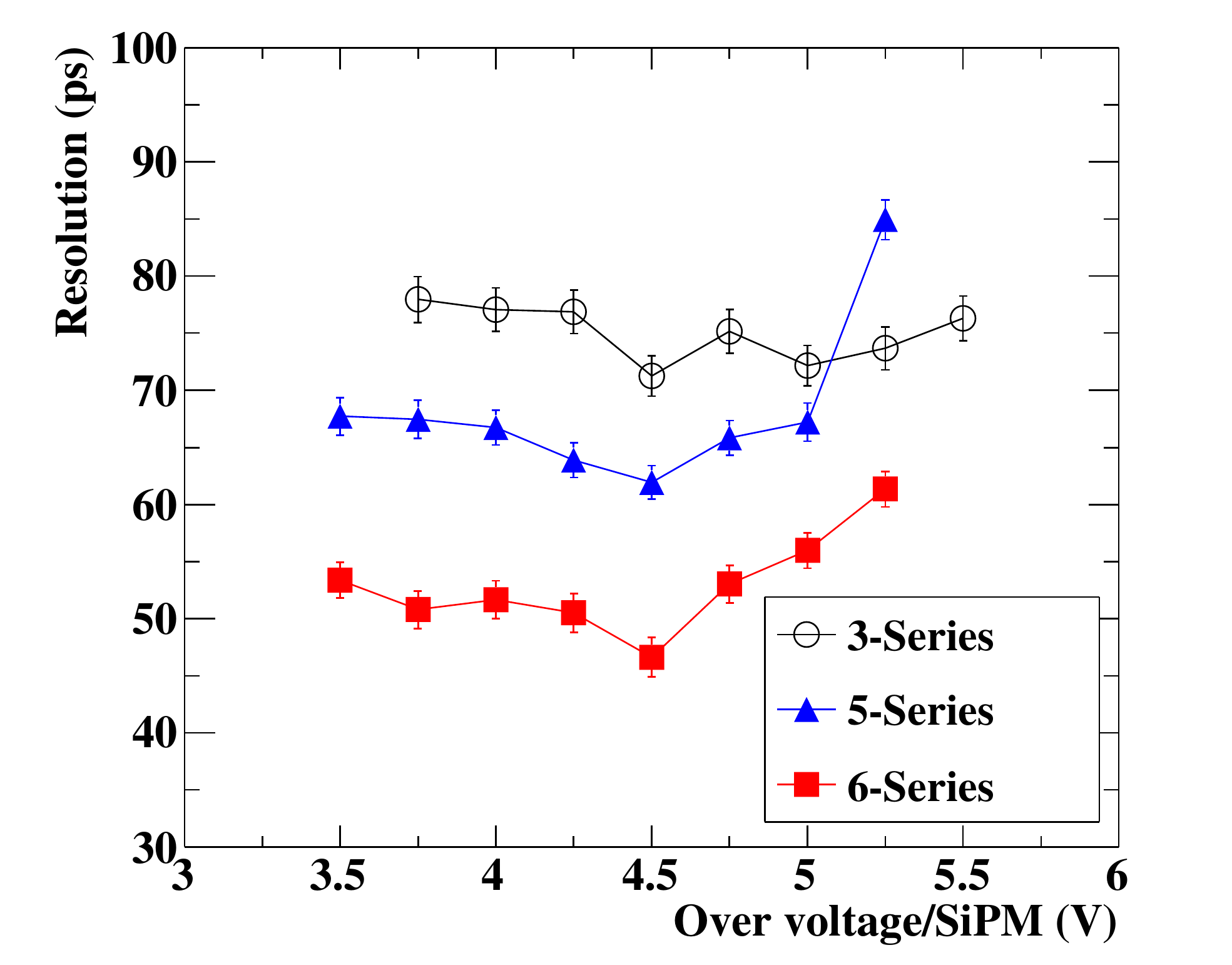}
\caption{Time resolution measured with different numbers of SiPMs. 3, 5, and 6 SiPMs (ASD-NUV3S-P-50) connected in series and coupled to each end of \SI[product-units=power]{90 x 40 x 5}{\mm} scintillator (from \cite{Nishimura:2015qev}).}
\label{fig:nSiPM}
\end{figure}

Another way is increasing the sensor coverage by using more SiPMs.
Figure~\ref{fig:nSiPM} shows the time resolution measured with different numbers of SiPMs.
In this study, SiPMs from ASD were used. A clear improvement with a larger number of SiPMs is observed, 
and a time resolution of \SI{50}{\ps} is achieved with 6 SiPMs at each end coupled to \SI[product-units=power]{90 x 40 x 5}{\mm} scintillator. This is better than that achieved with 3 HPK SiPMs (\SI{58}{\ps}).
The question as to how many sensors can be used depends on the final geometry of the detector 
and cost, so the decision of the SiPM model and the number was made after fixing those parameters.
We finally adopted the 6-series solution using ASD SiPMs, which gives the best performance within our budget constraint.

\begin{figure}[tb]
\centering
\includegraphics[width=1\linewidth]{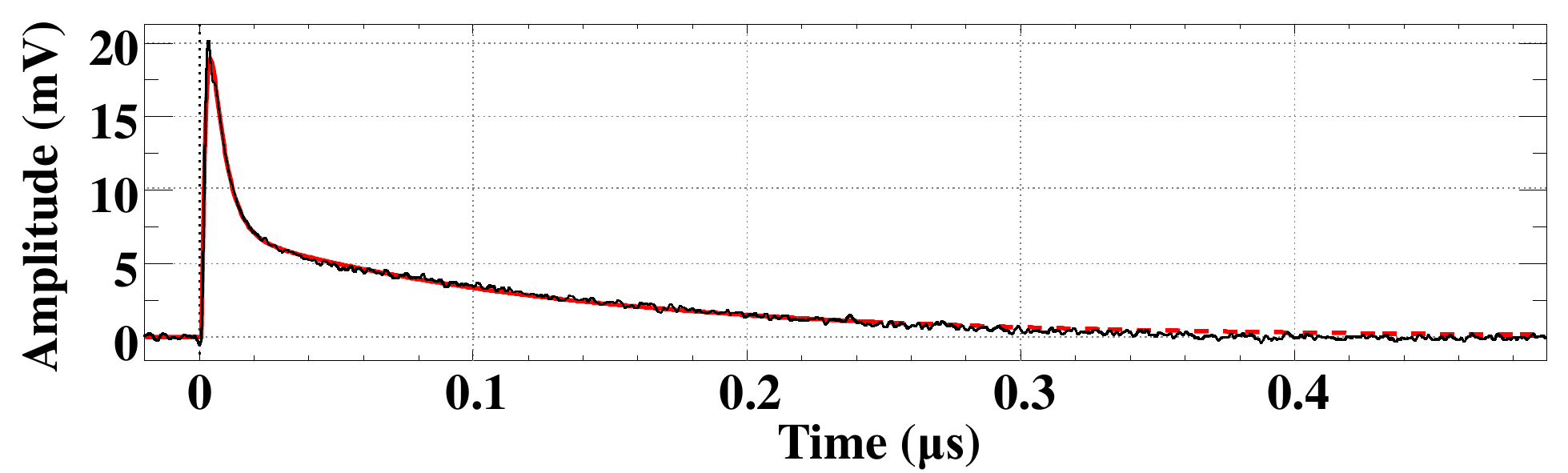}
\caption{Pulse shape of a single-cell-fired signal from an ASD-NUV3S-P-High-Gain (with a gain 60 amplifier). The black line shows the averaged pulse shape and the red curve is the best fit function.}
\label{fig:single_waveform}
\end{figure}

The model used in the pTC is the ASD-NUV3S-P-High-Gain; the specifications provided by AdvanSiD are listed in Table~\ref{tab:ASD}.
Figure~\ref{fig:single_waveform} shows the measured single-cell-fired signal. The SiPM's specific long exponential tail, with a time constant of \SI{124}{\ns}, is due to the recharge (recovery) current determined predominantly by the quench resistance and the cell capacitance, which are measured to be ${R_q} = \SI[separate-uncertainty]{1100(50)}{\kilo\ohm}$ and $C_D = \SI[separate-uncertainty]{100(10)}{\femto\farad}$, respectively.
\begin{table}[tb]
\centering
\caption{\label{tab:ASD}Specifications of AdvanSiD SiPM ASD-NUM3S-P-50-High-Gain.}
\begin{tabular}{lcc}
   \hline
   Parameter                       & Value & Unit\\
   \hline \hline
   Effective active area         & \num{3x3} & \si{\mm\squared}\\
   Cell size                     & \num{50x50}& \si{\um\squared}\\
   Cells number                  & 3600 &\\
   Spectral response range       & \numrange{350}{900} & \si{\nm}\\
   Peak sensitivity wavelength   & 420& \si{\nm} \\
   Breakdown voltage $V_\mathrm{BD}$  & $24\pm0.3$ & \si{\volt}\\
   Work voltage range            & $V_\mathrm{BD} + 2$ to $V_\mathrm{BD} + 3.5$&\si{V}\\
   Dark count                    & \num{<100} & $\si{kcps\per\mm^2}$\\
   Gain                          & \num{3.3e6} &\\
   $V_\mathrm{BD}$ temperature sensitivity & 26 &\si{\mV/\degreeCelsius}\\
   \hline
\end{tabular}
\end{table}

\begin{figure}[t]
\centering
\includegraphics[width=1\linewidth]{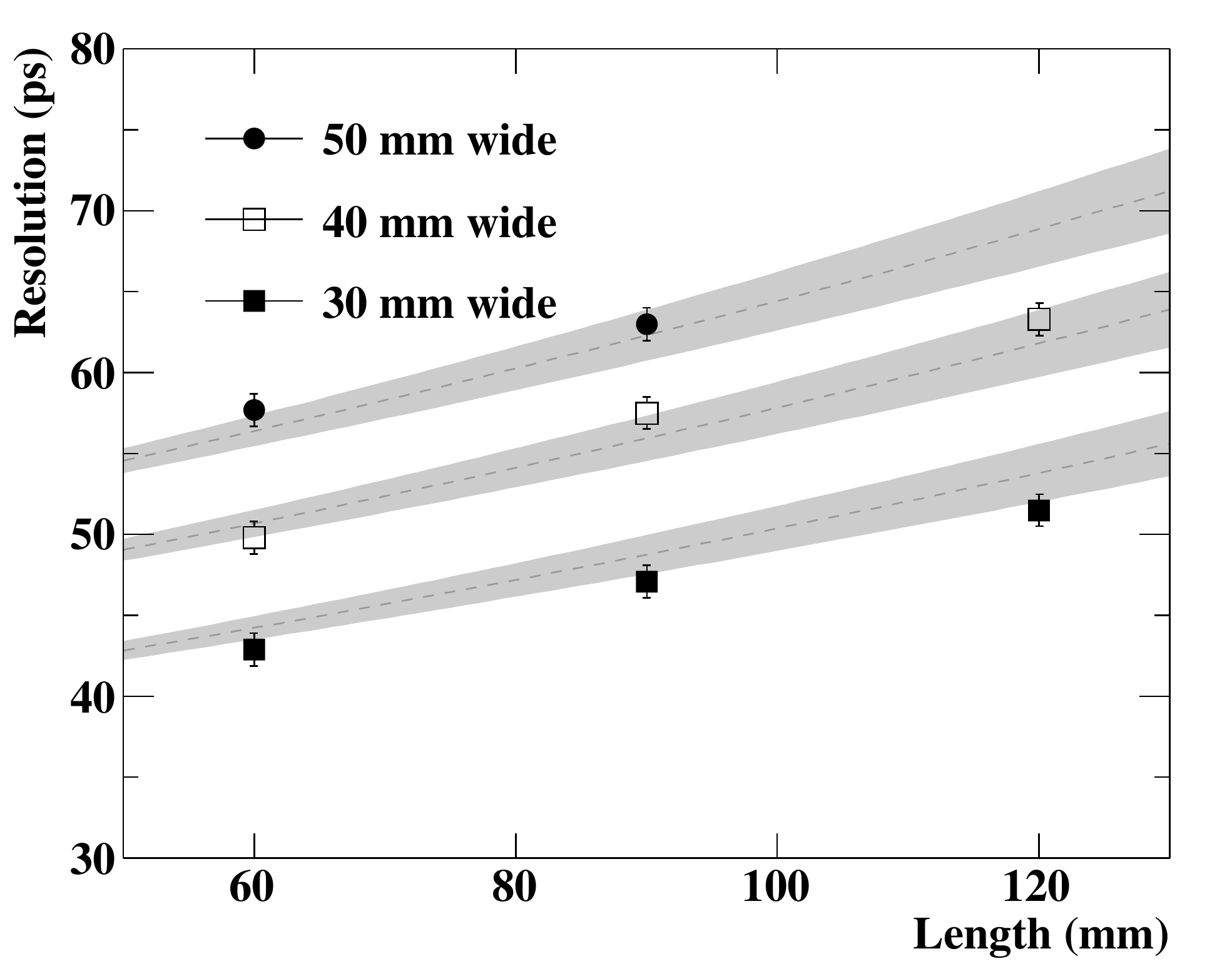}
\caption{Dependence of the counter time resolution on the size measured with 3 HPK SiPMs (S10362-33-050C) at each end.
The superimposed curves show the dependence expected from the detected photon statistics. The shaded bands show the uncertainty. See \cite{Cattaneo:2014uya} for the detailed description.}
\label{fig:sizeDependence}
\end{figure}

\paragraph{Geometry}
The single counter time resolution was measured for different sized scintillator tiles,
and the results are shown in Fig.~\ref{fig:sizeDependence}.
The size dependence is understandable from the photon statistics expected from the sensor coverage to the scintillator cross-section (dependent on $W$) and the light attenuation in the scintillator (on $L$).

The size has to be optimised by a balance between single-counter resolution (smaller is better) and hit multiplicity and detection efficiency (larger is better).
This optimisation is performed via a MC simulation study using the measured single-counter resolutions.
As a result, longer counters (up to the measured maximum $L = \SI{120}{\mm}$) are found to give a better performance.
Considering the hit rate and the double-hit probability, we did not test longer counters and fixed the length to be $L=\SI{120}{\mm}$. 
The optimal counter width $W$ is dependent on the longitudinal position because the radial spread of the signal \positron\ trajectories depends on the longitudinal position in the pTC region.
We adopt two different sizes: $W=\SIlist[list-units = single]{40;50}{\mm}$. The $W=\SI{50}{\mm}$ counters are assigned to the middle longitudinal position (see Fig.~\ref{fig:TCdesign}) where the radial spread becomes large.  
We observed a moderate dependence of the resolution on the thickness $T$ and decided for $T=\SI{5}{\mm}$, which is sufficiently thick to match the SiPM active area.
A \SI{5}{\mm} thick scintillator causes a deflection of \SI{50}{\MeV} positron direction for $\theta^\mathrm{RMS}_\mathrm{MS}\SI{\sim25}{\milli\radian}$, whose impact on the propagation time estimation is estimated to be \SI{\sim5}{\ps},
negligibly small compared to the counter resolution.

\paragraph{Final design of the counter module}
\begin{figure*}[tb]
\centering
\includegraphics[width=0.9\linewidth]{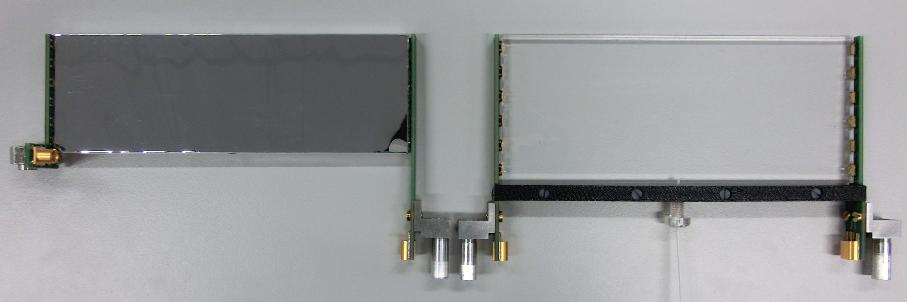}
\caption{Picture showing both types of counter modules. Left: $W=\SI{40}{\mm}$ counter wrapped in the reflector (example with the L-shaped PCB). Right: $W=\SI{50}{\mm}$ counter with optical fibre before wrapping in the reflector. }
\label{fig:counter_picture} 
\end{figure*}
Figure~\ref{fig:counter_picture} shows examples of the final counter modules.
A counter consists of a tile of BC-422 with dimensions of $L\times W\times T=120\times (40~ \mathrm{or}~50) \times 5$~\si{\mm\cubed} and 12 ASD SiPMs, 6 on each $(W\times T)$-side, directly coupled to the scintillator with optical cement (BC-600).
The scintillator is wrapped in \SI{32}{\um} thick ESR film, and then the module is wrapped in a \SI{25}{\um} thick black sheet of Tedlar for light tightness.

\begin{figure}[tb]
\centering
\includegraphics[width=18pc]{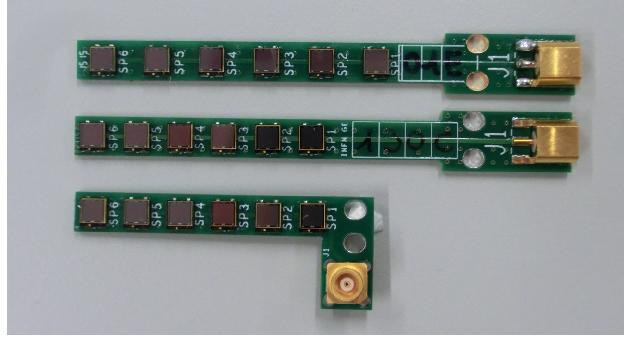}
\caption{The SiPM mounting PCBs. The top one is for the \SI{50}{\mm} counters and the others are for the \SI{40}{mm} ones.}
\label{fig:pcb_picture}
\end{figure}

Figure~\ref{fig:pcb_picture} shows the PCBs on which the SiPMs are soldered. The L-shaped PCBs are used for the counters at the inner (small $|z|$) location where the radial space is more restricted because of the smaller inner diameter of the magnet coils. 
Parts made of aluminium are attached to the PCBs and thermally coupled to one of the metal layers on the PCBs. They are used not only to mechanically fix the counters but also to thermally link the SiPMs to the main support structure whose temperature is controlled by a chiller system.

Each counter except for the counters using the L-shaped PCBs is equipped with an optical fibre for the laser calibration (described in Sect.~\ref{sec:cal-laser}).

\subsubsection{Read-out chain}
The basic idea of the read-out scheme is to send the raw SiPM-output signals directly to the WaveDREAM read-out boards (WDBs) (see Sect.~\ref{sec:tdaq}), on which the signals are amplified, shaped, and digitised.
Hence, the SiPMs and amplifiers are separated by long cables without any pre-amplification. 
This approach is adopted for both simplification and for space and power consumption reasons.
The reduction of the sensor capacitance by the series connection allows \SI{50}{\ohm} transmission without significant broadening of the pulse.

The counter modules are mounted on \SI{1}{\m} long custom PCBs (back-planes) placed on the mechanical support structures, allowing the signals to be transmitted outside the spectrometer. 
The back-planes have coaxial-like signal lines with a \SI{50}{\ohm} characteristic impedance. The ground lines are independent of each other to avoid possible ground loops.
The signals are then transmitted  to the WDBs on \SI{7}{\m} long non-magnetic RG-178 type coaxial cables (Radiall C291\,140\,087). 
MCX connectors are used for all connections.

The SiPM bias voltages, typically \SI{164}{\volt} for the six ASD SiPMs in series, are supplied from the WDBs through the signal lines; only one cable per channel.

The input signals are amplified by a factor 100 at the analogue part of WDBs.
It turns out to be very important to eliminate the long time constant component of the SiPM output pulse for a precise time measurement in order to suppress the effect of dark counts and obtain a stable baseline, especially after some radiation damage.
For this reason a pole-zero cancellation circuit is incorporated on the  WDB.

The amplified and shaped waveforms are digitised at a sampling frequency of 2~GSPS by the DRS4 chips on the WDBs for a detailed offline analysis of the pulses in order to compute the precise signal times.

\subsubsection{Mechanical support structure}
The mechanical support structures are made of aluminium cylinders with inner and outer radii of \SIlist{380;398}{\mm}, respectively.
The back-planes are fit to grooves machined on the structures.
A hole is drilled below the centre of each counter to pass an optical fibre from underneath.
Cooling-water pipes are laid on the outer side of the structure and connected to the chiller
to keep the temperature below \SI{30}{\degreeCelsius} with a stability better than \SI{1}{\degreeCelsius}.\footnote{The main heat source is the front-end electronics of CDCH (Sect.~\ref{sec:CDC_Electronics}).}

\subsection{Hit distribution \& rate}
\begin{figure*}[tb]
\centering
\includegraphics[width=.8\linewidth]{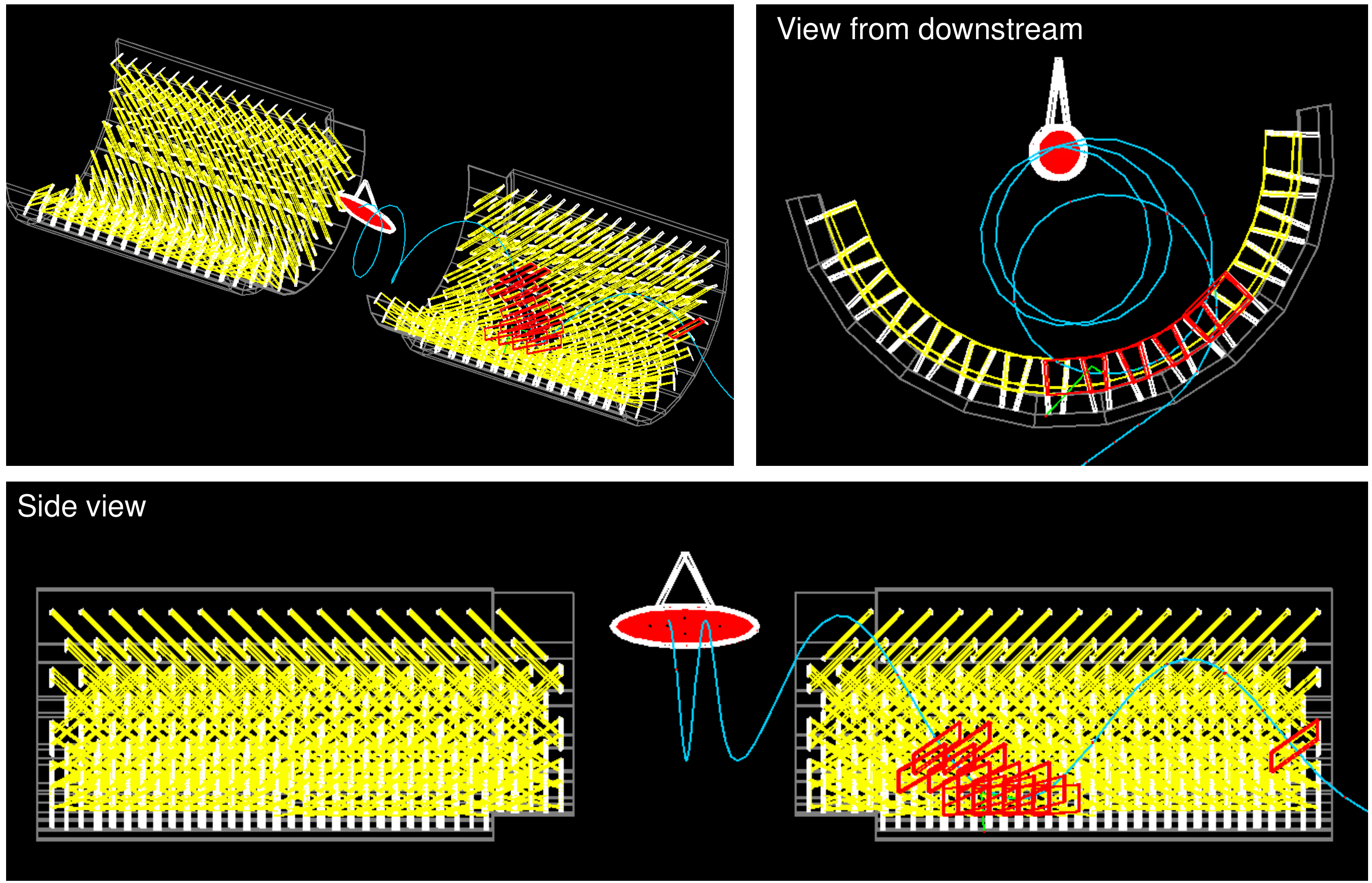}
\caption{An example of a hit pattern by a simulated signal \positron. CDCH is not drawn in these figures.}
\label{fig:mcdisplay}
\end{figure*}

\begin{figure}[tb]
\centering
\includegraphics[width=1.\linewidth]{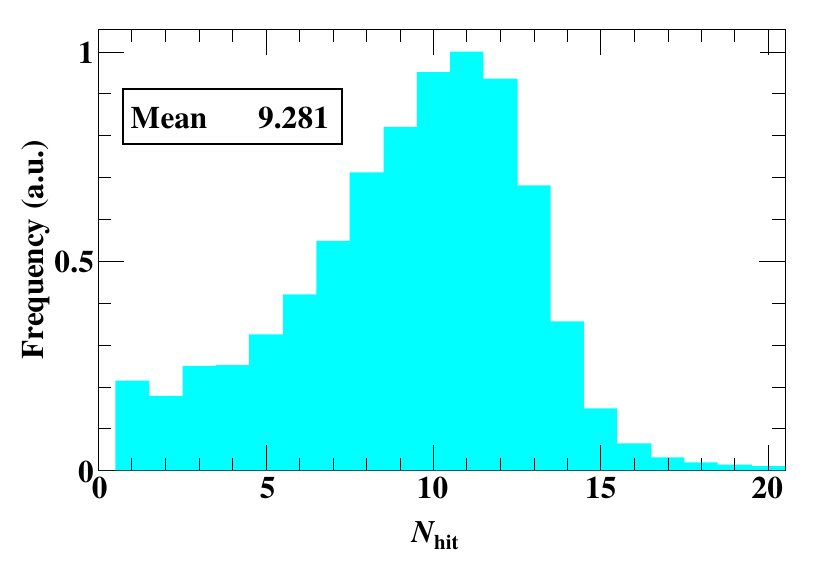}
\caption{Distribution of the expected number of hit counters for signal positrons, from a MC simulation.}
\label{fig:nhits}
\end{figure}

A MC simulation based on \textsc{Geant4} (version 10.0) \cite{Geant4,Geant4_2,Geant4_3} is performed with the final detector configuration to evaluate the hit distribution and hit rates.
Figure~\ref{fig:mcdisplay} shows an example of a hit pattern in the pTC by a \positron
from a $\megc$ decay.
Figure~\ref{fig:nhits} shows the distribution of the number of hit counters for signal positrons generated in the angular acceptance.\footnote{Defined so that the corresponding photon (with a direction opposite to the \positron) enters the fiducial volume of the LXe photon detector.}
The mean hit multiplicity is evaluated to be $\bar{N}_\mathrm{hit}=9.3$.

\begin{figure}[tb]
\centering
\includegraphics[width=1\linewidth]{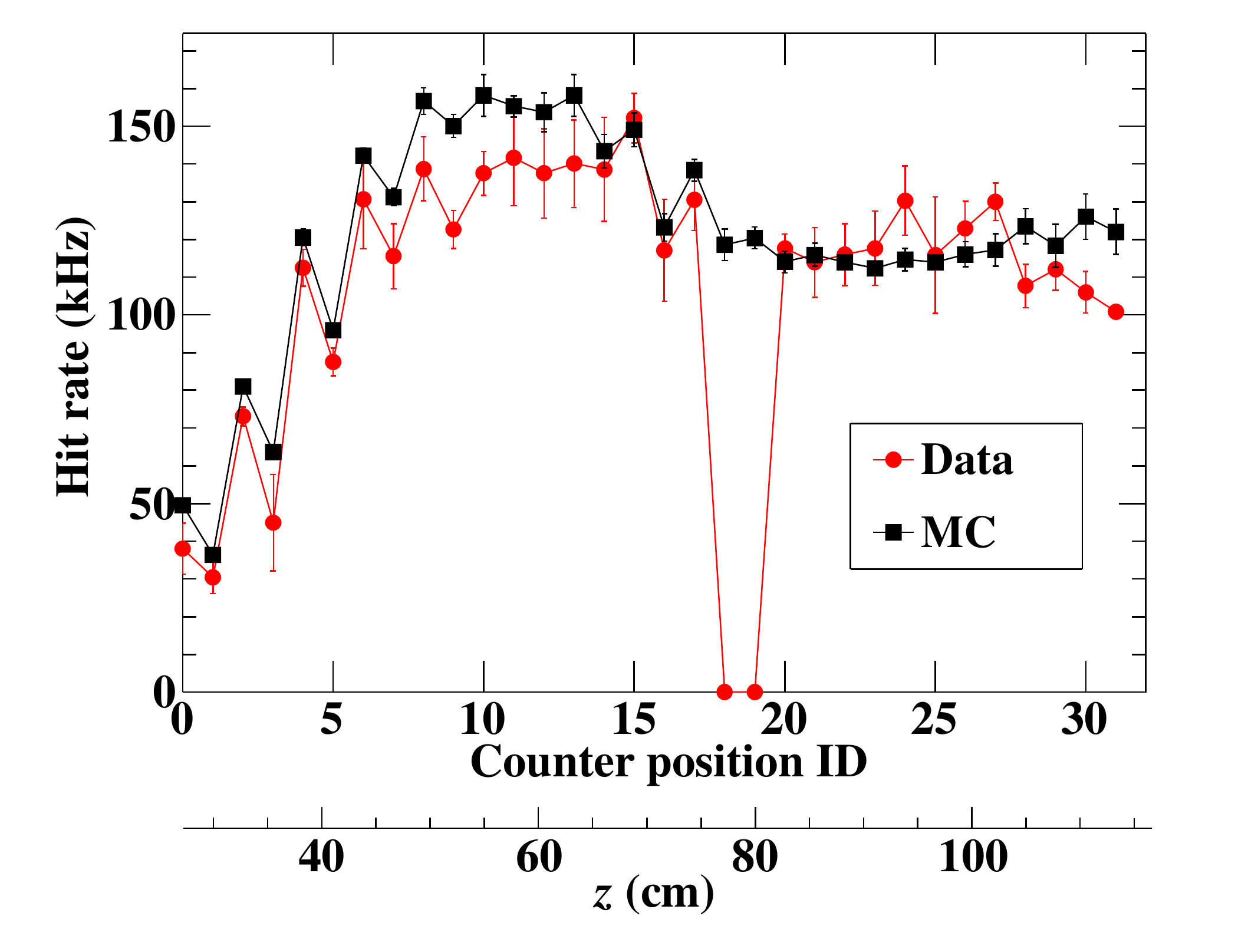}
\caption{Counter hit rate under MEG~II beam conditions, as a function of counter $z$-position. The black squares are from the MC simulation and the red circles are from the pilot run. The points with zero hit rate are due to dead channels in the readout electronics (from \cite{Uchiyama:2016vci}).}
\label{fig:hitrate}
\end{figure}

The hit rates at the individual counters are estimated by a simulation of a \muonp-beam 
(at a rate \SI{9e7}{\per\second})\footnote{It is necessary because Michel positrons from off-target decays (especially downstream of the target) have a non-negligible effect on the rates.} which then decay in accordance with the SM calculation.
The result is shown in Fig.~\ref{fig:hitrate} as a function of $z$-position of the counters. 
The rates are position dependent, and the maximum is \SI{160}{\kilo\hertz}.  This result is confirmed by measurements in the pilot run described in Sect.~\ref{sec:pTC-performance}.

\subsection{Calibration methods}
\label{sec:tc_calib}
%
It is also quite important to precisely synchronise all the counters, although the effect of the misalignment of the individual counter times can be diluted by taking the average over the multiple hit counters as seen from Eq.~(\ref{eq:resolution}).
Considering the dilution effect, $\sigma_{\tpositron}^\mathrm{inter\mathchar`-counter}\SI{\sim30}{\ps}$ is required for the precision of each counter time-alignment. 
Two schemes are under development for the inter-counter time-alignment.
They are complementary to each other and have independent systematic errors.

\subsubsection{Track-based method}
 \label{sec:track_calib}
  High momentum Michel positron pass through more than one counter as would signal one.
  Multiple hits allow time-alignment between adjacent counters after correcting for 
  the \positron\ travel time between hits.
  The track information analysed by the CDCH can be used for a precise correction of the travel time.
  Generally, the track-based method provides very precise results; $O(\mathrm{ps})$ is achieved in a
  study using a MC simulation. However, this method is subject to systematic position-dependent biases
  caused by small systematic errors in the travel time estimation. Such biases will be detected and
  corrected for by the laser-based method detailed in Sect.~\ref{sec:cal-laser}. Furthermore, this method cannot be used to synchronise the two super-modules.
     
\subsubsection{Laser-based method}
 \label{sec:cal-laser}
\begin{figure}[tb]
\centering
\includegraphics[width=1.\linewidth]{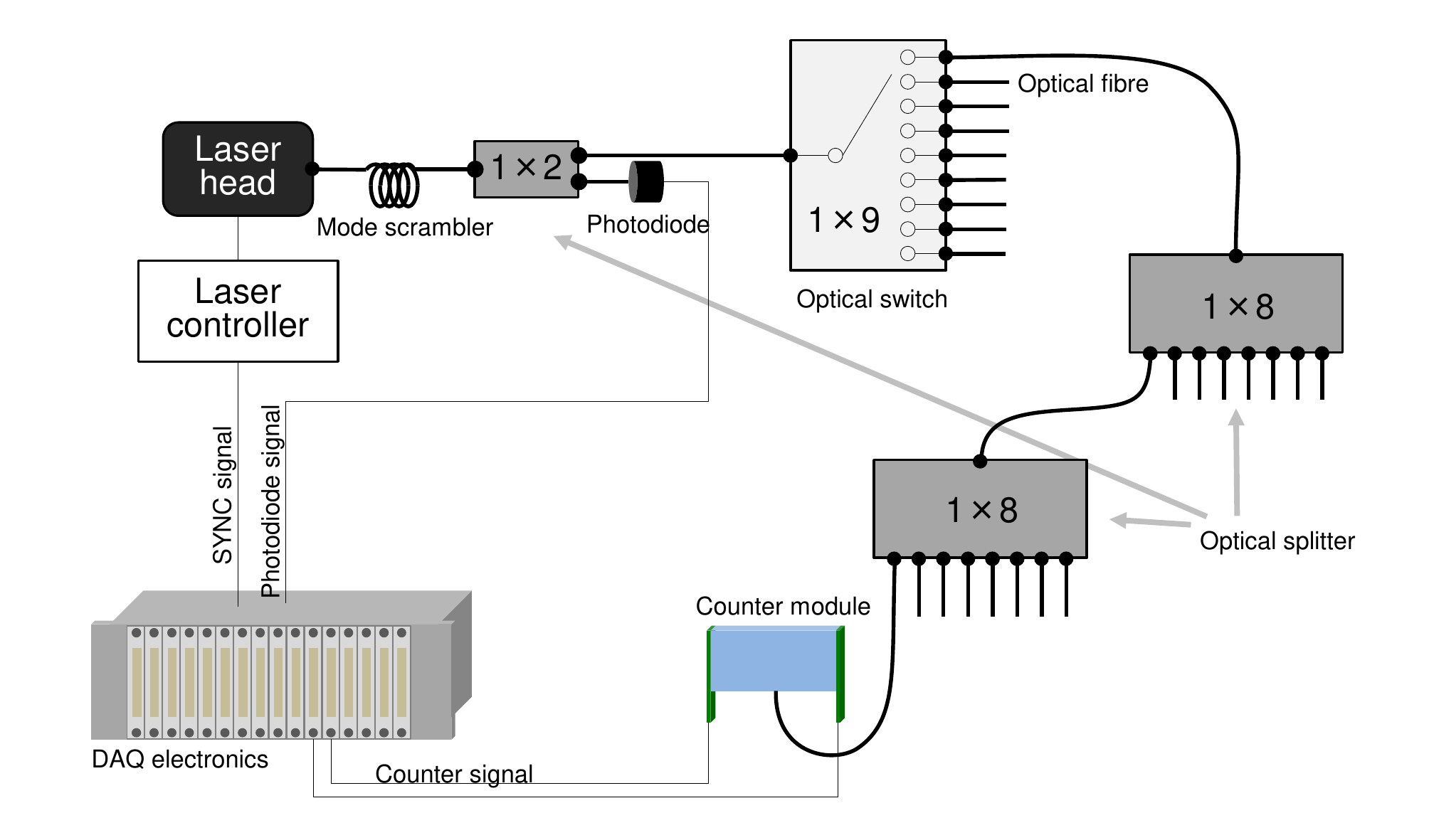}
\caption{Schematic of the pTC laser calibration system.}
\label{fig:laser_system}
\end{figure}

 The counters can also be time-aligned by distributing synchronous light pulse to all the counters through optical fibres.
 To this goal, we have developed a laser calibration system shown schematically in Fig.~\ref{fig:laser_system} (see also \cite{Bertoni_2016}).
 Ideally we should distributing lase light to all counters, however, it turned out to be impossible to install optical fibres to those at the innermost location 
 (in total 80 counters) due to space limitation. For those counters, we rely on the calibration by the track-based method detailed in 
      Sect.~\ref{sec:track_calib}.


The light source is a PLP10-040 \cite{HPK:PLP-10}, with an emission wavelength at \SI{401}{\nm}, pulse width of \SI{60}{\ps} (FWHM) and peak power of \SI{200}{\milli\watt}.
The fast light pulse is first split into two outputs; one is directed to a photodiode to gauge the signal amplitude, and the other 
serves as an input to an active optical multiplexer \cite{jena} with nine output channels and 
remotely controlled, such that the signal is outputted alternatively to each of them. 

\begin{figure}[tb]
\centering
\includegraphics[width=.5\linewidth]{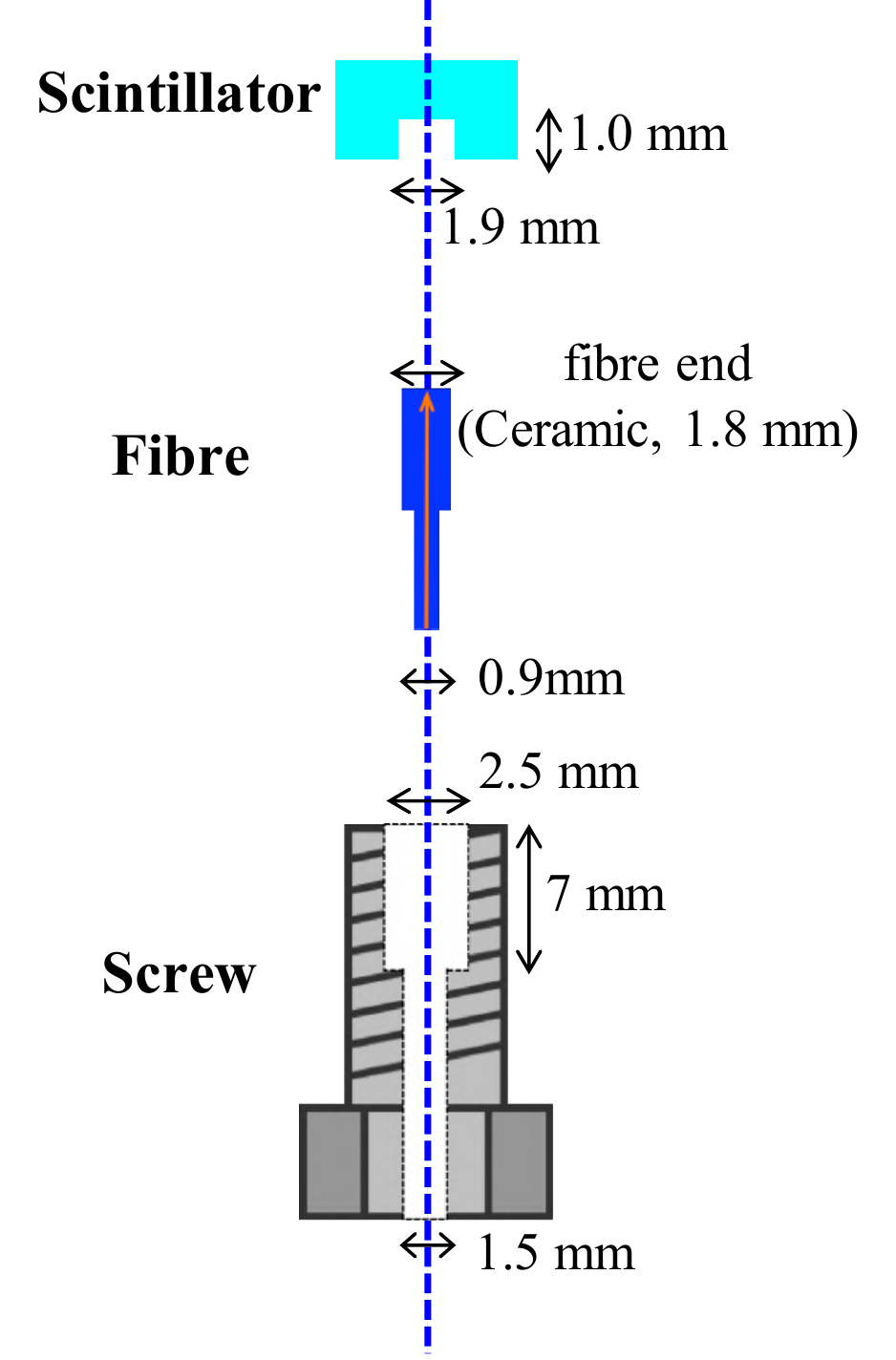}
\caption{Schematic of the fibre fixing method.}
\label{fig:fibre_fix_method}
\end{figure}

Each of the outputs of the multiplexer (except one used as a monitor) is then inputted to two cascaded stages of \num{1x8} optical splitters \cite{lightel}. 
each of which splits the input signal into eight signals of approximately equal output amplitudes. As a result, 64 channels become available in parallel with an
amplitude \num{\sim 1/64} of the original (actually smaller due to losses in the various stages). 
Finally, each output signal from the last stage of splitters is fed into a counter through an optical fibre.
Figure~\ref{fig:fibre_fix_method} shows how the optical fibre is fixed to the scintillator:
 to stably fix the fibre, a small hole (\SI{2.5}{\mm} diameter, \SI{1}{\mm} depth) is drilled into the bottom face of the scintillator,
and the ferrule of the fibre is inserted into the hole using a polycarbonate screw and a support bar (ABS resin) across the two PCBs.

\begin{figure}[tb]
\centering
\includegraphics[width=1.\linewidth]{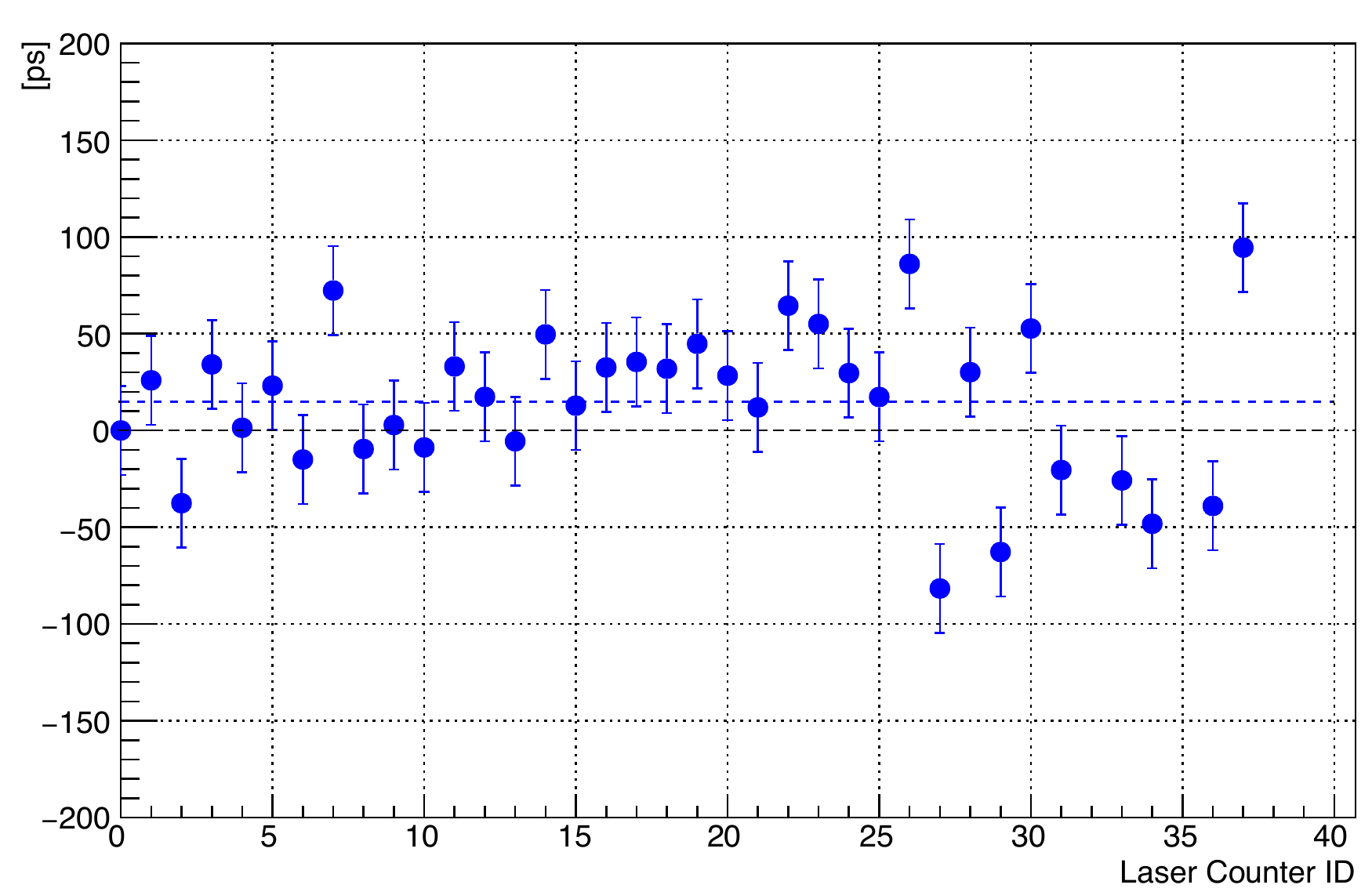}
\caption{\label{fig:timeoffset_difference}%
Difference of time offsets between the laser-based method and the track-based method at the beginning of the pilot run in 2016.
The difference was calculated only for the laser-installed-counters.
The standard deviation is \SI{39}{\ps}.
Each error bar includes systematic uncertainties of the two methods.
}
\end{figure}

As mentioned in Sect.~\ref{sec:pTC-performance}, we performed pilot runs using the MEG~II beam. 
In the 2016 run, we installed the laser calibration system for 40 counters and tested the system by examining the consistency with 
the track-based time calibration method detailed in Sect.~\ref{sec:track_calib}.
The time offset of each counter was calculated independently using two methods.
Figure~\ref{fig:timeoffset_difference} shows the difference between the results of the two methods. The dispersion (\SI{39}{\ps} in standard deviation) includes the systematic errors of both methods, 
therefore, the precision of each method is better. 
The time dependence was stable during the 3-week-run to a $\sigma = \SI{6}{\ps}$.
By combining the two methods, it is possible to calibrate all the counter offsets to a precision better than $\sigma_{\tpositron}^\mathrm{inter\mathchar`-counter}=\SI{30}{\ps}$.
The average contribution to the inter-counter calibration can be evaluated as $\sigma_{\tpositron}^\mathrm{inter\mathchar`-counter}/\sqrt{\bar{N}_\mathrm{hit}}=\SI{10}{\ps}$.

\subsection{Expected performance}
\label{sec:pTC-performance}
The single-counter performance is evaluated using electrons from a $^{90}\mathrm{Sr}$ source.
All the assembled counters were irradiated by the electrons at three positions (\SIlist[list-units = single]{-45;0;+45}{\mm} along $L$) 
to measure the time resolution, position resolution, light yield, and effective light speed $v_\mathrm{eff}$. 
The pulse time of each read-out side is picked-off by the digital-constant-fraction method ($t_1$ and $t_2$). The hit time is then reconstructed by averaging the two times, $(t_1+t_2)/2$, while the hit position along $L$ is reconstructed by the time difference, $(t_1-t_2)\times v_\mathrm{eff}/2$.
The mean time resolutions for all assembled counters are \SIlist{72;81}{\ps} for $W=$\SIlist{40;50}{\mm} counters, respectively. These are about 15\% worse than those obtained with the prototype counters in the R\&D phase because of the quality control of SiPMs and scintillators in the mass production phase.
The hit position resolution is $\sigma_{L}\SI{\sim10}{\mm}$.

The performance with multiple counters was studied in a series of beam tests carried out at 
the Beam Test Facility (BTF) at LNF and the $\mathrm{\pi E5}$ beam channel at PSI.
Six to ten prototype counters aligned as a telescope were irradiated by \SI{50}{\MeV} 
monochromatic positrons at the BTF or by Michel positrons at PSI.
The effects of multiple Coulomb scattering and secondary particles, 
such as $\delta$-rays, were examined, and the time resolutions was found to 
improve by use of multiple counters following closely Eq.~(\ref{eq:resolution}).
Detailed reports are available in \cite{Cattaneo:2015mcu, Nishimura:2015qev}.

Finally, we performed pilot runs in 2015 and 2016 using the MEG~II \muonp\ beam and the one-forth system of the pTC (consisting of 128 counters) installed in the COBRA spectrometer.
The system was thoroughly tested from the hardware point of view: the geometrical consistency, the installation procedure, and the operation under beam.
The laser calibration system was partially implemented and also tested.
Data from Michel positrons were also taken with a prototype of the WDBs, under various trigger conditions.

\begin{figure}[tb]
\centering
\includegraphics[width=1.\linewidth]{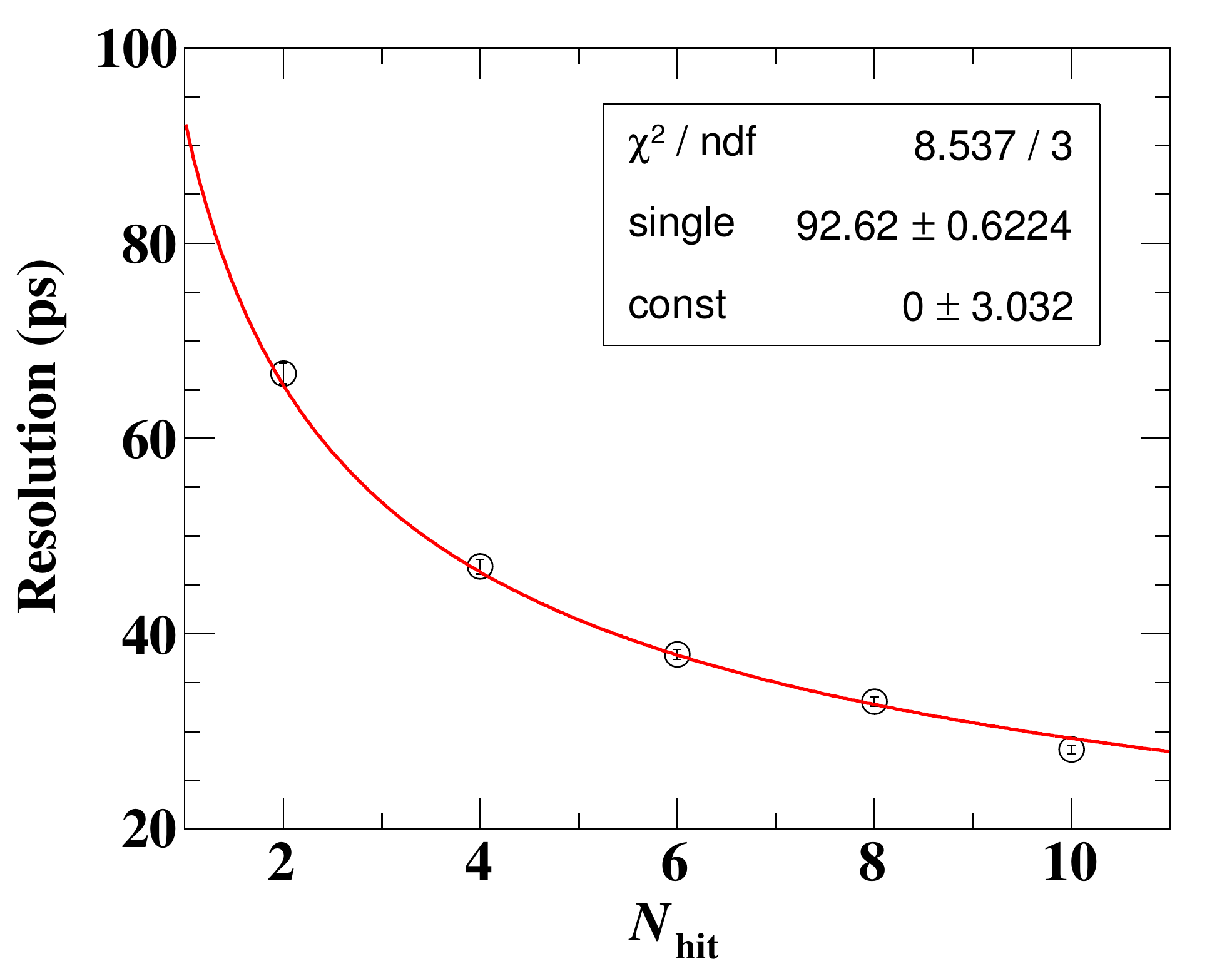}
\caption{The total time resolution vs. the number of hits measured by the \lq{}$\mathrm{odd}-\mathrm{even}$\rq{} analysis in the pilot run 2016. 
The points are the average of the 22 counter-sets weighted by the event fraction. 
The red curve is the best fit function of $\sigma_{\tpositron}(N_\mathrm{hit})=\sigma_{\tpositron}^\mathrm{single}/\sqrt{N_\mathrm{hit}} \oplus \sigma_{\tpositron}^\mathrm{const}$.}
\label{fig:nhit_resolution}
\end{figure}

The multi-counter time resolutions are evaluated by an \lq{}$\mathrm{odd}-\mathrm{even}$\rq{} analysis.
For a given set of hit counters, hits are alternately grouped into \lq{}odd\rq{} ($N_\mathrm{odd}$) and \lq{}even\rq{} ($N_\mathrm{even}$) by the order of the pixels
traversed by the positron, the time difference being defined as ($N_\mathrm{hit} = N_{\mathrm{even}} + N_{\mathrm{odd}}$)
\begin{align}
	t_\mathrm{odd-even}(N_\mathrm{hit}) = 
      \frac{1}{N_\mathrm{hit }}\left(\sum_{i=1}^{N_\mathrm{odd }} t_{\mathrm{hit} (2i-1) } - 
      \sum_{i=1}^{N_\mathrm{even}} t_{\mathrm{hit} (2i) }\right). \nonumber
	\label{eq:odd-even}
\end{align}
The standard deviation of $t_\mathrm{odd-even}(N_\mathrm{hit})$ is used as an estimator of the time resolution for $N_\mathrm{hit}$ hits and examined for 22 sets of counters.
Figure~\ref{fig:nhit_resolution} shows the result obtained in the pilot run 2016. The total time resolution improves as Eq.~(\ref{eq:resolution}) with $\sigma^\mathrm{single}_{\tpositron}$\SI{=93}{\ps}. 
At the mean $\bar{N}_\mathrm{hit} = 9$, $\sigma_{\tpositron}(\bar{N}_\mathrm{hit} =9)$\SI{=31}{\ps} was achieved.


\subsection{Other issues}

The modest radiation hardness of SiPMs is considered as a weak point of SiPMs.
Increase of the dark current and change of the gain of SiPMs are typical effects after substantial irradiation.
The SiPMs in the pTC will be irradiated by a high flux of Michel positrons during the experiment. 
The integrated fluence of the Michel positrons during three-years running is estimated to be 
\SI{\sim e11}{\positron\per\cm\squared}.\footnote{This estimation is based on the measured hit 
rate in the pilot run. It is twenty times higher than the previous estimation in \cite{Baldini:2013ke}.}

\begin{figure}[tb]                                                                                                                                                                  
\begin{center}
\includegraphics[width=1.\linewidth]{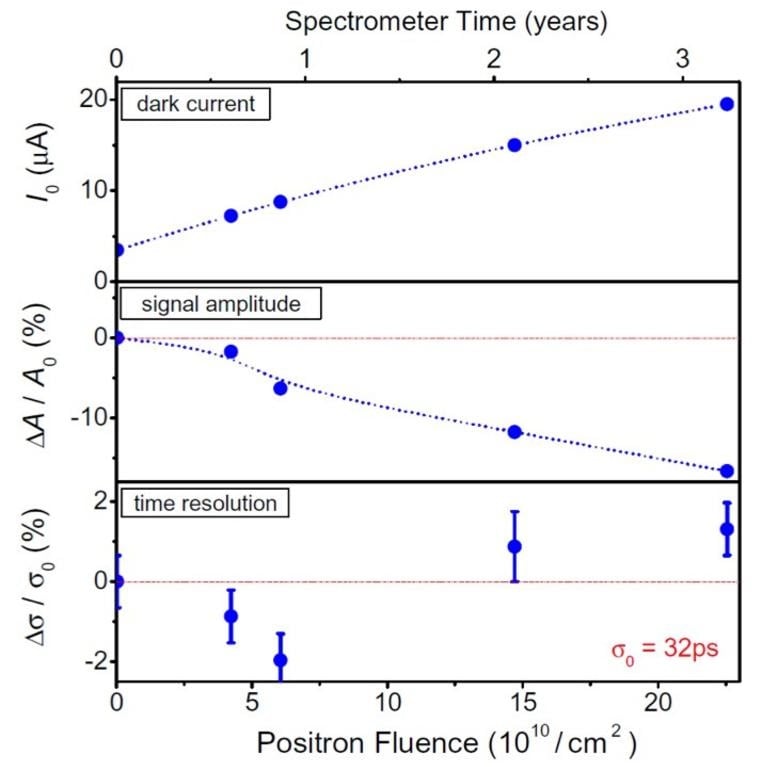}
\caption{\label{fig:muSR MPPC irradiation test}
Results from the irradiation tests of HPK SiPM (S10362-33-050C) performed by the PSI $\mu$SR group.
Significant increase of dark current (top) and 15\% gain decrease (middle) are observed, while the timing resolution is unchanged (bottom).
Courtesy from Dr. A.~Stoykov of Paul Scherrer Institut.
}
\end{center}
\end{figure}

The PSI $\mu$SR group performed irradiation tests using Michel positrons as shown in 
Fig.~\ref{fig:muSR MPPC irradiation test}\ \cite{muSR:irradiation-test}.
The SiPMs from HPK are irradiated by Michel positrons with fluences 
up to \SI{2.5e11}{\per\cm\squared}, which is more than twice higher than MEG~II expectation.
They observed a significant increase of the dark current by a factor of six and a 15\% gain decrease.   
Interestingly the timing resolution is unchanged even with the highest fluence.

During the pilot run, we observed increases in the SiPM current. 
By extrapolating the observed increase, the dark current of each channel would reach
$\mathcal{O}(\SI{100}{\micro\ampere})$ in the three years run.
This is higher than the expectation from the study above.
Further studies are necessary to assess the impact of the radiation damage on the timing performance.
We plan to carry out irradiation tests of our SiPMs and counter modules using high intensity $\beta$ sources and test beams such as BTF at LNF.

The SiPMs are also irradiated by neutrons and $\gamma$-rays in our experiment.
The effect is discussed in detail in Sect.~\ref{sec:LXe MPPC issues}
for the SiPMs planned to be used for the LXe photon detector
and it turns out not to influence the performance of SiPMs.

Another possible issue is the temperature stability of the SiPMs. 
The temperature coefficient of the breakdown voltage for ASD-NUM3S-P-50-High-Gain is 
\SI{26}{\milli\volt\per\degreeCelsius}; the gain at an over-voltage of \SI{2.5}{\volt} changes by 1\% for a temperature change of \SI{1}{\degreeCelsius}.
The temperature will be controlled and stabilised to within \SI{1}{\degreeCelsius} by an air-conditioning system of our detector hut and the cooling water system on the mechanical support structure.
Therefore, the temperature dependence of the SiPMs should not be an issue in our case. 

\clearpage
\newpage
\section{LXe photon detector}
\label{sec:LXe_Calorimeter}
\subsection{Upgrade concept }

The liquid xenon (LXe) photon detector is a key ingredient 
to identifying the signal and suppressing the background in the \megc\ search. 
It is, therefore, crucial to substantially improve its performance in MEG~II. 

\begin{figure}[b]
\centering
\includegraphics[width=1\linewidth]{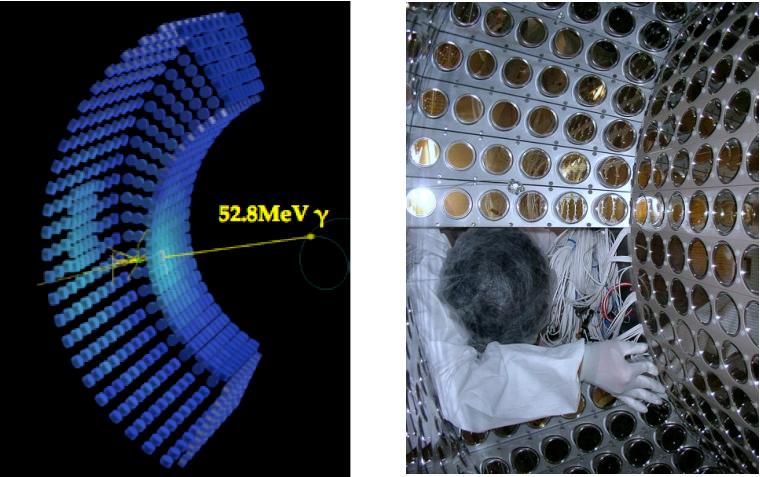}
\caption{\label{fig:current detector}%
   MEG LXe photon detector with \SI{900}{\litre} LXe surrounded by 846 UV-sensitive PMTs.
}
\end{figure}

The MEG LXe photon detector, shown in Fig~\ref{fig:current detector}, 
was one of the world's largest detectors based on LXe scintillation light
with \SI{900}{\litre} of LXe surrounded by 846 PMTs submerged in liquid to detect the scintillation light
in the VUV range ($\lambda = \SI[separate-uncertainty]{175(5)}{\nm}$).
The 2-\si{inch} PMT (Hamamatsu R9869) used in the detector is UV-sensitive 
with a photo-cathode of K-Cs-Sb and a synthetic quartz window.
The quantum efficiency (QE) was about 16\% for the LXe scintillation light at a 
LXe temperature of \SI{165}{\kelvin}. 

\begin{figure}[tb]                                                                                                                                                                  
\centering
\includegraphics[width=1\linewidth]{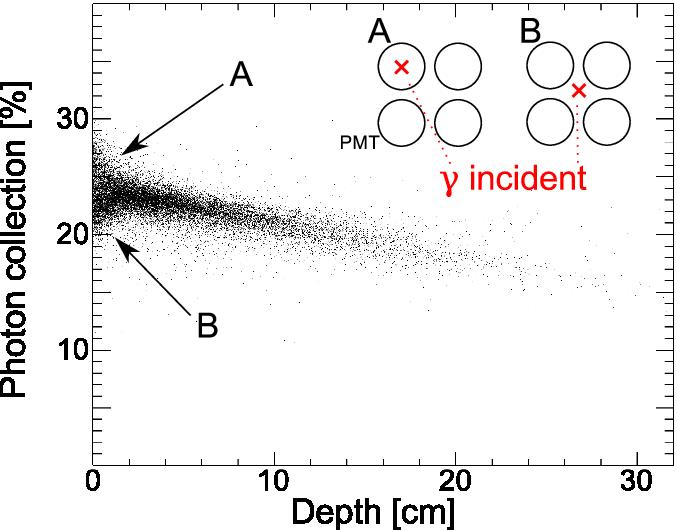}
\caption{\label{fig:current detector non-uniformity}%
   Efficiency of the scintillation light collection estimated by a MC simulation as a function 
      of the depth of the first interaction of a signal photon of \SI{52.8}{\MeV}. 
}
\end{figure}

The photon entrance inner face was covered by 216 PMTs with a minimum spacing between adjacent PMTs.
The photo-cathode of the PMT was, however, round-shaped with a diameter of \SI{46}{\mm} 
which was much smaller than the interval between adjacent PMTs of \SI{62}{mm}.
The performance of the MEG LXe photon detector was limited due to this non-uniform PMT coverage. 
Figure~\ref{fig:current detector non-uniformity} shows the efficiency of scintillation light collection 
as a function of the depth of the first interaction for signal photons of \SI{52.8}{\MeV}.
The collection efficiency strongly depended on the incident position.
The non-uniform response was partly corrected for in the offline analysis,
but it still deteriorated the energy and position resolutions due to event-by-event fluctuations of the shower shape,
    especially for shallow events.

\begin{figure*}[tb]                                                                                                                                                                  
\centering
\includegraphics[width=0.45\textwidth]{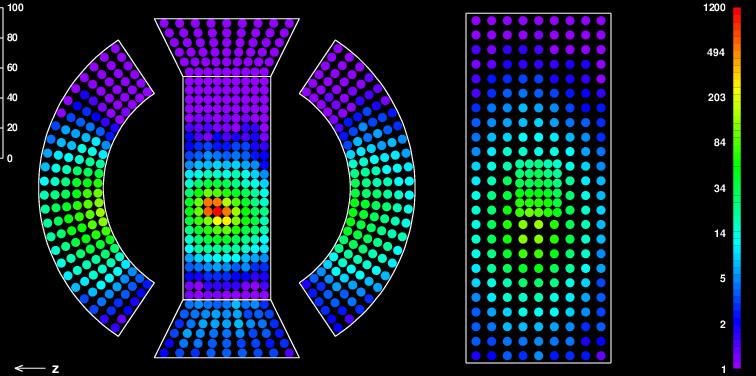}
\includegraphics[width=0.45\textwidth]{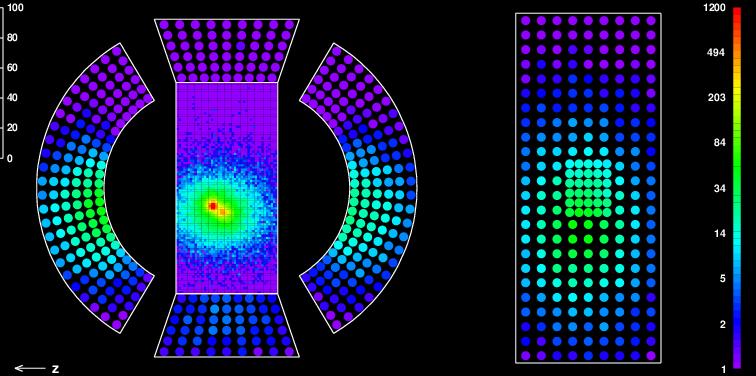}
\caption{\label{fig:comparison of imaging}%
   Example of scintillating light distributions detected by photo-sensors 
      in case of (left) PMTs and (right) smaller photo-sensors (\SI[product-units=single]{12x12}{\mm\squared}) 
      on the inner face for the same MC event.
}
\end{figure*}
\begin{figure}[tb]                                                                                                                                                                  
\centering
\includegraphics[width=1\linewidth]{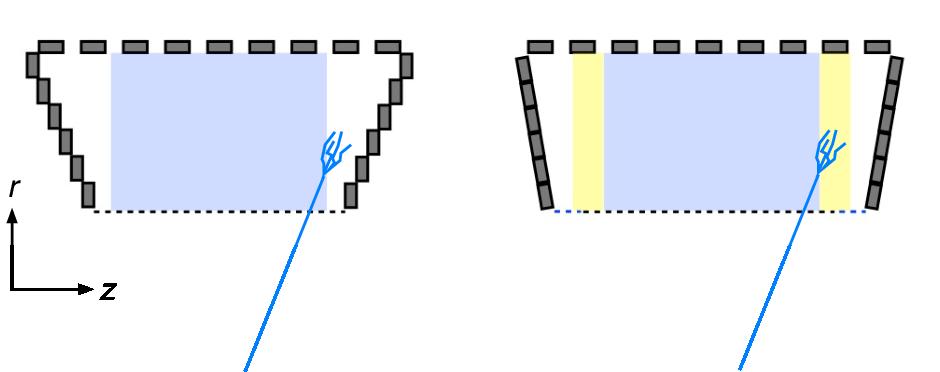}
\caption{\label{fig:PMT layout modification}%
MEG (left) and MEG~II (right) layouts of the PMTs viewed in an $r$-$z$ section.
}
\end{figure}

The main concept of the upgrade of the LXe photon detector for MEG~II is to reduce this non-uniform response 
by replacing the PMTs on the inner face with smaller photo-sensors.
Figure~\ref{fig:comparison of imaging} shows a comparison of how the same event would look for
the two cases with the current PMTs and smaller photo-sensors (\SI[product-units=single]{12x12}{\mm\squared})
on the inner face. 
The imaging power is greatly improved with smaller photo-sensors.
For example, two local energy deposits in the same shower are clearly separated in this event.
It turns out that both the energy and position resolutions greatly improve 
especially for shallow events as shown in Sect.~\ref{sec:expected performance}.\\
SiPMs are adopted as smaller photo-sensors for the inner face of the MEG~II LXe photon detector.
The motivation for choosing SiPM is discussed in detail in Sect.~\ref{sec:MPPC}.\\
The PMTs which were used on the inner face of the MEG LXe photon detector are re-used on the other faces.
It turns out by detailed MC studies that the best use of those PMTs
is achieved by modifying the layout of the PMTs on the lateral faces.
Figure~\ref{fig:PMT layout modification} illustrates the modified layout viewed on a $r$-$z$ section.
The inner face extends along $z$, outside the acceptance region by 10\% on each side.
The extended volume reduces the energy leakage for events near the lateral walls. 
The PMTs on the lateral faces are tilted such that all the photo-cathodes lie in the same plane.
This configuration minimises the effect due to shower fluctuations for events near the lateral walls.
The energy resolution is thus improved especially for those events.

\subsection{Development of VUV-sensitive MPPC}
\label{sec:MPPC}
\subsubsection{MPPC advantage}

The MPPC\textsuperscript{\textregistered} (Multi-Pixel Photon Counter), a new type of photon counting device produced 
by Hamamatsu Photonics K.K., is a kind of SiPM device.
The MPPC has many excellent features suited for the MEG~II experiment. 
It is insensitive to magnetic fields and is sensitive to single photons, 
which enables an easier and more reliable calibration of the detector.
Moreover, a finer read-out granularity of the scintillation light with MPPCs
allows for a more precise reconstruction of shallow events.
Less material budget before the LXe active region results in a 9\%\ higher detection efficiency,
as discussed in Sect.~\ref{sec:expected performance}.
The typical bias voltage is less than \SI{100}{\volt}. 

\subsubsection{Issues}
\label{sec:LXe MPPC issues}
There are several issues to be addressed concerning the detection of LXe scintillation light by MPPCs.

The first issue is the photon detection efficiency (PDE) for VUV light.
There are two types of layer structures for the SiPM, p-silicon on an n-substrate (p-on-n) and n-on-p. 
In general, since the ionisation coefficient for electrons is higher than that for holes, 
the breakdown initiation probability of electrons is always higher than that of holes. 
Blue light is absorbed close to the SiPM surface and electrons initiate the avalanche breakdown in the p-on-n case, 
which results in a higher sensitivity in the blue light region.
Our MPPC uses the p-on-n structure, which is suitable to detect the blue light.
The PDE of standard MPPC for VUV light is, however, nearly zero 
since VUV photons can not reach the sensitive layer due to a protection coating layer made of epoxy resin or silicon rubber.
Furthermore, an anti-reflection (AR) coating layer is not optimised to 
the refractive index of LXe at the scintillation light wavelength.

The second issue is the MPPC size.
The current largest single MPPC commercially available is \SI[product-units=power]{6 x 6}{\mm}, 
which is still too small to cover the inner face of the LXe photon detector with an affordable
number of read-out channels.
It is desirable to develop a large-area MPPC with \SI[product-units=power]{10x10}{\mm} or larger.
However, the larger size of MPPCs could cause a larger dark count rate, larger gain non-uniformity, 
and larger capacitance (longer tail in the waveform, larger noise etc.) \cite{Ootani:2015cia}.  

A large-area UV-sensitive MPPC has been developed in collaboration with Hamamatsu Photonics 
to be used in the upgraded LXe photon detector.
We will describe its characteristics in the following sections.

\subsubsection{PDE}
\label{sec:PDE}
Many prototypes optimised for VUV detection have been produced by Hamamatsu Photonics,
which have no protection coating, a thinner contact layer
or optimised AR coating with different parameters.

\begin{figure}
   \centering
      \includegraphics[width=1\columnwidth,angle=0, clip, trim=0 0 0 20pt]{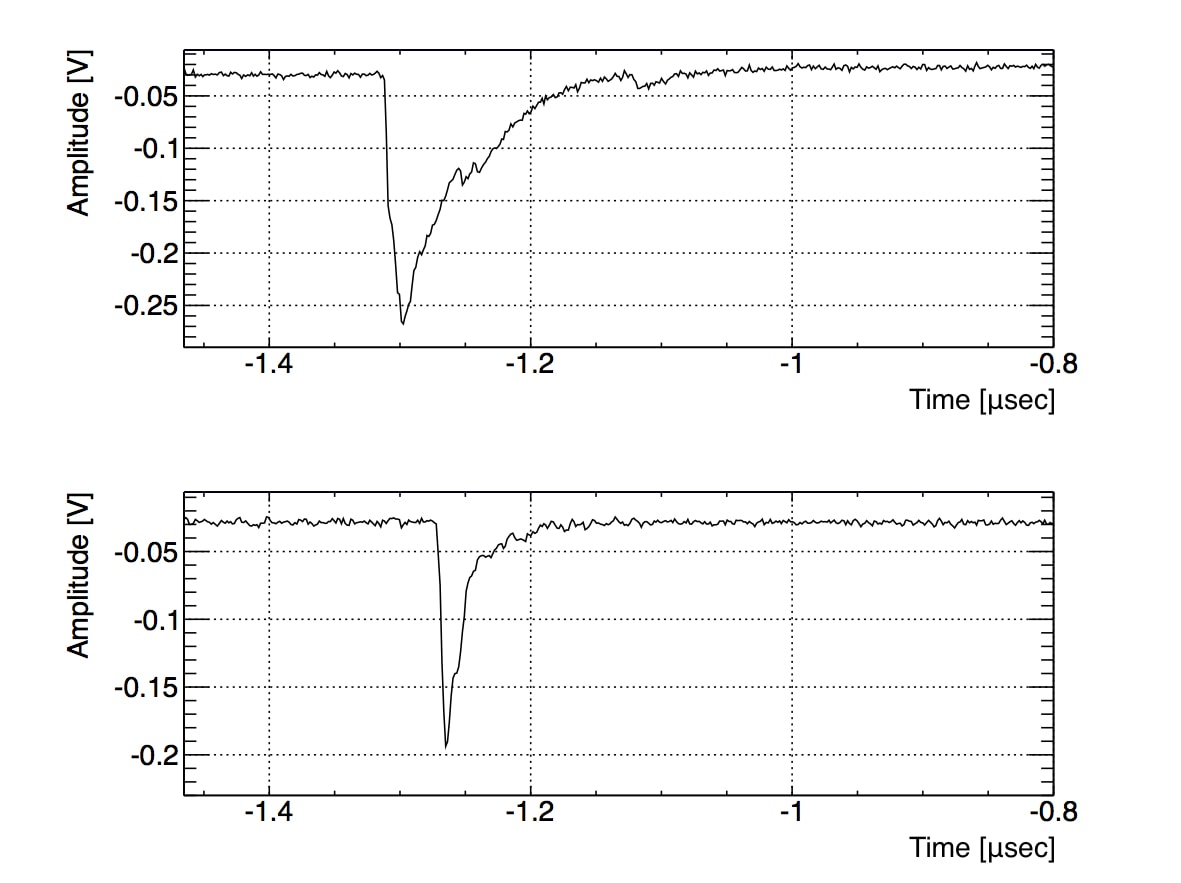}
      \caption[]{An MPPC signal waveform (upper) and a PMT signal waveform (lower) for the same $\alpha$-event digitised
        at a sampling frequency of 700~MSPS.}
      \label{fig:mppc_pmt_waveform}
\end{figure}

We succeeded in detecting the LXe scintillation light from $\alpha$-events by using one such prototype sample.
Figure~\ref{fig:mppc_pmt_waveform} shows signal waveforms 
from the MPPC sample (upper figure) and a UV-sensitive PMT (lower one)
for the same $\alpha$ event. 


The number of detected photoelectrons for $\alpha$-event is calculated from the ratio of the observed charge to that obtained 
for a single photoelectron event.
The PDE is then estimated from a ratio of the detected number of photoelectrons 
to the expected number of incoming scintillation photons from $\alpha$-events.
This PDE still contains contributions from cross-talk, after-pulses, and the infrared component of the LXe scintillation light.
The contribution from the infrared component is estimated to be $\sim$1\%\ indirectly, by using the signal observed with a commercial MPPC (S10362-33-100C),
which is supposed to be insensitive to the VUV component.
The cross-talk + after-pulse components are estimated using a flashing LED in such a way that the MPPC detects less than 1~p.e. on average. 
The expected 1~p.e. probability ($p_\mathrm{1~p.e. expected}$) is calculated 
from the Poisson distribution with the mean estimated from the observed probability of 0~p.e. events.
We can estimate the cross-talk+after-pulse probability 
by comparing this with the measured probability of 1~p.e. events ($p_\mathrm{1~p.e.~measured}$)\cite{Yokoyama:2008hq}.
This method yields a
cross-talk + after-pulse probability
= $(p_\mathrm{1~p.e. expected}-p_\mathrm{1~p.e. measured})/p_\mathrm{1~p.e. measured}$.
of between \numrange[range-phrase=--]{10}{50}\%, depending on the over-voltage.

\begin{figure}
   \centering
      \includegraphics[width=.45\textwidth,angle=0]{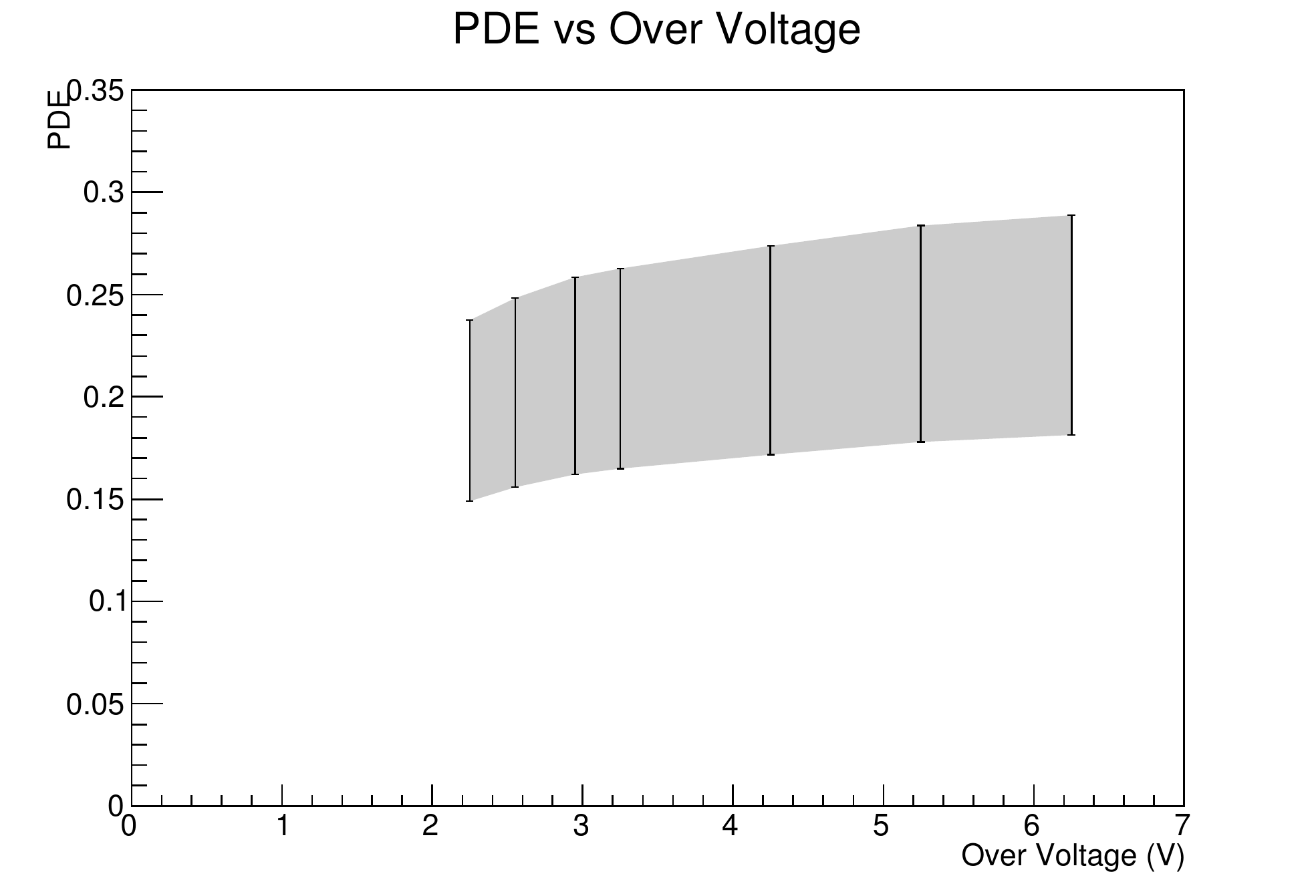}
      \caption[]{Measured PDEs as a function of the over-voltage.
      The large uncertainty mainly stems from the different measurement setups.
      }
      \label{fig:MPPC_PDE}
\end{figure}

Figure~\ref{fig:MPPC_PDE} shows the measured PDEs for four MPPC samples after correcting for the contributions from cross-talk and after-pulses.
There is roughly a 30\% uncertainty in the PDE value, that is estimated from the variation of the PDE measured in different setups.
The result shows that the PDE is higher than the 15\%\ PDE measured in LXe, which is similar to the QE for the UV-sensitive PMT of the current detector ($\sim$16\%).
Since the sensor coverage on the inner face is increased by 50\%\ using MPPCs, the total photoelectron statistics would be increased.


\subsubsection{Temperature dependence}

\begin{figure}[tb]
   \centering
      \includegraphics[width=.45\textwidth,angle=0]{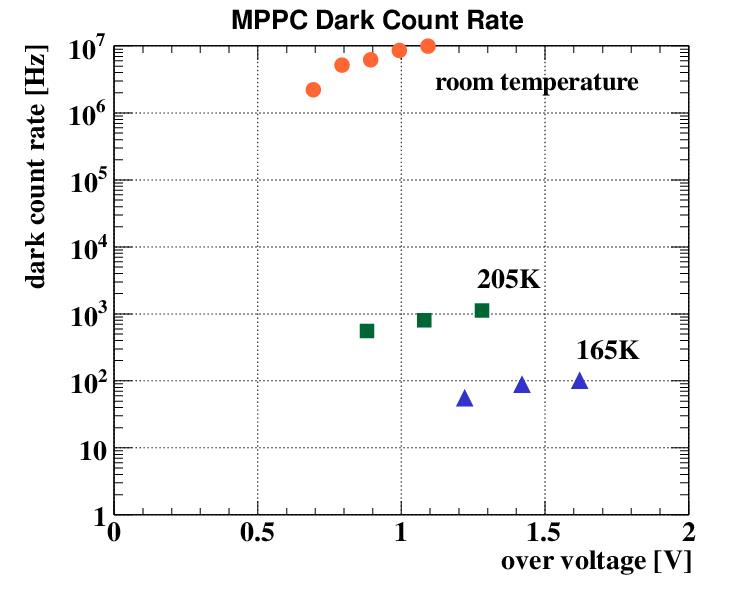}
      \caption[]{The dark count rate measured at different temperatures (room temperature, 205~K and 165~K). }
      \label{fig:MPPC_dark}
\end{figure}

Thermally generated free carriers in a depleted layer produce dark counts.
The typical dark count rate is \SIrange[range-phrase=--,range-units=single]{0.1}{10}{\MHz\per\mm\squared} at room temperature.
The dark count rate is known to be suppressed by five orders of magnitude at 
LXe temperature (\SI{165}{\kelvin}) \cite{JanicskoCsathy:2010bh}.
Our test measurements confirm that the dark count rate is reduced down to 
\SIrange[range-phrase=--,range-units=single]{1.}{100}{\hertz} for \SI[product-units=single]{3x3}{\mm\squared} samples 
at LXe temperature as shown in Fig.\ref{fig:MPPC_dark}.

Poly-silicon was used in the previous versions of MPPCs as quenching resistors, but now switched 
to metal resistors are more common.

However, the resistivity of the poly-silicon increases when the temperature decreases, for example, 
the resistance at LXe temperature is measured to be more than a factor of two higher than at room temperature.
A metal quenching resistor,
which has 1/5 of the temperature coefficient of a poly-silicon resistor,
is more suitable in our application 
in order to keep the quenching resistance low, which can avoid long fall time of MPPC signal.

The breakdown voltage of MPPCs is known to have a relatively large temperature coefficient (\SI{56}{\mV\per\degreeCelsius})
and the gain and PDE can, therefore, easily shift depending on the temperature, 
influencing the stability of the detector performances.

The LXe temperature stability of the MEG LXe photon detector has been measured to be smaller 
than \SI{0.15}{\kelvin} (RMS), most likely dominated by the precision of the temperature measuring device.
The fluctuation of the MPPC gain at an over voltage around \SI{7}{\volt} is expected to be smaller than 0.1\% (RMS). 
The PDE at around \SI{7}{\volt} over voltage is already saturated, and no fluctuations are expected from 
temperature variation.
The voltage dependence of the cross-talk and the after-pulse of the MPPC should be smaller than 30\%/V, which corresponds to
0.23\%. 
These fluctuations are smaller than the expected energy resolution of the MEG~II LXe photon detector described in Sect.~\ref{sec:expected performance}.

\subsubsection{Radiation hardness}

Radiation produces defects in the silicon bulk or at the Si/SiO$_{2}$ interface of SiPMs. 
As a result, some parameters of SiPMs such as the breakdown voltage, leakage current, 
dark count rate, gain, and PDE may change after irradiation. 
There have been many studies on the radiation hardness of SiPMs 
irradiated by photons, neutrons, protons, or electrons. These studies show the following.

An increased dark count rate was observed at more than \SI{e8}{n\per\cm\squared}, 
and loss of single p.e. detection capability was observed at more than 
\SI{e10}{n\per\cm\squared}~\cite{Matsumura:2006zz}. 
From the neutron flux measured in the MEG experimental area, the total neutron 
fluence is estimated to be less than \SI{1.6e8}{n\per\cm\squared} in MEG~II.

Increased leakage current was observed with a photon irradiation of \SI{200}{\gray}~\cite{Matsubara:2006zz}, 
while the photon dose in the MEG~II is estimated to be \SI{0.6}{\gray}. 

The radiation damage by photons, or neutrons should not be an issue for the MPPCs in MEG~II.

\subsubsection{Linearity}

\begin{figure}[tb]
   \centering
      \includegraphics[width=1\linewidth,angle=0]{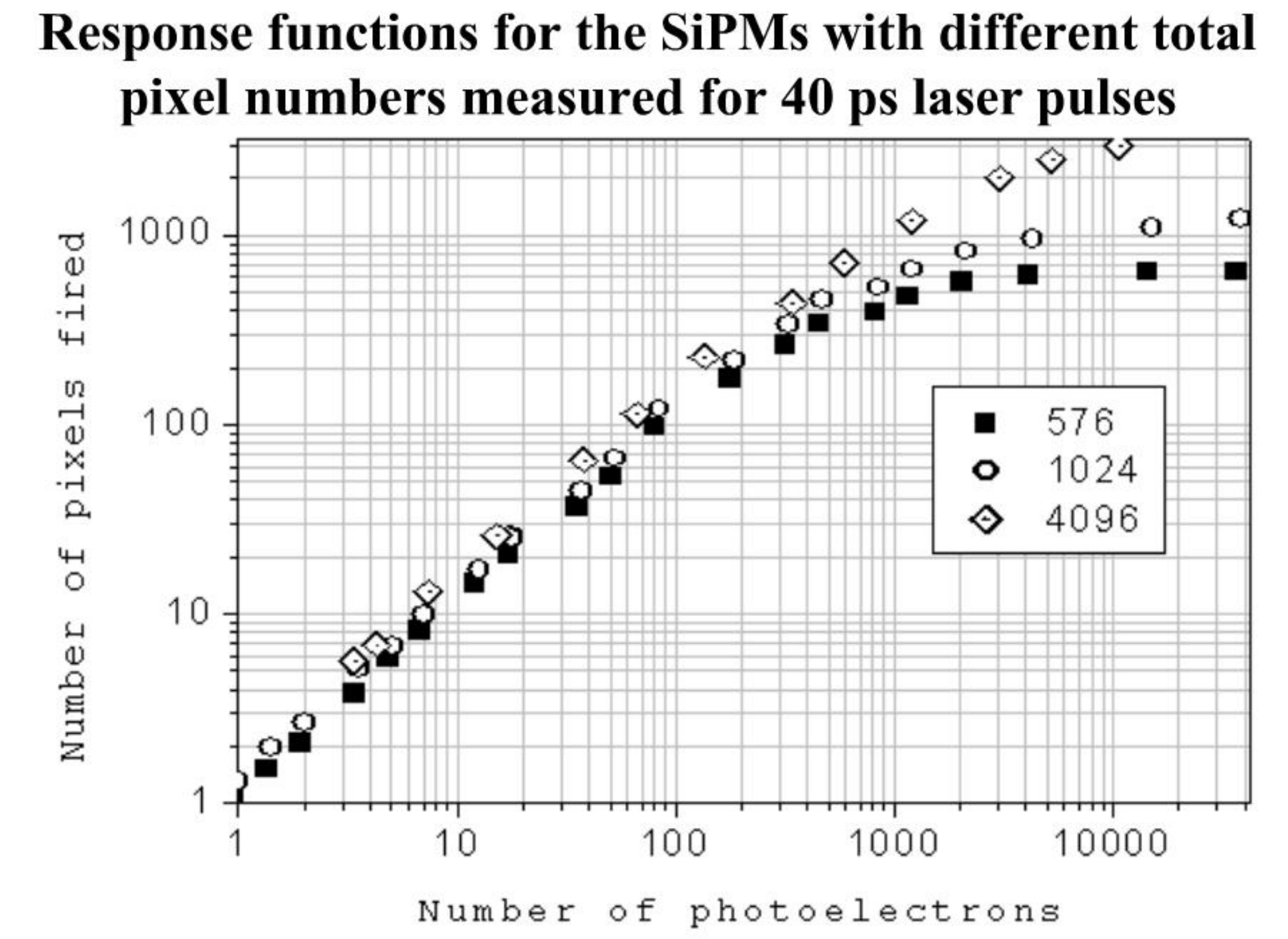}
      \includegraphics[width=.45\textwidth,angle=0]{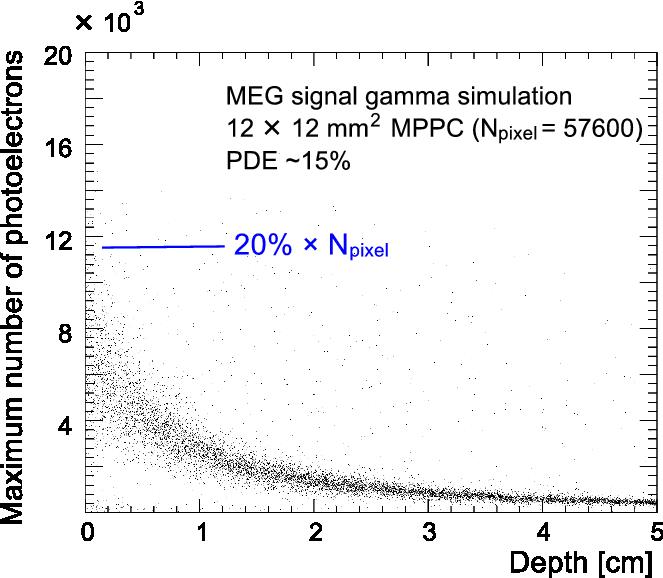}
      \caption[]{(Top) Response functions for the SiPMs with different total pixel numbers measured for a 40~ps laser pulses~\cite{Andreev:2004uy}.
        (Bottom) The number of photoelectrons expected from a \SI[product-units=single]{12x12}{\mm\squared} MPPC versus conversion depth in the MEG~II MC simulation.}
      \label{fig:MPPC_linearity}
\end{figure}

SiPMs show a non-linear response when the number of incident photons is comparable to or 
larger than the number of pixels of the device.
The optimal condition is that the number of incident photons be much smaller than the number 
of pixels without any localisation.
Figure~\ref{fig:MPPC_linearity} (top) shows the measured response functions for 
\SI[product-units=single]{1x1}{\mm\squared} SiPMs with different total numbers of pixels 
illuminated by a \SI{40}{\ps} laser pulse~\cite{Andreev:2004uy}. 
For the MEG~II LXe detector, the expected number of photoelectrons reaches up to \num{12000} p.e. 
on \SI[product-units=single]{12x12}{\mm\squared} sensor area for very shallow signal events as shown in Fig~\ref{fig:MPPC_linearity} (Bottom), which 
is only 20\%\ of the total number of pixels, \num{57600}. Considering also that some of the fired 
pixels are recovered during the emission time of the scintillation light, the expected 
non-linearity is, therefore, small and can be corrected for by a careful calibration.

\subsubsection{Large area MPPC}
\label{sec:largeMPPC}

The current largest MPPC (\SI[product-units=single]{6x6}{\mm\squared}) is still too small for the MEG~II LXe photon detector,
and we need at least \SI[product-units=single]{10x10}{\mm\squared} to replace the PMTs. 
For a larger size sensor, we have to pay attention to a possible increase
in the dark count rate, an increase of the sensor capacitance, and gain non-uniformity over the sensor area. 

The increase of the dark count rate is not an issue in MEG~II due to the LXe temperature. 
To reduce the sensor capacitance, the large area of \SI[product-units=single]{12x12}{\mm\squared} is formed by 
connecting in series four smaller MPPCs (\SI[product-units=single]{6x6}{\mm\squared}) each).
In this configuration, the decay constant of the signal waveform becomes \SIrange[range-phrase=--,range-units=single]{40}{50}{\ns} from \SI{130}{\ns}.
To equalise the gain in the large area sensor, four MPPCs are selected
with similar breakdown voltages.

Instead of a simple series connection, each sensor chip is decoupled with a capacitor 
to enable the bias voltage to be supplied via a parallel connection. 
In this way, we can still extract signals from the series connection, 
and the common bias voltage such as \SI{\sim 55}{\volt} can be supplied to the four sensor chips.

\subsection{Detector design}
\subsubsection{Design of sensor package and assembly}
\label{sec:Design of Sensor Package and Assembly}%
\begin{figure}[tb]
\centering
\includegraphics[width=1\linewidth]{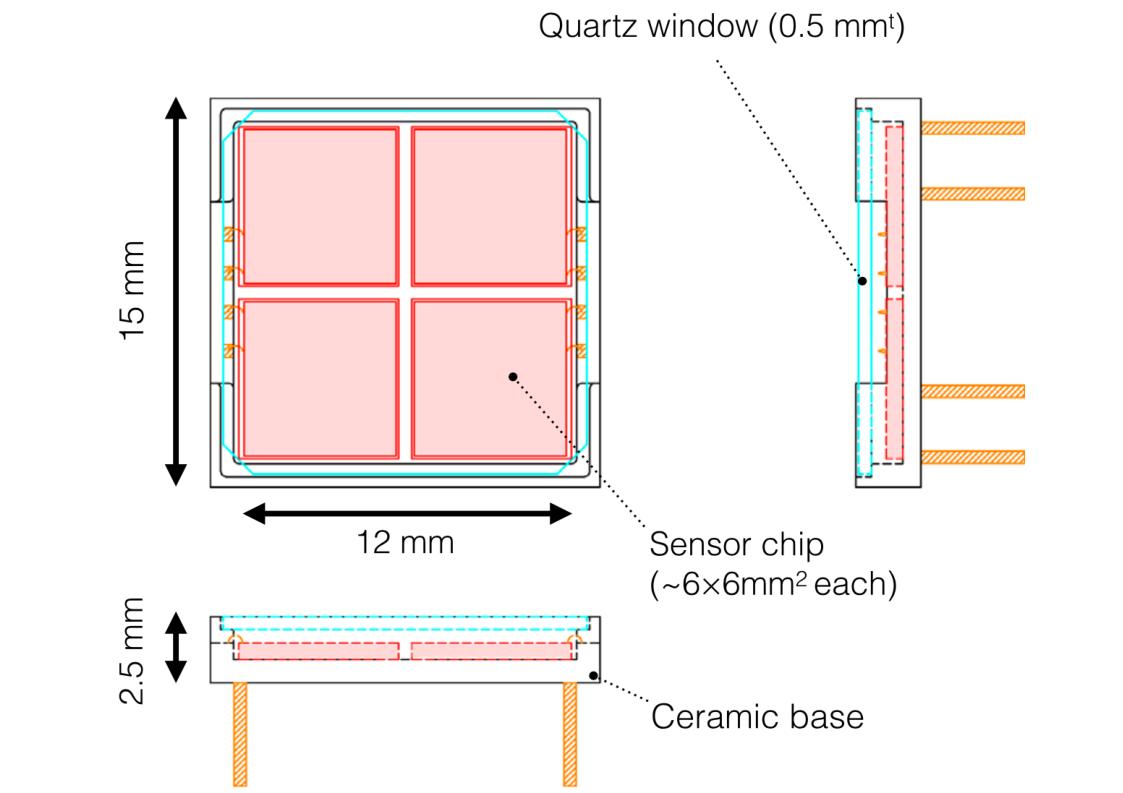}
\caption{\label{fig:MPPC package design}%
   MPPC package design.
}
\end{figure}

Figure~\ref{fig:MPPC package design} shows a design of the UV-enhanced MPPC package used for the MEG~II LXe photon detector.
Four sensor chips with a total active area of \SI[product-units=single]{12x12}{\mm\squared} 
are glued on a ceramic base of \SI[product-units=single]{15x15}{\mm\squared}.
The ceramic is chosen as a base material because the thermal expansion rate is close to that of silicon at LXe temperatures. 

\begin{figure}[tb] 
\centering
\includegraphics[width=1\linewidth]{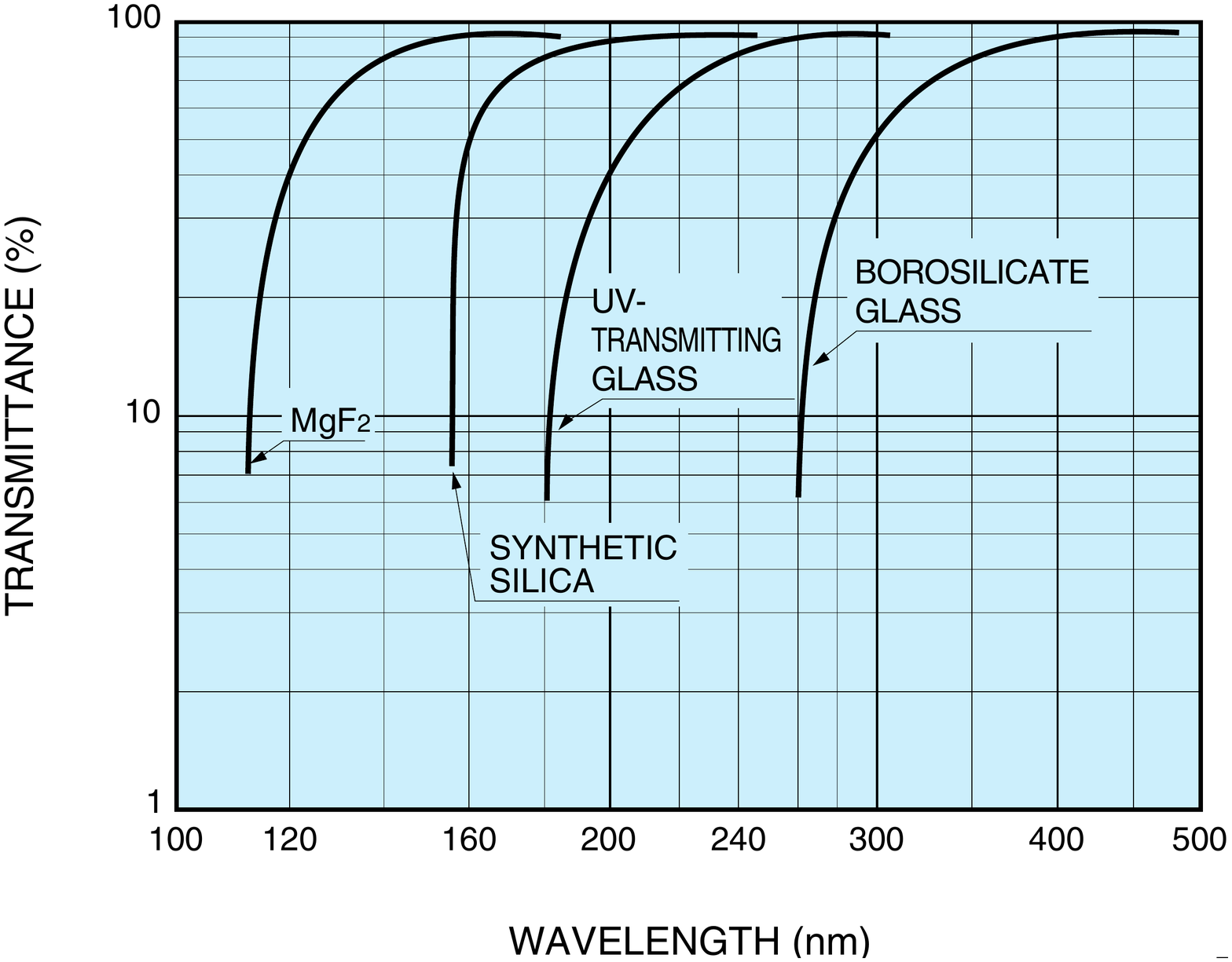}
\caption{\label{fig:quartz transmission}%
   Transmission efficiency as a function of wavelength for high quality VUV-transparent quartz.
}
\end{figure}

The sensor active area is covered with a thin 
high quality VUV-transparent quartz window for protection.
The window is not hermetic; there is a gap between the sensor
and the window in which LXe penetrates.
Figure~\ref{fig:quartz transmission} shows the transmission efficiency of different window materials 
as a function of wavelength~\cite{quartz_transmittance},
showing that the transmittance of the synthetic silica, which is used in our MPPC, is quite high for the LXe scintillation light (\SI{175}{\nm}).
The reflection loss is small since both sides of the quartz window touch LXe
whose refractive index is close to that of the quartz window
($n_\mathrm{LXe}=1.64$, $n_\mathrm{quartz} = 1.60$).  

\begin{figure}[tb] 
\centering
\includegraphics[width=1\linewidth, clip, trim=0 50pt 0 50pt]{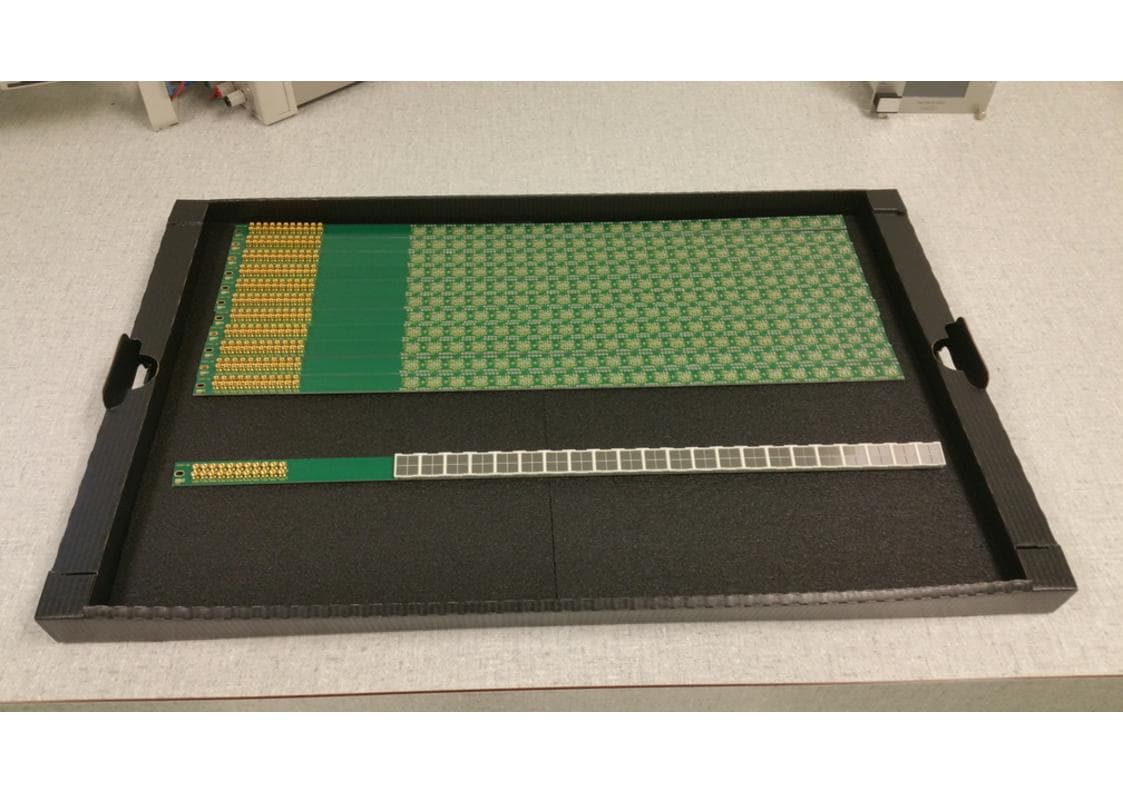}
\caption{\label{fig:MPPC mounting PCB}%
PCBs for MPPC mounting. One PCB in the front has already 22 MPPCs mounted.}
\end{figure}

\begin{figure}[tb]
\centering
\includegraphics[width=1\linewidth, clip, trim=0 50pt 0 50pt]{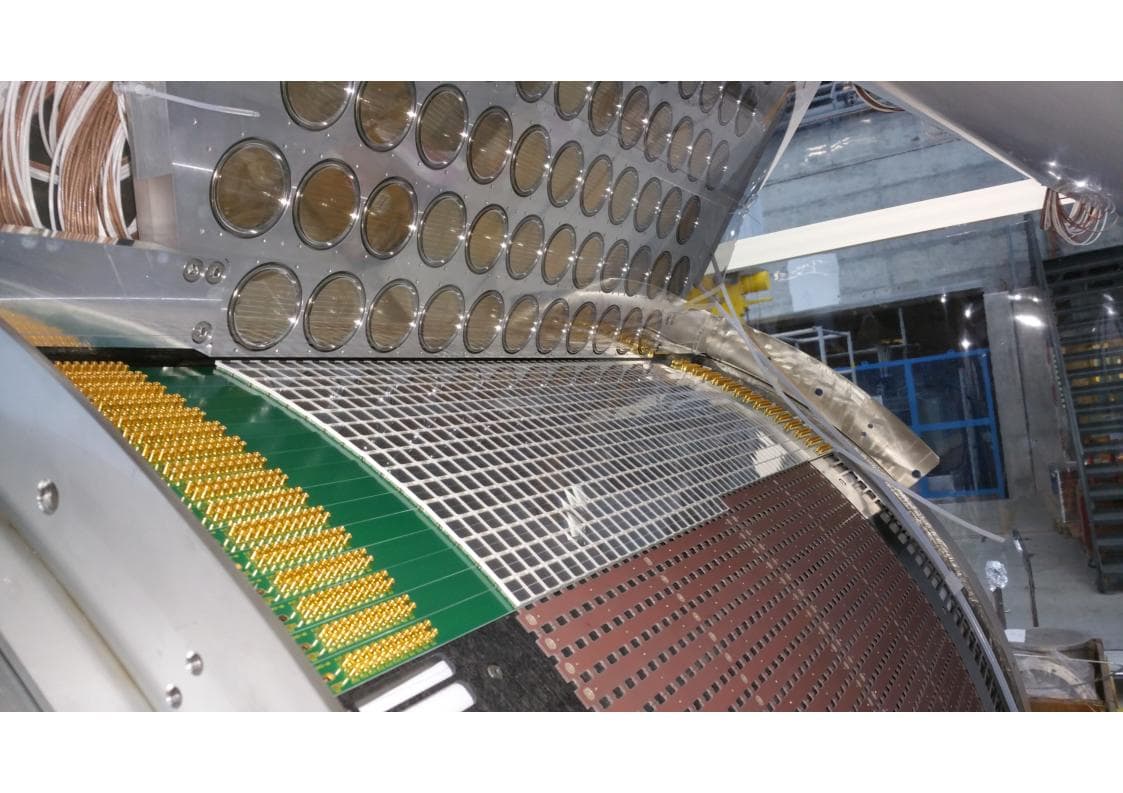}
\caption{\label{fig:MPPC slab assembly}%
   Installing PCB strips onto the inner face.
}
\end{figure}

The MPPCs are mounted on a PCB strip as shown in Fig.~\ref{fig:MPPC mounting PCB}.
Each PCB strip has 22 MPPCs, and two PCBs are mounted in a line along the $z$-direction
with 93 lines (186 strips) covering the $\phi$-direction 
on the inner wall of the detector cryostat as shown in Fig.~\ref{fig:MPPC slab assembly}.
The total number of MPPCs is 4092.
One MPPC package has eight electrode pins (an anode and a cathode from each sensor chips)
 which are plugged into the corresponding sockets on the PCB.
 This mounting scheme allows easy replacement of the MPPC module if necessary.
The PCB has additional circuit parts of capacitances and resistors to realise the signal line in series
and the bias line via parallel connection as described in Sect.~\ref{sec:largeMPPC}.

The signals from the MPPCs are transmitted on the signal lines of the PCB 
which are designed to be well shielded from both outside and the adjacent channels and have a \SI{50}{\ohm} impedance.
Similar PCBs are used in the feed-throughs of the cryostat as described in Sect.~\ref{sec:cable and feed-through}.

\begin{figure}[tb]
\centering
\includegraphics[width=1\linewidth]{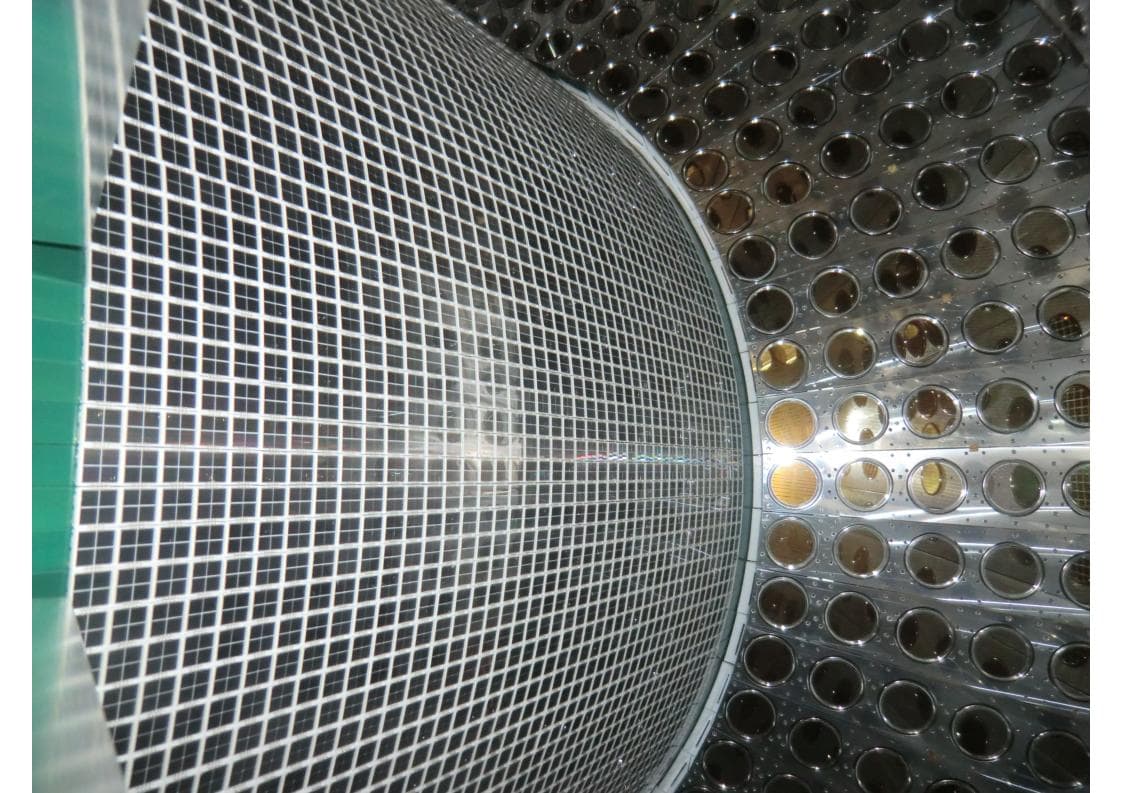}
\caption{\label{fig:MPPC Photo}%
   The inside of the LXe photon detector after the MPPCs and PMTs are assembled.
}
\end{figure}

It is important to precisely align the PCB strips on the inner wall of the detector cryostat
and to minimise the gap between the strips and the wall 
since LXe in this gap deteriorates the photon detection efficiency
and causes an undesirable low energy tail in the energy response function of the detector.
Figure~\ref{fig:MPPC Photo} shows the inside of the LXe photon detector after the MPPCs and PMTs are mounted. 

\subsubsection{Design of PMT support structures}
\begin{figure}[tb]
\centering
\includegraphics[width=0.35\textwidth]{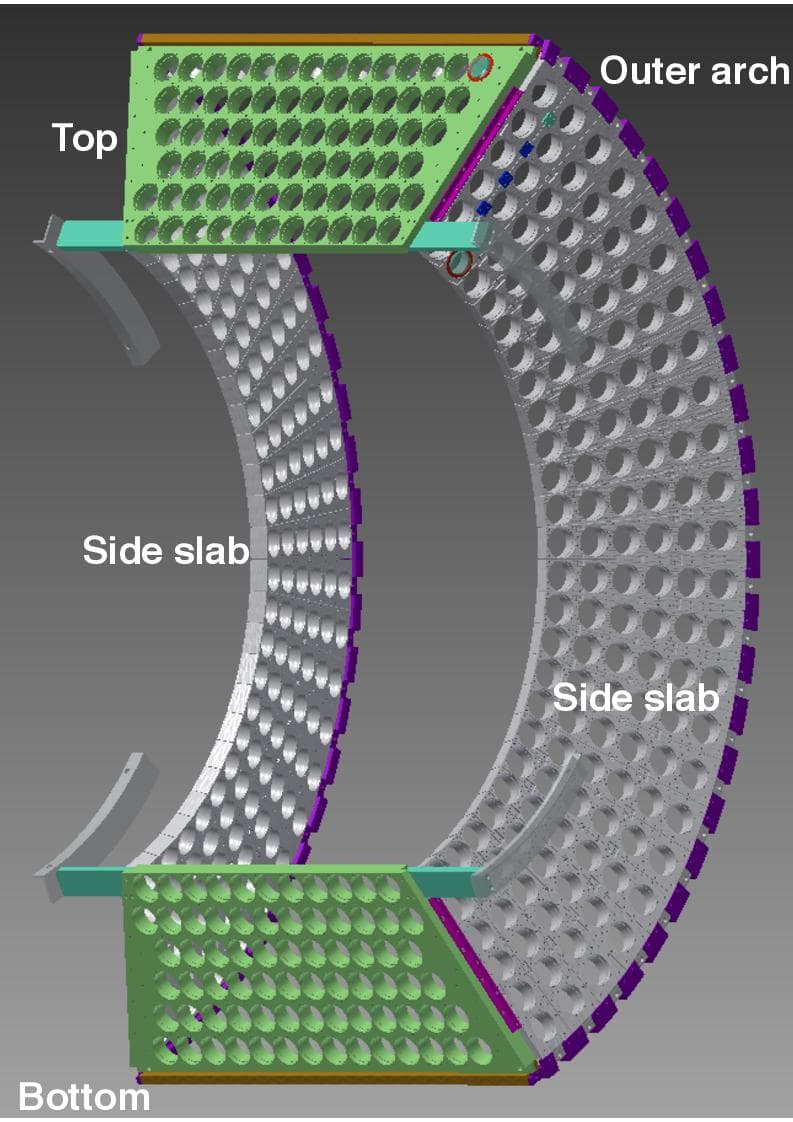}
\caption{\label{fig:PMT support}%
   3D CAD design of the LXe photon detector PMT support structure.
}
\end{figure}

Figure~\ref{fig:PMT support} shows the 3D CAD design of the PMT support structure. 
There is no support structure at the inner side, and only the outer part has screw 
holes to an arch-shaped support structure repaired by bolts. 
Joint brackets are used to fix two adjacent side slabs. 
The side and outer faces of the PMT support structure are re-used from the MEG LXe photon detector, 
while the top and bottom panels are modified to fit the larger number of PMTs, 73 instead of 54. 
In total, 668 PMTs are installed in five faces except for the inner one.

\subsubsection{Signal transmission}
\label{sec:cable and feed-through}

The transmission of 4092 MPPC signals to the DAQ electronics without introducing noise or distortion is challenging. 
We have to pay attention to pickup noise, cross-talk, and limited space in the cryostat as well as the feed-throughs etc.
In order to overcome such issues, we have developed a multi-layer PCB with coaxial-like signal line structure.
It is used for both the PCBs for MPPC mounting and the vacuum feed-through of the cryostat. 

\begin{figure}[tb]
\centering
\includegraphics[width=1\linewidth]{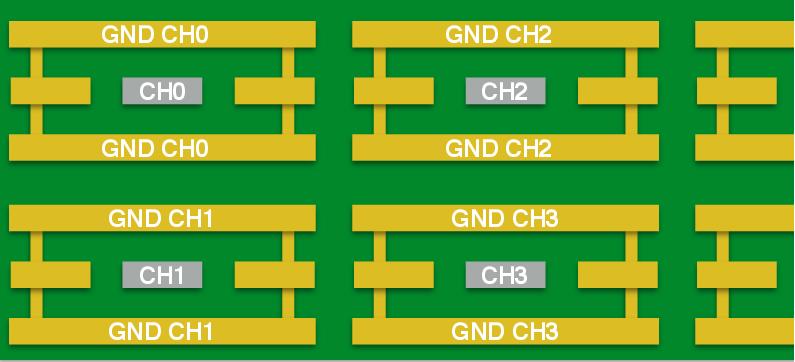}
\caption{\label{fig:PCBlayout}%
   A cross-sectional view of the PCB schematic drawing in which a signal line is shielded by surrounding ground lines and ground layers. 
  The total thickness of the PCB board is \SI{1.6}{\mm}.}
\end{figure} 

As described in Sect.~\ref{sec:Design of Sensor Package and Assembly},
22 MPPCs are mounted on a PCB strip and signal lines embedded in the strip transmit signals to an end.
The total length of the signal lines is about \SI{35}{\cm}, and the width of the PCB is \SI{15}{\mm}.
MPPCs are plugged into socket pins on the PCB and 22 MMCX (micro-miniature coaxial) connectors are used at the end of the signal lines.
The signal lines on the PCB strip are connected to (real) thin coaxial cables
by means of connectors at the edge of PCBs. 
Then the signals are transmitted to feed-throughs using 
the thin coaxial cables, with a length of \SIrange{2.5}{4.9}{\m} depending on their $\phi$ positions.
The coaxial cables (RG178-FEP) are produced by JYEBAO\cite{JYEBAO}. An MMCX connector is assembled on one end, and the other end is directly soldered on the feed-through PCB.

\begin{figure}[tb]
\centering
\includegraphics[width=1\linewidth, clip, trim=0 2cm 0 2cm]{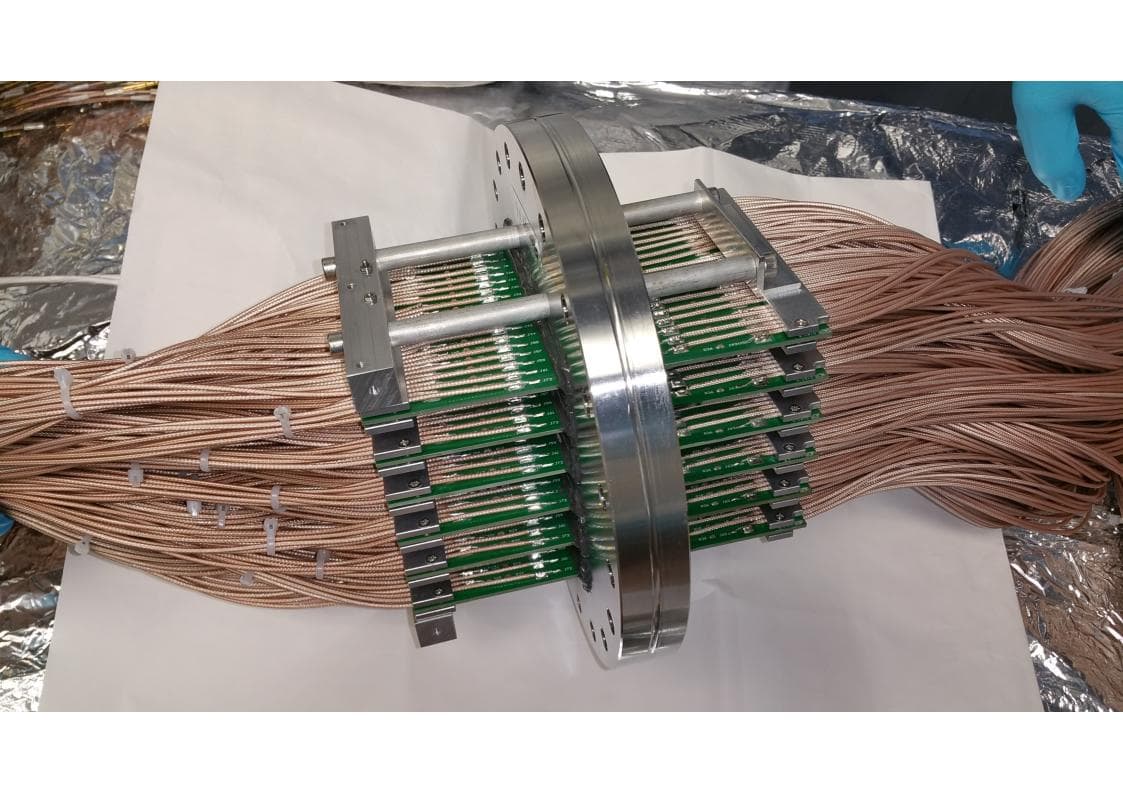}%
\caption{\label{fig:feedthru} PCB-type vacuum feed-through for the MEG~II LXe photon detector.  }
\end{figure}

Figure~\ref{fig:PCBlayout} shows the layer structure of the PCB used as our feed-through PCB.
Each signal line is surrounded by different ground patterns to minimise cross-talk and to shield from the outside. 
To avoid any ground loop, different ground patterns for different signal lines are separated. 
In total, six layers (two layers of signal, and four layers of ground) are used.
The dimensions of the layers are adjusted to have a \SI{50}{\ohm} impedance. 
The MEG LXe photon detector had in total 10 DN160CF flanges for the signal and HV cables of 846 PMTs. 
Since the number of readout channels has increased, more feed-through ports are necessary. 
A PCB-type feed-through shown in Fig.~\ref{fig:feedthru} similar to the PCB for MPPC mounting has been developed,
which allows a high density signal transmission through vacuum walls and a low-noise environment.
On both sides of a PCB, 72 cables are directly soldered, and six PCBs are glued by Stycast 2850 FT + Catalyst 24 LV into a DN160CF flange.
In total, 10 DN160 CF flanges are used for MPPC signals (up to 4320 channels) and 2 flanges for PMT signals, 
while 4 flanges are used for PMT HV cables.
The signal from the feed-through is transmitted to the readout electronics via 10~m coaxial cables.

\subsubsection{Read-out electronics}

Both the PMTs and MPPCs signals are read out by WaveDREAM boards (WDBs).
Amplifiers are mounted on the boards with switchable gain settings from 0.5 to 100 (see Sect.~\ref{sec:wavedream} in detail).
The different gain stages can then be switched at any time.
The higher gain mode is used to detect single photo-electrons for the calibration of the MPPCs,
while the low gain mode is used to take physics data 
where a large dynamic range is needed.

No amplifier is installed between MPPC and WDB.
The bias voltage for MPPC, which is typically \SI{50}{\volt}, is supplied from the WDBs through the signal cable.

\subsubsection{Cryogenics}
The MEG LXe cryostat is re-used for the MEG~II LXe photon detector. 
In order to cover the increase of the external heat inflow due to \num{\sim 4000} extra signal cables for the MPPCs, 
the cooling power of the refrigerator is increased by adding another Gifford-McMahon (GM) refrigerator, model AL300
produced by CRYOMECH \cite{cryomech}.
The new refrigerator will produce more than \SI{400}{\watt} of cooling power which should be sufficient to cool the MEG~II LXe photon detector.

\begin{figure}[tb]
\centering
\includegraphics[width=1\linewidth, clip, trim=0 0 0 25pt]{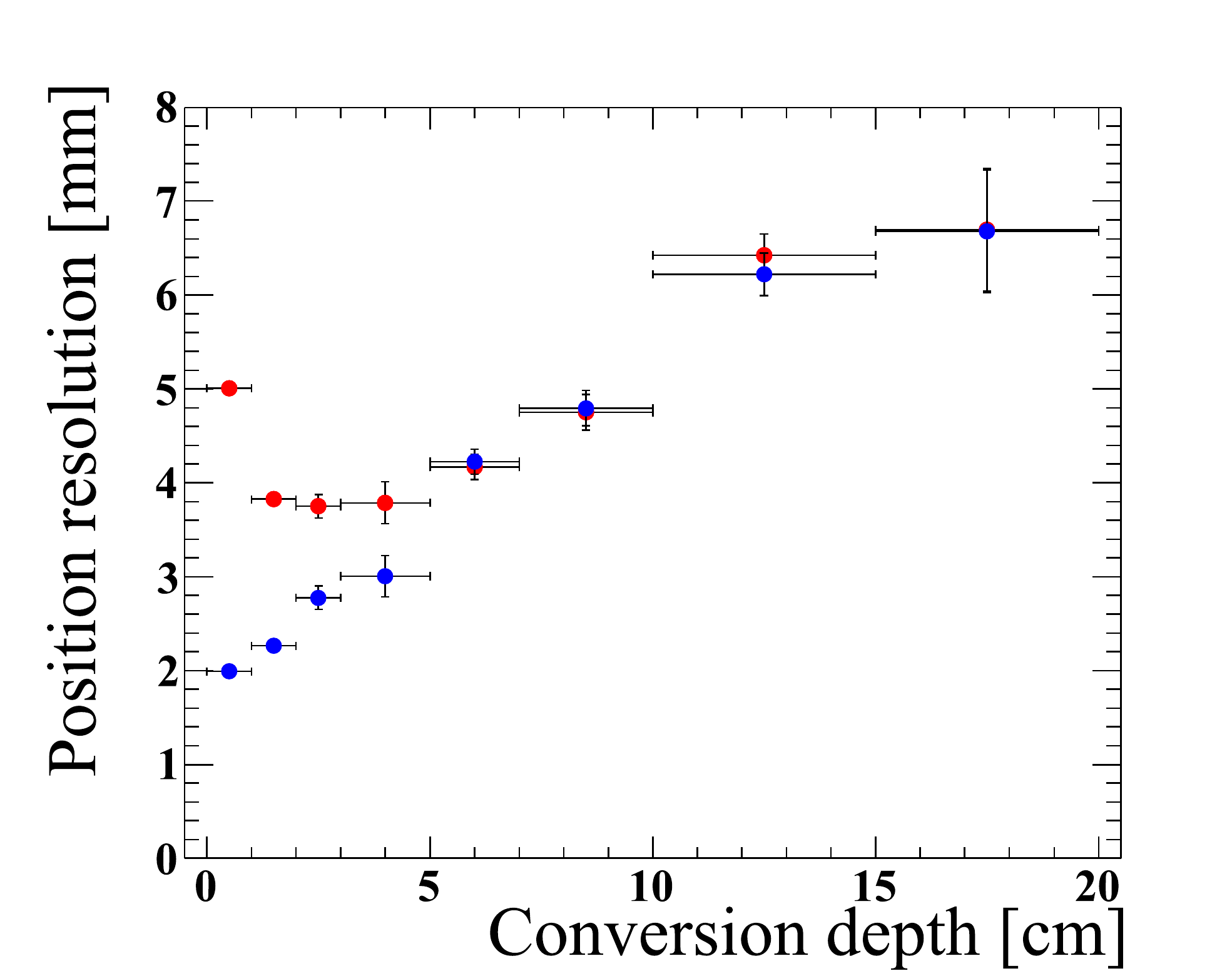}\\ 
\includegraphics[width=1\linewidth, clip, trim=0 0 0 25pt]{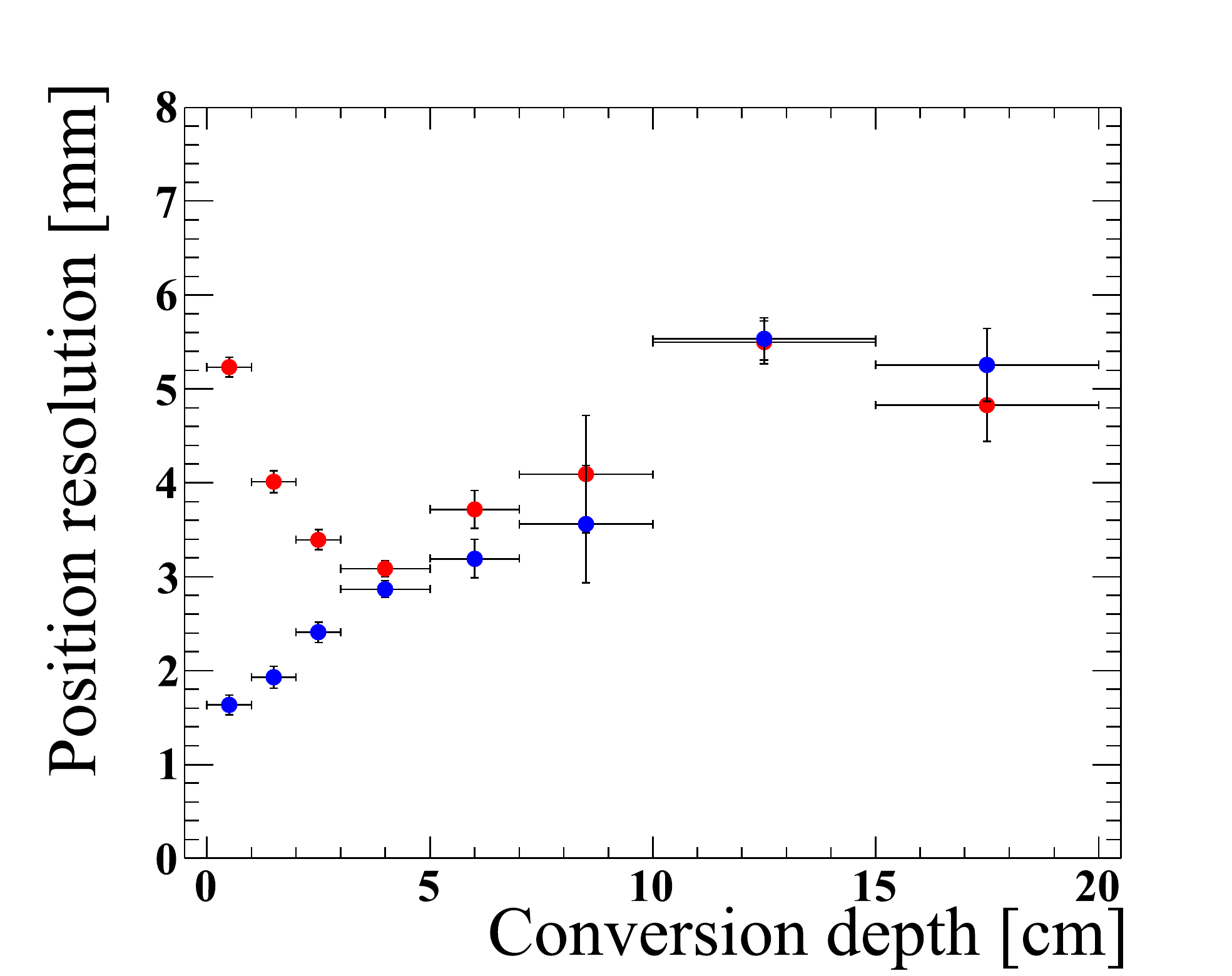} 
\caption{\label{fig:xec_posreso}%
Position resolution in the horizontal (top) and vertical (bottom) directions as
a function of the first conversion depth. The resolutions in MEG
are shown with red markers, and those in MEG~II
are shown with blue markers.}
\end{figure}

\subsection{Calibration and monitoring}
\label{sec:LXe_Calorimeter_Calibration}

The LXe detector necessitates careful calibration and monitoring of the energy scale over its full
energy range. That requires several methods that have already been introduced and commissioned in MEG and will be inherited by MEG~II with some modifications to match the upgrade. They are listed in Table~\ref{tab:ListOfCalibrations} and summarised in the following (see \cite{megdet} for more details):

\begin{enumerate}
\item The behaviour of the LXe photon detector is checked in the low-energy region using \SI{4.4}{\MeV} $\gamma$-rays from an AmBe source, placed in front of the inner face, and
\SI{5.5}{\MeV} $\alpha$-particles from $^{241}$Am sources deposited on thin wires, mounted inside the active volume of the detector. The $\alpha$-signals are also used to evaluate and
monitor in-situ the PMT quantum efficiencies (QEs) and measure the Xe optical properties on a daily basis.
In addition, \SI{9.0}{\MeV} $\gamma$-rays from capture by $^{58}$Ni of thermalised neutrons produced by a neutron
generator are also available. This is the only method which allows to check the response of the LXe photon detector with and without the particle flux associated with the muon beam and/or the other beams.
\item The performance of the LXe photon detector in the intermediate-energy region is measured two/three times per week
using a Cockcroft--Walton accelerator,
by accelerating protons, in the energy range \SIrange[range-phrase=--,range-units=single]{400}{1000}{\keV}, onto a Li$_2$B$_4$O$_7$ target.
$\gamma$-rays of \SI{17.6}{\MeV} energy from $^7 \mathrm {Li} (\mathrm{p}, \gamma) ^8 \mathrm {Be}$ are used to monitor the energy scale, resolution and the uniformity of the detector, while time-coincident 4.4 and \SI{11.6}{\MeV} $\gamma$-rays
from $^{11} \mathrm {B} (\mathrm{p}, \gamma \gamma) ^{12} \mathrm{C}$ are used to inter-calibrate the relative timing of the LXe photon detector
with the pTC detector.
\item The response of the LXe photon detector around the \megc\ signal region and above is measured once/twice a year using photons from $\pi^0$ decays produced by the $\pi^-$
charge exchange reaction (CEX) in a liquid hydrogen target, $\mathrm{p}(\pi^- , \pi^0)\mathrm{n}$.
Photons with energy of \SI{129}{\MeV} are also produced via the radiative capture reaction, $\mathrm{p}(\pi^- ,\gamma)\mathrm{n}$, with a relative probability of
$\Gamma(\mathrm{p}(\pi^-,\pi^0)\mathrm{n})/\Gamma(\mathrm{p}(\pi^- ,\gamma)\mathrm{n}) = 1.546 \pm 0.009$ \cite{spuller77} (Panofsky ratio).
\item The RMD can be used as well for calibration purposes with dedicated triggers. In particular the selection of the $\egammapair$ pair represents a strong quality check of the complete apparatus and a straightforward way to extract the global time resolution (the resolution of the timing difference between the positron and the photon) and the relative offset.
\end{enumerate}

\begin{table*}
\caption{\label{tab:ListOfCalibrations}%
The calibration tools of the LXe detector for the MEG~II experiment.}
\centering
{\begin{tabular*}{\textwidth}{@{\extracolsep{\fill}}lllll@{}}
\hline
{\bf Process} &  & {\bf Energy} & {\bf Main Purpose} & {\bf Frequency} \\
\hline
Cosmic rays & $\mu^{\pm}$ from atmospheric showers & Wide spectrum $\cal O$(GeV) & LXe--CDCH relative position & Annually \\
& & & LXe purity & On demand\\
Charge exchange & $\pi^- \mathrm{p}  \to \pi^0 \mathrm{n} $  & $55, 83, 129$~MeV photons & LXe energy scale/resolution & Annually \\
& $ \pi^0 \to \gamma \gamma$ & & & \\
Radiative $\mu-$decay  & $\radiative$ & Photons \SI{> 40}{\MeV},   & LXe--pTC relative timing & Continuously \\
& & Positrons \SI{> 45}{\MeV} & & \\
Proton accelerator & $^7 {\rm Li} (\mathrm{p}, \gamma) ^8 {\rm Be}$ & 14.8, 17.6~MeV photons & LXe uniformity/purity& Weekly \\
& $^{11} {\rm B} (\mathrm{p}, \gamma) ^{12} {\rm C}$ & 4.4, 11.6, 16.1~MeV photons & LXe--pTC timing & Weekly \\
Neutron generator & $^{58} {\rm Ni}(\mathrm{n},\gamma) ^{59}{\rm Ni}$ & 9~MeV photons & LXe energy scale & Weekly \\
Radioactive source & $^{241}{\rm Am}(\alpha,\gamma)^{237}{\rm Np}$ & 5.5~MeV $\alpha$'s & LXe PMT/SiPM calibration & Weekly \\
& & & LXe purity& \\
Radioactive source & $^9\mathrm{Be}(\alpha_{^{241} {\rm Am}}, \mathrm{n})^{12}\mathrm{C}^{\star}$ & 4.4~MeV photons & LXe energy scale & On demand \\
                   & $^{12}\mathrm{C}^{\star}(\gamma)^{12}\mathrm{C}$ & & & \\
Radioactive source & $^{57}\mathrm{Co}(\rm{EC},\gamma)^{57}\mathrm{Fe}$ & 136 (11 $\%$), 122 keV (86 $\%$) X-rays & LXe--spectrometer alignment & Annually \\ 
LED & & UV region & LXe PMT/SiPM calibration & Continuously \\
\hline
\end{tabular*}}
\end{table*}

\subsection{Alignment }

Precise relative alignment of the photon detector and the positron magnetic spectrometer is important 
to ensure that the angular acceptance criteria for \megc\ signal events are not compromised. 
For example, a \SI{5}{\mm} error in the measured position of the photon in the LXe photon detector would 
result in a signal event possibly being missed because it would not be consistent with being emitted 
opposite to the direction of the positron. The relative alignment of the LXe photon detector and the 
spectrometer is implemented using optical survey techniques. For the LXe photon detector, the survey is 
complicated by the fact that the photo-sensors (SiPMs) are not visible once the LXe photon detector is 
closed and that their positions relative to the external survey markers change due to thermal contraction and 
buoyant forces as the LXe photon detector is cooled and filled with liquid xenon.

\subsubsection{The X-ray alignment system for the LXe photon detector}

A newly introduced technique will measure the position of each SiPM using the novel technique of X-ray imaging 
each sensor. The technique uses a well collimated and precisely aligned X-ray beam in the radial direction 
originating from the axis of the COBRA magnet (at $x =y =0$ in the MEG coordinate system) at precisely known 
axial ($z$) and azimuthal ($\phi$) coordinates. The X-rays are collimated to produce a ribbon-like beam, 
narrow (\num{\approx 10}\% of the dimension of a SiPM at its face) in one dimension ($\phi$ or $z$). 
The energy of the X-rays is chosen such that they penetrate the COBRA and LXe cryostats with significant probability, 
yet interact within \SI{\approx 1}{\mm} of liquid xenon, primarily by photo-absorption. 
Scintillation light produced by the photo-electrons in the liquid xenon is detected by the SiPM directly 
in front of the interaction. The $z$-coordinate of each SiPM is deduced by orienting the narrow 
(\SI{1.5}{\milli\radian}) beam dimension in the axial direction and then scanning it in that direction. 
The axial extent of a given SiPM is given by the axial extent of the X-ray beam position for which light 
is detected in that element. The $\phi$-coordinate is similarly determined by rotating the collimator so 
the beam is narrow in the azimuthal direction and scanning in azimuth.

The X-rays are produced by decay of a $\rm{^{57}Co}$ source, producing X-ray lines at \SI{122}{\keV} (\num{\approx 80}\%) and \SI{136}{\keV} (\num{\approx 10}\%). They penetrate the COBRA magnet and the front of the LXe cryostat with \num{\approx 30}\% probability. We use a commercial point source with an activity of \SI{\approx 3e10}{\becquerel} and collimate the beam to \SI[product-units=single]{1.5 x 50}{\milli\radian\squared} with a brass collimator. The $z$-coordinate of the origin of the beam and its $\phi$-direction are set using precise linear and rotary translation stages. The signal induced in the SiPM is about 30\% of that induced in a single SiPM by a typical shower of a \SI{\sim53}{\MeV} photon from \megc\ events. Data are collected by implementing a trigger on the signal detected in a limited number of SiPMs in the region to which the X-ray beam points.

\begin{figure}
\centering
\includegraphics[width=0.99\linewidth]{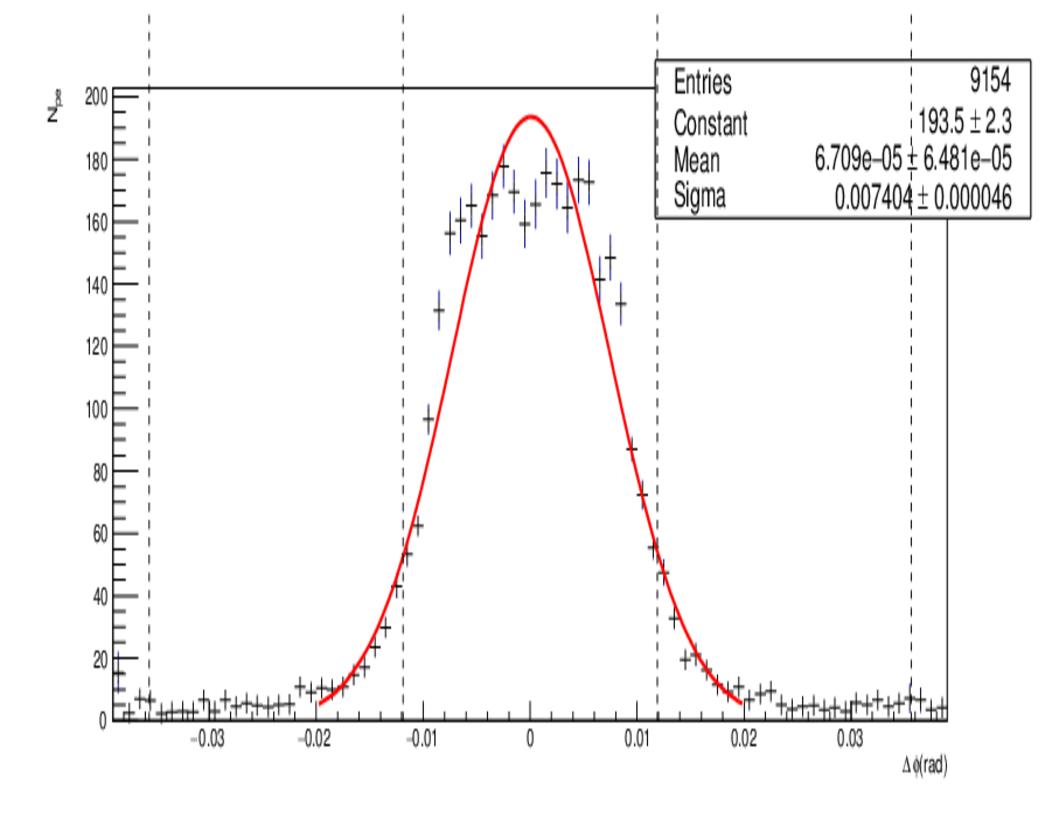}
\caption{Mean number of photoelectrons vs. $\Delta\phi$ for X-rays events fitted with a Gaussian distribution. 
The dashed lines show the boundaries of the neighbouring SiPMs.}
\label{fig:Xraymethod}
\end{figure}

The expected performances is studied with a \textsc{Geant4} MC simulation of the X-ray 
beam and the MEG~II detector. X-rays are generated in the beam solid angle, propagated 
through the COBRA cryostat and into the LXe. Scintillation light is produced from the 
electron produced by the X-ray interaction and the SiPM response is simulated. 
Figure~\ref{fig:Xraymethod} shows a plot of the average number of detected photoelectrons 
per interaction in a SiPM as a function of the difference $\Delta\phi$ between the 
$\phi$-coordinate of the beam with respect to the $\phi$-coordinate of the SiPM centre. 
Each bin contains \num{\approx 90} detected X-rays, corresponding to an exposure time of 
\SI{\sim 2.5}{\second} per position. An approximate estimation of the precision with 
which the SiPM centre $\phi$-coordinate can be measured is obtained by fitting the 
distribution with a Gaussian; the statistical uncertainty is $\sigma_{\overline{\Delta\phi}} 
\simeq$\SI{0.06}{\milli\radian}. Similar precision is obtained fitting the distribution 
with a rectangular function smeared with error functions.

Systematic uncertainties in the position determination will be due to the uncertainty in our 
knowledge of the direction and origin of the X-ray beam. The position and angle alignment 
of the collimator is made by an optical survey to a precision of \SI{< 100}{\um} and 
\SI{0.2}{\milli\radian}. Variations in the beam direction as it moves along the translation 
stage (of the order of \SI{0.5}{\milli\radian} from the device specifications and our measurements) 
will be monitored with a laser attached to the translation stage and projected to a quadrant 
photodiode, as well as with a spirit-level on the translation stage. In addition, a cross-check 
of the optical survey of the cryostat and the X-ray beam is made by mounting small 
LYSO\footnote{Lutetium-yttrium oxyorthosilicate Lu$_\mathrm{2(1-x)}$Y$_\mathrm{2x}$SiO$_5$.} crystals 
scintillators and thin lead absorbers in well-surveyed positions just in front of the LXe cryostat. 
The X-ray beam should be detected in the LYSO detectors at the calculated X-ray beam 
$\phi$- and $z$-coordinates and the signal in the LXe should be shadowed at the calculated X-ray 
beam $\phi$- and $z$-coordinates of the thin lead absorbers.

\subsection{Expected performance\label{sec:expected performance}}
The expected performance of the upgraded LXe photon detector is evaluated using a MC simulation.

\subsubsection{Simulation}
A full MC simulation code based on \textsc{Geant4} was developed to compare the performance of the MEG and the MEG~II design.
In the simulation, scintillation photon propagation is simulated by \textsc{Geant4}.
The reflection of scintillation photons on the MPPC surface was simulated using the complex refractive index of a pure silicon crystal. 
The reflectance is typically about 60\%. 
In the simulation, the index-number of hit pixel and the arrival time of each scintillation photon are recorded. 
They are used to form avalanche distributions in each MPPC. 
The dark-noise, optical cross-talk, after-pulsing, saturation and recover are modelled based on real measurements and incorporated in the simulation.  
The waveform of the MPPC is simulated by convolving the single photo-electron pulse and the time distribution of avalanches. 
A simulated random electronics noise is added assuming the same noise level as the MEG read-out electronics.

The event reconstruction analyses are basically the same as those for the MEG detector, 
while the parameters, such as waveform integration window and corrections for light collection efficiency 
depending on the conversion position, are optimised for the new design.
The non-linear response of the MPPC due to pixel saturation (see
Fig.~\ref{fig:MPPC_linearity}), resulting in a non-linear energy
response of the detector, is taken into account.
However the effect on the energy reconstruction is negligible
because the fraction of the total number of photoelectrons observed by each MPPC is small.

\subsubsection{Results}

Figure~\ref{fig:xec_posreso} shows the position resolutions for signal photons as a
function of the reconstructed conversion depth ($w$). In MEG, the
position resolution is worse in the shallow depth part than in the deeper part because of the PMT size. 
The position resolution in the shallow part is much better in MEG~II due to smaller size of the photo-sensors.

\begin{figure*}[tb]
\centering
\begin{minipage}{1\textwidth}
\subfigure[MEG ($w < \SI{2}{\cm}$)]{
\includegraphics[width=0.45\textwidth, clip, trim=0 0 0 40pt]{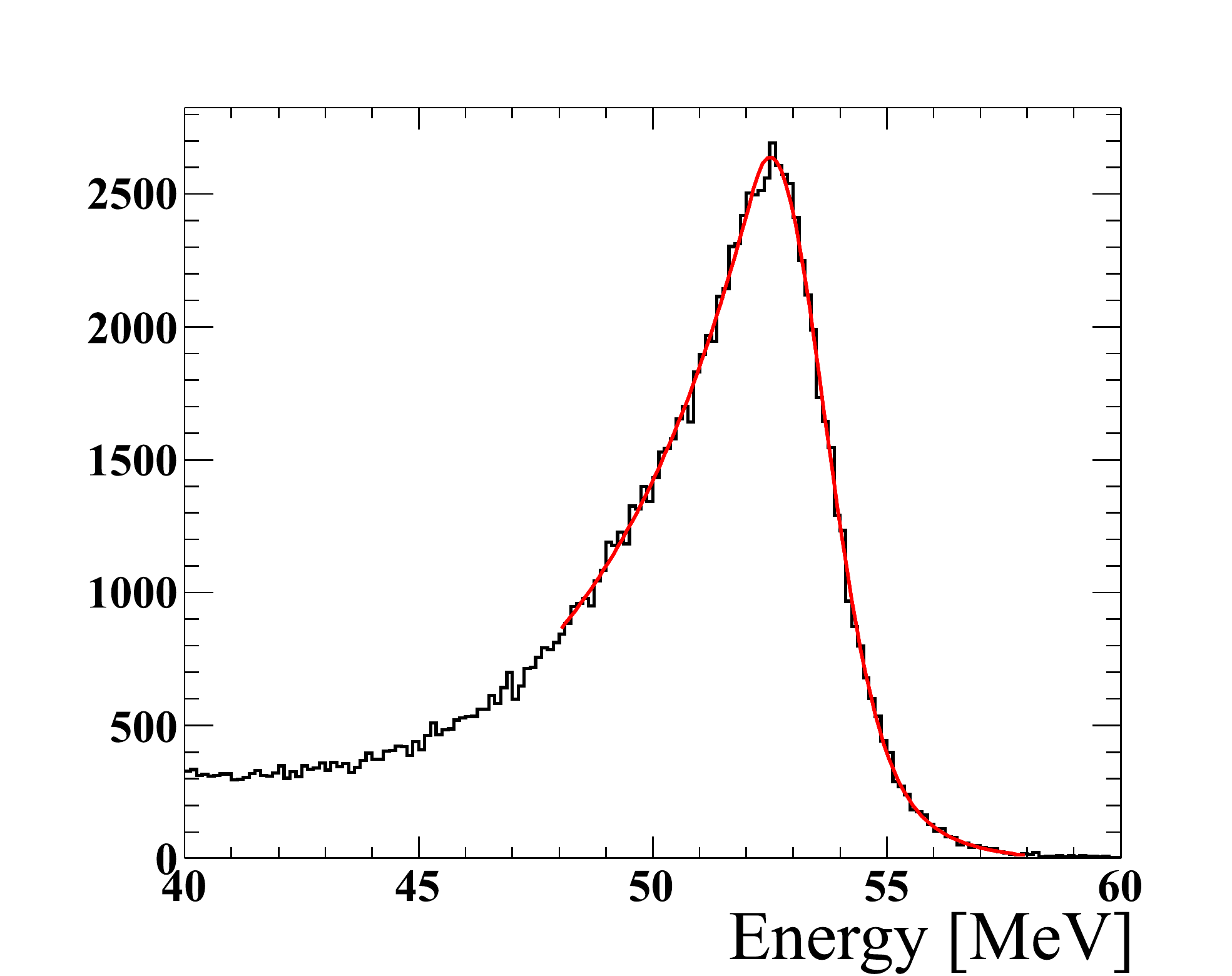}
}
\subfigure[MEG~II ($w < \SI{2}{\cm}$)]{
\includegraphics[width=0.45\textwidth, clip, trim=0 0 0 40pt]{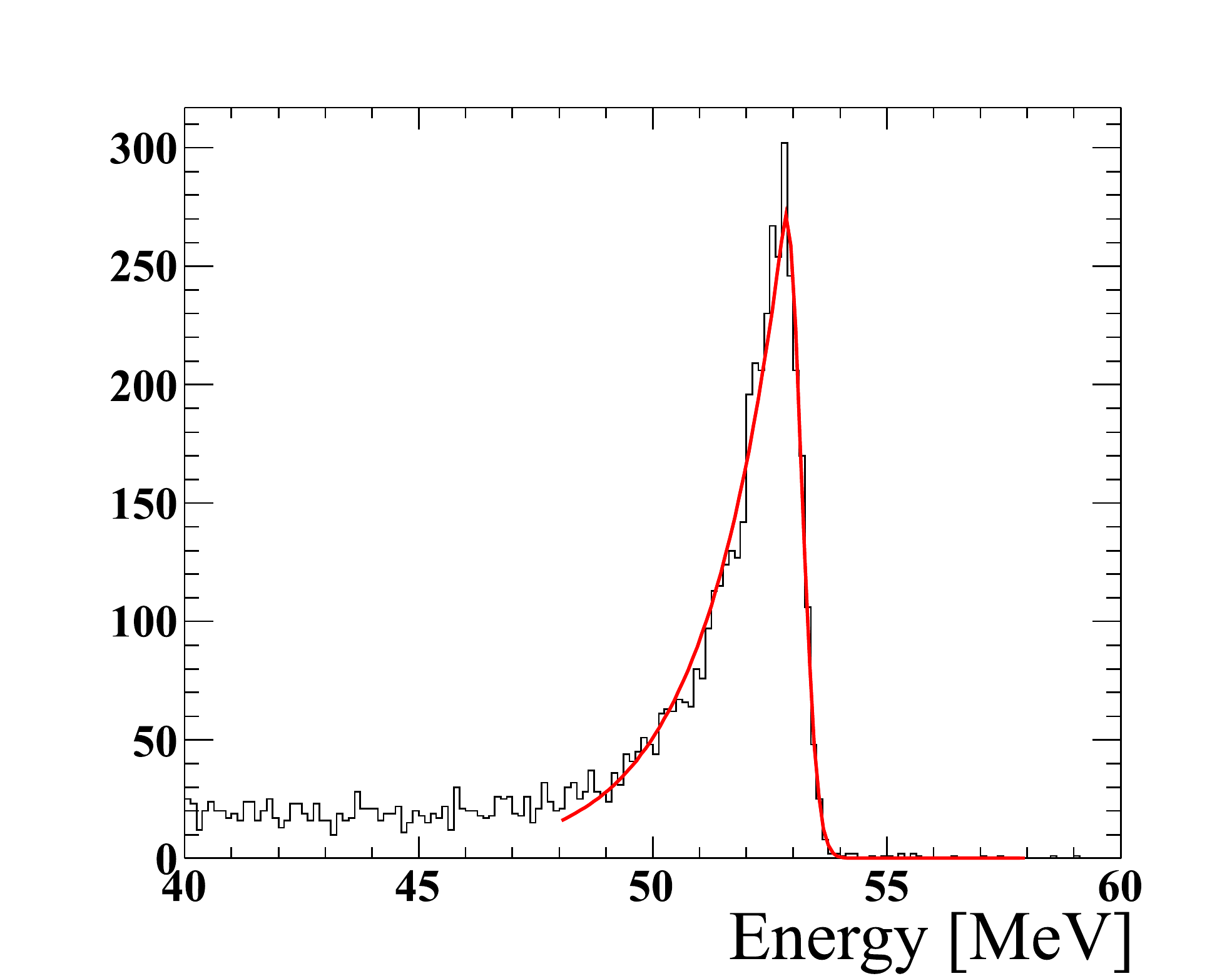} 
}
\end{minipage}
\begin{minipage}{1\textwidth}
\subfigure[MEG ($w \geq \SI{2}{\cm}$)]{
\includegraphics[width=0.45\textwidth, clip, trim=0 0 0 40pt]{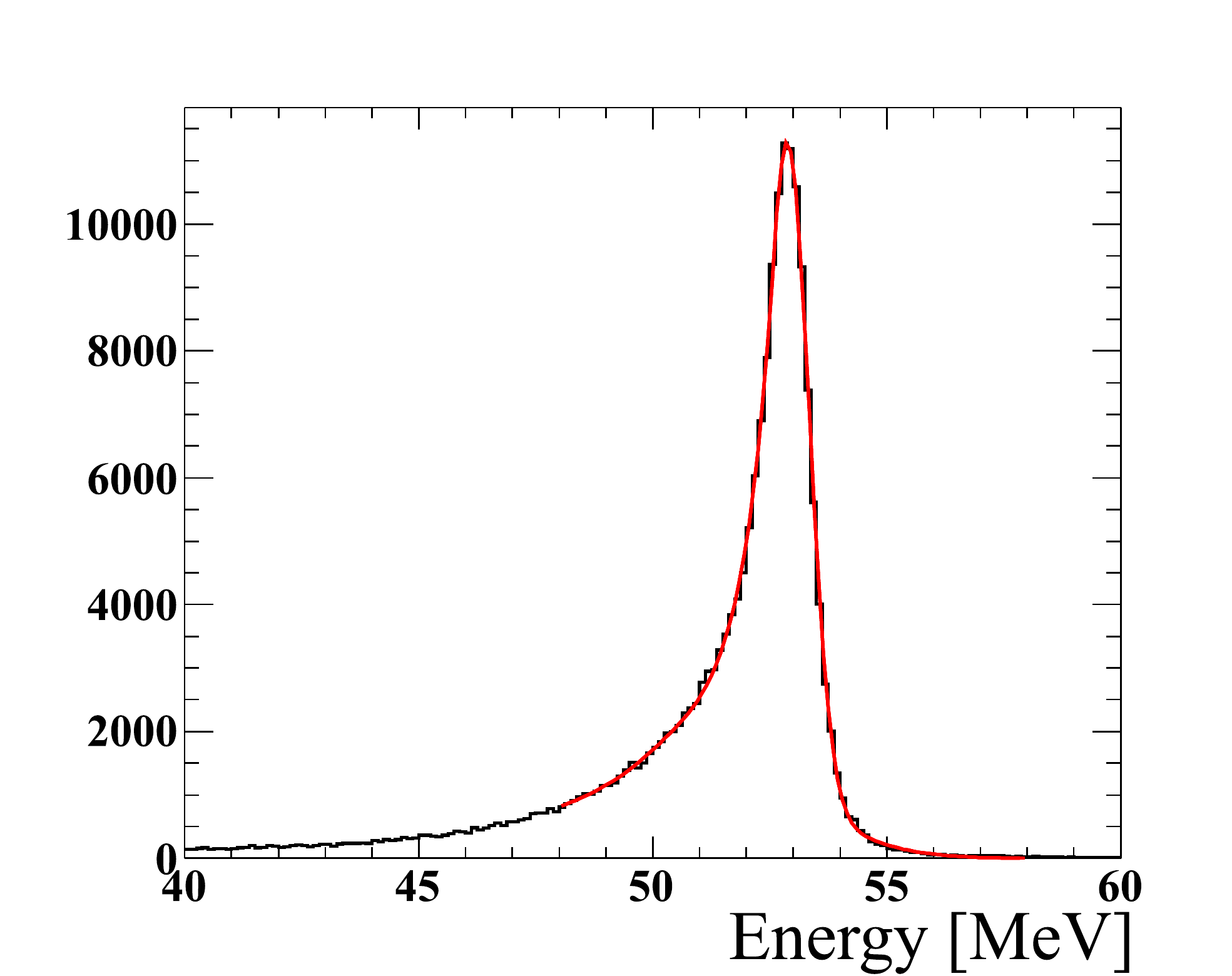} 
}
\subfigure[MEG~II ($w \geq \SI{2}{\cm}$)]{
\includegraphics[width=0.45\textwidth, clip, trim=0 0 0 40pt]{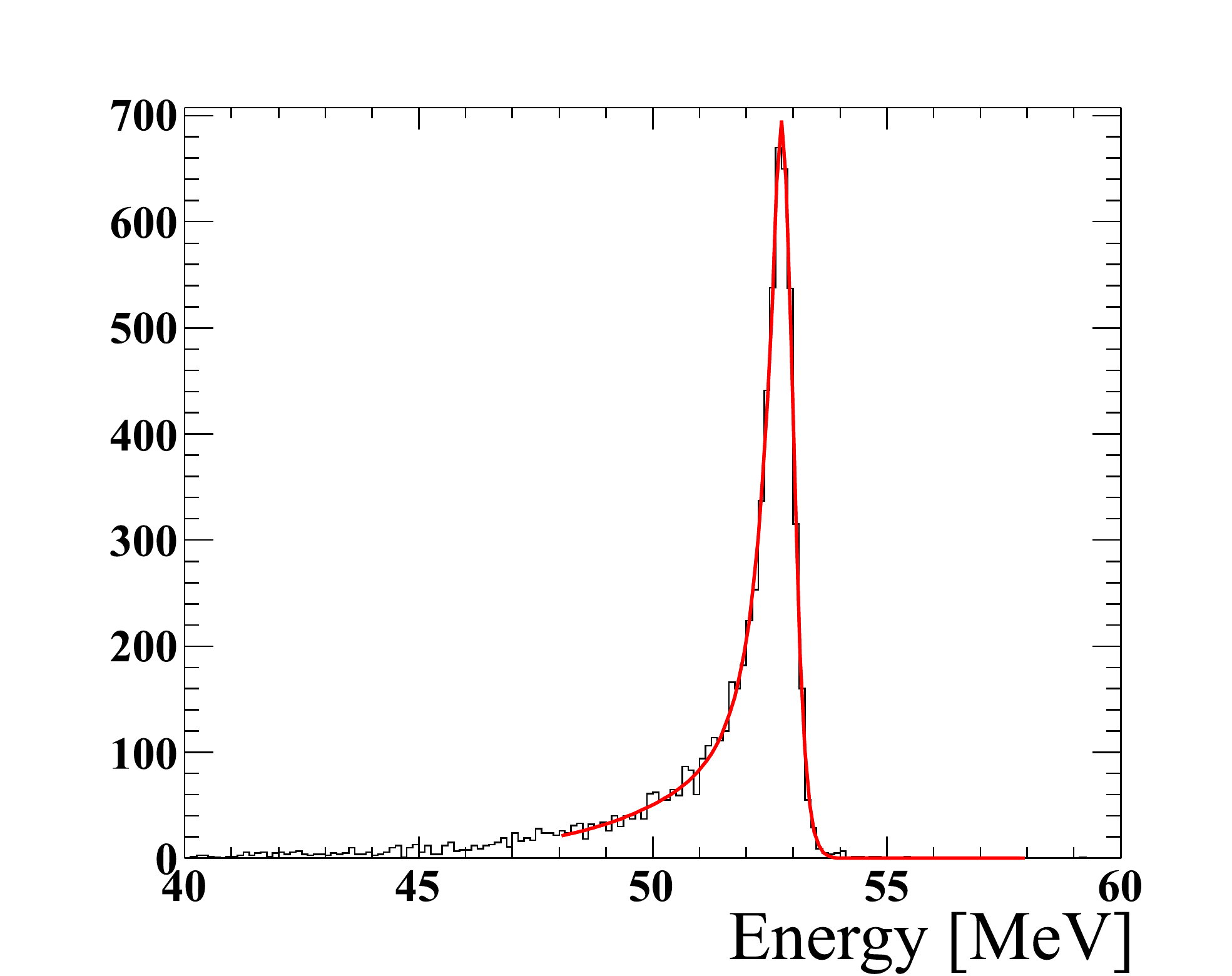} 
}
\end{minipage}
\caption{\label{fig:xec_enereso}%
   Energy PDFs for $\egamma=\SI{52.83}{\MeV}$ photons converting in the 
   MEG (left) and the MEG~II (right) LXe photon detectors. 
   The response to shallow (top) and deep (bottom) events are shown separately.}
\end{figure*}

The energy resolution is also much better in the shallow part with the MEG~II
design than that of MEG as shown from the probability density function (PDF) for $\egamma=\SI{52.83}{\MeV}$ photons
in Fig.~\ref{fig:xec_enereso} due mainly to a more uniform photon collection efficiency.
The low energy tail is smaller because of the lower
energy leakage at the acceptance edge with the improved layout of the lateral PMTs. 
The resolution is also better in the deeper part because of the modification
of the angle of the lateral PMTs.

\begin{figure}[tb]
\centering
\subfigure[$w < \SI{2}{\cm}$]{
\includegraphics[width=0.45\textwidth, clip, trim=0 0 0 30pt]{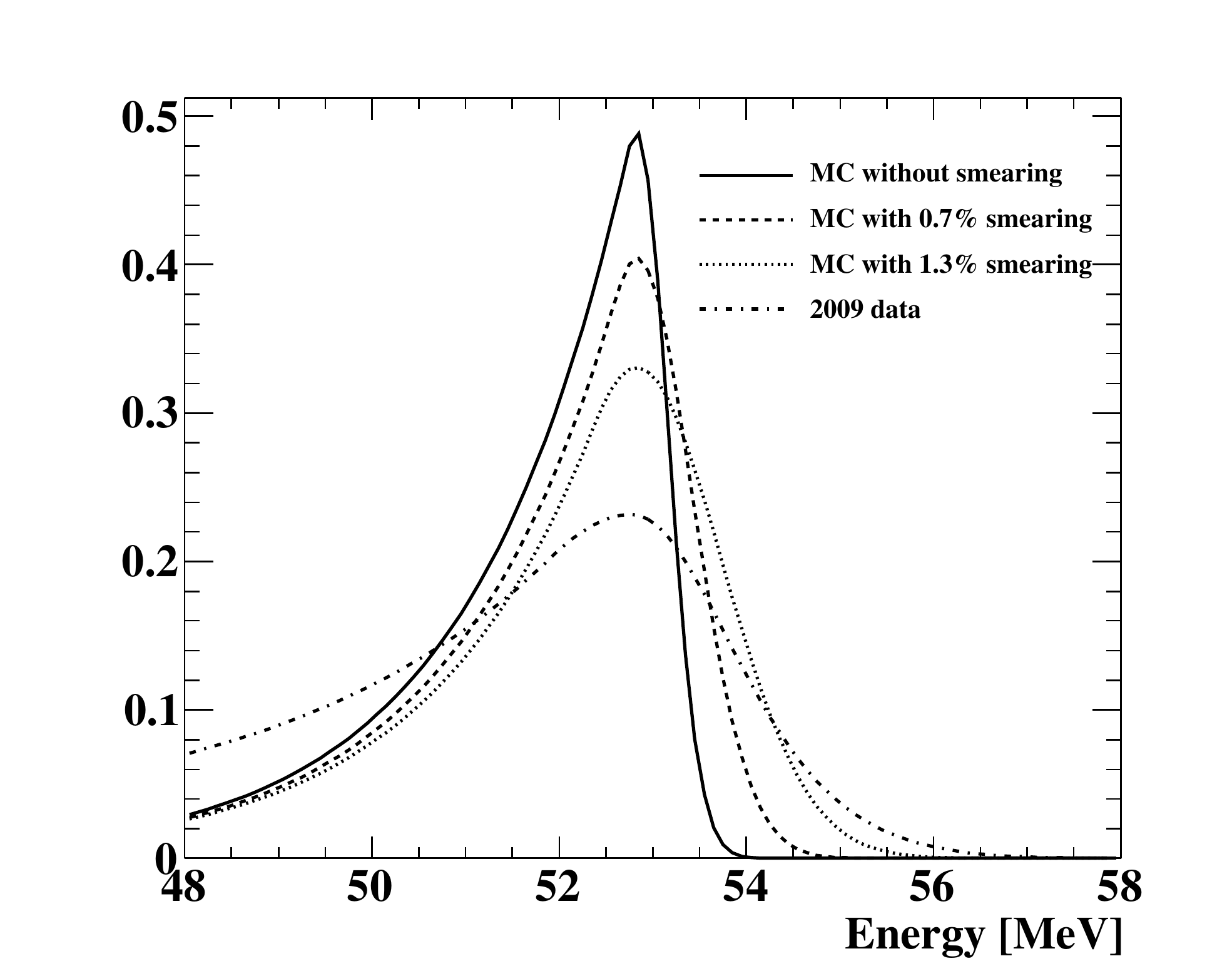}
}
\subfigure[$w \geq \SI{2}{\cm}$]{
\includegraphics[width=0.45\textwidth, clip, trim=0 0 0 30pt]{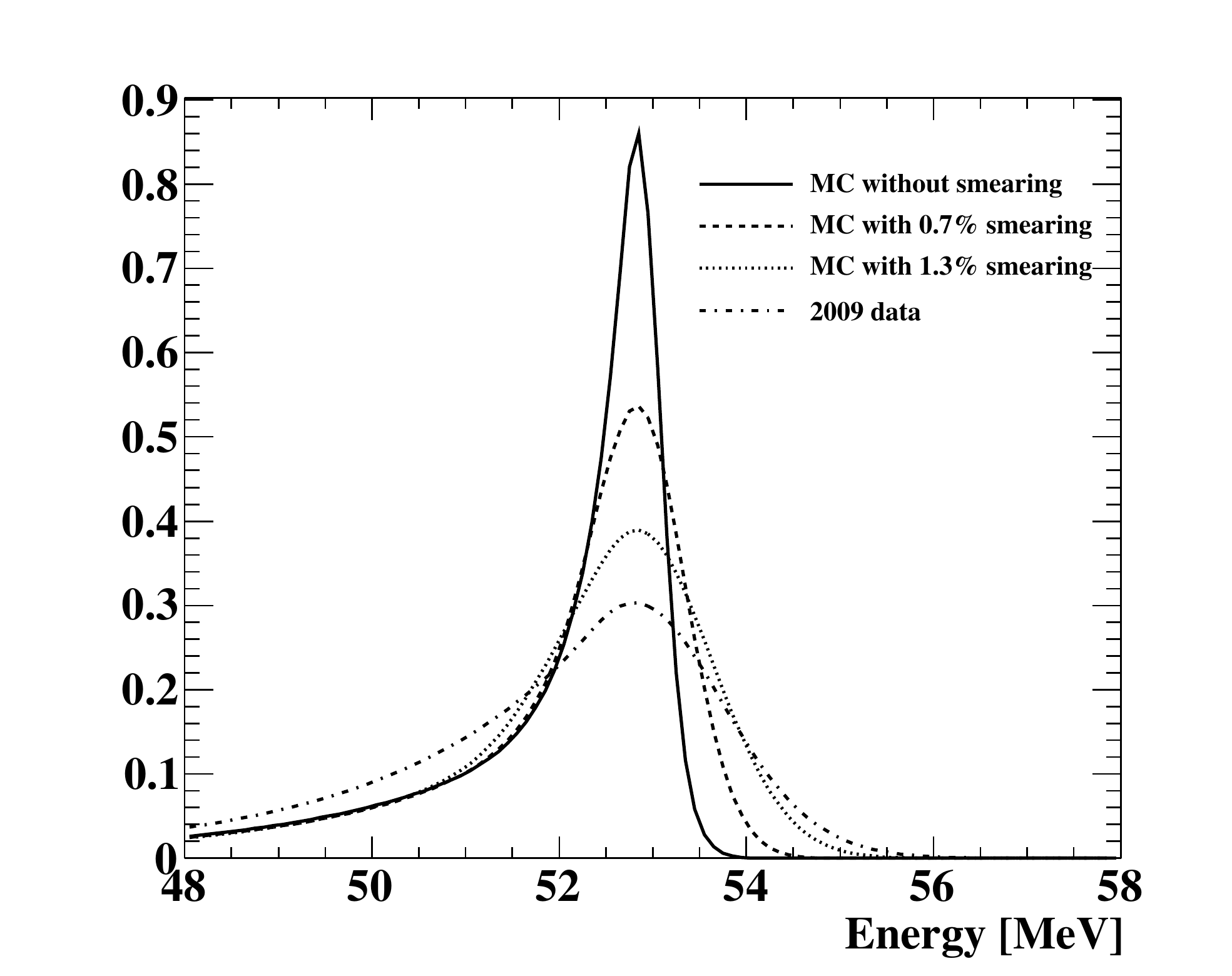}
}
\caption{\label{fig:xec_energy_smear}%
Energy response functions with various assumptions of additional fluctuation (0, 0.7 and
1.3\%) and that of the 2009 data.
}
\end{figure}

The measured energy resolution of MEG (1.7\% for $w>\SI{2}{\cm}$) was worse than
that in the simulation (1.0\% for $w>\SI{2}{\cm}$). 
The reason is not fully understood, while the source of the difference could be related
to the behaviour of the PMTs (e.g.~gain stability, angular dependence and so on) or the optical
properties of liquid xenon (e.g.~effect of convection). In the former case, the
difference can become smaller in the upgraded configuration.
On the other hand, in the latter case, the difference could remain.
Figure~\ref{fig:xec_energy_smear} shows the energy response under different assumptions:
\begin{enumerate}
\item the additional fluctuation completely vanish in MEG~II, 
\item a part of fluctuation remains
which corresponds to 1.2\% resolution in MEG (the
resolution achieved with the MEG LXe large prototype detector), 
\item the fluctuation remains making the resolution of MEG 1.7\%.
\end{enumerate}
We will use the assumption 2 for the sensitivity calculation in Sect.~\ref{sec:sensitivity}.

The time resolution of the LXe photon detector $\sigma_{\tgamma}$ can be separated into six components;
the transit time spread (TTS) of the photo-sensors, the statistical fluctuation of scintillation photons, 
the timing jitter of the read-out electronics, the electronics noise,
the resolution of the photon conversion point and the finite size and the fluctuation of the energy deposits in the LXe. 
Most of these are common to both MEG and MEG~II, but the
effect from the TTS and electronics noise are different because of the different
photo-sensors.
The effect of the TTS is negligible because it scales as a function of the number of
photoelectrons, and the light output of liquid xenon is large.
The effect of the electronics is larger in MEG~II than in MEG because the leading 
time of an MPPC pulse for liquid xenon scintillation signal is slower than that of a PMT pulse.
In order to estimate the effect, the time resolution of the upgraded detector for signal
photons is measured in the simulation. 
The evaluated time resolution with preliminary
waveform and reconstruction algorithms is $\sigma_{\tgamma}\simeq\SI{50}{\ps}$ assuming 
a noise level up to \SI{1}{\mV}. 
The main improvements come from the better time of flight estimate,
deriving from the better position reconstruction, and higher photon statistics.
Since parameters such as the rise time of the waveform and the noise components may not 
be correctly considered in the simulation, the time resolution in the worse case might 
still be at the MEG level, hence a conservative estimation 
is $\sigma_{\tgamma}$\SIrange{\sim50}{70}{\ps}.

\subsubsection{High intensity}

The higher background photon rate due to the higher muon intensity in MEG~II should not be a
problem for the photo-sensor operation. 
On the other hand, the background rate in the analysis photon energy region
would be increased due to pile-up. In the MEG analysis, the energies of pile-up photons are 
unfolded using the waveform and light distribution on the inner face. 

\begin{figure}
\centering
\subfigure[Before unfolding pile-up photons]{
\includegraphics[width=0.45\textwidth, clip, trim=0 0 0 35pt]{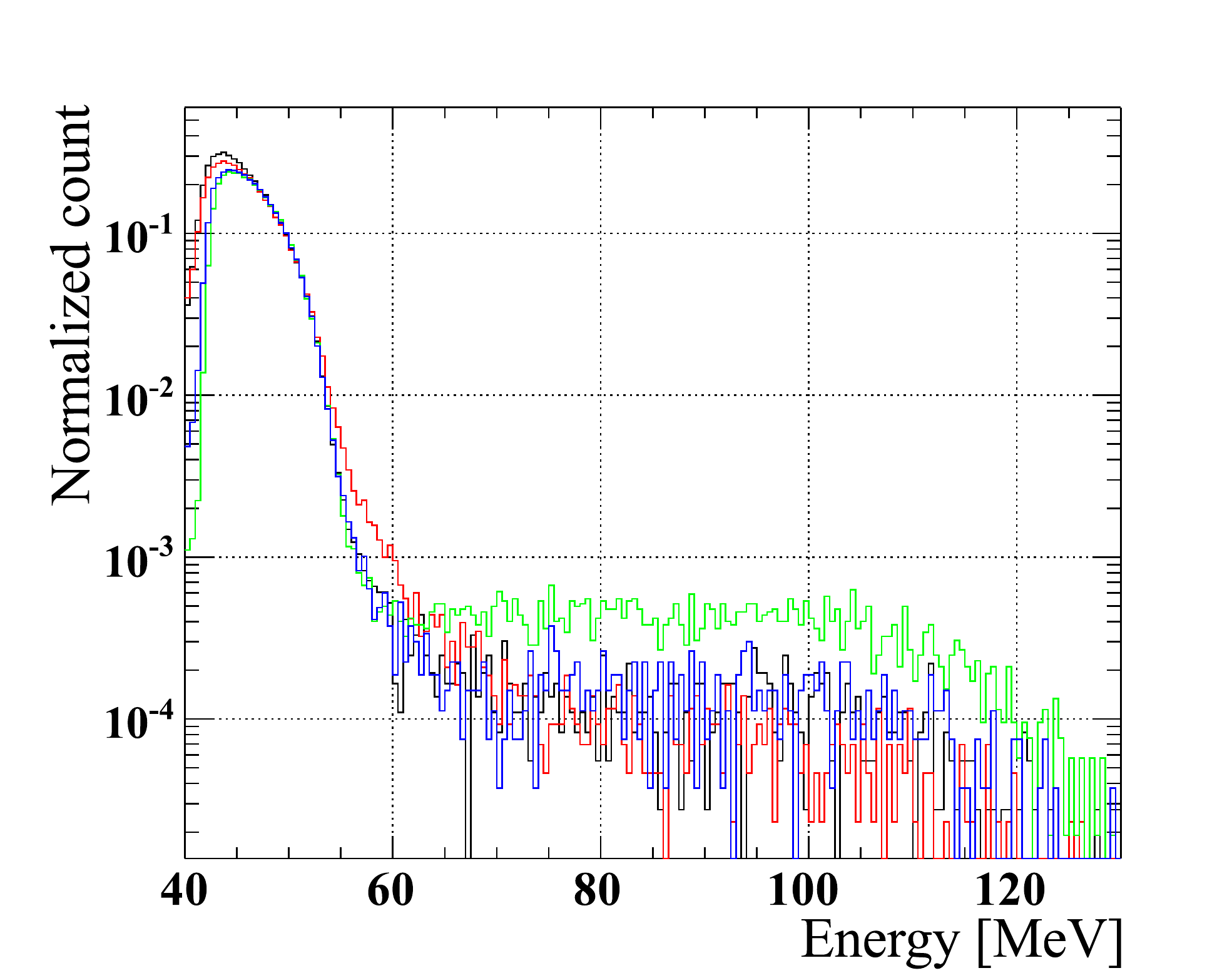}
}
\subfigure[After unfolding pile-up photons]{
\includegraphics[width=0.45\textwidth, clip, trim=0 0 0 35pt]{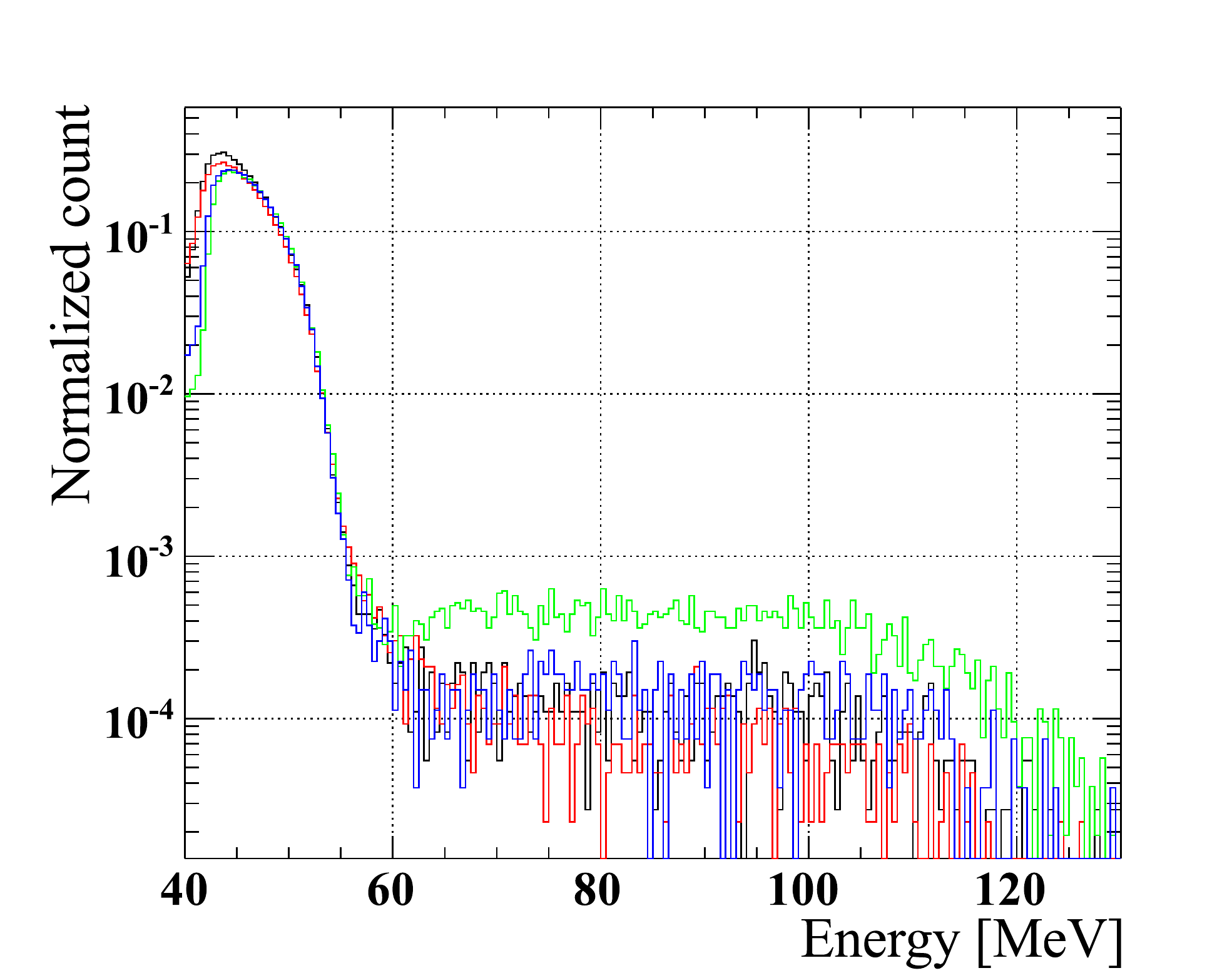}
}
\caption{\label{fig:xec_gammabg_intensity}%
Reconstructed energy spectrum obtained for different beam intensities.
The horizontal axis shows energies in GeV without unfolding pile-up photons (a) and the same after
unfolding and subtracting the energy of pile-up photons (b).
Green, black, blue and red lines show the spectrum at muon stopping rates of 1.0, 3.0, 3.3 and
\SI{8e7}{\muonp\per\second}, respectively.
The spectra are normalised by the number of events in the range \SIrange[range-phrase=--,range-units=single]{48}{58}{\MeV}; the scaling factors are consistent with the muon
stopping rate on the target.
A difference in the low energy part below \SI{45}{\MeV} is due to different effective trigger
thresholds; a difference in the high energy part is due to the different
ratio between the photons backgrounds and the cosmic ray background.
}
\end{figure}

In 2011 we took data with the MEG LXe detector at different beam
intensities: 1.0, 3.0, 3.3 and \SI{8.0e7}{\muonp\per\second}.
Figure~\ref{fig:xec_gammabg_intensity} shows the photon spectrum normalised to the
number of events from \SIrange[range-units=single]{48}{58}{\MeV}; the scaling factors are consistent with to the muon
stopping rate on the target.
The shapes of spectra are almost identical in the analysis region 
after subtracting the energies of pile-up photons.
Since the same analysis can be used also for the MEG~II upgraded detector, a higher beam rate
is not expected to cause an additional background rate due to pile-up.

\clearpage
\newpage
\section{Radiative Decay Counter}
\label{sec:RDC}

The Radiative Decay Counter (RDC) is an additional detector to be installed in
MEG~II. It is capable of identifying a fraction of the low-energy positrons from RMD decays 
having photon energies close to the kinematic limit, which are the dominant source of 
photons for the accidental coincidence background. 
This section describes the concept and the
design of the detector, as well as the results of the pilot run and the
expected performances. 
  \begin{figure}[b]
   \centering
    \includegraphics[width=1\linewidth]{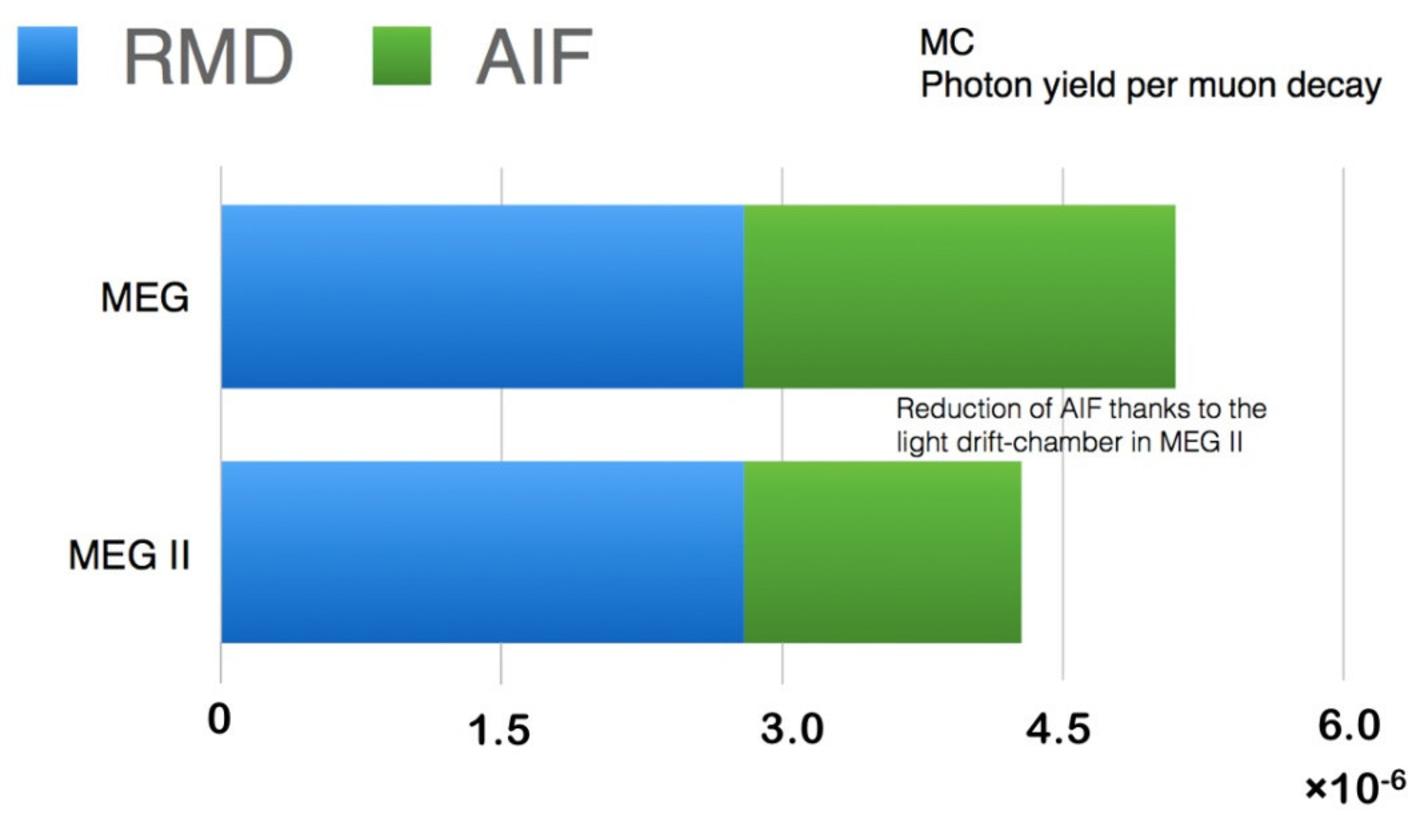}
    \caption{
    \label{fig:RMD_frac}%
    Sources of the background photons ($\egamma > \SI{48}{\MeV}$) in
    accidental background events for MEG and MEG~II.
    }
  \end{figure}

  \subsection{Identification of the RMD photon background} 

  As mentioned in Sect.~\ref{sec:Introduction}, RMD and accidental
  coincidences are the backgrounds in \megc\ search. In the case of the accidental background, 
  which is dominant in MEG~II, photons are produced from either RMD or positron 
  AIF. Figure~\ref{fig:RMD_frac} shows the fraction of 
  background photons expected in MEG and MEG~II from different sources. 
  The AIF background decreases in MEG~II thanks to the reduced mass of
  the CDHC compared with the MEG drift chambers and it is possible to decrease it further by
  looking for a disappearing positron track in the analysis. On the other
  hand, the RMD photon background does not change. Therefore, it is important to
  identify these events. According to simulations, the RDC can detect 
  \num{\sim42}\%\ of the RMD photon background events ($\egamma >\SI{48}{\MeV}$), 
  (the product of the fraction of positrons going downstream (\num{\sim48}\%) 
  and the RDC positron detection efficiency 
  (\num{\sim88}\%, see Table~\ref{tab:rdc_ds_performance})) 
  thus improving the sensitivity of the \megc\ search by 15\%.
  
\begin{figure}[tb] 
   \centering
    \includegraphics[width=1\linewidth]{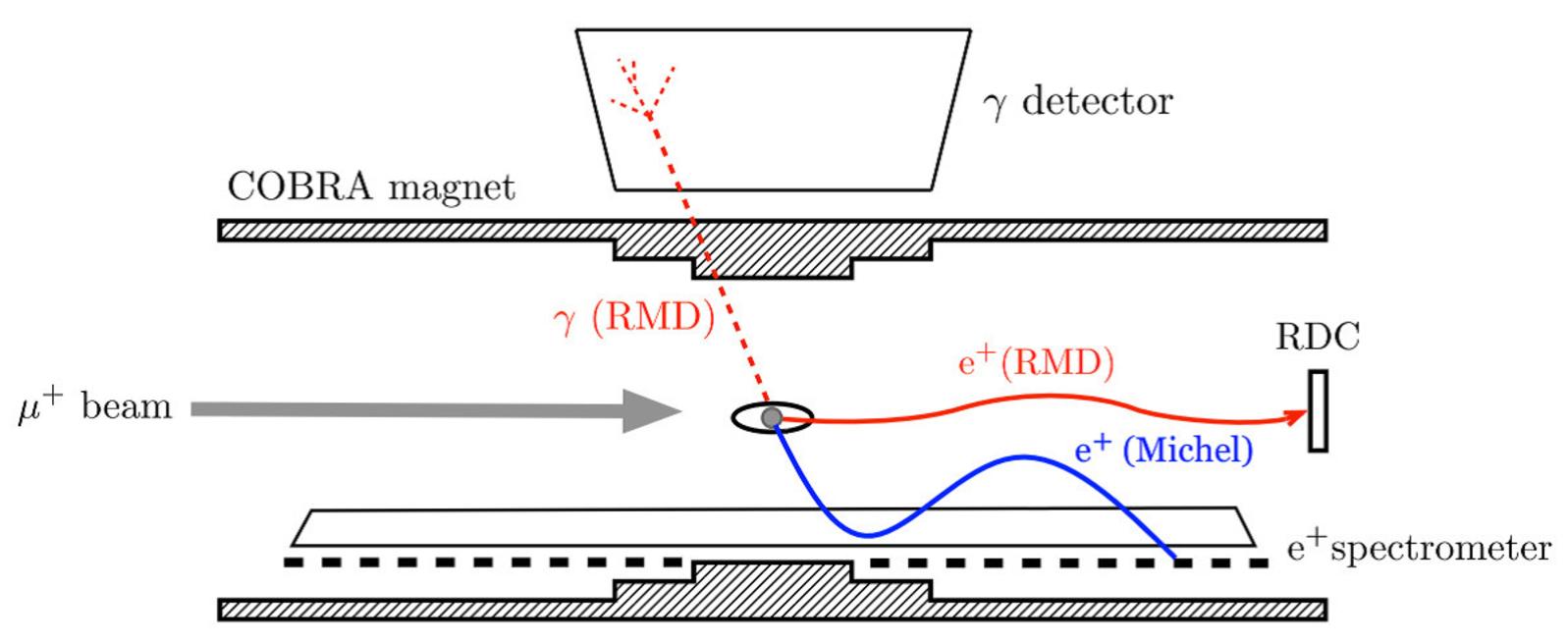}
    \caption{
    \label{fig:rdc_overview}%
    Schematic view of the detection of RMD with the RDC.
    }
  \end{figure}

  The RDC will be installed downstream the \muonp\ stopping
  target as shown in Fig.~\ref{fig:rdc_overview}. A fraction of the RMD events 
  can be identified by
  tagging a low-energy positron in time coincidence with the detection of a high energy
  photon in the LXe detector. This low-energy positron of \SIrange[range-phrase=--,range-units=single]{1}{5}{\MeV} 
  (with $\egamma >\SI{48}{\MeV}$) follows an almost helical trajectory with small radius
  around the B-field lines. Therefore, it can be seen by a small detector with a radius of
  only \SI{\sim 10}{\cm}, placed on the beam axis. There is an option to
  install a detector also upstream, as described in
  Sect.~\ref{sec:rdc_us}. 

  \subsection{Detector design} \label{sec:rdc_design}

\begin{figure}[tb] 
   \centering
    \includegraphics[width=0.4\textwidth]{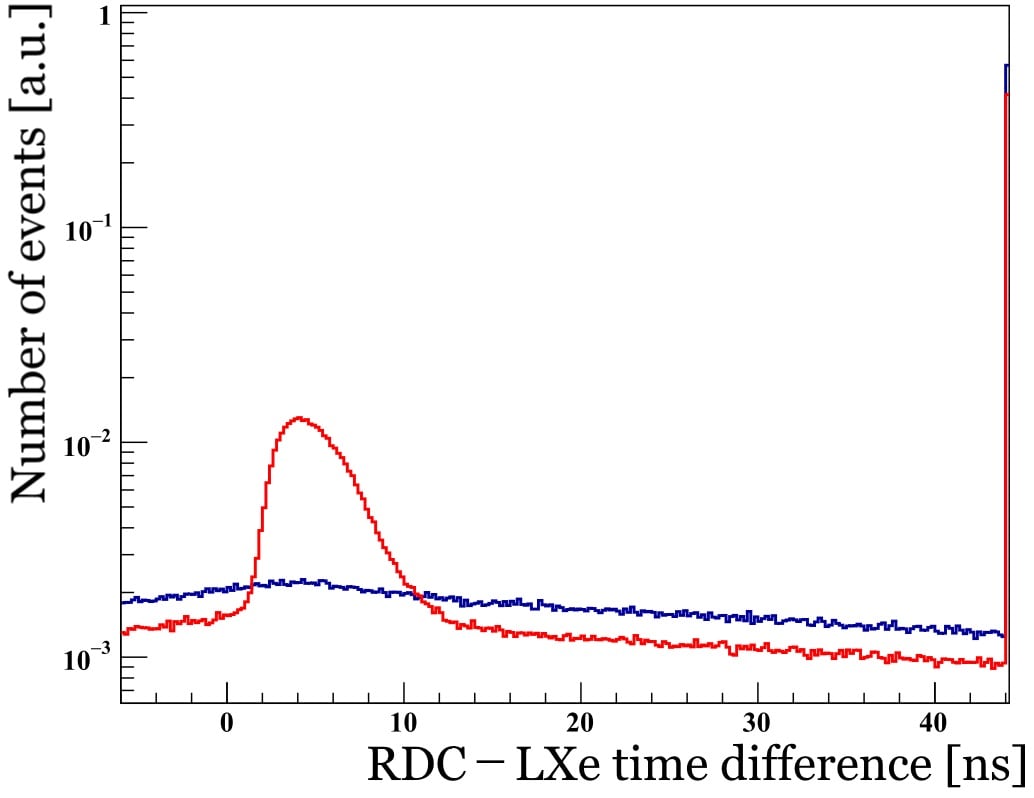}
    \caption{
    \label{fig:rdc_tdiff}
    Simulated time differences between the RDC and LXe photon detectors for
    accidental background events (red) and \megc\ signal events (blue). 
    }
  \end{figure}

\begin{figure}[tb] 
   \centering
    \includegraphics[width=0.4\textwidth]{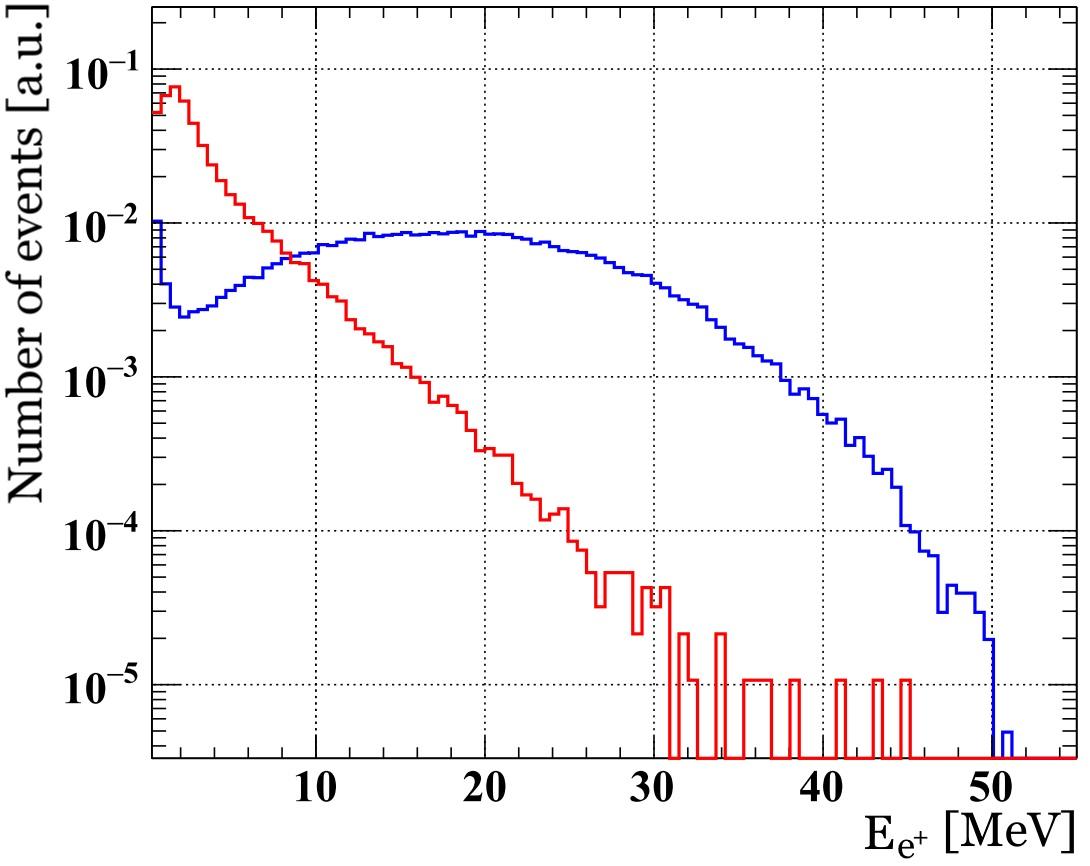}
    \caption{
    \label{fig:rdc_energy}%
    Expected energy distribution at the RDC for RMD events
    with $\egamma > \SI{48}{\MeV}$ (red) and for the Michel events (blue).
    }
  \end{figure}

  The red histogram in Fig.~\ref{fig:rdc_tdiff} shows the expected 
  distribution of the time difference between RDC and the LXe photon detector 
  for accidental background events (with photons from RMD or AIF), while the 
  blue histogram is the distribution due to \megc\ signal events. The peak in the red
  histogram corresponds to the RMD events, while the flat region in both histograms
  corresponds to background Michel positrons.
  As the detector is placed on the beam-axis, there are many background
  Michel positrons (\SI{{\sim}e7}{\positron\per\second}). They can 
  be distinguished from RMD positrons by measuring their energy
  since they typically have higher energies
  as shown in Fig.~\ref{fig:rdc_energy}. Hence, the RDC consists of
  fast plastic scintillator bars (PS) for timing and a LYSO crystal
  calorimeter for energy measurements.

\begin{figure}[tb] 
   \centering
    \includegraphics[width=0.35\textwidth]{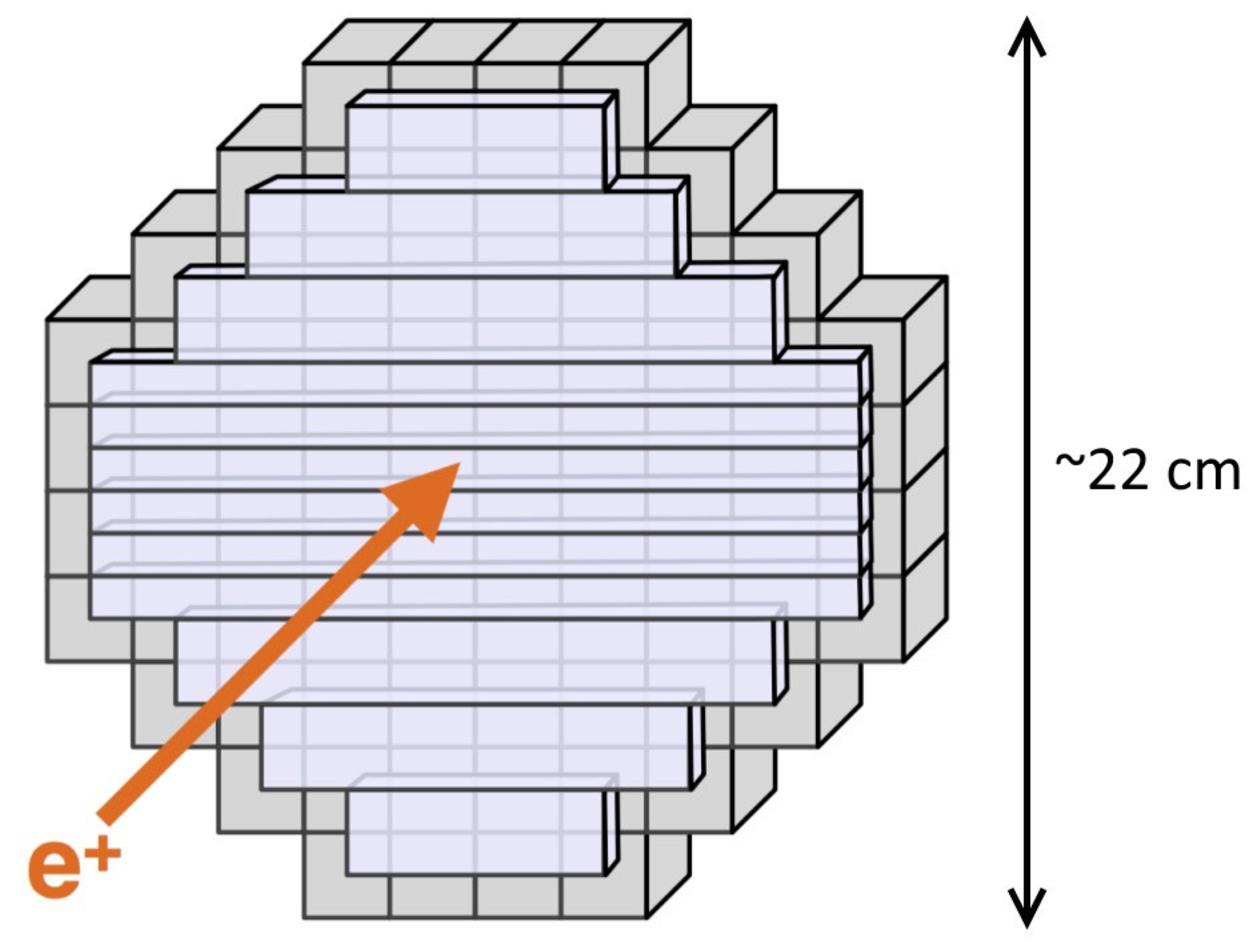}
    \caption{
    \label{fig:rdc_sketch}%
    Schematic view of the RDC. The horizontal long plates in
    front are the plastic scintillator bars, and the cubes behind
    are the LYSO crystals.
    }
  \end{figure}

  Figure~\ref{fig:rdc_sketch} shows a schematic view of the RDC
  detector: 12 plastic scintillator bars in the front detect the timing of
  the positrons, and 76 LYSO crystals behind are the calorimeter
  for energy measurement. In order to distinguish RMD positrons
  from Michel ones, both the PS and the LYSO calorimeter are finely segmented. Because the
  background rate is larger close to the beam axis, the width of
  the PS in the central region is \SI{1}{\cm} while it is \SI{2}{\cm} at the outer
  part. The size of each LYSO crystal is \SI[product-units=single]{2x2x2}{\cm\cubed}.  

\begin{figure}[tb] 
   \centering
    \includegraphics[width=1\linewidth]{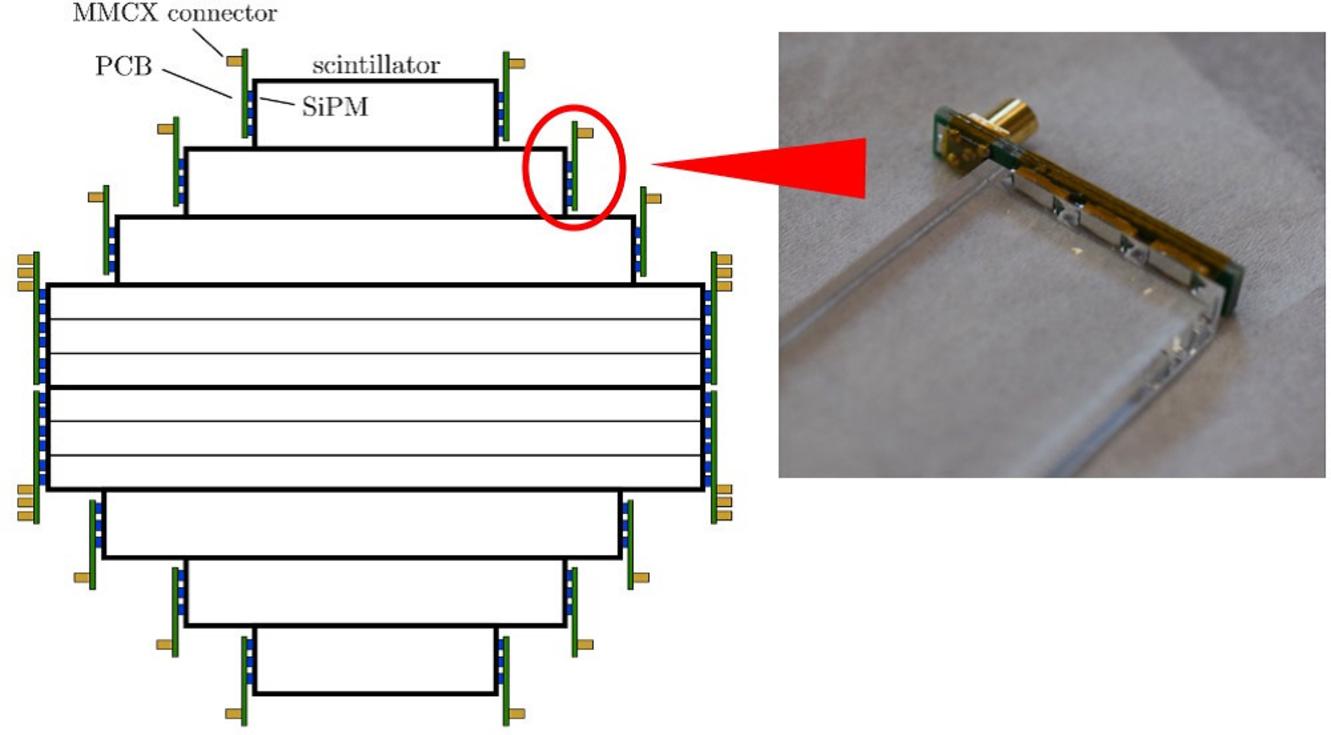}
    \caption{
    \label{fig:rdc_ps}%
    Plastic scintillator bars of the RDC. SiPMs are connected to the scintillator bars at both ends.
    }
  \end{figure}

  The PS shown in Fig.~\ref{fig:rdc_ps} consists of plastic scintillators read out by SiPMs.
  The design of the PS is very similar to that of the
  pTC (Sect. \ref{sec:Pixelated_Timing_Counter}). In
  order to have good timing resolution, 
  scintillators must have a high light yield and short rise time.
  BC-418 from Saint-Gobain~\cite{Saint-Gobain-BC4XX} was selected as it satisfies these requirements. The
  scintillation light is read out by SiPMs at both ends of each
  scintillator. SiPMs are compact and operate in high magnetic fields, and so
  are suitable for the RDC having many readout channels in
  a limited space. The MPPC S13360-3050PE from Hamamatsu Photonics~\cite{hamamatsu-MPPC-RDCPS} 
  was selected for the SiPM
  for PS because of its high gain and high photon detection
  efficiency. In order to detect as much scintillation light as
  possible, multiple SiPMs (two for the central part and three for the outer
  part) are attached to both ends of the scintillators. The
  SiPMs are connected in series on the readout printed
  circuit boards (PCBs) to reduce the number of readout channels and to reduce
  the rise time of the signal due to the reduced capacitance. 
  They are glued to the scintillators
  by optical cement. Each scintillator is wrapped with
  a \SI{65}{\um} thick reflective sheet (ESR from 3M) 
  to increase light yield and to provide optical separation as
  well as black sheets of Tedlar for light shielding.

\begin{figure}[tb] 
   \centering
    \includegraphics[width=1\linewidth]{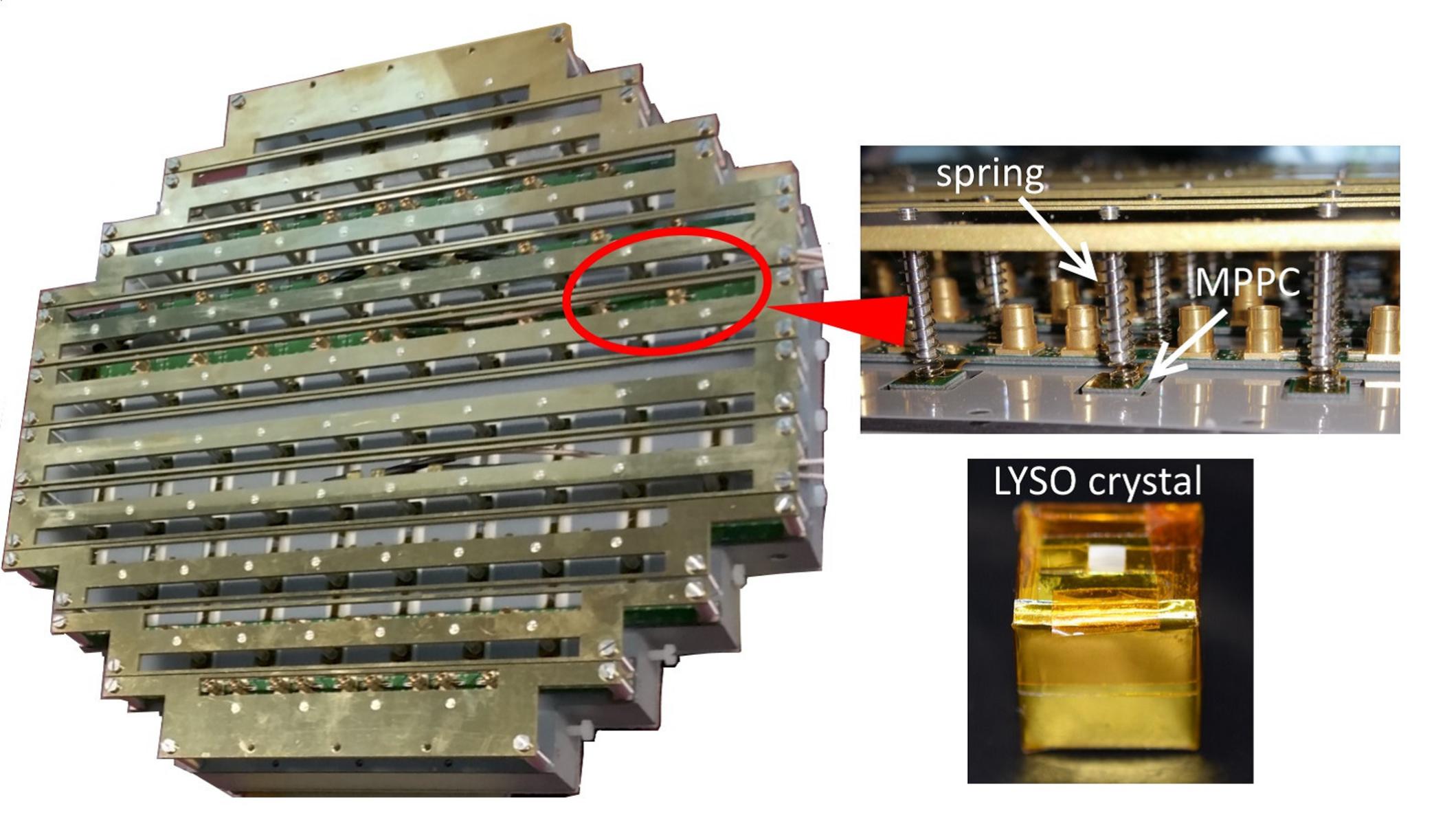}
    \caption{
    \label{fig:rdc_lyso}%
    LYSO crystals with the SiPMs attached with springs.
    }
  \end{figure}

  The calorimeter is made of 76 LYSO crystals (Shanghai Institute of Ceramics). 
  LYSO crystals have a high light yield (\SI{3e4}{photon\per\MeV}) and a short 
  decay time constant (\SI{42}{\ns}). These characteristics are suitable for the
  measurement of positron energy in a high rate environment. 
  LYSO contains the radio isotope $^{176}$Lu, which decays to
  $^{176}$Hf with emission of a $\beta^-$ (with end-point energy of \SI{596}{\keV}
  and half life of \SI{3.78e10}{years}), followed by a cascade
  of \SIlist{307;202;88}{\keV} $\gamma$-rays. As described in
  Sect. \ref{sec:rdc_beamtest}, this intrinsic radioactivity can be
  used for an energy calibration. The decay rate is measured to be
  small (\SI{\sim 2}{\kHz}), therefore not affecting the detection of positron
  from RMD.
  Each LYSO crystal is connected to one SiPM at the
  downstream side (see Fig.~\ref{fig:rdc_lyso}). A SiPM with a small pixel size
  of \SI{25}{\um} (S12572-025 from Hamamatsu Photonics~\cite{hamamatsu-MPPC-RDCLYSO}) 
  was selected as it
  has good linearity for high intensity of incident scintillation light. The SiPM 
  has spring-loaded contact to the crystal, using optical grease, instead of being
  glued. Therefore, it is possible to replace the SiPM or the crystal. 
  
  \subsection{Tests and construction in the laboratory} \label{sec:rdc_construction}

  The characteristics of each SiPM for the PS are measured before
  construction. The breakdown voltage is obtained for each SiPM from the
  measurement of the current--voltage response curve. SiPMs with the
  breakdown voltages close to each other are grouped together and
  connected in series. After the construction of the PS, the timing resolution of each
  counter is measured to be less than \SI{90}{\ps} by using a $^{90}$Sr source. 

  The LYSO crystals are also tested individually. We measured the light yield
  and the energy resolution of all the crystals by using a $^{60}$Co
  source. The energy resolution was measured to be \num{\sim6}\% at
  $\egamma = \SI{1}{\MeV}$ for all the crystals.
  In a high rate environment, energy
  resolution can be worsened by the ``afterglow'' effect of LYSO.
  Afterglow is a delayed light emission of crystals with very
  long time constant (typically few hours). This effect was
  studied by exposing the crystals to a $^{90}$Sr source. The increase of
  the current due to afterglow was measured with the SiPMs attached to the crystals. 
  According to this measurement, the expected increase of
  the current in the MEG~II beam environment is estimated to be \SI{\sim10}{\uA} at maximum. The influence on the energy 
  resolution is expected to be less than 1\% at $\egamma = \SI{1}{\MeV}$.

  The support structures of the PS and of the LYSO calorimeter are constructed with
  non-magnetic materials such as aluminium. The front part of the PS is not
  covered with metal, in order to minimise the amount of material. In
  order to absorb the stress of the springs 
  (\SI{\sim 2.5}{\kg} in total) with the minimum amount of material, a \SI{3.3}{\mm}
  Rohacell plate sandwiched with two 
  CFRP (Carbon Fibre Reinforced Polymer) plates (\SI{0.2}{mm}
  each) is inserted between the PS and the LYSO calorimeter. In addition, a \SI{0.1}{\mm} thin
  aluminium plate is inserted for better light shielding. The back side of
  the crystals is covered by two Delrin\textsuperscript{\textregistered} plates and one CFRP plate.

\begin{figure}[tb] 
   \centering
    \includegraphics[width=1\linewidth]{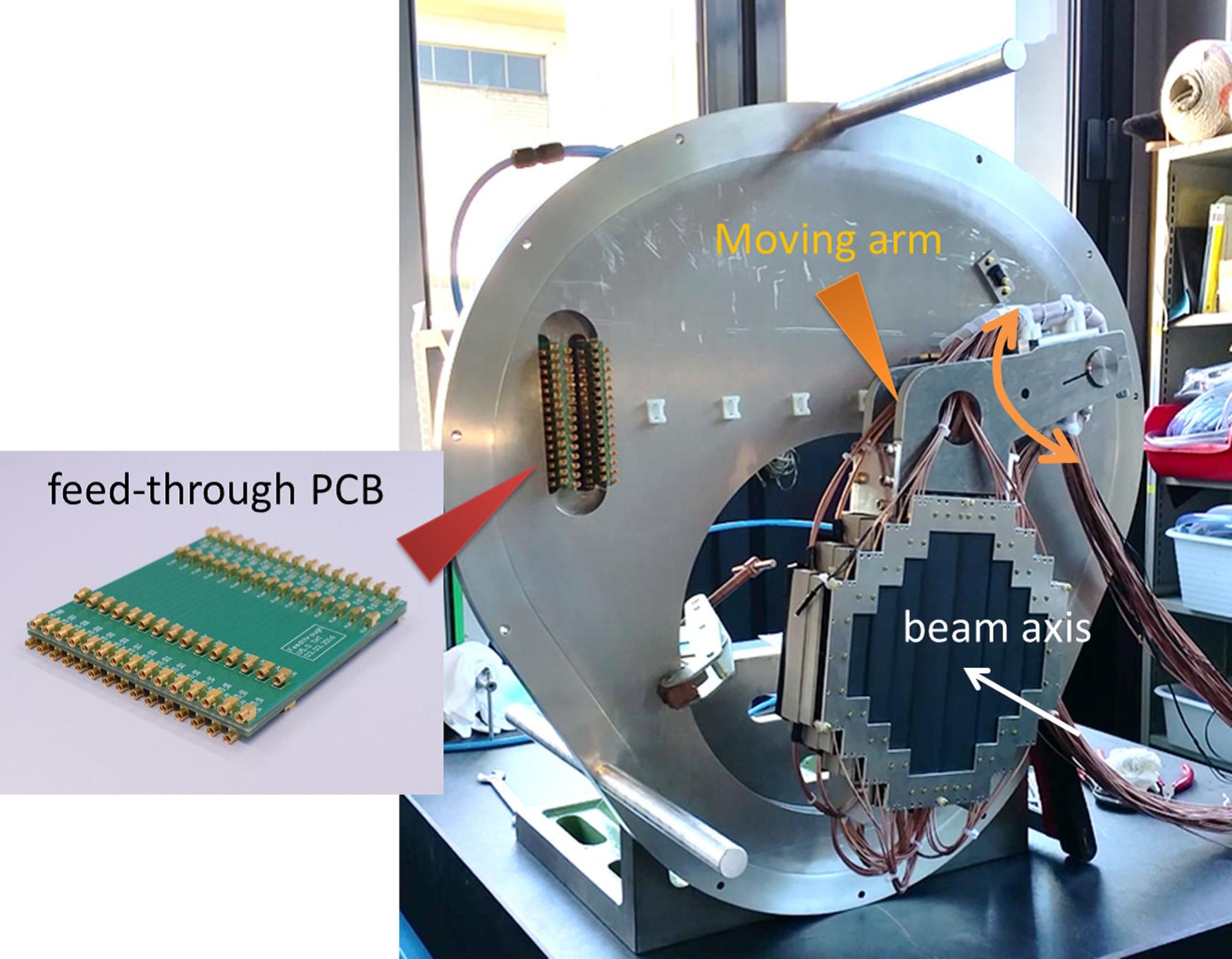}
    \caption{
    \label{fig:rdc_mover}%
    Downstream RDC mounted on a moving arm. 
    }
  \end{figure}

  Figure \ref{fig:rdc_mover} shows the RDC mounted on a moving arm
  system attached to the end-cap of the COBRA magnet. The RDC can be remotely moved
  away from the beam-axis when the calibration target for the LXe photon detector is
  inserted from the downstream side. The moving arm is controlled by
  water pistons made of plastic, which work in a magnetic field. The
  supporting mechanics are made of aluminium except for the titanium shaft, which works under 
  heavy loads. The end-cap of COBRA separates the inner volume 
  (filled with helium) from the outside. SiPM signals are transmitted
  through the end-cap by using feed-through PCBs attached to the
  end-cap. The design of the feed-through is essentially the same as
  used for the LXe photon detector (see Sect.~\ref{sec:cable and feed-through}). 

  \subsection{Pilot run with a muon beam} \label{sec:rdc_beamtest}

\begin{figure}[tb] 
   \centering
    \includegraphics[width=0.45\textwidth, clip]{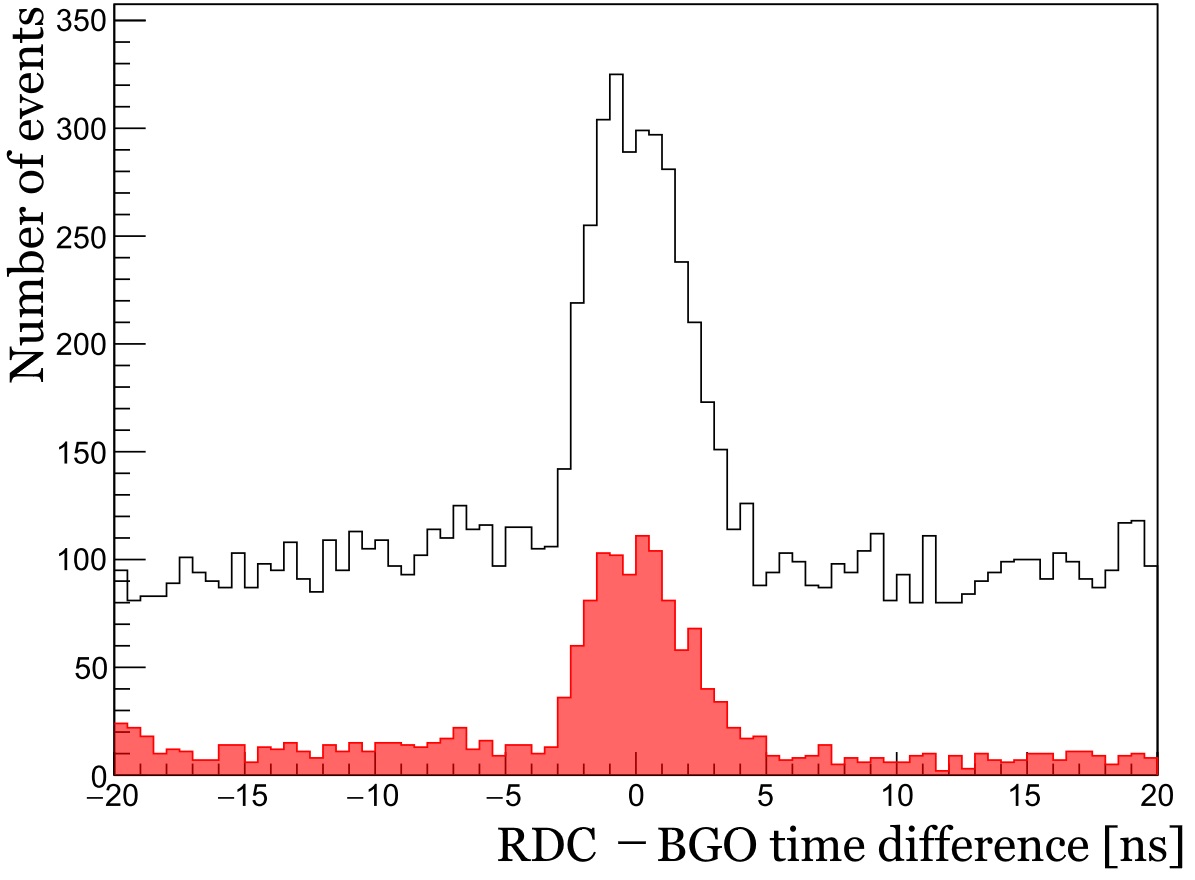}
    \caption{
    \label{fig:rdc_beamtest_tdiff}%
    Time difference of the RDC PS and BGO hits, from the beam
    test. Black (red) histogram shows the distribution before (after)
    applying a cut to the energy deposit in the LYSO calorimeter.
    }
  \end{figure}

  The full detector system was tested in the $\pi$E5 beam line at PSI
  with a beam intensity of \SI{\sim 1e8}{\muonp\per\second}.
  The RDC was mounted at the downstream end of the COBRA magnet. 
  For the detection of photons from RMD, a BGO detector consisting of 16
  crystals (\SI[product-units=power]{4.6x4.6x20}{\cm} each) was used as a substitute of the LXe photon detector.
   RMD events were acquired by requiring an energy deposit larger than
   \SI{\sim35}{\MeV} in the the BGO.
   After event selection to reject cosmic rays, \num{\sim 15000} events
   remained.
   The distribution of the time difference between the BGO hit
   and the PS hit is shown in Fig.~\ref{fig:rdc_beamtest_tdiff}. A clear
   peak corresponding to RMD events is successfully observed.
   
\begin{figure}[tb] 
   \centering
    \includegraphics[width=0.45\textwidth, clip]{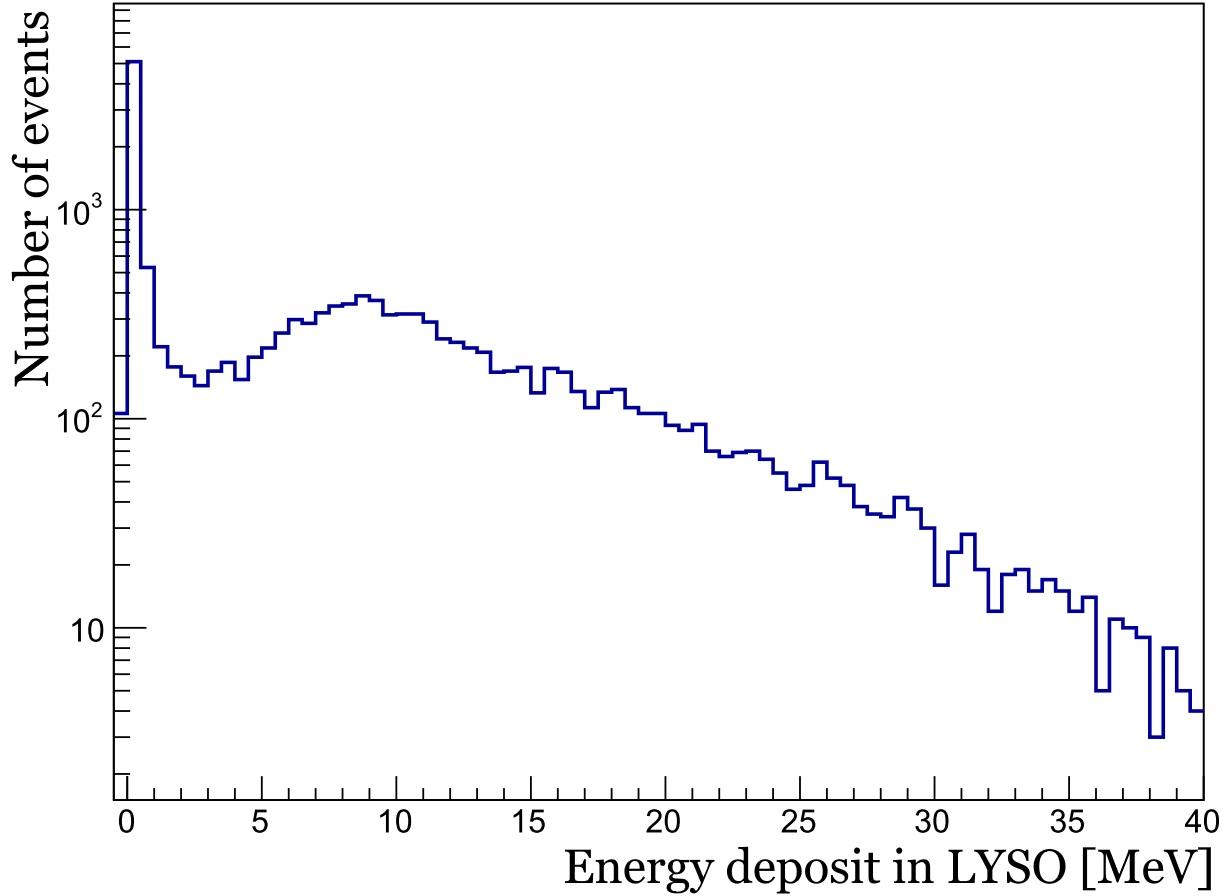}
    \caption{
    \label{fig:rdc_beamtest_lyso}%
    Energy distribution observed in the LYSO calorimeter.
    }
  \end{figure}

  Figure \ref{fig:rdc_beamtest_lyso} shows the measured distribution of the energy loss in 
  the LYSO calorimeter.
  Higher energy tail events are mainly Michel positron backgrounds,
  while the low energy part (\SI{<5}{\MeV}) corresponds to RMD.
  For a demonstration, we applied an event selection to reject events
  with an energy release in the calorimeter above \SI{4}{\MeV}. 
  The red histogram in Fig.~\ref{fig:rdc_beamtest_tdiff}
  shows the timing distribution after the calorimeter LYSO energy cut. The flat
  region which corresponds to backgrounds is reduced to \num{\sim 1/10} by
  this cut. The peak region (i.e. RMD events) is also reduced to
  \num{\sim 1/3} because the energy threshold for the BGO trigger was 
  low and therefore the energy of the RMD positron could be high.

  \subsection{Expected performance} \label{sec:rdc_performance}

  The sensitivity of the \megc\ search in MEG~II including the RDC is
  calculated by using the expected timing difference distribution of the RDC
  and LXe photon detectors (see Fig.~\ref{fig:rdc_tdiff}) and the expected energy
  distribution in the LYSO calorimeter (see Fig.~\ref{fig:rdc_energy}) (see Sect.~\ref{sec:sensitivity} for the details). 
  In the MEG physics analysis \cite{meg2010, meg2013, meg2016}, the
  likelihood depends on the number of events (signal and background) and
  the probability density functions (PDF) based on 
  the energy, timing and relative angles of positron and photon.
  The RDC observables can be added in the likelihood analysis by using the
  PDFs of the PS--LXe timing difference and of the LYSO calorimeter energy. 
  Table~\ref{tab:rdc_ds_performance} summarises the performances of
  the RDC assumed in the calculation.
  By using the RDC, the sensitivity of MEG~II is expected to improve by 15\%.

  \begin{table}[tb]
   \caption{Performances of the RDC assumed in the sensitivity
   calculation. RMD acceptance is the probability to detect 
   RMD positrons going downstream, for $\egamma > \SI{48}{\MeV}$.
   RMD detection efficiency is the probability of detecting a positron 
   falling in the geometrical acceptance range.
   Accidental probability is the probability of
   observing a Michel positrons in the RDC uncorrelated to the 
   photon at $R_\muonp = \SI{7e7}{\per\second}$.}
   \label{tab:rdc_ds_performance}
   \centering
    \begin{tabular}{lc}
     \hline
     Parameter              & Value    \\ 
     \hline 
     LYSO energy threshold       & \SI{30}{\keV}   \\ 
     RMD detection efficiency   & 100\%    \\
     LYSO energy resolution & 8\%      \\
     Time resolution        & \SI{100}{\ps}   \\
     Accidental probability & 9\%      \\
     RMD acceptance         & 88\%     \\
     \hline
    \end{tabular}
  \end{table}

  \subsection{Further background reduction with an upstream RDC} \label{sec:rdc_us}

  Because half of the positrons from RMD go upstream, 
  it is possible to further improve the sensitivity by adding 
  an additional RDC in the beam line upstream the muon
  stopping target, near the end of the COBRA magnet. 
  The upstream RDC has to be very different from the downstream
  RDC as it must be placed on the beam path. First of all, the material thickness
  must be small enough to minimise the impact on the beam which prevents
  the use of a calorimeter. Secondly, the detector
  must be able to distinguish the RMD positrons from beam muons. This could be
  possible with a fast response, finely segmented detector.

  A possible candidate is a layer of scintillation fibres with
  SiPM readout. Fibres can be bundled at both ends to reduce the
  number of readout channels. A fibre candidate is BCF-12 (Saint-Gobain
  \cite{SaintGobainFibres}), a double-clad square shaped fibre
  \SI{250}{\um} wide. With this thickness, the effect to the
  muon beam optics is expected to be negligibly small. However, 
  radiation damage on fibres and pile-up of the beam muon signals
  (after-pulse of SiPMs increases the pile-up probability) may affect the
  detector performance. 

  Another candidate is a synthetic diamond detector. Diamond detectors
  have fast signal, and can be manufactured in a thin layer. They are 
  also known to be radiation hard. The drawback is their low signals, which
  requires high gain amplifiers with low noise.

  The estimated improvement of the sensitivity with the upstream
  RDC is 10\% when the detection efficiency is 100\%.

\clearpage
\newpage
\section{Trigger and DAQ}
\label{sec:tdaq}
%

This section describes an innovative integrated trigger and data acquisition system designed for the MEG~II detector. 
After a description of the main requirements, the designed circuit characteristics and their interplay are described. 
We conclude with the latest results from the research and development phase.

\subsection{Requirements}

The MEG~II sensitivity goal requires a substantial detector and read-out electronics 
redesign to deal with a factor of two increase in muon stopping rate 
with respect to MEG. 
As a consequence we replaced many of the PMTs of the LXe and timing counter detectors with SiPMs and MPPCs; 
similarly the new CDCH design requires more read-out channels compared to the MEG drift chambers. 
In summation this has led to an almost tripling of read-out channels with respect to MEG.
The requirement for an efficient offline pile-up reconstruction and rejection is the availability of full waveform information;
thus the DAQ waveform digitiser has to provide state-of-the-art time and charge resolution and a sampling speed in the GSPS range.

In addition, SiPMs have a lower gain than PMTs and require electronic signal amplification. 
Using SiPMs in LXe prevents us from placing preamplifiers directly next to the photo-sensors because of cooling problems;
it is therefore mandatory that the new electronics contains flexible amplification stages for small signals (single photo-electrons for calibration) as well as large signals ($\gamma$-showers).

\begin{figure}[b]
\centering
\includegraphics[width=0.95\columnwidth]{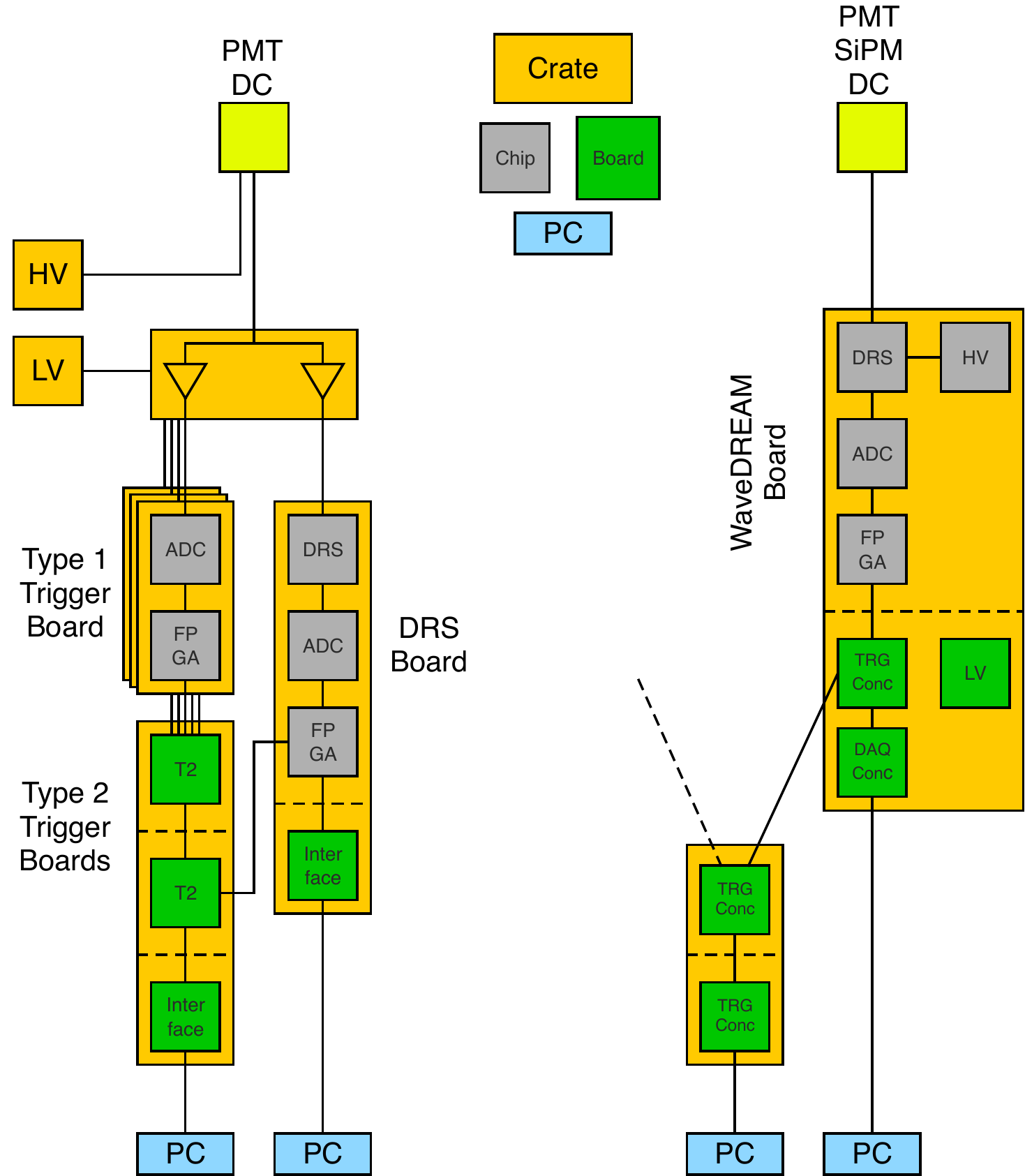}
\caption{\label{fig:tdaqschema}%
Comparison of the old (left) vs. the new (right) TDAQ electronics designs; 
the active splitter system present in the old version is integrated, together with the \lq\lq{}Type1\rq\rq{} and \lq\lq{}DRS Board\rq\rq{} functionalities in the WaveDREAM board, making the MEG~II TDAQ system extremely compact.}
\end{figure}

As shown in Fig.~\ref{fig:tdaqschema}, the detector signals in MEG were actively split and then sent to the dedicated VME-based trigger and DAQ systems; 
the limited space for the electronics in the experimental area prevents us from adopting such a scheme with the increased number of channels expected in MEG~II. 

\subsection{The WaveDREAM Board}
\label{sec:wavedream}
\begin{figure*}[tb]
\centering
\includegraphics[width=1\linewidth]{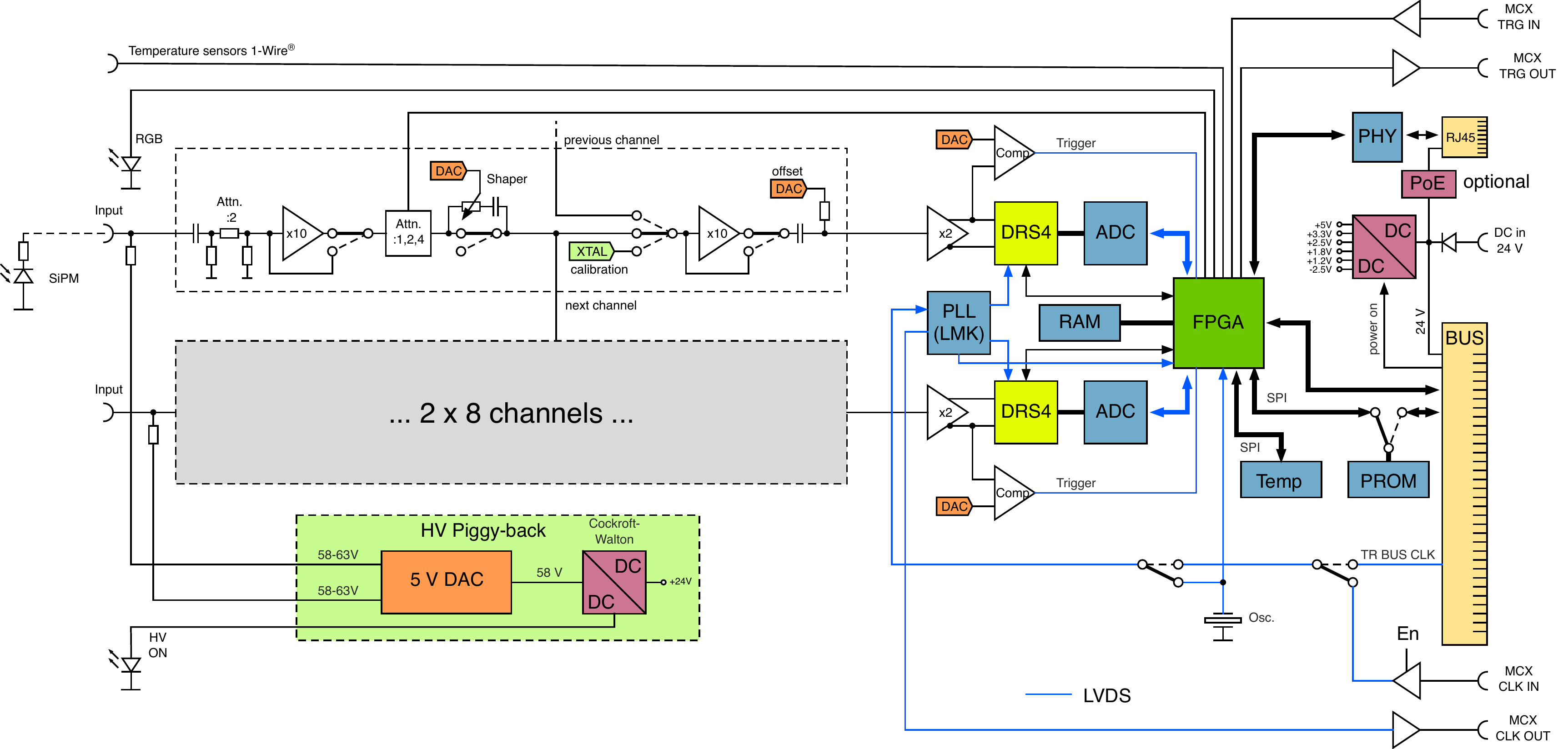}
\caption{\label{fig:WDBSchematics}%
Simplified schematics of the WaveDREAM board. 
It contains 16 variable gain input amplifiers, two DRS4 chips, 16 ADC channels and a Spartan 6 FPGA. 
A optional high voltage generator for SiPM biasing can be mounted as a piggy-back board.}
\end{figure*}

The new system integrates the basic trigger and DAQ (TDAQ) functionalities onto the same electronics board, the WaveDREAM board (WDB).
A simplified schematics of the WDB is shown in Fig.~\ref{fig:WDBSchematics}. 

It contains 16 channels with variable gain amplification and flexible shaping through a programmable pole-zero cancellation.
Switchable gain-10 amplifiers and programmable attenuators allow an overall input gain from 0.5 to 100 in steps of two.
A multiplexer can be used to send one input signal to two channels simultaneously which can be set at different gains, at the expense of only having 8 channels per board.
Two DRS4 chips~\cite{ritt_2004_nim} are connected to two 8-channel ADCs, which are read out by a Field-Programmable Gate Array (FPGA).
In normal operation, the DRS4 chips work in \lq\lq{}transparent mode\rq\rq{}, 
where they sample the input signals continuously at a speed up to 5~GSPS in an analogue ring buffer.
At the same time, a copy of the input signal is sent to the DRS4 output, where it is digitised continuously by the ADCs at 80~MSPS with a resolution of \SI{12}{\bit}.

The output stream of the ADCs is used in the FPGA to perform complex trigger algorithms such as a threshold cut on the sum of all input channels.
Interpolation of the ADC samples via look-up tables allows time coincidence decisions with resolutions of a few nanoseconds to be made, much less than the ADC sampling speed.
In case a trigger occurs, the DRS4 chip is stopped and the internal 1024-cell analogue memory is digitised through the same ADCs previously used for the trigger.
With this technique, both complex triggering and high speed waveform sampling is possible on the same board.

\begin{figure}
\centering
\includegraphics[width=0.7\columnwidth]{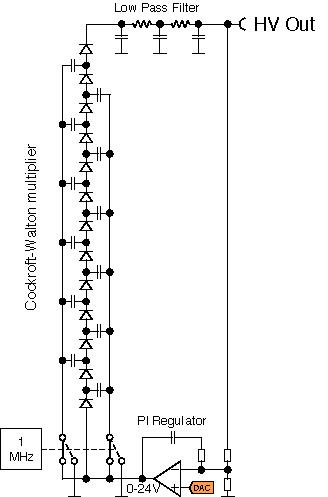}
\caption{\label{fig:CW}%
Simplified schematics of the Cockcroft--Walton voltage multiplier.}
\end{figure}

The SiPMs of the MEG~II experiment require bias voltages in the range of \SIrange[range-units=single,range-phrase=--]{30}{60}{\volt}.
Some detectors use six SiPMs in series, which requires a maximum voltage up to \SI{240}{\volt}.
This voltage can be supplied through the signal cables with capacitive de-coupling of the signal into the amplifiers as shown in Fig.~\ref{fig:WDBSchematics}.
An ultra-low noise bias voltage generator has been designed to accommodate these needs.
A Cockcroft--Walton (CW) stage (also known as Greinacher multiplier) generates a high voltage output of \SI{24}{\volt} at a switching frequency of \SI{1}{\MHz} (see Fig.~\ref{fig:CW}).

\begin{figure}
\centering
\includegraphics[width=1\columnwidth]{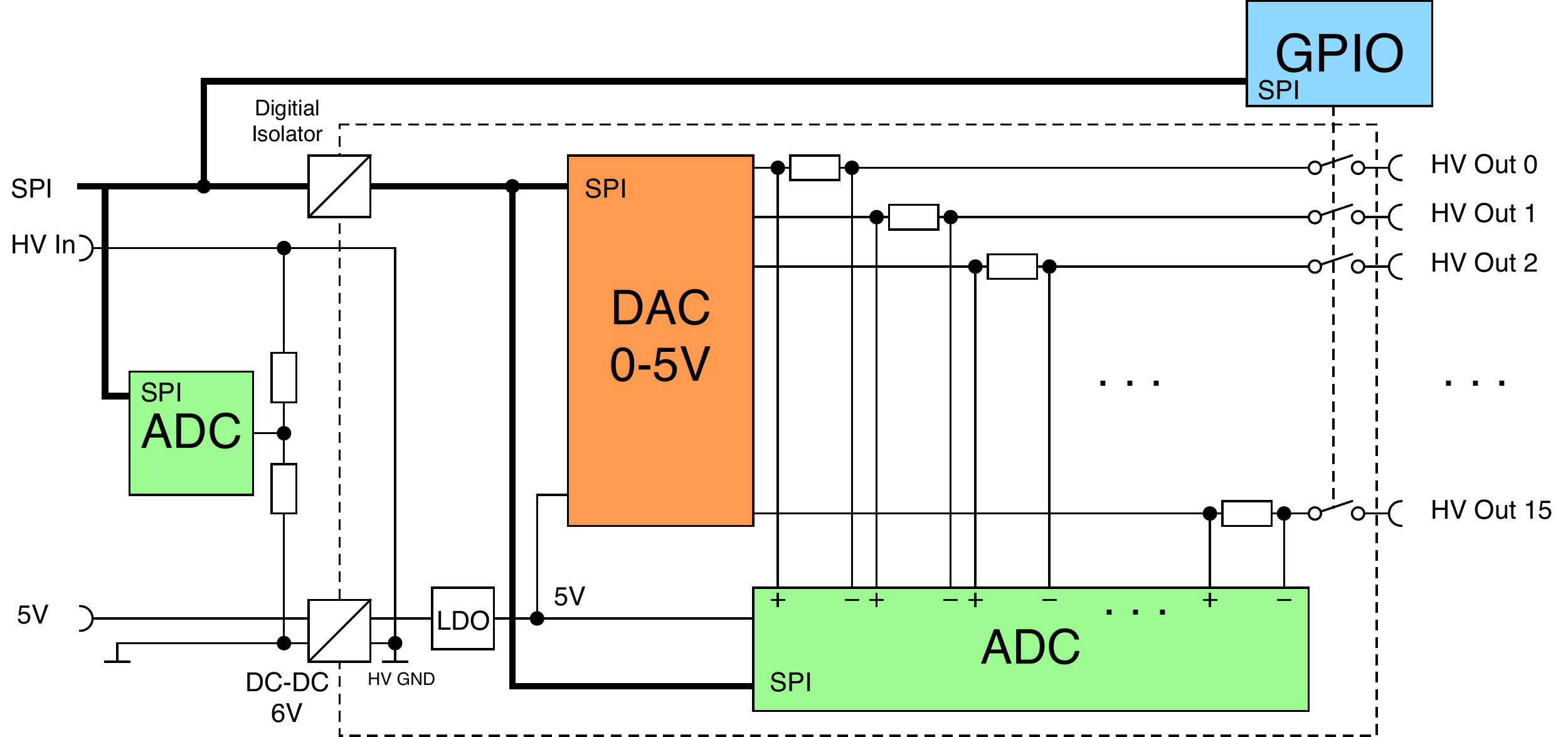}
\caption{\label{fig:HV}%
 Simplified schematics of the bias voltage control.}
\end{figure}

A Proportional-Integral (PI) regulator keeps the output voltage stable by comparing it through a voltage divider with a demand voltage given by a DAC.
An elaborate low pass filter reduces the output ripple to below \SI{0.1}{\milli\volt}, so that it cannot be seen, even with an amplifier gain of 100, at the input of the WDB.
Since SiPMs require slightly different bias voltages, a simple \SI{5}{\volt} DAC \lq\lq{}sitting\rq\rq{} at the high voltage potential can add between \SIrange{0}{5}{\volt} to the output voltage on a channel-by-channel basis (see Fig.~\ref{fig:HV}).

The \SI{5}{\volt} DAC and the ADC for current measurements are placed at a high voltage ground defined by the CW generator.
An isolated DC-DC converter generates, together with a low drop-out (LDO) regulator, the \SI{5}{\volt}  power supply voltage required
by the DAC and the ADC. They are interfaced through a SPI bus via a digital isolator. 
A separate \SI{24}{\bit} ADC measures the CW voltage through a precision voltage divider.

At a CW voltage of \SI{58}{\volt} for example, output voltages from \SIrange{58}{63}{\volt} can be generated for each channel, 
which is sufficient to accommodate variations between different SiPMs.
The output current is measured via shunt resistors and a \SI{16}{\bit} ADC with differential inputs.
The voltage drop across the shunt resistor is measured and converted into a current by the control software running on the soft core processor in the FPGA.
The DAC is adjusted according to the voltage drop to keep the output voltage stable independent of the current,
while high voltage CMOS switches (IXYS CPC7514) are used to turn off individual channels.
Different CW generators have been developed for different output voltages and powers, 
reaching up to \SI{240}{\volt} and \SI{50}{\milli\ampere}.
Alternatively, a single high voltage can be distributed throughout the crate backplane, 
reducing costs by eliminating individual CW generators for each WDB.

\begin{figure}
\centering
\includegraphics[width=0.95\columnwidth]{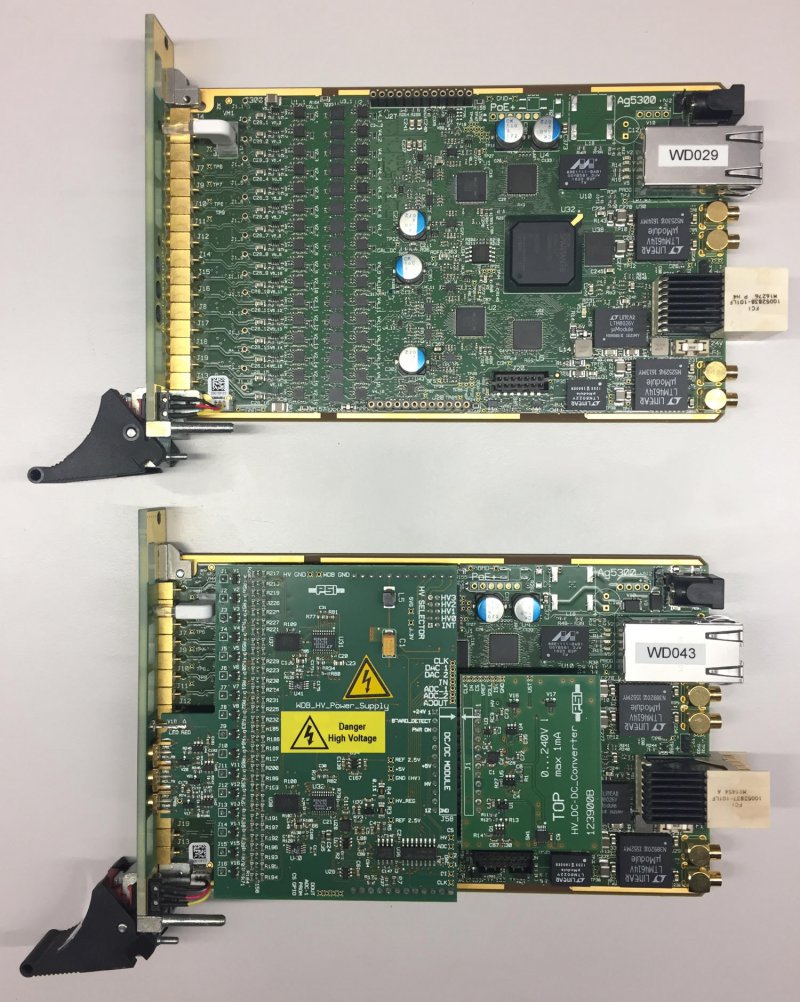}
\caption{\label{fig:WDB}%
Two WaveDREAM boards without (top) and with a high-voltage piggy-back board (bottom).}
\end{figure}

Using this scheme, a cost effective and highly precise bias generator has been realised.
The absolute voltage accuracy (as measured with an external multimeter) is below \SI{1}{\milli\volt} at a maximum current of \SI{2.5}{\milli\ampere}.
The current measurement has a resolution of \SI{1}{\nano\ampere} at a full range of \SI{50}{\micro\ampere} with an accuracy of 0.1\%.
The high voltage bias generator is implemented as an optional piggy-back PCB placed on top of the WDB
(see Fig.~\ref{fig:WDB}), so it can be omitted for channels which do not need biasing (such as PMT channels which have a separate high voltage supply), thus reducing costs. 

\begin{figure}
\centering
\includegraphics[width=1\linewidth]{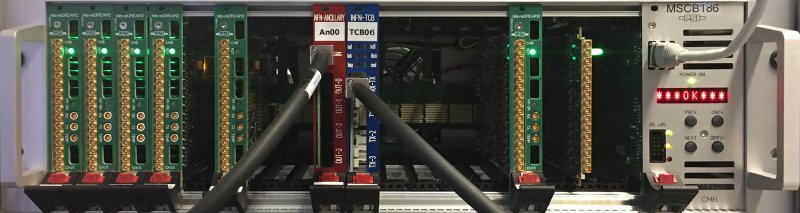}
\caption{\label{fig:WDCrate}%
 WaveDREAM crate shown with 7 WDB (green), one ancillary board (red), one TCB (blue) and the CMB (right).}
\end{figure}

The WDB can be used in stand-alone mode, where it is read out through Gigabit Ethernet and powered through Power-over-Ethernet (PoE+).
For MEG~II, it has been decided to house 16 boards in a compact crate. 
This crate requires \si{\giga\bit} links for the simultaneous read-out of waveform and trigger data, 
a common high voltage for the SiPM biasing,
an integrated trigger distribution and
an ultra-low jitter clock with a few picoseconds precision.
Since such a crate is not available on the market, a new standard has been developed.
The WaveDAQ crate is a 3~HE 19" crate with 16+2 slots and a custom backplane as can be seen in Fig.~\ref{fig:WDCrate}.
The Crate Management Board (CMB) contains the 220~V power supply together with a shelf management unit and is placed to the right side of the crate.
The power supply generates a \SI{24}{\volt} crate power of \SI{350}{\watt} and a \SI{5}{\volt} standby power for the shelf manager.
Cooling is achieved by fans on the rear-side blowing air from the back to the front, 
where it exits through holes in the various boards.
This topology allows stacking of crates directly on top of each other, making the whole system very compact.

The CMB contains an \SI{8}{\bit} micro-controller programmed in the C-language. 
It is connected to a dedicated Ethernet network for remote control and monitoring, 
and has a LED display and buttons for local control.
Current and temperature sensors reflect the state of the crate, and each of the 18 slots can be powered on and off individually.
The micro-controller is connected to all slots via a Serial Peripheral Interface (SPI) bus.
This allows detection of individual boards in each slot, communication with all WDBs as well as remote firmware updates through the backplane.
A physical select line for each slot allows geographic addressing as in the ``good old CAMAC days''.

The WaveDAQ crate contains 16 slots for WDBs, which provide 256 input channels.
The flexibility of the WDBs allows the readout of SiPMs, PMTs and drift chamber channels. 
The MEG~II experiment will use a total of 37 such crates for the data acquisition of all detectors.
The global trigger as well as the trigger and clock distribution is also housed in WaveDAQ crates,
increasing the total number to 39.

In addition, the WaveDAQ crate contains two slots for so-called \lq\lq{}concentrator\rq\rq{} boards. 
The Trigger Concentrator Board (TCB) receives 8 \si{\giga\bit} serial links from each WDB and is described later.
The Data Concentrator Board (DCB) received two separate \si{\giga\bit} serial links from each WDB for waveform readout.
The dual star topology allows the operation of both the trigger and the DAQ system simultaneously without interference.

An integrated trigger bus allows the distribution of trigger signals through the backplane.
Busy signals from each slot are connected via a \lq\lq{}wired-or\rq\rq{} and are used to re-arm the trigger after an event.
A low jitter clock with skew corrected PCB traces is distributed through the backplane.
Measurements show a slot-to-slot variation below \SI{50}{\ps} and a jitter below \SI{5}{\ps}.
All backplane communication signals except the busy line use the LVDS standard.

Each WDB supports hot-swap functionality. 
During hot insertion, an inrush current controller ramps up the board's capacitors gently, avoiding connector sparks and backplane power supply glitches. 
A switch at the handle latch switches off the internal power before the board is extracted.

\subsection{Data read-out: the Data Concentrator Board}

The DCB is responsible for the configuration of all boards inside the crate through the SPI links,
the distribution of the master clock and trigger signals, 
the readout of waveform data from each slot through dedicated serial links, 
the merging and formatting of the data, and the interface to the global DAQ computers through Gigabit Ethernet. 
It uses a Xilinx Zynq-7030 chip which contains a dual-core ARM Cortex-A9 processor embedded 
in the FPGA fabric and running at \SI{1}{\giga\hertz}.
This chip is complemented with a SD card to store the Linux operating system, 
\SI{512}{\mega\byte} of DDR3 RAM, 
and a Small Form-Factor Pluggable (SFP) transceiver for 1 or \SI{10}{\giga\bit/\second} Ethernet.
A dedicated clock distributor with integrated jitter cleaner (LMK03000~\cite{clockcond}) 
receives an internal or external clock and distributes it through the backplane to all slots via a star topology.

A dedicated font-end program runs on the ARM processors which collects waveform data from all 16 WDBs, 
merges them into one event, and sends it to the central DAQ computers via Gigabit Ethernet (optional \SI{10}{\giga\bit}). 
The event format is compatible to the MIDAS DAQ system used in MEG~II. 
In addition, the front-end program configures and monitors all WDBs and the TCB through the SPI links.
It can communicate to the CMB to reboot individual slots in case of problems or firmware upgrades.

\subsection{Trigger processing: Trigger Concentrator Board}
\label{sec:trg}

The trigger processing includes suppressing the background by almost six orders of magnitude resulting in an acquisition rate of about \SI{10}{\hertz}. 
The real time reconstruction algorithms rely on the fast response detectors: 
the LXe detector for the photon observables and the pTC for the positron ones. 
The ionisation drift time in the CDCH cells prevents the trigger system from using any information from the track reconstruction from trigger level 0.

Event selection relies on an on-line reconstruction of decay product observables, 
such as momenta, relative timing, and direction. 
Logic equations are mapped in FPGA cells and implemented at 80~MHz so as to be synchronous with the FADC data flow. 
An estimate of the photon energy is obtained by the linear sum of pedestal-subtracted signal amplitudes of LXe photo-sensors, 
each weighted according to their own gain, 
which is efficiently implemented by using digital signal processor (DSP) units in the FPGA. 
An increased ADC resolution (12 vs. \SI{10}{\bit}) coupled with the improved single photoelectron 
response of the new sensors will allow to achieve a resolution better than that of MEG 
(7\% FWHM at the signal energy $\egamma = \SI{52.8}{\MeV}$), though
the final resolution will depend on running conditions. 
Concerning the relative timing, this will benefit from using WDB comparators coupled to each input signal (on both the LXe detector and pTC), 
whose latch time can be further refined by implementing look-up tables on the FPGA to correct for time-walk effects. 
Also in this case we expect the resolution to be significantly improved from the \SI{3}{\ns} achieved in MEG; 
some results of the expected online time resolutions are reported in Sect.~\ref{sec:tdaqres}. 
Moreover, the enhanced imaging capability due to the finer detector segmentation (smaller LXe photo-sensors and pTC counters) 
permits tighter angular constraints on the decay kinematics.

\begin{figure}
\centering
\includegraphics[width=0.95\columnwidth]{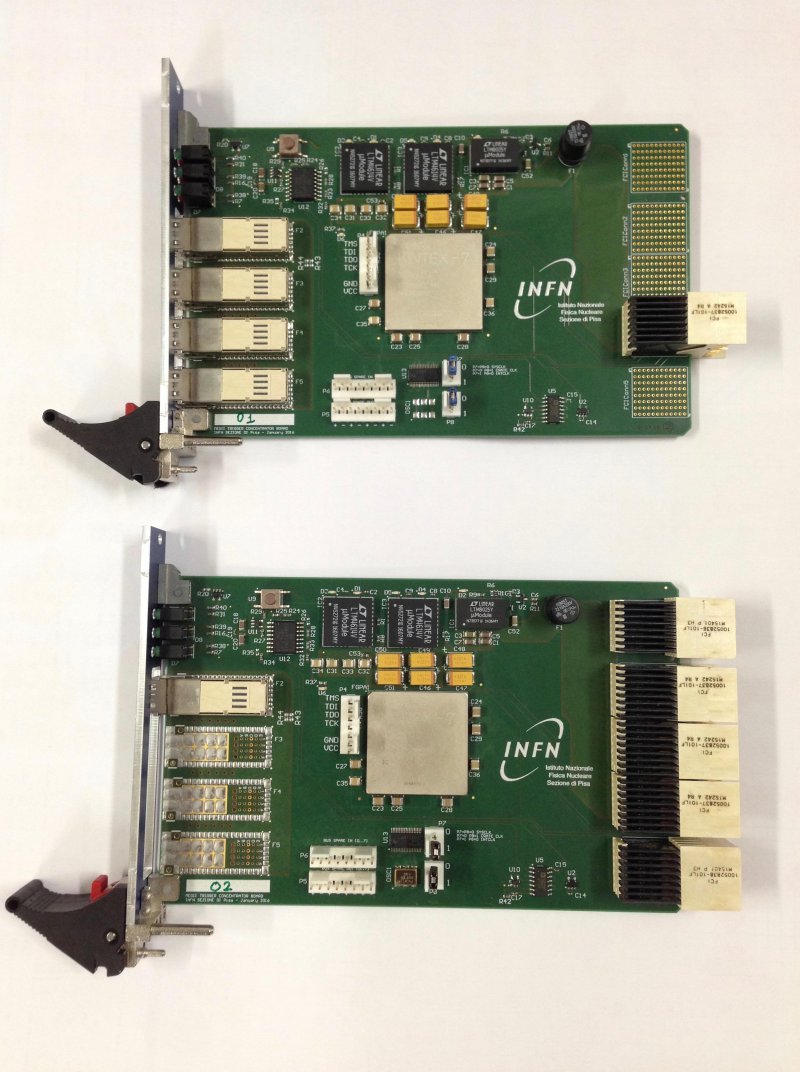}
\caption{\label{fig:TCB}%
 TCB configured as slave (top) and master (bottom); 
in case of a slave board the data-flow is from the left (front panel) to the right (back plane) and vice versa for a master.}
\end{figure}

The boards designed for the online data processing are called Trigger Concentrator Boards (TCB). 
A TCB gathers the information from a lower level trigger board, which could be a WDB via 
back-plane connections or another TCB which could be in another or in the same crate. 
In the first case the connection is provided via the back-plane in the second by a cable 
connected on the front panel. 
In order to minimise design and production costs, 
we decided to use the same 12-layer layout for all TCBs, 
independent of the role each one plays in the trigger hierarchy. 
TCBs differ from each other by the firmware operating on an on-board Xilinx Kintex7 FPGA~\cite{fpga}. 
Apart from reconstruction algorithms, which depend on individual sub-detectors, 
other features might depend on the slot assignment. 
For instance, the direction of I/O data lines is set from the back-plane to the FPGA if the TCB is located at the centre of the crate 
(Master position in all the crates), while it is the other way round for higher level TCBs hosted in a Slave position in the trigger crate, 
the two configuration are shown in Fig.~\ref{fig:TCB}.

\subsection{System synchronisation: Ancillary board}

The main task of the Ancillary system is to provide the TDAQ boards with an ultra-low jitter clock signal to be used as the experiment time reference. 
We selected a low jitter 80~MHz oscillator~\cite{pericom} and a low jitter fan-out from Maxim~\cite{maxim}, 
as a result we measure an overall jitter better than \SI{10}{\ps} at the WaveDREAM input. 
The distribution is arranged on a master-to-slave fan-out and implemented on a board, 
the Ancillary board, which can be configured as both master or slave: as a Master it generates the low jitter 
clock signal and receives the control signals, 
such as the trigger and synchronisation pulses, 
from the master TCB and forwards them to all the other TDAQ modules through the Slave modules, the link is provided by the backplane. 
The other way round the busy signal is distributed from the DAQ crates 
to the trigger crates and used as a veto for any trigger signal generation.

\subsection{Slow Control}

Each experiment has quantities that must be monitored or controlled \lq\lq{}slowly\rq\rq{}.
Examples are temperatures, power supply currents, and environmental values such
as humidity and pressure. 
This is the task of the slow control system.
MEG~II relies on the Midas Slow Control Bus (MSCB) which has been
successfully used in the MEG experiment over the past decade.
It uses the RS-485 standard for communication and a set of optimised
commands~\cite{mscb} for effective and quick exchange of data representing physical values.

The MSCB protocol has been implemented in the CMB,
which allows the control and monitoring of the WaveDAQ crates directly from
the MEG~II slow control system.
In addition, an MSCB communication line has been added to the WaveDAQ crate backplane,
so the CMB can forward any MSCB command to individual slots in the crate.
Each WDB implements an MSCB core for the control and monitoring of the bias high voltage for each channel.
This core is implemented in the FPGA soft-core processor (Xilinx MicroBlaze),
and connected to the DACs and ADCs of the high voltage piggy back board. 
Individual channels can be switched on and off, demand values set
and currents can be read back through the slow control system.

In addition, a connector has been placed on the front panel of the WDB, 
which implements the 1-Wire\textsuperscript{\textregistered} bus system~\cite{onewire}.
This system allows the connection of virtually any number of sensors to a single line.
Each sensor has a unique address under which it can be accessed.
In addition to the serial communication, also the sensor power is delivered through the same line,
hence the name 1-Wire.
This scheme allows each of the 16 SiPMs connected to each WDB to be equipped with an
individual temperature sensor. All 16 sensors are connected to this 1-Wire bus
and are accessible by the bias voltage control program inside the FPGA and the MSCB slow controls system.
This allows the implementation of an algorithm which adjusts the bias voltage of each SiPM
to keep the breakdown voltage and therefore the gain constant even 
with temperature drifts.
 
\subsection{Performance}
\label{sec:tdaqres}

The TDAQ efficiency, defined as the product of the trigger efficiency to select
candidate signal events and the experiment live-time fraction, 
affects the experiment sensitivity (cf. Eq.~(\ref{eq:signal})).

The read-out scheme guarantees a data transfer dead time of about \SI{1}{\ms} leading 
to a possible trigger rate of about \SI{100}{\hertz} with irrelevant dead time, 
such a value is however not sustainable by the offline infrastructure since the 
overall data size would increase by much more than a factor 10 with respect to MEG. 
As a consequence a maximum trigger rate of about \SI{10}{\hertz}, associated with an online selection efficiency close to unity is sought.

In order to accomplish this task the online event reconstruction algorithms have to be 
refined as described in Sect.~\ref{sec:trg}, in particular for $\egamma$; 
the trigger resolution on the photon energy reconstruction was estimated by using the 
MC generated events reconstructed with an emulator on the FPGA firmware (FW) written in C{}\verb!++!. 
The projected resolution is more than a factor 2 better than in MEG,
 $\sigma_{\egamma}/\egamma = 1.5\%$ at \SI{45}{\MeV} (it was 3.5\% in MEG). 

\begin{figure}
\centering
\includegraphics[width=0.95\columnwidth]{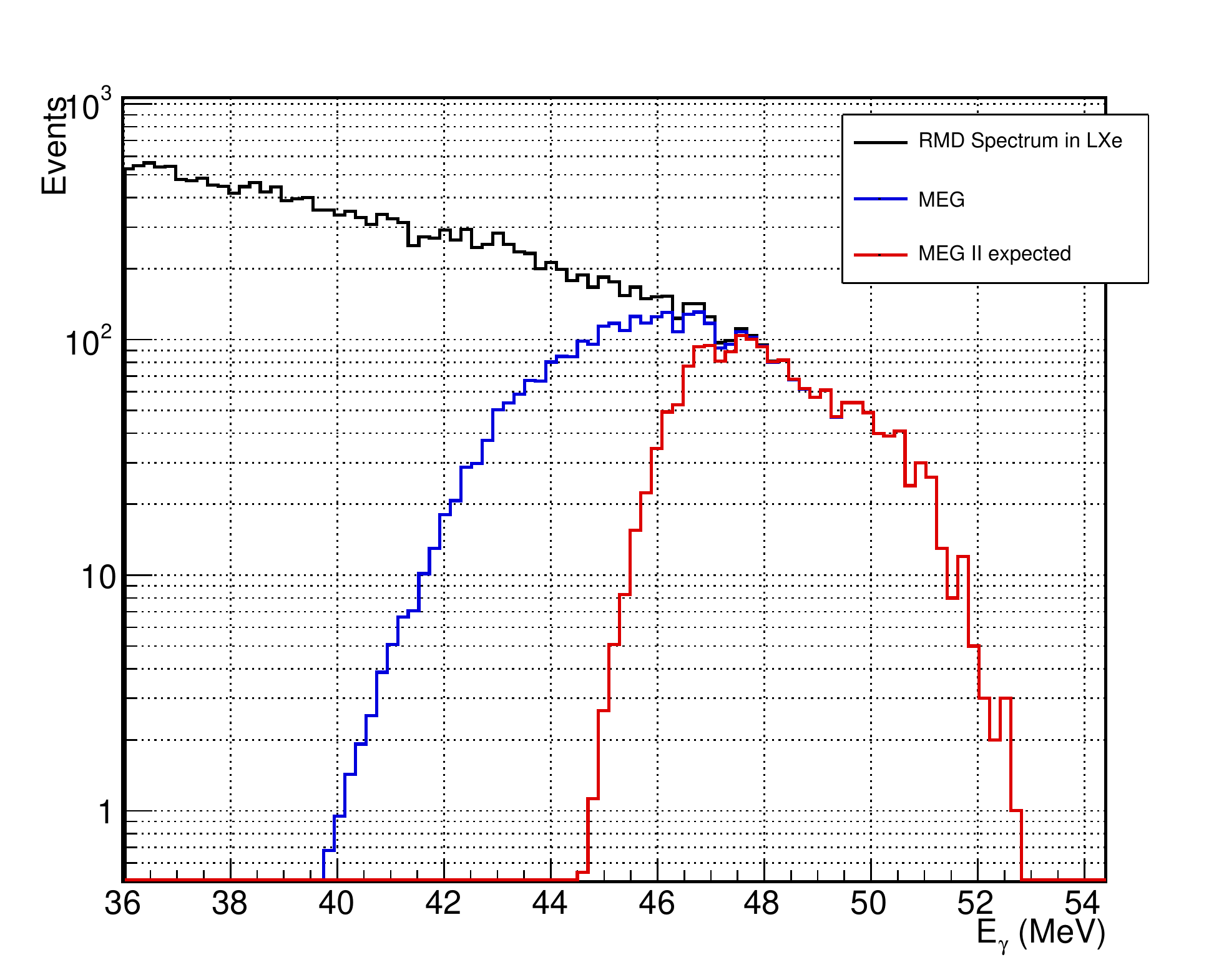}
\caption{\label{fig:egammacut}%
 ${\egamma}$-spectrum in the LXe photon detector shown in black and the effective spectra by applying 
an online threshold at \SI{42}{\MeV} with a relative resolution of 3.5\%\ (blue) and \SI{45}{\MeV} with a resolution of 1.5\%\ (red).}
\end{figure}

The improved resolution will allow an increase in the online $\egamma$-threshold without loss of efficiency in the analysis region, 
i.e. over \SI{48}{\MeV}, as reported in Fig.~\ref{fig:egammacut}. 
The results indicate that we will be able to increase the online threshold by at least \SI{2}{\MeV}, 
from 42 to \SI{45}{\MeV}, leading to a trigger rate reduction of about a factor 2.

The online time measurement will be extracted by sampling the WDB discriminator 
output at \SI{800}{\MHz} and intercepting the first sample over threshold, 
all the TDCs will be relatively synchronised by the clock signal distributed by the ancillary system. 
The intrinsic resolution of this TDC is the clock period divided by $\sqrt{12}$, being \SI{\approx 350}{\ps}.

The method was tested during a beam test with a pTC prototype at PSI. 
The time resolution was measured by comparing the measured times of two adjacent pixels, 
where the transit time spread of the positron along that path is \SI{\approx50}{\ps}, much lower that the expected resolution. 
The measured time resolution on a single pixel is \SI{\approx500}{\ps}, close to the intrinsic limit; 
the origin of the difference has been studied and found to be due to two main factors: 
time walk on the discriminator and electronics jitter on FPGA processing. 
The former factor will be corrected in the final system and we estimate reaching a single channel time resolution of about \SI{450}{\ps}.

\begin{figure}
\centering
\includegraphics[width=0.95\columnwidth]{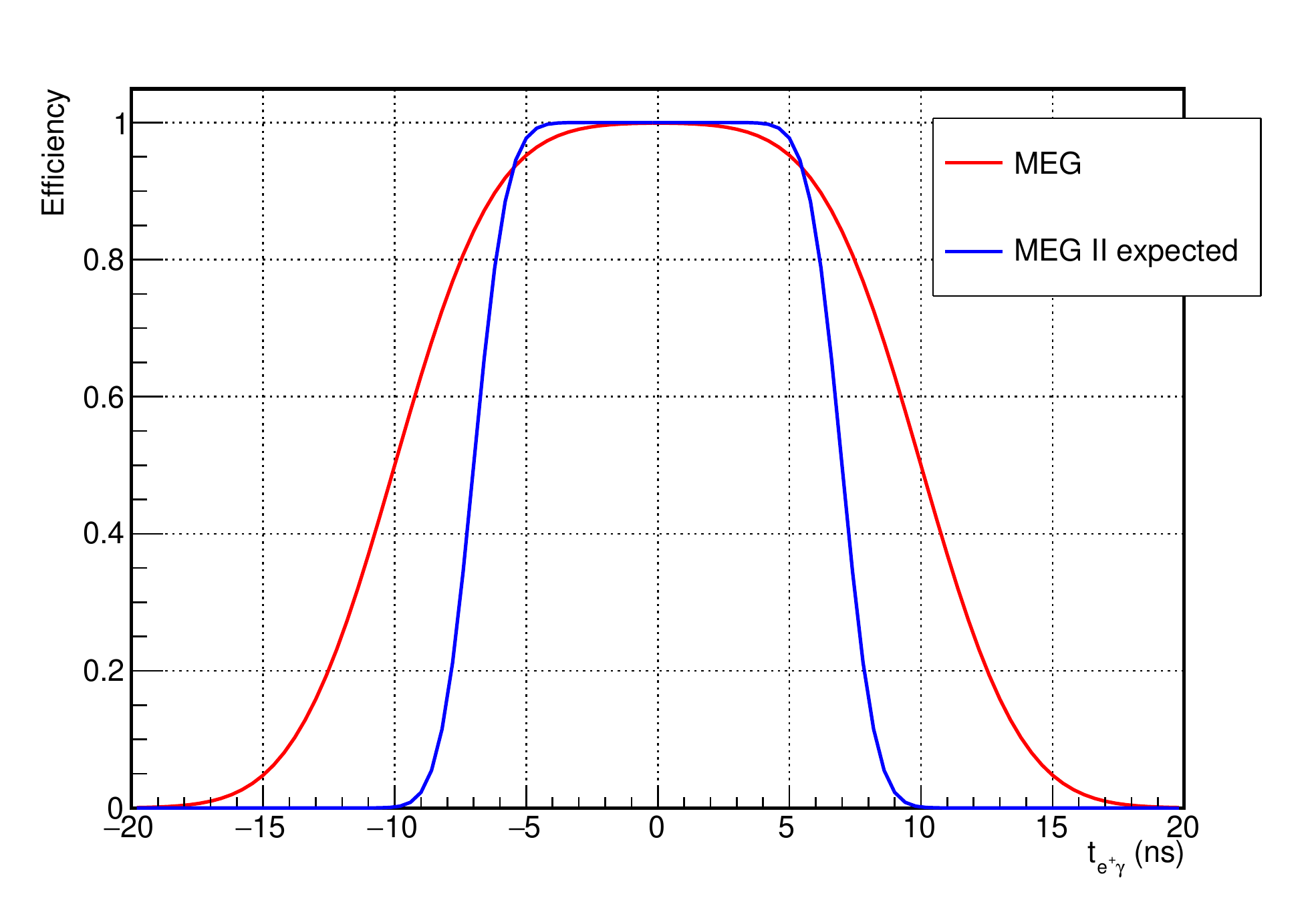}
\caption{\label{fig:ontim}%
 Comparison of the online positron--photon timing trigger selection efficiency of MEG~II (blue) 
to MEG (red); the selection width (FWHM) for MEG~II is \SI{14}{\ns} while previously it was \SI{20}{\ns} for MEG.}
\end{figure}

The online time resolution of the positron--photon coincidence is then expected to be better than \SI{1}{\ns}, 
more than a factor 3 better than in MEG. Figure~\ref{fig:ontim} shows the effective trigger coincidence window for MEG~II superimposed on that of the MEG. Thanks to the improved time resolution it will be possible
a substantial reduction in the coincidence width (FWHM) from \SI{20}{\ns} to at least \SI{14}{\ns}, leading to a trigger rate reduction of a factor \numrange[range-phrase=--]{1.5}{2} with no
efficiency loss on signal.

The MEG~II trigger rate is then expected to be \SI{\approx10}{\hertz}, a value comparable with that of MEG. 
The improvements on the online reconstruction resolutions are expected to compensate for the increased muon stopping rate.

\clearpage
\newpage
\section{Expected sensitivity}
\label{sec:sensitivity}

The estimation of the MEG~II sensitivity follows the approach exploited in MEG \cite{meg2016}. 
A detailed MC simulation of the beam and the detector is implemented together with a reconstruction
of the particle's observables. The probability density functions (PDFs) of the observables relevant 
for discriminating signal from background are generated with the help of simulation and prototype data. 
Then, an ensemble of simulated experiments (\emph{toy MC}) are generated from the PDFs and analysed 
extracting a set of upper limits (UL).
Finally, the sensitivity is estimated.

\subsection{Simulation and reconstruction}

We developed a full simulation of the detector based on \textsc{Geant4}, adding information, 
where necessary, from measurements (e.g. light propagation properties in LXe) or dedicated simulations 
(e.g. ionisation density in the drift chamber from \textsc{Garfield} \cite{garfield}). The \textsc{Geant4}
hits are then converted into simulated electronic signals, making use of waveform templates 
extracted from data collected with prototypes or with the final detectors. 
At this stage, we also mix different \textsc{Geant4} events in order to simulate the pile-up of 
multiple muon decays within the same DAQ time window.

Both data and simulated events go through the same reconstruction chain. For each sub-detector, 
a waveform analysis is performed in order to extract raw observables, such as the signal time and charge. 
A hit reconstruction procedure is then applied to translate them into calibrated physical observables. The following
variables are extracted:
\begin{enumerate}
\item the drift time of the ionisation electrons in the drift chamber and the hit position along the $z$-coordinate,
\item the hit time and position in each pTC and RDC PS tile and
\item the number of collected photons in each photo-sensor of the LXe photon detector and RDC calorimeter.
\end{enumerate}
Several reconstruction algorithms are then applied to extract the single particle's observables. 
Most notably, dedicated pattern recognition algorithms and a Kalman filter technique are used to extract the 
positron track parameters; the positron is tracked 
through the pTC tiles to extract the best estimate of the positron time;
number and timing of collected scintillation photons 
of each photo-detector in the LXe photon detector are used to extract the photon time and conversion vertex 
as well as the photon energy.

Finally, these observables are combined to extract the kinematic variables characterising a \megc~decay 
allowing the discrimination from background events: the photon energy $\egamma$, the positron energy 
$\epositron$, the relative timing $\tegamma$, and the relative polar and azimuthal angles ($\thetaegamma$, $\phiegamma$). 

The probability density functions (PDFs) describing the distributions 
of each kinematic variable for the signal and the backgrounds 
are generated relying on MC simulated events or on data collected from prototypes.

A representative scenario for MEG~II resolutions and efficiencies is summarised in Table~\ref{tab:scenario} 
and compared to the MEG performance. 
The efficiency of the positron reconstruction is greatly improved to that of MEG, thanks to the high efficiency of the tracking system 
and to the optimised geometry of CDCH and pTC.
The resolution on the relative time between the \positron\ and the $\gamma$ is estimated 
to be $\sigma_{\tegamma}\simeq\SI{84}{\ps}$ by adopting the most conservative estimation for the LXe 
photon detector timing resolution of $\sigma_{\tgamma} \simeq\SI{70}{\ps}$ and an error 
on the positron timing due to the pTC resolution of $\sigma_{t_\positron^\mathrm{pTC}} 
\simeq\SI{31}{\ps}$, which includes an inter-counter calibration contribution
$\sigma_{t_\positron}^\mathrm{inter\mathchar`-counter}/\sqrt{\bar{N}_{hit}}\simeq\SI{10}{\ps}$, 
a synchronisation contribution between WDBs of
$\sigma^\mathrm{WDB}_{t_\positron}\simeq\SI{25}{\ps}$ and a contribution due to the track
extrapolation along the CDCH measured trajectory of $\sigma^\mathrm{CDCH}_{t_\positron}\simeq\SI{20}{\ps}$.
  
\begin{table}
\caption{ \label{tab:scenario}Resolutions (Gaussian $\sigma$) and efficiencies of MEG~II compared 
with those of MEG}
\centering
\newcommand{\minu}{\hphantom{$-$}}
\newcommand{\cc}[1]{\multicolumn{1}{c}{#1}}
\begin{tabular}{@{}lll}
\hline
  {\bf PDF parameters }  & \minu MEG & \minu  MEG~II \\ 
\hline\noalign{\smallskip}
$\epositron$ (keV)                & \minu  380    & \minu 130   \\
$\thetae$ (mrad)                  & \minu 9.4     & \minu 5.3     \\
$\phie$ (mrad)                    & \minu 8.7     & \minu 3.7     \\
$\zpos/\ypos$ (mm) core           & \minu 2.4/1.2 & \minu 1.6/0.7  \\
$\egamma$(\%)  ($w\SI{>2}{\cm}$)/($w\SI{<2}{\cm}$)     & \minu 2.4/1.7 & \minu 1.1/1.0 \\
$\ugamma,\vgamma,\wgamma$ (mm)          & \minu 5/5/6   & \minu 2.6/2.2/5  \\
$\tegamma$ (ps)   & \minu 122     & \minu 84 \\
\hline
{\bf  Efficiency (\%)} & &  \\ 
\hline
Trigger             & \minu $\approx$ 99  & \minu  $\approx$ 99 \\  
Photon        & \minu 63            & \minu  69 \\
\positron (tracking $\times$ matching)              & \minu 30            & \minu  70  \\
\hline
\end{tabular}
\end{table}

\begin{figure}
\centering
\includegraphics[width=1\linewidth, clip, trim=0 0 0 4pc]{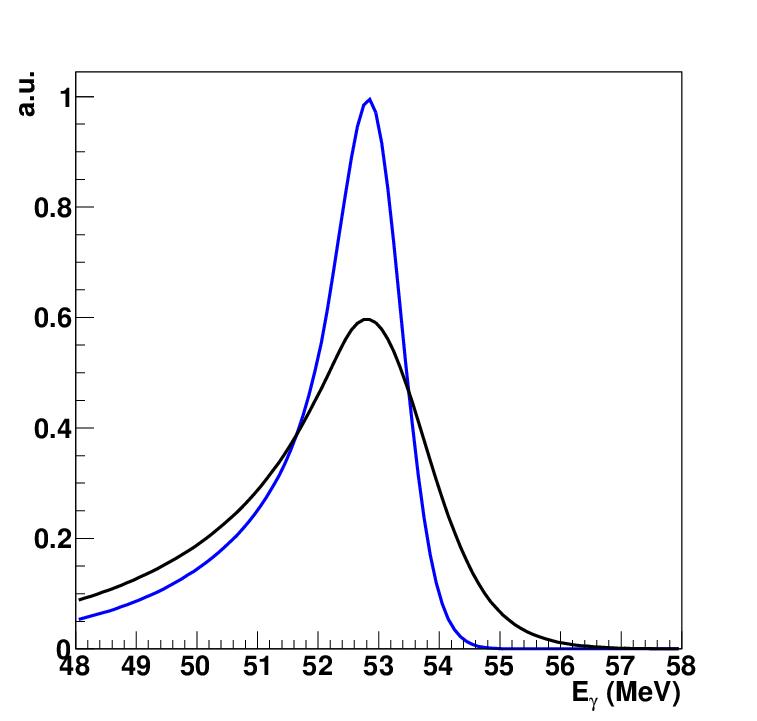}
\caption{Comparison of the $\egamma$ PDFs for signal events 
based on the resolutions obtained in 2010 data (black) and on 
the projected value for the upgrade (blue). \label{fig:senspdf1}} 
\end{figure}

\begin{figure}
\centering
\includegraphics[width=1\linewidth, clip, trim=0 0 0 4pc]{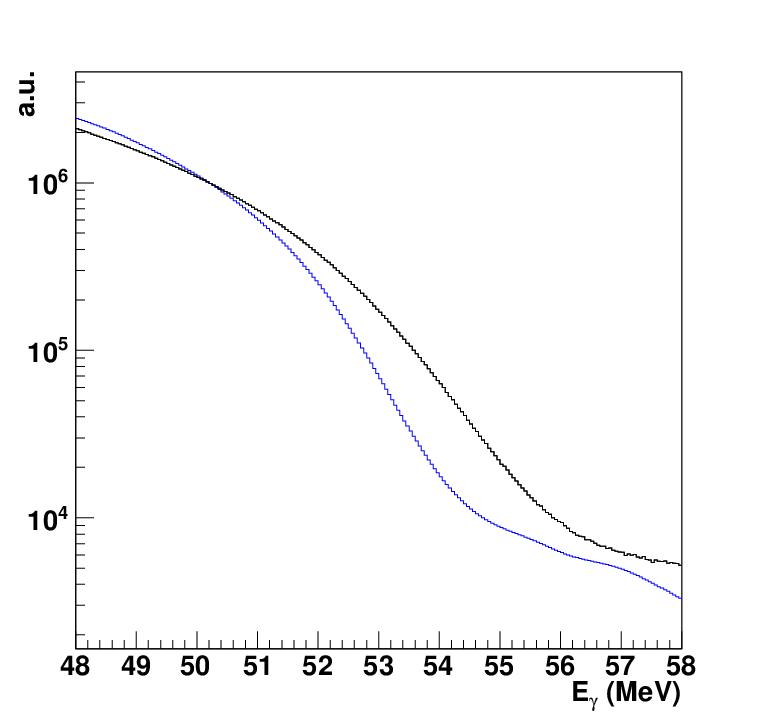}
\caption{\label{fig:senspdf2}Comparison of the $\egamma$ PDFs for accidental 
background events based on the resolutions obtained in 2010 data (black) and on 
the projected value for the upgrade (blue). Differences in relative background contributions between 
RMD, AIF and pile-up are also taken into account.} 
\end{figure}

As an example we show the $\egamma$ PDFs for signal (see Fig.~\ref{fig:senspdf1})
and accidental background events (see Fig.~\ref{fig:senspdf2}). The expected improvement in MEG~II is visible 
by comparing these PDFs (blue) with the 2010 MEG data PDFs (black). In the 
$\egamma$ background PDFs various contributions are taken into account: RMD, 
photons from positron AIF and from bremsstrahlung on materials in the detector, pile-up events, as well as resolution effects. 
The configuration of the CDCH, with a smaller amount of material close to the 
LXe photon detector, reduces the AIF contribution, which is dominant for 
$\egamma>\SI{52}{\MeV}$, by about $20\%$ with respect to the MEG detector. 
The combined effect of the increased 
resolution and of the lower high energy background is clearly visible in 
Fig.~\ref{fig:senspdf2}.

\subsection{Analysis }

Each toy MC is analysed using the maximum likelihood analysis technique developed 
following the MEG data analysis \cite{meg2010, meg2013, meg2016}
to extract an UL at 90\% CL on the number of signal events, following the prescription
of \cite{feldman_1998}, that is converted 
to an UL on $\BR(\megc)$ by using the appropriate normalisation factor.
This technique is more efficient and 
reliable than a box analysis, since all types of background are correctly folded in the global 
likelihood function and taken into account with their own statistical weights. 
The enhanced precision of the MEG~II detectors allows a much better separation of the signal from the background and 
reduces significantly the spill of the photon and positron background distributions into the signal region, 
which is due to experimental resolution effects. 


\subsection{Sensitivity estimate}

An ensemble of simulated experiments (\emph{toy MC}) with a statistics comparable to the expected number of events
during MEG~II data taking are generated from the 
PDFs assuming zero signal events and an average number of radiative and accidental 
events obtained by extrapolating the results of the MEG experiment and taking into account the new detector 
performances. The numbers of RMD and accidental events are then left free to fluctuate, according to 
Poisson statistics. 
For each toy MC we extract an UL on the $\BR(\megc)$. Following \cite{meg2016}, we define as 
\emph{sensitivity} the median of the distribution of the ULs obtained from the toy MCs.

\begin{figure}
\centering
\includegraphics[width=0.5\textwidth, clip, trim=0 1pc 1pc 5pc]{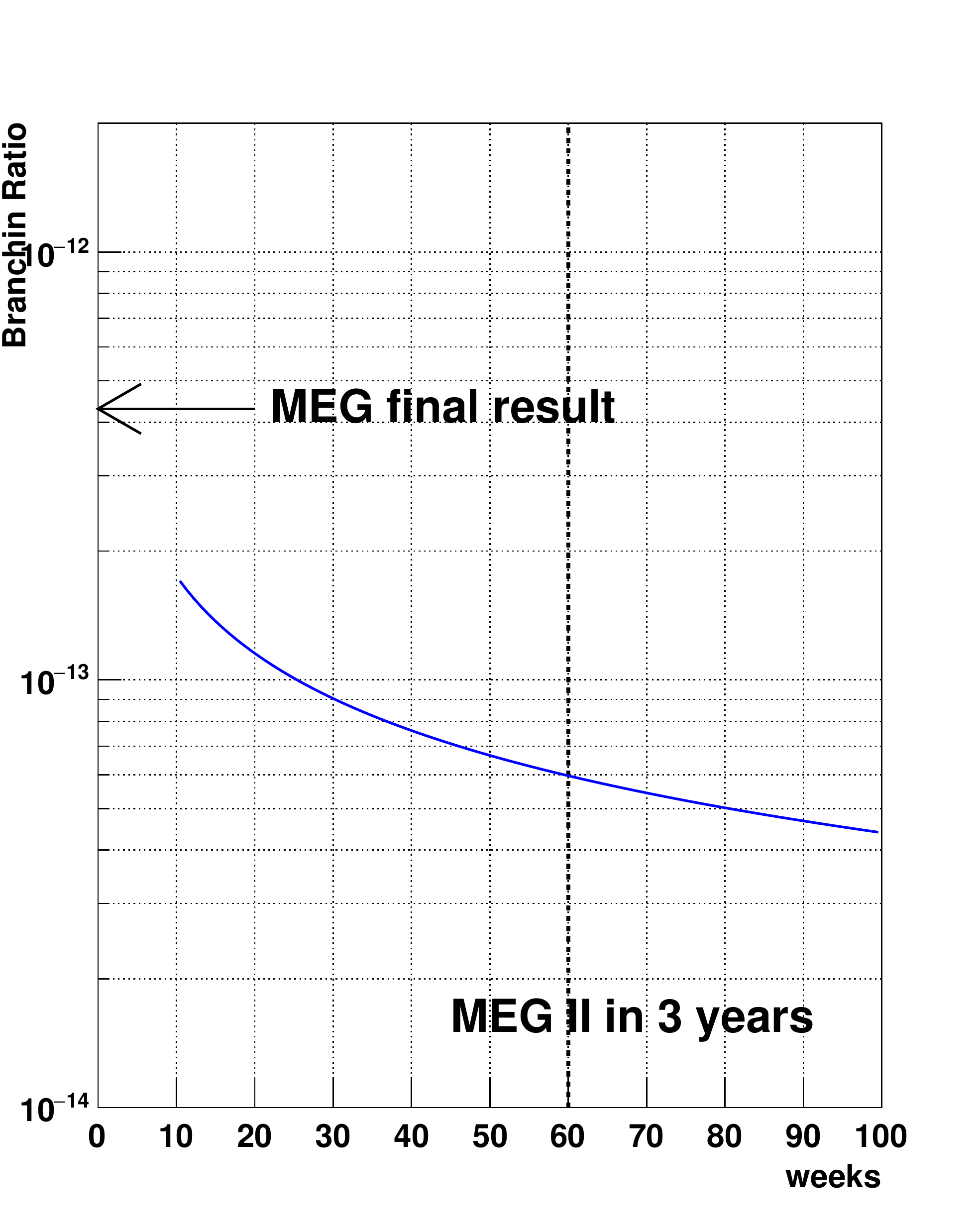}
\caption{\label{fig:sensEv}Expected sensitivity of MEG~II as a function of 
the DAQ time compared with the bounds set by MEG \cite{meg2016}. 
Assuming conservatively $20$ DAQ weeks per year, we expect a branching ratio sensitivity of
\num{6e-14} in three years.}
\end{figure}

In Fig.~\ref{fig:sensEv} we show the evolution of the sensitivity as a function of the 
DAQ time (in weeks). 
Assuming conservatively 140 DAQ days per year, we can reach a sensitivity of \num{6e-14} in three years.
The sensitivity has been re-evaluated since the proposal \cite{Baldini:2013ke} according to the updated 
estimations of the expected detector performances, the inclusion of the downstream RDC and a more 
conservative assumption on the DAQ time.

\clearpage

\newpage
\section{Conclusions }
\label{sec:cConclusions}

We have presented the detailed design of the components of the MEG~II detector, together with a presentation of the scientific merits of the experiment.
The MEG~II detector results from a mixture of upgraded components of the MEG experiment (beam line, target, calibration,
LXe photon detector) 
and of newly designed components (CDCH, pTC, RDC, trigger and DAQ). The design has been
completed and construction and commissioning are ongoing.

The resolutions on the relevant physical variables are expected to improve by about a factor of 2, as suggested 
by simulation and preliminary results from laboratory and beam tests. Those improvements, together with an 
increase by more than a factor of 2 both in muon decay rate and signal detection efficiency, 
are expected to bring the sensitivity to the $\megc$ decay rate down to \num{6e-14} in three years of data taking.
In term of discriminating power of parameters of models beyond the Standard Model, 
this limit is comparable to those achievable by the next generation of cLFV experiments exploiting other channels.

\clearpage

\section*{Acknowledgements}

We are grateful for the support and co-operation provided 
by PSI as the host laboratory and to the technical and 
engineering staff of our institutes. This work is supported by 
Schweizerischer Nationalfonds (SNF) Grant 200021 137738
(Switzerland), 
DOE DEFG02-91ER40679 (USA), INFN (Italy)
and MEXT/JSPS KAKENHI Grant Numbers JP22000004, JP25247034, JP26000004,
JP15J10695, JP17J03308, JP17J04114, JP17K14267 and 
JSPS Overseas Research Fellowships 2014-0066 (Japan).
Partial support of the Italian 
Ministry of University and Research (MIUR) Grant 
No. RBFR138EEU 001 is acknowledged.

\bibliographystyle{unsrt}
\bibliography{MEG}
\end{document}